\documentclass[twocolumn]{aastex631}
\usepackage{gensymb} 
\usepackage{amsmath} 
\usepackage{multirow} 
\usepackage{bm} 
\usepackage{xspace} 
\usepackage{float}
\usepackage{customcommands} 
\usepackage{newunicodechar,graphicx}
\PassOptionsToPackage{hyphens}{url}\usepackage{hyperref}

\DeclareRobustCommand{\okina}{%
  \raisebox{\dimexpr\fontcharht\font`A-\height}{%
    \scalebox{0.8}{`}%
  }%
}
\newunicodechar{ʻ}{\okina}

\begin{document}
\title{The TESS-Keck Survey. XVI. Mass Measurements for 12 Planets in Eight Systems\footnote{Based on observations obtained at the W. M. Keck Observatory, which is operated jointly by the University of California and the California Institute of Technology.}}
\correspondingauthor{Joseph M. Akana Murphy}
\email{joseph.murphy@ucsc.edu}

\author[0000-0001-8898-8284]{Joseph M. Akana Murphy}
\altaffiliation{NSF Graduate Research Fellow}
\affiliation{Department of Astronomy and Astrophysics, University of California, Santa Cruz, CA 95064, USA}

\author[0000-0002-7030-9519]{Natalie M. Batalha}
\affiliation{Department of Astronomy and Astrophysics, University of California, Santa Cruz, CA 95064, USA}

\author[0000-0003-3623-7280]{Nicholas Scarsdale}
\affiliation{Department of Astronomy and Astrophysics, University of California, Santa Cruz, CA 95064, USA}

\author[0000-0002-0531-1073]{Howard Isaacson}
\affiliation{Department of Astronomy, University of California, Berkeley, CA 94720, USA}
\affiliation{Centre for Astrophysics, University of Southern Queensland, Toowoomba, QLD, Australia}

\author[0000-0002-5741-3047]{David R. Ciardi}
\affiliation{NASA Exoplanet Science Institute-Caltech/IPAC, Pasadena, CA 91125, USA}

\author[0000-0002-9329-2190]{Erica J. Gonzales}
\altaffiliation{NSF Graduate Research Fellow}
\affiliation{Department of Astronomy and Astrophysics, University of California, Santa Cruz, CA 95064, USA}

\author[0000-0002-8965-3969]{Steven Giacalone}
\affiliation{Department of Astronomy, University of California, Berkeley, CA 94720, USA}

\author[0000-0002-6778-7552]{Joseph D. Twicken}
\affiliation{NASA Ames Research Center, Moffett Field, CA 94035, USA}
\affiliation{SETI Institute, Mountain View, CA 94043, USA}

\author[0000-0002-1092-2995]{Anne Dattilo}
\affiliation{Department of Astronomy and Astrophysics, University of California, Santa Cruz, CA 95064, USA}

\author[0000-0002-3551-279X]{Tara Fetherolf}
\altaffiliation{UC Chancellor's Fellow}
\affiliation{Department of Earth and Planetary Sciences, University of California, Riverside, CA 92521, USA}

\author[0000-0003-3856-3143]{Ryan A. Rubenzahl}
\altaffiliation{NSF Graduate Research Fellow}
\affiliation{Department of Astronomy, California Institute of Technology, Pasadena, CA 91125, USA}

\author{Ian J. M. Crossfield}
\affiliation{Department of Physics and Astronomy, University of Kansas, Lawrence, KS 66045, USA}

\author[0000-0001-8189-0233]{Courtney D. Dressing}
\affiliation{Department of Astronomy, University of California, Berkeley, CA 94720, USA}

\author[0000-0003-3504-5316]{Benjamin Fulton}
\affiliation{NASA Exoplanet Science Institute/Caltech-IPAC, Pasadena, CA 91125, USA}

\author[0000-0001-8638-0320]{Andrew W. Howard}
\affiliation{Department of Astronomy, California Institute of Technology, Pasadena, CA 91125, USA}

\author[0000-0001-8832-4488]{Daniel Huber}
\affiliation{Institute for Astronomy, University of Hawai\okina i, Honolulu, HI 96822, USA}

\author[0000-0002-7084-0529]{Stephen R. Kane}
\affiliation{Department of Earth and Planetary Sciences, University of California, Riverside, CA 92521, USA}

\author[0000-0003-0967-2893]{Erik A. Petigura}
\affiliation{Department of Physics and Astronomy, University of California, Los Angeles, CA 90095, USA}

\author[0000-0003-0149-9678]{Paul Robertson}
\affiliation{Department of Physics and Astronomy, University of California, Irvine, CA 92697, USA}

\author[0000-0001-8127-5775]{Arpita Roy}
\affiliation{Space Telescope Science Institute, 3700 San Martin Drive, Baltimore, MD 21218, USA}
\affiliation{Department of Physics and Astronomy, Johns Hopkins University, Baltimore, MD 21218, USA}

\author[0000-0002-3725-3058]{Lauren M. Weiss}
\affiliation{Department of Physics and Astronomy, University of Notre Dame, Notre Dame, IN 46556, USA}

\author[0000-0001-7708-2364]{Corey Beard}
\affiliation{Department of Physics and Astronomy, University of California, Irvine, CA 92697, USA}

\author[0000-0003-1125-2564]{Ashley Chontos}
\altaffiliation{Henry Norris Russell Fellow}
\affiliation{Department of Astrophysical Sciences, Princeton University, Princeton, NJ 08540, USA}
\affiliation{Institute for Astronomy, University of Hawai\okina i, Honolulu, HI 96822, USA}

\author[0000-0002-8958-0683]{Fei Dai}
\altaffiliation{NASA Sagan Fellow}
\affiliation{Division of Geological and Planetary Sciences, California Institute of Technology, Pasadena, CA, 91125, USA}
\affiliation{Department of Astronomy, California Institute of Technology, Pasadena, CA 91125, USA}

\author[0000-0002-7670-670X]{Malena Rice}
\altaffiliation{Heising-Simons 51 Pegasi b Postdoctoral Fellow}
\affiliation{Department of Physics, Massachusetts Institute of Technology, Cambridge, MA 02139, USA}
\affiliation{Kavli Institute for Astrophysics and Space Research, Massachusetts Institute of Technology, Cambridge, MA 02139, USA}
\affiliation{Department of Astronomy, Yale University, New Haven, CT 06511, USA}

\author[0000-0002-4290-6826]{Judah Van Zandt}
\affiliation{Department of Physics and Astronomy, University of California, Los Angeles, CA 90095, USA}

\author[0000-0001-8342-7736]{Jack Lubin}
\affiliation{Department of Physics and Astronomy, University of California, Irvine, CA 92697, USA}

\author[0000-0002-3199-2888]{Sarah Blunt}
\affiliation{Department of Astronomy, California Institute of Technology, Pasadena, CA 91125, USA}

\author[0000-0001-7047-8681]{Alex S. Polanski}
\affil{Department of Physics and Astronomy, University of Kansas, Lawrence, KS 66045, USA}

\author[0000-0003-0012-9093]{Aida Behmard}
\altaffiliation{NSF Graduate Research Fellow}
\affiliation{Division of Geological and Planetary Science, California Institute of Technology, Pasadena, CA 91125, USA}

\author[0000-0002-4297-5506]{Paul A.\ Dalba}
\altaffiliation{Heising-Simons 51 Pegasi b Postdoctoral Fellow}
\affiliation{Department of Astronomy and Astrophysics, University of California, Santa Cruz, CA 95064, USA}
\affiliation{SETI Institute, Carl Sagan Center, Mountain View, CA 94043, USA}

\author[0000-0002-0139-4756]{Michelle L. Hill}
\affiliation{Department of Earth and Planetary Sciences, University of California, Riverside, CA 92521, USA}

\author[0000-0001-8391-5182]{Lee J.\ Rosenthal}
\affiliation{Department of Astronomy, California Institute of Technology, Pasadena, CA 91125, USA}

\author[0000-0002-4480-310X]{Casey L. Brinkman}
\affiliation{Institute for Astronomy, University of Hawai\okina i, Honolulu, HI 96822, USA}

\author[0000-0002-7216-2135]{Andrew W. Mayo}
\affiliation{Department of Astronomy, University of California, Berkeley, CA 94720, USA}

\author[0000-0002-1845-2617]{Emma V. Turtelboom}
\affiliation{Department of Astronomy, University of California, Berkeley CA 94720, USA}

\author[0000-0002-9751-2664]{Isabel Angelo}
\affiliation{Department of Physics and Astronomy, University of California, Los Angeles, CA 90095, USA}
\affiliation{Mani L. Bhaumik Institute for Theoretical Physics, University of California, Los Angeles, CA 90095, USA}

\author[0000-0003-4603-556X]{Teo Mo\v{c}nik}
\affiliation{Gemini Observatory/NSF's NOIRLab, Hilo, HI 96720, USA}

\author[0000-0003-2562-9043]{Mason G. MacDougall}
\affiliation{Department of Physics and Astronomy, University of California, Los Angeles, CA 90095, USA}

\author[0000-0001-9771-7953]{Daria Pidhorodetska} 
\affiliation{Department of Earth and Planetary Sciences, University of California, Riverside, CA 92521, USA}

\author[0000-0003-0298-4667]{Dakotah Tyler}
\affiliation{Department of Physics and Astronomy, University of California, Los Angeles, CA 90095, USA}

\author[0000-0002-6115-4359]{Molly R. Kosiarek}
\altaffiliation{NSF Graduate Research Fellow}
\affiliation{Department of Astronomy and Astrophysics, University of California, Santa Cruz, CA 95064, USA}

\author[0000-0002-5034-9476]{Rae Holcomb}
\affiliation{Department of Physics and Astronomy, University of California, Irvine, CA 92697, USA}

\author[0000-0003-3179-5320]{Emma M. Louden}
\affiliation{Department of Astronomy, Yale University, New Haven, CT 06511, USA}

\author[0000-0001-8058-7443]{Lea A.\ Hirsch}
\affiliation{University of Toronto, Mississauga, ON L5L 1C6, Canada}

\author{Jay Anderson}
\affiliation{Space Telescope Science Institute, Baltimore, MD 21218, USA}

\author[0000-0003-3305-6281]{Jeff A. Valenti}
\affiliation{Space Telescope Science Institute, Baltimore, MD 21218, USA}

\begin{abstract}
With \jwst's successful deployment and unexpectedly high fuel reserves, measuring the masses of sub-Neptunes transiting bright, nearby stars will soon become the bottleneck for characterizing the atmospheres of small exoplanets via transmission spectroscopy. Using a carefully curated target list and more than two years' worth of \apflevy and \keckhires Doppler monitoring, the TESS-Keck Survey is working toward alleviating this pressure. Here we present mass measurements for 11 transiting planets in eight systems that are particularly suited to atmospheric follow-up with \jwst. We also report the discovery and confirmation of a temperate super-Jovian-mass planet on a moderately eccentric orbit. The sample of eight host stars, which includes one subgiant, spans early-K to late-F spectral types (\teff $=$ 5200--6200 K). We homogeneously derive planet parameters using a joint photometry and radial velocity modeling framework, discuss the planets' possible bulk compositions, and comment on their prospects for atmospheric characterization.
\end{abstract}

\section{Introduction} \label{sec:intro}
The \kepler spacecraft \citep{borucki10} taught us about the Milky Way Galaxy's intrinsic planet radius distribution for planets interior to 1 AU \citep{howard12, batalha13, fressin13, petigura13, fulton17, bryson21}. However, the long-stare, single-field nature of the survey precluded ground-based Doppler mass measurements for all but the brightest host stars. NASA's \tess mission \citep{ricker14}, on the other hand, with its detections of transiting exoplanets orbiting bright, nearby stars across the full sky, is allowing us to better understand the observed exoplanet mass and radius distribution. Of particular interest to theories of planet formation and evolution are the masses of sub-Neptunes, whose seeming diversity in bulk density has challenged our post-\kepler interpretations of small planet formation and evolution (e.g., \citealt{luque22}). Thus far, \tess is responsible for discovering nearly 100 planets smaller than 4 \rearth that also have robust mass measurements.\footnote{Data accessed via the NASA Exoplanet Archive on 2023-Mar-28. For planets with better than 50\% and 15\% fractional measurement precision in mass and radius, respectively.}

Measurements of atmospheric properties are key to understanding the interior composition of sub-Neptunes \citep{rogersSeager10} as they lie at the confluence of theoretical isocomposition curves in the mass-radius plane \citep{valencia07, adams08, zeng19, otegi20b}. Constraints of atmospheric composition can also inform theories of their formation and evolution histories (e.g., \citealt{madhusudhan19} and references therein; \citealt{kite20}). For these reasons, transit spectra of sub-Neptunes are extremely valuable. However, precise knowledge of the planets' surface gravities is required in order to interpret these data due to the degeneracy between surface gravity and atmospheric mean molecular weight \citep{batalha19}. Space-based transit photometry delivers precise planet radii, making planet mass the dominant source of uncertainty in the surface gravity calculation. Therefore, precise mass measurements and the substantial investments of ground-based resources that they require remain the critical first step in the effort to understand the physical drivers of sub-Neptune diversity.

\subsection{The TESS-Keck Survey: Planet atmospheres}
The TESS-Keck Survey (TKS; \citealt{chontos22}), a multi-semester Doppler monitoring campaign of promising \tess planet candidates with the \keckhires and \apflevy spectrographs, is working to provide the precise planet mass measurements required by future efforts in atmospheric characterization, among other investigations (e.g., \citealt{scarsdale21, lubin22}). TKS science falls along four main axes: (1) planet bulk composition, (2) system architectures and dynamics, (3) planet atmospheres, and (4) evolved systems. The systems presented in this work were all observed as members of science case three (SC3), planet atmospheres.

The goal of the TKS SC3 program is to measure precise masses for transiting planets in \tess systems that are particularly amenable to atmospheric follow-up. The SC3 target list was constructed using input from a quantitative selection function in addition to hand-tuning based on results from the \tess Follow-up Observing Program (TFOP; e.g., a target would be dropped despite a favorable selection function value if reconnaissance spectroscopy revealed the system to be an eclipsing binary).\footnote{TFOP contributions to the target selection process are acknowledged in \cite{chontos22}. This work does not make use of proprietary TFOP information beyond what was acknowledged by \cite{chontos22}.} Details of the target selection procedure can be found in \cite{scarsdale21} and \cite{chontos22}. The latter contains the complete TKS target list. 

In short, the quantitative selection function we used to identify potential SC3 targets strikes a balance between favorable prospects for atmospheric characterization and Doppler observing cost. The function is the ratio of a planet candidate's expected transmission spectroscopy metric (TSM; \citealt{kempton18}), a \jwst signal-to-noise ratio (\snr) proxy, and the estimated \keckhires exposure time required to achieve a 5$\sigma$ mass measurement. TSM is defined as
\begin{equation} \label{eqn:tsm}
\text{TSM} = (\text{scale factor}) \times \frac{R_\mathrm{p}^3   T_\mathrm{eq}}{M_\mathrm{p}  R_*^2} \times 10^{-J/5},
\end{equation}
where ``scale factor" is a normalization constant that depends on planet radius,
\begin{equation*}
    \text{scale factor} = 
    \begin{cases} 
          0.19 & \text{for } R_\mathrm{p} < 1.5 \text{ } R_\mathrm{\oplus} \\
          1.26 & \text{for } 1.5 < R_\mathrm{p} < 2.75 \text{ } R_\mathrm{\oplus} \\
          1.28 & \text{for } 2.75 < R_\mathrm{p} < 4 \text{ } R_\mathrm{\oplus} \\
          1.15 & \text{for } 4 < R_\mathrm{p} < 10 \text{ } R_\mathrm{\oplus}. \\
    \end{cases}
\end{equation*}
\rplanet and \mplanet are in Earth units, \teq is in Kelvin (and assumes zero Bond albedo and full day-night heat redistribution), \rstar is in solar units, and $J$ is the host star's apparent magnitude in $J$-band. The scale factor values account for all unit conversions. Since TSM depends on planet mass, we used the mass-radius relation from \cite{chenKipping17} to translate planet radius values from the \tess object of interest (TOI) catalog \citep{guerrero21} into preliminary mass estimates. The \keckhires exposure time required to achieve a 5$\sigma$ mass measurement was estimated using the methods in \cite{plavchan15}. A high value for this ratio indicates a more favorable target. 

In order to encourage a sample of planets that was spread evenly over parameter space, TOIs were divided into bins in stellar \teff, planet radius, and planet instellation flux. Selection function values for planets in the same \teff-\rplanet-\sincplanet bin were compared against one another. The top five highest ranking planets in each bin were then considered as candidates for the final TKS SC3 target list. Though our binning technique attempted to select a sample that spanned a wide range of host star \teff, since \keckhires is not optimized for observing cooler stars,\footnote{\keckhires measures stellar radial velocities using a warm cell of molecular iodine \citep{butler96}, which imprints absorption lines on the stellar spectrum between $\sim$5000--6000 \AA. Therefore, \keckhires is generally less efficient at measuring the radial velocities of M dwarfs compared to G dwarfs, for example.} our final SC3 target list is comprised primarily of planets orbiting G dwarfs. Our target list also focuses on sub-Neptunes since they offer reasonable expected Doppler observing costs (compared to super-Earths) but are still not giant planets, for which the literature already contains numerous atmospheric measurements. 

At the start of the survey in 2019B, we identified 20 \tess systems with at least one high-value planet candidate for atmospheric characterization according to our sample selection procedure. After more than two years of Doppler monitoring, the majority of these systems are now either already published, e.g., HD 63935 \citep{scarsdale21}, HD 191939 \citep{lubin22}, or the subject of publications in preparation by TKS collaborators. The eight systems presented in this work constitute the remaining systems of the TKS SC3 target list.

\subsection{Targets in this work} \label{sec:intro_targets_this_work}
In order of increasing TOI number, the systems presented in this work are:
\begin{itemize}
    \item \sysI (\toiI): a G dwarf hosting two sub-Neptunes.
    \item \sysII (\toiII): an early-K dwarf hosting one sub-Neptune.
    \item \sysIII (\toiIII): a late-F dwarf hosting a sub-Neptune and a super-Earth on opposite sides of the radius valley \citep{fulton17, vaneylen18}.
    \item \toiIV: a G dwarf hosting one sub-Neptune.
    \item \sysV (\toiV): an early-K dwarf hosting one sub-Neptune.
    \item \sysVI (\toiVI): a G dwarf hosting two sub-Neptunes, each with $P > 20$ d. The system is also host to a massive, distant companion as seen by a linear trend in the radial velocities. The nature of the companion is uncertain.
    \item \sysVII (\toiVII): an early-G dwarf hosting one sub-Neptune. The host star also appears to be gravitationally bound to a mid-M dwarf companion (\sysVIIcomp). The two stars have a sky-projected separation of about 200 AU.
    \item \sysVIII: a slightly evolved G star hosting one sub-Neptune and one nontransiting, super-Jovian-mass planet on a moderately eccentric orbit. The system is also host to a massive, distant companion as seen by a linear trend in the radial velocities. The nature of the companion is uncertain.
\end{itemize}

Why are these systems attractive targets for atmospheric observations? While not every planet presented here has an extraordinarily high TSM value---\cite{kempton18} suggest that ``good'' targets for atmospheric characterization have TSM $> 50$, which is not true for three of the 11 transiting planets in this work---\cite{batalha23} make clear that the most informative samples for inferring population-level characteristics are not necessarily composed of the best individual targets for atmospheric characterization. Furthermore, \cite{batalha23} note that planets are often chosen for Doppler and subsequent atmospheric follow-up because they are extreme in some way. This novelty bias systematically disfavors planets which are in fact the Galaxy's most common products of planet formation. To this end, six of the 11 transiting planets presented here land on the mode of the sub-Neptune mass-radius distribution. As noted above, the majority of the planets in the TKS SC3 program orbit G dwarfs, stellar hosts which are currently underrepresented in the set of atmospheric targets for \jwst, as much focus remains on small planets orbiting cool stars.\footnote{While planets transiting cool stars are typically more efficient targets for transmission spectroscopy (owing in part to the larger planet-star radius ratio), the radius distribution of planets orbiting M dwarfs is distinct from that of FGK dwarfs \citep{dressing13}. This implies differences in the dominant channel(s) of planet formation and evolution as a function of stellar properties, and, consequently, that the atmospheric characteristics of planets around M dwarfs may not be representative of planets with Sun-like hosts.} Finally, this work presents a large sample of planets with homogeneously derived physical properties, mitigating the effects of potential systematic biases from the data analysis. We discuss the planets' prospects for atmospheric characterization further in \S\ref{sec:atmo_char}.

\vspace{5mm}
The paper is organized as follows: We summarize the \tess 2-min cadence observations in \S\ref{sec:tess_obs}. We present high resolution imaging of the host stars in \S\ref{sec:imaging} and describe our stellar characterization in \S\ref{sec:stellar}. We discuss our Doppler observations and data reduction in \S\ref{sec:rvs}. We discuss our light curve inspection, cleaning, and initial transit modeling in \S\ref{sec:lightcurve_inspection}. We search for radial velocity trends and nontransiting companions in \S\ref{sec:nt_search}. We examine stellar activity in \S\ref{sec:stellar_activity_considerations}. In \S\ref{sec:joint_model} we describe our joint photometry, radial velocity, and stellar activity modeling framework. We present the results of this modeling in \S\ref{sec:results}. In \S\ref{sec:bulk_comp} we discuss possible bulk compositions for the planets and place them in the mass-radius diagram. In \S\ref{sec:atmo_char} we discuss the planets' prospects for atmospheric characterization. We conclude in \S\ref{sec:conclusion}. \revision{We note that the times of observations labeled in Barycentric Julian Date (BJD) or Barycentric \tess Julian Date (BTJD; $\mathrm{BTJD} = \mathrm{BJD} - 2457000$; i.e., the \tess and Doppler observations) were measured using the using the Barycentric Dynamical Time standard \citep[TDB; e.g.,][]{eastman10}.}

\section{\tess photometry} \label{sec:tess_obs}

\begin{deluxetable}{lcc}
\tablecaption{Summary of 2-min cadence \tess observations \label{tab:tess_summary}}
\tablehead{\colhead{System} & \colhead{Sectors} & \colhead{Observing start/end} \\ \colhead{} & \colhead{} & \colhead{(UT)}}
\vspace{-0.5cm}
\startdata
\sysI & 3, 30 & 2018-Sep-20/2020-Oct-21\\
\sysII & 6, 33 & 2018-Dec-11/2021-Jan-13\\
\sysIII & 5, 32, 43, 44 & 2018-Nov-15/2021-Nov-06\\
\sysIV & 9, 35 & 2019-Feb-28/2021-Mar-07\\
\sysV & (14 total) & 2019-Jul-18/2023-Jan-18\\
\sysVI & 17, 42, 43 & 2019-Oct-07/2021-Oct-12\\
\sysVII & 17, 57 & 2019-Oct-07/2022-Oct-29\\
\sysVIII & (6 total) & 2019-Nov-02/2022-Dec-23\\
\enddata
\tablecomments{According to data available on MAST as of 2023-Mar-07. Systems are listed in order of increasing TOI number, starting with \sysI (TOI-266). The start and end dates of the \tess observing baseline are listed in the ``Observing start/end'' column, but the systems were not necessarily observed continuously during this period. \sysV has 2-min cadence \tess light curves from a total of 14 sectors: 14, 15, 16, 20, 21, 22, 26, 40, 41, 47, 49, 53, 56, and 60. \sysVIII has 2-min cadence \tess light curves from a total of six sectors: 18, 19, 25, 52, 58, and 59.}
\end{deluxetable}

Of the 12 planets characterized in this work, 11 are detected in transit by \tess. Table \ref{tab:tess_summary} summarizes the 2-min cadence \tess observations for each system as they were available on the Mikulski Archive for Space Telescopes (MAST) on 2023-Mar-07 (i.e., up to and including \tess Sector 60). Each system was observed in at least two sectors, with \sysV being observed in 14. The photometry was processed by the \tess Science Processing Operations Center pipeline \citep[SPOC;][]{jenkins16}. \revision{All of the \tess data used in this paper can be found in MAST: \dataset[10.17909/y06k-3f04]{http://dx.doi.org/10.17909/y06k-3f04}.}

For all of our targets, there are no individual sources from \gaia Data Release 3 \citep[DR3;][]{gaia, gaiadr3arxiv} within 20\arcsec\ that cause $> 1\%$ dilution, nor does the combined flux of all DR3 sources within that radius cause $>1 \%$ dilution for any target. Furthermore, the SPOC data products we use are already corrected for dilution from \gaia sources per \gaiadrtwo \citep{gaiadr2}. In \S\ref{sec:imaging} we present high resolution imaging observations that rule out significant dilution from unresolved companions.\footnote{The results of TFOP imaging data, in addition to the results of other reconnaissance observations, were used to inform TKS target selection, as acknowledged by \cite{chontos22}. This paper formally reports the results of imaging observations of targets without such data already in the literature.} We discuss our light curve inspection and cleaning in \S\ref{sec:lightcurve_inspection} and our transit modeling in \S\ref{sec:transit_modeling}.

\section{High resolution imaging} \label{sec:imaging}
\begin{figure*}
    \centering
    \includegraphics[width=0.7\textwidth]{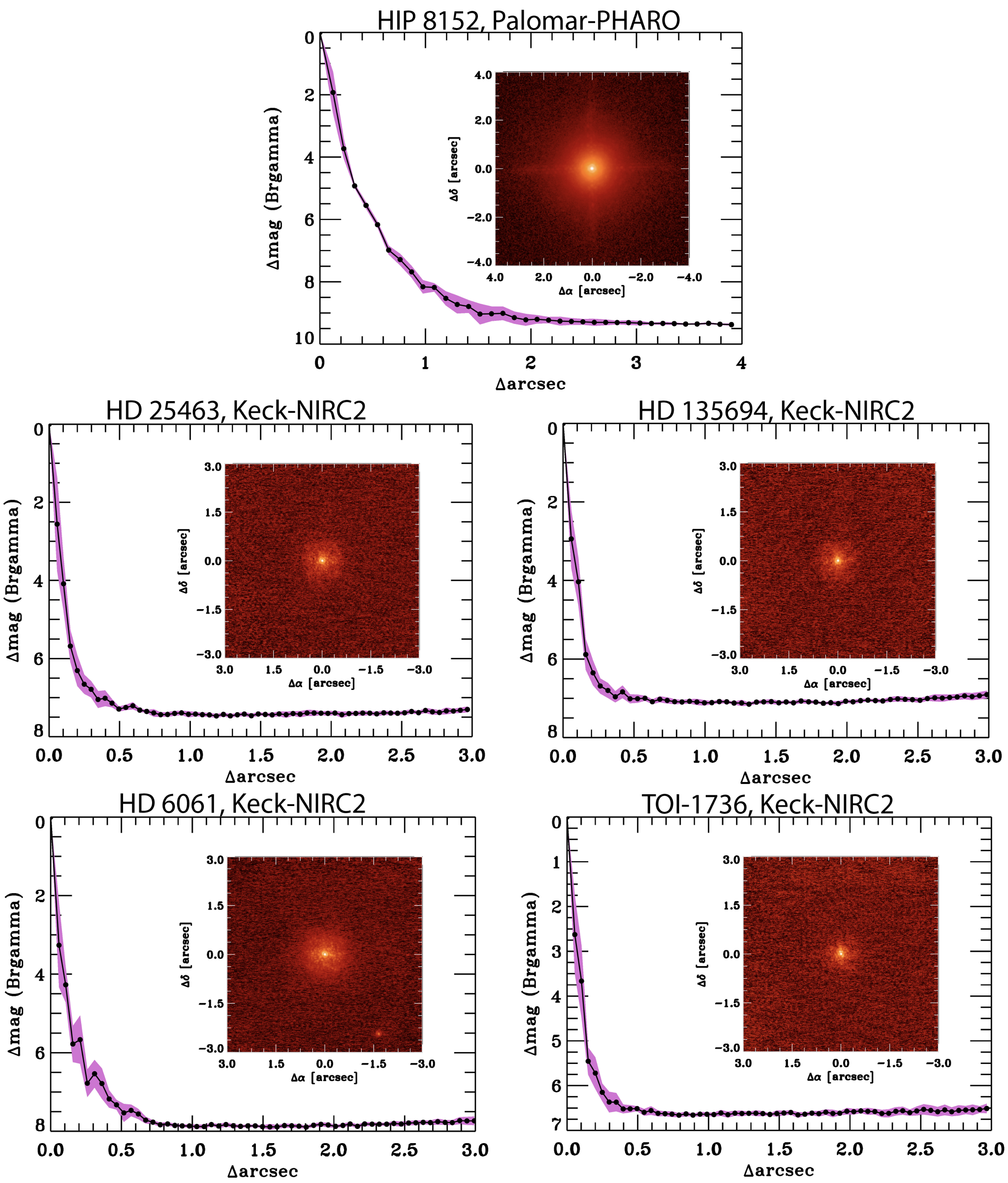}
    \caption{AO imaging results for \sysI, \sysIII, \sysV, \sysVII, and \sysVIII from our observations with \palomarpharo and \kecknirctwo. Contrast curves are shown in black with $1\sigma$ error envelopes in purple. The images themselves are shown as the postage stamp insets. \sysI, \sysIII, \sysV, and \sysVIII all appear single. \kecknirctwo observations of \sysVII were taken in both Br-$\gamma$ and $J$-cont, but only the former is shown here. In the image of \sysVII, \sysVIIcomp can be seen in the lower right corner at a separation of $\approx 3\arcsec$. \sysVIIcomp is fainter than \sysVII by $6.2$ mag in the \tess bandpass, meaning that its dilution of \sysVII b's transits is about a 0.1\% effect (i.e., much less than the uncertainty on the stellar radius). \sysVII and \sysVIIcomp have consistent distances and proper motions according to \gaiadrthree, meaning that the two stars are almost certainly gravitationally bound. At a distance of 67 pc, their on-sky separation of $3\arcsec$ translates to a sky-projected separation of about 200 AU.}\label{fig:imaging}
\end{figure*}

To ensure that the planet transits were not subject to dilution from sources not resolved by \gaia, we used high resolution imaging (HRI) to place contrast limits on potential nearby companions. \sysII, \sysIV, and \sysVI all have high resolution images in the literature that rule out dilution from nearby companions. We summarize the results of these observations in \S\ref{subsec:HRILit}. For the remaining five systems, \sysI, \sysIII, \sysV, \sysVII, and \sysVIII, we present new observations from \palomarpharo \citep{pharo} and \kecknirctwo \citep{nirc2}. The \palomarpharo and \kecknirctwo observations were obtained under the programs of PIs D. R. Ciardi and E. J. Gonzales, respectively. A summary of the imaging observations from this work can be found in Table \ref{tab:ao_summary} and sensitivity curves are shown in Figure \ref{fig:imaging}.

\begin{deluxetable*}{lccccccc}
\tablecaption{Imaging observations from this work \label{tab:ao_summary}}
\tablehead{\colhead{System} & \colhead{Instrument} & \colhead{Observation date} & \colhead{Filter} & \colhead{$t_\mathrm{exp}$} & \colhead{$N_\mathrm{exp}$} & \colhead{Resolution} & \colhead{Contrast at $0\arcsec.5$}\\ \colhead{} & \colhead{} &  \colhead{(UT)} & \colhead{} & \colhead{(s)} & \colhead{} & \colhead{(FWHM)} & \colhead{($\Delta$ mag)}}
\vspace{-0.5cm}
\startdata
\sysI    & \palomarpharo & 2018-Dec-22 & Br-$\gamma$ &   9.9 & 15 &  $0\arcsec.11$ & 6.0 \\ \hline
\sysIII  & \kecknirctwo  & 2020-Sep-09 & Br-$\gamma$ &   0.2 &  9 &  $0\arcsec.05$ & 7.5 \\ \hline
\sysV    & \kecknirctwo  & 2020-May-28 & Br-$\gamma$ &   1.0 &  9 &  $0\arcsec.05$ & 7.0 \\ \hline
\sysVII  & \kecknirctwo  & 2020-May-28 &    $J$-cont &   1.2 &  9 &  $0\arcsec.04$ & 7.0 \\
\sysVII  & \kecknirctwo  & 2020-May-28 & Br-$\gamma$ &   1.5 & 18 &  $0\arcsec.05$ & 7.5 \\
\sysVII  & \palomarpharo & 2020-Dec-05 &    $H$-cont &   1.4 & 15 &  $0\arcsec.08$ & 7.1 \\
\sysVII  & \palomarpharo & 2020-Dec-05 & Br-$\gamma$ &   1.4 & 15 &  $0\arcsec.09$ & 6.7 \\ \hline
\sysVIII & \kecknirctwo  & 2020-Sep-09 & Br-$\gamma$ &   0.5 &  9 &  $0\arcsec.05$ & 6.5 \\
\enddata
\tablecomments{$J$-cont: $\lambda_0 = 1.213$ $\mu$m and $\Delta \lambda = 0.020$ $\mu$m. $H$-cont: $\lambda_0 = 1.668$ $\mu$m and $\Delta\lambda = 0.018$ $\mu$m. Br-$\gamma$: $\lambda_0 = 2.169$ $\mu$m and $\Delta\lambda = 0.0323$ $\mu$m. Systems are listed in order of increasing TOI number.}
\end{deluxetable*}

\begin{figure}
    \centering
    \includegraphics[width=\columnwidth]{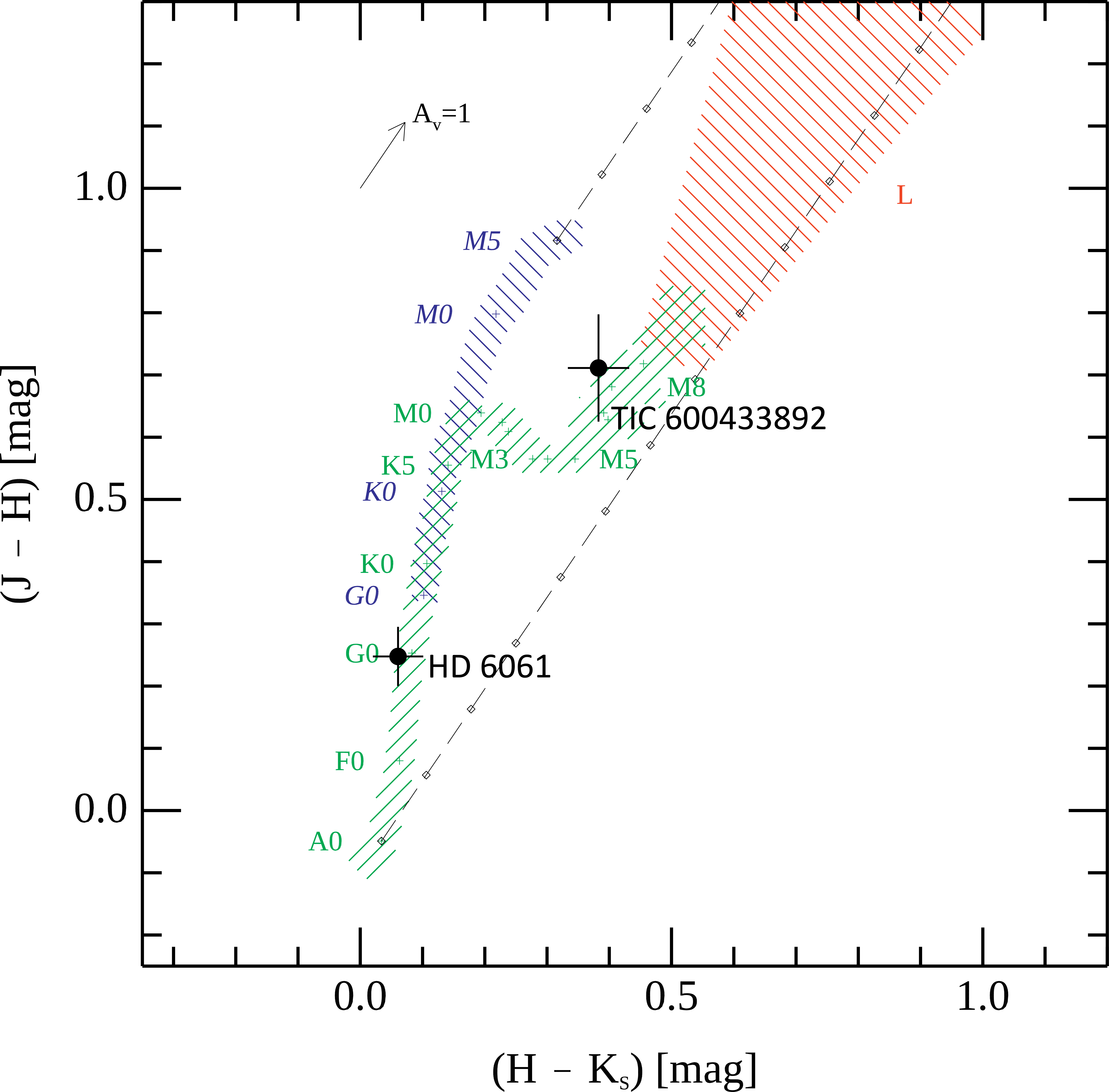}
    \caption{A \twomass \emph{JHK$_s$} color-color diagram. The dwarf branch, giant branch, and brown dwarf loci are shown with green, blue, and red hashes, respectively. Black dashed lines represent the direction of reddening induced by extinction ($A_V$). \sysVII and \sysVIIcomp are overplotted as the black circles with 1$\sigma$ error bars. \sysVIIcomp is consistent with being an M4/5V dwarf. We find that \sysVII is consistent with being a G0 dwarf, which agrees with the classification from \cite{cannon93}.}\label{fig:sysVII_color_color}
\end{figure}

\subsection{Literature Observations} \label{subsec:HRILit}
\sysII and \sysIV have \kecknirctwo observations from \cite{schlieder21}. \sysII was observed with \kecknirctwo on 2019-Mar-25 using the narrow-band Br-$\gamma$ filter ($\lambda_0 = 2.169$ $\mu$m and $\Delta\lambda = 0.032$ $\mu$m) and an integration time of 30 s. The star appears single and the \kecknirctwo observation rules out companions of $\Delta 7.4$ mag at $5\sigma$ confidence at a separation of 0\arcsec.5. The \sysIV \kecknirctwo observation was taken on 2019-Jun-09 using the $K$ filter ($\lambda_0 = 2.196$ $\mu$m and $\Delta\lambda = 0.336$ $\mu$m) and an integration time of 10 s. The star appears single, and the \kecknirctwo observation rules out companions of $\Delta 7.6$ mag at $5\sigma$ confidence at a separation of 0\arcsec.5.

\osbornMMXXIII reported the discovery and confirmation of \sysVI b and c using space-based photometry from \tess and \cheops \citep{cheops} along with radial velocity observations from CAFE, HARPS-N, and SOPHIE \citep{cafe, harpsn, sophie}. To rule out dilution from nearby sources, the authors observed \sysVI with a variety of optical speckle and near-infrared (NIR) adaptive optics (AO) instruments, including \kecknirctwo. The HRI shows no evidence of stellar companions within 1\arcsec.

\subsection{\palomarpharo and \kecknirctwo observations}
\subsubsection{\sysI (\toiI)} \label{sec:sysI_ao}
Palomar Observatory HRI observations of \sysI were made with the PHARO instrument on the 5.1 m Hale telescope. \palomarpharo has a pixel scale of $0\arcsec.025$ pix$^{-1}$ for a total field of view of about $25\arcsec$. Observations of \sysI were taken on 2018-Dec-22 in the narrow-band Br-$\gamma$ filter. Observations were acquired using the natural guide star AO system P3K \citep{dekany13} in the standard 5-point \texttt{quincunx} dither pattern with steps of $5\arcsec$. Each dither position was observed three times, with $0\arcsec.5$ positional offsets between each observation, for a total of 15 frames. For \sysI, each frame had an integration time of 9.9 s, amounting to a total on-source time of 149 s. No stellar companions were detected.

\subsubsection{\sysIII (\toiIII)} \label{sec:sysIII_ao}
Keck Observatory HRI observations of \sysIII were made with the NIRC2 instrument on the 10 m Keck II telescope. \kecknirctwo was used in the narrow-angle mode with a pixel scale of approximately $0\arcsec.01$ pix$^{-1}$ and a full field of view of about $10\arcsec$. Observations of \sysIII were taken on 2020-Sep-09 in the narrow-band Br-$\gamma$ filter. Observations were acquired using the natural guide star AO system in the standard three-point dither pattern to avoid the lower left quadrant of the detector, which is typically noisier than the other three quadrants. The dither pattern has a step size of $3\arcsec$. Each dither position was observed three times, with $0\arcsec.5$ positional offsets between each observation, for a total of nine frames. For \sysIII, each frame had an integration time of 0.2 s, amounting to a total on-source time of 1.8 s. No stellar companions were detected.

\subsubsection{\sysV (\toiV)}
\kecknirctwo observations of \sysV were taken on 2020-May-28 following the methods described in \S\ref{sec:sysIII_ao}. Images were taken in the narrow-band Br-$\gamma$ filter. Each frame had an integration time of 1.0 s, amounting to a total on-source time of 9 s. No stellar companions were detected.

\subsubsection{\sysVII (\toiVII)}
\kecknirctwo observations of \sysVII were taken on 2020-May-28 following the methods described in \S\ref{sec:sysIII_ao}. Images were taken in both the $J$-continuum ($\lambda_0 = 1.213$ $\mu$m and $\Delta\lambda = 0.020$ $\mu$m) and Br-$\gamma$ narrow-band filters. Observations of \sysVII were taken in multiple filters due to the visual observation of a nearby diluting source (\sysVIIcomp, separation of $\approx 3\arcsec$) in order to further ascertain colors and the likelihood of the nearby stellar object being bound. For the Br-$\gamma$ observations, six images were taken at each dither position, for a total of 18 frames. Each frame had an integration time of 1.2 s in $J$-cont and 1.5 s in Br-$\gamma$, amounting to a total on-source time of 11 s in $J$-cont and 27 s in Br-$\gamma$.

\sysVII was also observed with \palomarpharo on 2020-Dec-05 following the methods described in \S\ref{sec:sysI_ao}. \sysVII was observed in both the Br-$\gamma$ and $H$-continuum ($\lambda_0 = 1.668$ $\mu$m and $\Delta\lambda = 0.018$ $\mu$m) narrow-band filters. Each frame (in both Br-$\gamma$ and $H$-cont) had an integration time of 1.4 s, amounting to a total on-source time of 21 s in each filter. We discuss the nature of the stellar companion in \S\ref{sec:ao_sysVIIcomp}. Other than \sysVIIcomp, no other stellar companions were detected.

\subsubsection{\sysVIII}
\kecknirctwo observations of \sysVIII were taken on 2020-Sep-09 following the methods described in \S\ref{sec:sysIII_ao}. Images were taken in the narrow-band Br-$\gamma$ filter. Each frame had an integration time of 0.5 s, amounting to a total on-source time of 4.5 s. No stellar companions were detected.

\subsection{\palomarpharo and \kecknirctwo reduction}
Both the \palomarpharo and the \kecknirctwo data were reduced using the same methods. The science frames were flat-fielded and sky-subtracted. The flat fields were generated from a median average of dark subtracted flats taken on-sky. The flats were normalized such that the median value of the flats is unity. The sky frames were generated from the median average of the dithered science frames; each science image was then sky-subtracted and flat-fielded. The reduced science frames were combined into a single co-added image using an intrapixel interpolation that conserves flux, shifts the individual dithered frames by the appropriate fractional pixels, and median-coadds the frames. The final resolutions of the combined dithers were determined from the full-width half-maximum (FWHM) of the point spread functions in the corresponding filter. 

The sensitivities of the final combined AO image were determined by injecting simulated sources azimuthally around the primary target every 20\degree\ at separations of integer multiples of the central source's FWHM \citep{furlan17}. The brightness of each injected source was scaled until standard aperture photometry detected it with $5\sigma$ significance. The resulting brightness of the injected sources relative to each target set the contrast limits at that injection location. The final $5\sigma$ limit at each separation was determined from the average of all of the determined limits at that separation. The uncertainty on the limit was set by the root-mean-square (RMS) dispersion of the azimuthal slices at a given radial distance. The final sensitivity curves are shown in Figure \ref{fig:imaging}. For all targets, no stellar companions were detected within $1\arcsec$.

\subsection{\sysVIIcomp: a stellar companion to \sysVII} \label{sec:ao_sysVIIcomp}
\sysVII was observed in multiple filters with both \kecknirctwo and \palomarpharo due to the presence of a nearby stellar companion. In Br-$\gamma$, the companion, \sysVIIcomp, has a separation of $3\arcsec.06 \pm 0\arcsec.20$ and a position angle of $213 \pm 1\degree$ E of N. \sysVIIcomp is fainter than \sysVII by $6.2$ mag in the \tess bandpass, meaning that its dilution of \sysVII b's transits, approximately a 0.1\% effect, is negligible. According to \gaiadrthree, \sysVII and \sysVIIcomp have consistent distances to 1$\sigma$ ($67.69 \pm 0.07$ pc and $66.3 \pm 1.4$ pc, respectively) and consistent proper motions to 3$\sigma$ ($\mu_\alpha = -9.72 \pm 0.01$ mas/yr and $\mu_\delta = -10.09 \pm 0.01$ mas/yr for \sysVII, $\mu_\alpha = -10.7 \pm 0.5$ mas/yr and $\mu_\delta = -10.6 \pm 0.2$ mas/yr for \sysVIIcomp). This implies that the two stars are almost certainly gravitationally bound. At a distance of 67 pc, $3\arcsec$ translates to a sky-projected separation of about 200 AU. 

Following the methods of \cite{ciardi18}, relative photometry was conducted on the \kecknirctwo $J$-cont image and the \palomarpharo $H$-cont and Br-$\gamma$ images to deblend the infrared magnitudes of the two stars (where Br-$\gamma$ is taken to have a central wavelength that is sufficiently close to $K_s$). The resulting Two Micron All Sky Survey \citep[\twomass;][]{2mass} $JHK_s$ color-color diagram suggests that \sysVIIcomp is an M4/5V dwarf (Figure \ref{fig:sysVII_color_color}). Following the reasoning in \cite{ciardi18}, it is unlikely that \sysVIIcomp is a heavily reddened ($A_V > 6$ mag using an $R = 3.1$ extinction law) early-F or late-A background star, given that the entire line-of-sight extinction through the Galaxy is only $A_V \approx 2$ mag \citep{schlafly11}.

\section{Determination of stellar properties} \label{sec:stellar}
\subsection{Stellar template observations} \label{sec:template_obs}
We used the High Resolution Echelle Spectrometer (HIRES; \citealt{vogt94}) on the 10 m Keck I telescope at the W. M. Keck Observatory on Maunakea to obtain iodine-free spectra of each system at high resolution and \snr, which were used to produce a deconvolved stellar spectral template (DSST) for each host. The exposure parameters for each template are summarized in Table \ref{tab:keckTemplates}. Triple-shot exposures of rapidly rotating B stars were taken with the iodine cell in the light path immediately before and after the high-resolution templates were collected in order to precisely constrain the instrumental point-spread function (PSF). The data collection and reduction followed the methods of the California Planet Search (CPS) as described in \cite{howard10}.

\subsection{Stellar characterization} \label{sec:stellar_characterization}
We performed an initial stellar characterization of each host star using \specMatchEmp \citep{yee17} to constrain stellar effective temperature (\teff), metallicity (\feh), and stellar radius (\rstar) directly from the iodine-free \keckhires template spectra. \specMatchEmp fits stellar spectra between 5000 and 5800 \AA\ in 100 \AA\ segments using a linear combination of spectral templates from a library of over 400 precisely characterized FGKM stars.

To estimate the posteriors of the fundamental stellar parameters, we used \isoclassify (\citealt{huber17}; \citealt{berger20}) in \texttt{grid} mode with the \texttt{allsky} dust map, which is an extinction model obtained via a combination of the models from \cite{drimmel03}, \cite{marshall06}, and \cite{green19}. \isoclassify infers marginal posteriors for stellar properties by integrating over a grid of MIST isochrones \citep{choi16}. To inform the \isoclassify analysis, we input priors stemming from our \specMatchEmp results, parallaxes from \gaiadrthree, and \twomass $JHK_s$ magnitudes.\footnote{In the case of \sysVII, the \twomass $JHK_s$ magnitudes were deblended to account for the flux from \sysVIIcomp (see \S\ref{sec:ao_sysVIIcomp}).} Following \cite{tayar22}, to account for model-dependent systematic uncertainties, we inflated the errors on each host star's mass and radius by adding an additional 5\% and 4\% uncertainty, respectively, in quadrature with the measurement error reported by \isoclassify. The final stellar parameters are summarized alongside planet parameters in Appendix \ref{appendix:joint_models}.

\begin{deluxetable*}{lcccccc}
\tablecaption{\keckhires template observations \label{tab:keckTemplates}}
\tablehead{\colhead{System} & \colhead{Date} & \colhead{$t_\mathrm{exp}$} & \colhead{Decker} & \colhead{Airmass} & \colhead{\snr} & \colhead{$N_\mathrm{exp}$} \\
\colhead{} & \colhead{(UT)} & \colhead{(s)} & \colhead{} & \colhead{} & \colhead{(pix$^{-1}$)} & \colhead{}}
\vspace{-0.5cm}
\startdata
\sysI    & 2019-Aug-18 &  476 & B3 & 1.34 & 210 & 1 \\
\sysII   & 2019-Oct-31 &  314 & B1 & 1.21 & 217 & 2 \\
\sysIII  & 2019-Aug-18 &   26 & B3 & 1.52 & 200 & 3 \\
\sysIV   & 2020-Jan-04 & 1692 & B3 & 1.28 & 214 & 1 \\
\sysV    & 2020-Mar-09 &  180 & B1 & 1.66 & 211 & 2 \\
\sysVI   & 2020-Jan-30 &  263 & B3 & 1.42 & 210 & 2 \\
\sysVII  & 2019-Dec-28 &  180 & B3 & 1.15 & 213 & 2 \\
\sysVIII & 2020-Aug-11 &  187 & B3 & 1.59 & 211 & 1 \\
\enddata
\tablecomments{B1 decker: 3\arcsec.5 $\times$ 0\arcsec.574, $R=$ 60,000. B3 decker: 14\arcsec $\times$ 0\arcsec.574, $R=$ 60,000. \snr measured at 5500 \AA. $N_\mathrm{exp} > 1$ means that consecutive exposures were taken and then combined to produce the final template spectrum. For these cases, the $t_\mathrm{exp}$, airmass, and \snr reported in this table are the median values across the $N_\mathrm{exp}$ observations. All template observations were acquired with a moon separation of $> 30\degree$.
}
\end{deluxetable*}

\section{Doppler follow-up} \label{sec:rvs}
\subsection{\keckhires}
We obtained high-resolution spectra of each target with \keckhires to measure precise radial velocities (RVs). RVs were determined following the procedures of \cite{howard10}. In brief, a warm cell of molecular iodine was placed at the entrance slit during the RV observations \citep{butler96}. The superposition of the iodine absorption lines on the stellar spectrum provides both a fiducial wavelength solution and a precise, observation-specific characterization of the instrument's point spread function (PSF). As part of a forward model, the spectrum is divided into about 700 pieces between $\sim$5000--6000 \AA, with each piece being 2 \AA\ in width. For each piece, the product of the DSST and the Fourier Transform Spectrograph (FTS) iodine spectrum is convolved with the PSF to match the iodine-in observation. As one of the free parameters, an RV for each piece of spectrum is produced. The pieces are weighted using all observations of the star to produce a single RV for each observation. Our \keckhires Doppler observations are summarized in Table \ref{tab:rv_summary} \revision{and the RV measurements can be found in Table \ref{tab:all_rvs}.}

\subsection{\apflevy} \label{sec:rvs_apf}
\subsubsection{Data reduction and cleaning}
For the brighter targets in our sample ($V < 9.25$ mag) we also obtained high-resolution spectra with the Levy spectrograph mounted on the 2.4 m Automated Planet Finder telescope (APF; \citealt{vogt14}) at Lick Observatory on Mt. Hamilton near San Jos\'e, California. Though mounted on a much smaller telescope, \apflevy is complementary to Keck-HIRES in both latitude and observing cadence. In the case of \sysVIII, \apflevy observed periastron passage for the giant planet (\sysVIII c) while the system was inaccessible from Maunakea. With its queue-based observing schedule and lower oversubscription rate compared to Keck, \apflevy can also typically observe targets with higher cadence than \keckhires.

The standard reduction pipeline used to compute RVs from \apflevy spectra follows the methods of \cite{howard10}. As with our \keckhires observations, spectra were obtained with a warm cell of molecular iodine in the light path. We used the \keckhires DSSTs to compute RVs instead of acquiring independent iodine-free template spectra with \apflevy. \keckhires DSSTs have been shown to serve as effective replacements for \apflevy templates in the CPS Doppler reduction pipeline (e.g., \citealt{dai20, macdougall21, dalba22, lubin22}) and provide an efficient alternative to the long exposures that would otherwise be required to achieve similar \snr on an iodine-free \apflevy template. 

To avoid using low quality \apflevy RVs in our analysis, for each system we inspected the distribution of \apflevy RV errors as a function of \snr at 5500 \AA. We placed a conservative maximum RV error threshold of three times the median RV error for each target. Observations that landed above the error threshold were removed and the \apflevy RVs were recomputed using the cleaned data set. For \sysV, this resulted in removing four \apflevy spectra, all with $\sigma_\mathrm{RV} > 6.3$ \mps. For \sysVI, we removed six spectra, all with $\sigma_\mathrm{RV} > 5.8$ \mps. For \sysVII, we removed three spectra, all with $\sigma_\mathrm{RV} > 21.9$ \mps. For \sysVIII, we removed 13 spectra, all with $\sigma_\mathrm{RV} > 6.6$ \mps. The \apflevy Doppler observations used in our analysis are summarized in Table \ref{tab:rv_summary} \revision{and the RV measurements can be found alongside the \keckhires RVs in Table \ref{tab:all_rvs}.}

\subsubsection{The case of \sysIII} \label{sec:rvs_apf_igrand}
For one system, \sysIII, the reduction methods we used to measure velocities from the \apflevy spectra were slightly different from the methods of \cite{howard10} due to the star's rapid rotation (for \sysIII, we measure a sky-projected stellar rotational velocity of \vsini $=$ \vsiniIII km/s using \specMatchSynth; \citealt{specmatchsynth}). For ease of reference, we will refer to the methods in \cite{howard10} as the ``default'' reduction pipeline. To measure the radial velocity of a star from a spectrum, the default Doppler pipeline breaks the spectrum into small chunks and fits stellar absorption lines chunk-by-chunk. The size of each chunk is determined by a fixed pixel width. For \keckhires, this pixel width translates to a chunk width of about 2 \AA\ in wavelength space. However, because \apflevy has higher spectral resolution than \keckhires, this fixed pixel width translates to a smaller chunk width in wavelength space. For reference, the W decker on \apflevy has $R = 95,000$ \citep{vogt14} while the B5 decker on \keckhires has $R = 45,000$ \citep{vogt94}, where these deckers are typical for observations of \sysIII. Using the default Doppler reduction pipeline on \apflevy spectra therefore results in less spectral information being contained in each chunk than when it is applied to \keckhires spectra.

This difference in the wavelength space width of each chunk is typically not an issue for inactive, slowly-rotating stars (as is evident in the consistency between the \apflevy and \keckhires RVs for a representative system such as \sysVI). However, for more rapid rotators (\vsini $\gtrsim 10$ km/s), line broadening can conspire with the smaller chunk width to cause catastrophic errors in the \apflevy RV measurement process. This failure happens because single stellar absorption lines become too broad to fit within a single chunk. We observe this failure mode for \sysIII when trying to measure RVs from the \apflevy spectra via the default method. To circumvent this failure, for \sysIII's \apflevy spectra we compute RVs by fitting entire echelle orders simultaneously instead of fitting small chunks in series. This method also does not depend on an iodine-free template spectrum. Instead, we simultaneously solve for the stellar template using all of the iodine-in spectra. Save for these changes, the rest of the reduction is similar to the default method. 

We refer to this alternative method of computing the \apflevy RVs as the \igrand method. For completeness, Appendix \ref{appendix:apf_grand} contains figures comparing the default \apflevy RVs to the \igrand RVs for \sysIII. It is clear that the default \apflevy RVs are inconsistent with the contemporaneous \keckhires measurements (the default \apflevy RVs show nearly 100 m/s of scatter). In contrast, the spread and uncertainties of the \igrand velocities are more in line with expectations for a star of this magnitude ($V = 6.9$ mag) and spectral type (\teff $= 6200$ K). 

As we did for the other targets that were observed with \apflevy, we removed low quality \apflevy spectra of \sysIII by setting a maximum RV error threshold of three times the median \apflevy \igrand RV error. This resulted in removing 10 \apflevy spectra, all with $\sigma_\mathrm{RV} > 18.8$ \mps.

\begin{deluxetable*}{lcccccc}
\tablecaption{Summary of RV observations \label{tab:rv_summary}}
\tablehead{\colhead{System} & \colhead{Instrument} & \colhead{First/last observation} & \colhead{$N$ RVs (unbinned)} & \colhead{Median $t_\mathrm{exp}$} & \colhead{Median \snr} & \colhead{Typical decker} \\ \colhead{} & \colhead{} & \colhead{(UT)} & \colhead{} & \colhead{(s)} & \colhead{(pix$^{-1}$)} & \colhead{}}
\vspace{-0.5cm}
\startdata
\sysI & \keckhires & \sysIhiresjObsStartEnd & \sysIhiresjNBinNoBinRVs & 683 & \sysIhiresjMedRVSNR & C2\\
\hline \sysII & \keckhires & \sysIIhiresjObsStartEnd & \sysIIhiresjNBinNoBinRVs & 429 & \sysIIhiresjMedRVSNR & B5\\
\hline \multirow{ 2}{*}{\sysIII} & \keckhires & \sysIIIhiresjObsStartEnd & \sysIIIhiresjNBinNoBinRVs & 37 & \sysIIIhiresjMedRVSNR & B5\\
& \apflevy & \sysIIIapfObsStartEnd & \sysIIIapfNBinNoBinRVs & 592 & \sysIIIapfMedRVSNR & W\\
\hline \sysIV & \keckhires & \sysIVhiresjObsStartEnd & \sysIVhiresjNBinNoBinRVs & 897 & \sysIVhiresjMedRVSNR & C2\\
\hline \multirow{ 2}{*}{\sysV} & \keckhires & \sysVhiresjObsStartEnd & \sysVhiresjNBinNoBinRVs & 295 & \sysVhiresjMedRVSNR & B5\\
& \apflevy & \sysVapfObsStartEnd & \sysVapfNBinNoBinRVs & 1200 & \sysVapfMedRVSNR & W\\
\hline \multirow{ 2}{*}{\sysVI} & \keckhires & \sysVIhiresjObsStartEnd & \sysVIhiresjNBinNoBinRVs & 290 & \sysVIhiresjMedRVSNR & B5\\
& \apflevy & \sysVIapfObsStartEnd & \sysVIapfNBinNoBinRVs & 1800 & \sysVIapfMedRVSNR & W\\
\hline \multirow{ 2}{*}{\sysVII} & \keckhires & \sysVIIhiresjObsStartEnd & \sysVIIhiresjNBinNoBinRVs & 205 & \sysVIIhiresjMedRVSNR & B5\\
& \apflevy & \sysVIIapfObsStartEnd & \sysVIIapfNBinNoBinRVs & 1200 & \sysVIIapfMedRVSNR & W\\
\hline \multirow{ 2}{*}{\sysVIII} & \keckhires & \sysVIIIhiresjObsStartEnd & \sysVIIIhiresjNBinNoBinRVs & 226 & \sysVIIIhiresjMedRVSNR & B5\\
& \apflevy & \sysVIIIapfObsStartEnd & \sysVIIIapfNBinNoBinRVs & 1800 & \sysVIIIapfMedRVSNR & W\\
\enddata
\tablecomments{RVs are binned by 8 hrs. \keckhires B5 decker: 3\arcsec.5 $\times$ 0\arcsec.861, $R=$ 45,000. \keckhires C2 decker: 14\arcsec $\times$ 0\arcsec.574, $R=$ 45,000. \apflevy W decker: 1\arcsec $\times$ 3\arcsec, $R=$ 95,000. \snr is measured at 5500 \AA. All observations were acquired with a moon separation of $> 30\degree$.
}
\end{deluxetable*}


\begin{deluxetable*}{lcccccc} \label{tab:all_rvs}
\tablecaption{Radial Velocities and $S_\mathrm{HK}$ Values}
\tablehead{
  \colhead{System name} &
  \colhead{Time} & 
  \colhead{RV} & 
  \colhead{RV Unc.} & 
  \colhead{$S_\mathrm{HK}$} &
  \colhead{$S_\mathrm{HK}$ Unc.} &
  \colhead{Inst.} \\
  \colhead{}      &
  \colhead{(BJD)} & 
  \colhead{(m/s)} & 
  \colhead{(m/s)} &
  \colhead{}      &
  \colhead{}      &
  \colhead{}
}
\startdata
HIP 8152 & 2458710.099141 & -4.38   & 1.41    & 0.164   & 0.002   & HIRES   \\
\nodata  & \nodata       & \nodata & \nodata & \nodata & \nodata & \nodata \\
\enddata
\tablecomments{The RV and \shk measurements presented in this paper. \revision{Only the first row of the table (which is sorted by system and then by observation date) is shown here to inform its contents and format. BJD is reported using the TDB standard \citep[e.g.,][]{eastman10}.} Model-specific instrumental offsets have not been applied to the RV values. The RV errors listed here represent measurement uncertainty and have not been added in quadrature with the corresponding instrument jitter values resulting from our models of the data (see Appendix \ref{appendix:joint_models}). This table is available in its entirety online in machine-readable format.}
\end{deluxetable*}

\section{Light curve inspection and cleaning} \label{sec:lightcurve_inspection}
Before applying our joint analysis of the photometry and RVs, we first inspected and cleaned the \tess data. Using \texttt{lightkurve} \citep{lightkurve}, we downloaded all of the \tess Presearch Data Conditioning Simple Aperture Photometry \citep[PDCSAP;][]{smith12, stumpe12, stumpe14} 2-min cadence data for each target, excluding data with \texttt{NaN} values or data quality flags. We then normalized the data on a sector-by-sector basis. We also applied the following analysis to the Simple Aperture Photometry (SAP; \citealt{twicken10, morris20}) light curves for each target. While the best-fitting transit parameters were nearly identical between fits to the PDCSAP and SAP data, we generally found the SAP data contained obvious spacecraft systematics and required more outliers to be rejected.

\subsection{Transit search}
For each system, we searched for transits in the \tess PDCSAP light curve using the box least squares method (BLS; \citealt{kovacs02}). The signals reported by the SPOC were identified in the transiting planet search pipeline component, which employs an adaptive, noise-compensating matched filter \citep{jenkins02, jenkins10, jenkins20}. We recovered all SPOC-reported signals in the TOI catalog as of 2022-Oct-04, with a median \snr of 23 across all of our BLS detections and with each detection having \snr $\gtrsim 10$. After recovering the SPOC-reported signals, we masked the planet transits and re-ran our BLS search but failed to find any other candidates.

In the case of TOI-554.02, our BLS search recovers the candidate's shallow transits with a slightly lower significance (\snr $\approx 8$), motivating, in part, a more thorough investigation of the purported transit signal (see \S\ref{sec:validating_HD25463c}). In addition, there are two instances where we identify transiting planet candidates whose properties disagree with entries in the TOI catalog. These are the so-called ``duotransit" planets orbiting \sysI and \sysVI, which we discuss in \S\ref{sec:duo_transit_systems}.

\subsubsection{Statistical validation of TOI-554.02 (\sysIII c)} \label{sec:validating_HD25463c}
In the hierarchy of exoplanet detection, statistical validation is typically an intermediate step taken between planet candidacy and confirmation\footnote{For the purposes of this discussion, we take planet ``confirmation'' to mean that the planet's mass has been measured to some fiducial precision. For the mass-radius diagram in Figure \ref{fig:mr_diagram}, we show planets from the NASA Exoplanet Archive whose mass measurements have better than 50\% fractional precision.} where astrophysical false positive scenarios are systematically ruled out \citep[e.g.,][]{borucki12, morton16}. In the case of transiting planet candidates, validation is used to statistically exclude the possibility that the purported transit signal is in fact, for example, a background eclipsing binary star system. Since most of the transiting planets in this work are at least marginally ($\gtrsim 2.5\sigma$) detected with RVs, we bypass the statistical validation step as the measurement of their host star's Doppler signal confirms their planetary nature. However, in two cases, we measure only an upper limit on the planet mass. The first, \sysVI c, was externally validated and confirmed by \osbornMMXXIII using a combination of \tess and \cheops photometry and CAFE, HARPS-N, and SOPHIE RVs (see \S\ref{results:sysVI_comparison} for details). The second, TOI-554.02 (\sysIII c) has not yet been confirmed, so we take additional measures to statistically validate this planet.

A new Threshold Crossing Event (TCE) with $P = 3.04$ d was \revision{detected by the transit search of the SPOC Sectors 1--46 2-min light curve} for \sysIII (aka TOI-554). An initial limb-darkened transit model was fitted \citep{li19} and a suite of diagnostic tests were conducted to help \revision{determine whether or not the signal was planetary in nature} \citep{twicken18}. The transit signature passed all the diagnostic tests presented in the SPOC Data Validation reports. The \tess Science Office (TSO) reviewed the vetting information and issued an alert for TOI-554.02 on 2022-Apr-20.

The planet candidate is small (\rplanet $\approx 1.3$ \rearth from the SPOC report), and the pipeline only detects its transit with \snr $=8.5$. However, due to the candidate's short orbital period and the system's four sectors of photometry, \tess has observed 29 purported transits. After removing \sysIII b's transits from the light curve (as identified by our initial BLS search of the system), we re-ran BLS, but TOI-554.02's transit signal was not immediately apparent. We narrowed the BLS period grid to look for signals short of 10 d (down from 100 d) and increased the number of grid points (by a factor of 2). We identified a peak in the BLS power spectrum with \snr $\approx 8$ that corresponded to the SPOC-reported signal for TOI-554.02. While each individual transit is not entirely obvious by eye, the phase-folded transit shows a clear decrease in flux. We masked transits associated with TOI-554.02 and re-ran the BLS search but found no additional transit-like events. 

As discussed above, since our RV observations only place an upper limit on the mass of TOI-554.02 (see \S\ref{results:sysIII}), we independently analyzed the \tess photometry and our \kecknirctwo HRI with the planet validation framework \triceratops \citep{giacalone21} to rule out astrophysical false positive scenarios that might be responsible for TOI-544.02's purported transit signal. \triceratops validates planets by simulating astrophysical false positives arising from gravitationally bound stellar companions, chance-aligned foreground or background stars, and known nearby stars that are blended with the target in the \tess data. The marginal likelihoods of these false positive scenarios are calculated and compared to that of the scenario where the signal is caused by a planet transiting the target star. This calculation yields two quantities: the false positive probability (FPP; the overall probability that the signal is caused by something other than a planet transiting the target star) and the nearby false positive probability (NFPP; the probability that the signal is caused by a known nearby star that is blended with the target in the \tess data). In order for a planet to be considered validated, it must achieve ${\rm FPP} < 0.015$ and ${\rm NFPP} < 0.001$. To account for the intrinsic stochasticity in its calculation, we ran \triceratops 50 times on the same dataset, obtaining ${\rm FPP} = (4.6 \pm 0.2) \times 10^{-4}$ and ${\rm NFPP} = (5.7 \pm 0.9) \times 10^{-6}$. We find that the dominant contributor to FPP is the STP scenario, which involves a gravitationally bound stellar companion that hosts a transiting planet, although we note that this scenario is unlikely due to the absence of evidence for a stellar companion in our iodine-free spectra or RV data. Regardless, these values are sufficiently small to consider the planet statistically validated. We also note that these results are independent of the fact that we confirm \sysIII b using RVs, which makes it even more likely that TOI-554.02 is a true planet \citep{lissauer12, guerrero21}. Hereafter we refer to TOI-554.02 as \sysIII c.

\subsubsection{Duotransit systems} \label{sec:duo_transit_systems}
\sysI (TOI-266) and \sysVI (TOI-1471) both host two transiting planets, with the sub-Neptunes \sysI c and \sysVI c each having two transits in widely time-separated sectors \citep[these planets constitute a ``duotransit" scenario;][]{osborn22}. \cheops, in tandem with our \keckhires RVs of the systems, recently confirmed the correct period of \sysI c ($P = 19.61$ d; via private communication with the \cheops team, point of contact H. Osborn) and \sysVI c \citep[$P = 52.56$ d;][]{osborn23arxiv}. SPOC did not correctly identify the transit signals of these planets in the \tess data: For \sysI, TOI-266.02 is spuriously reported with $P = 6.19$ d and $T_\mathrm{c} = 1392.10$ BTJD. For \sysVI, as last updated on 2022-Apr-20, TOI-1471.02 is listed as having $P = 683.33$ d (the time difference between the transit in Sector 17 and the transit in Sector 42) and $T_\mathrm{c} = 1779.19$ BTJD (correct). After masking the transits of \sysI b and \sysVI b as identified by BLS, we re-ran our BLS search for both systems. In each case, the BLS power spectrum contained peaks with comparable significance (\snr $\approx 10$) at the aliases of the period allowed by the two widely time-separated transits. We masked the planet c transits by hand in each system and ran another BLS search, but found no additional transit-like events.

For \sysVI, the SPOC pipeline originally excluded all \revision{data points} in the Sector 17 light curve beyond 1787.72 BTJD due to a high level of scattered light from Earth, resulting in the exclusion of a second transit of \sysVI b in Sector 17 near 1788 BTJD. This initially caused the pipeline to match the first Sector 17 transit of \sysVI b with the Sector 17 transit of \sysVI c, and to report that \toiVI.01 had $P = 11.8$ d. It was not until later that TFOP follow-up revealed these two transits were actually of different depth and duration. To include the 1788 BTJD transit of \sysVI b in their analysis, \cite{osborn23arxiv}, hereafter \oXXIII, re-extract aperture photometry for \sysVI starting from the 2-min cadence target pixel files. In place of the PDC algorithm, they then use a custom light curve detrending method similar to \cite{vanderburg19} in order to remove spacecraft systematics. For the sake of homogeneity in our analysis of each system, we forgo replicating their custom light curve extraction and detrending, meaning that this work does not include the second transit of \sysVI b in Sector 17. We note that our measured transit parameters for \sysVI b are all consistent with the values reported by \oXXIII, and the primary reason for the difference in the size of our uncertainties on the radius of \sysVI b ($\pm 0.04$ \rearth from \oXXIII and $\pm 0.13$ \rearth from this work) is the difference in the reported uncertainty on our stellar radius measurements ($\pm 0.005$ \rsun from \oXXIII and $\pm 0.03$ \rsun from this work, where our error estimate has been inflated according to \citealt{tayar22}). We compare our results for \sysVI with those from \oXXIII in detail in \S\ref{results:sysVI_comparison}. 

\subsection{Light curve cleaning and initial transit fitting} \label{sec:phot_cleaning_init_transits}
\begin{figure}
    \centering
    \includegraphics[width=\columnwidth]{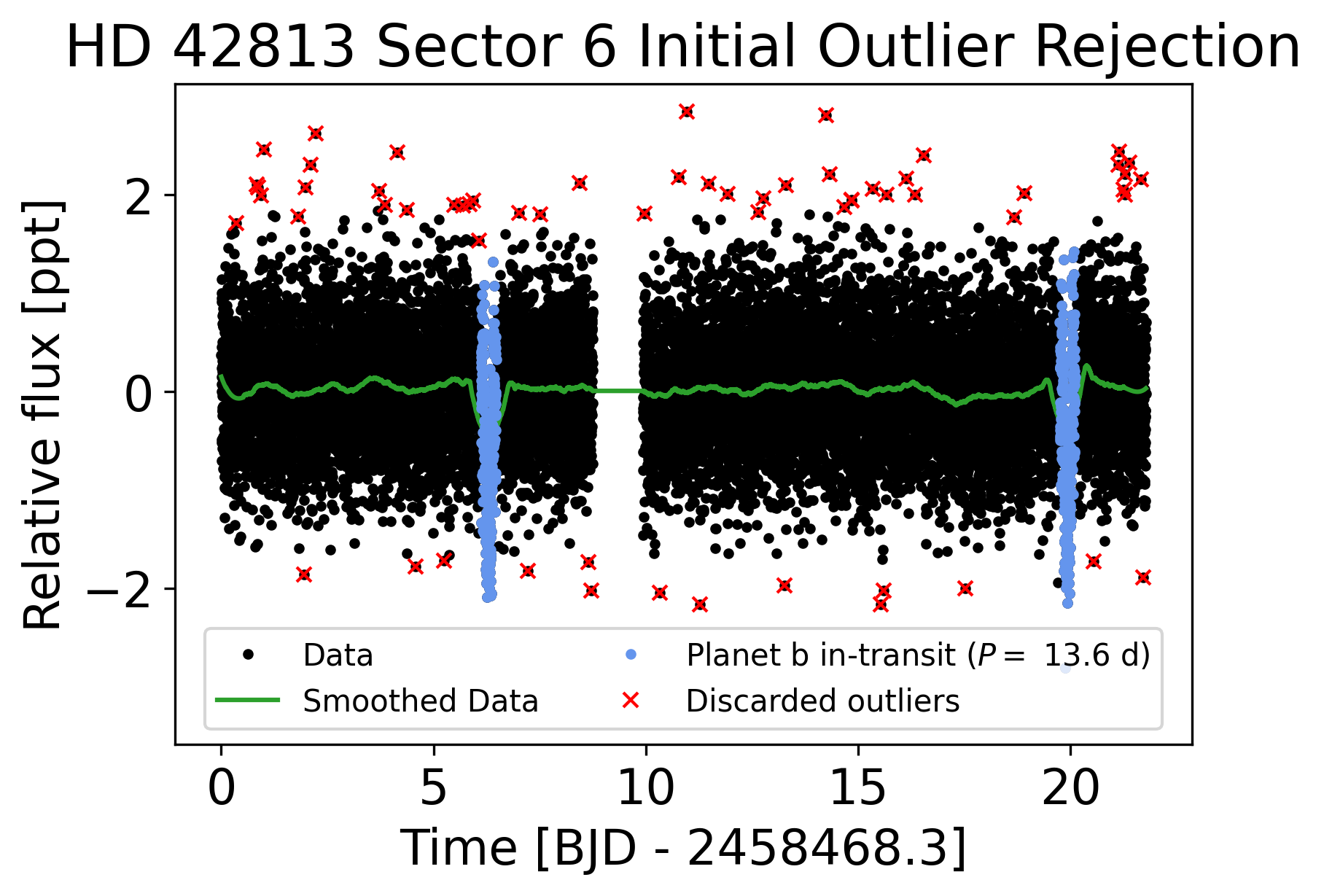}
    \caption{An example of our Savitzky-Golay filtering procedure for \sysII's Sector 6 PDCSAP photometry. The black points are the PDCSAP data, the green line is the data after being smoothed by the Savitzky-Golay filter, and the blue points are the in-transit data which are not subjected to the outlier rejection. Outliers are marked in red.}
    \label{fig:sg_filtering}
\end{figure}

After inspecting the \tess data for planet transits, we cleaned the photometry with an outlier rejection scheme. First, for each sector we smoothed the normalized \tess PDCSAP data in bins of 0.3 days with a cubic Savitzky-Golay filter \citep{savitzky64} and iteratively removed out-of-transit, $>3\sigma$ outliers until convergence. We used the SPOC-reported orbital period, time of transit, and transit duration to mask the planet transits, save for \sysI c and \sysVI c, since their orbital properties are incorrect in the TOI catalog (see \S\ref{sec:duo_transit_systems}). For \sysI c, we used the transit duration of a photometry-only fit to the \tess data (the same as the model described below) with a narrow Gaussian prior on the externally-confirmed period of $P = 19.61$ d. For \sysVI c, we used the transit duration from \oXXIII. Figure \ref{fig:sg_filtering} illustrates the results of the Savitzky-Golay filtering for \sysII's Sector 6 PDCSAP data. Across all systems, the number of outliers removed per sector by the Savitzky-Golay filtering was $74 \pm 14$. For each system, this outlier rejection excluded $\lesssim 0.5\%$ of all of the available \tess data. In each case, our iterative Savitzky-Golay filtering routine converged in three iterations, save for \sysVII, which converged in four iterations.

Next, we performed an additional outlier rejection step by fitting an initial, photometry-only transit plus Gaussian process model \citep[GP; e.g.,][]{rasmussen06} to the data and iteratively removing 7$\sigma$ outliers about the fit. The transit model was implemented with a quadratic limb darkening law \citep{exoplanet:kipping13} from \texttt{starry} \citep{starry} and the GP, used to remove low-frequency stellar variability and instrumental systematics, was constructed in \celeritetwo \citep{celerite2}. Following \cite{exoplanet:kipping13}, the limb darkening coefficients are parameterized as $q_1 \equiv (u_1 + u_2)^2$ and $q_2 \equiv 0.5 u_1(u_1 + u_2)^{-1}$, where $u_1$ and $u_2$ are the usual quadratic limb darkening coefficients. The transit model is parameterized using $\ln P$, $T_\mathrm{c}$, $\ln R_\mathrm{p}/R_*$, $b$, and $\ln T_\mathrm{dur}$. The parameters and priors of this initial, photometry-only model are generally the same as for the final, joint model of the photometry and RVs (Table \ref{tab:joint_model}). The main difference between the two is that the joint model does not assume a circular orbit and explicitly uses $\sqrt{e}\cos\omega$ and $\sqrt{e}\sin\omega$ instead of $\ln T_\mathrm{dur}$. For this initial, photometry-only model, we placed a broad Gaussian prior on $\ln T_\mathrm{dur}$, the center of which was the logarithm of the transit duration as reported in the TOI catalog when accessed on 2022-Oct-04, and whose width was $\ln 10$ d. This initial transit model also assumed no information about the stellar mass, since by employing a circular orbit and fitting in terms of $T_\mathrm{dur}$ we imply a stellar density. We elaborate on the differences between this initial, photometry-only model and our joint model in \S\ref{sec:transit_modeling}.

The kernel of the GP used to flatten the light curve is in the form of an overdamped stochastic harmonic oscillator (SHO). The power spectral density (PSD) of the SHO kernel can be written as
\begin{equation} \label{eqn:sho_kernel}
    S(\omega_f) = \sqrt{\frac{2}{\pi}} \frac{S_0 \omega_0^4}{(\omega_f^2 - \omega_0^2)^2 + \omega_0^2 \omega_f^2 / Q},
\end{equation}
where \omegaSHO is the angular frequency, $\omega_0$ is the undamped fundamental angular frequency, $S_0$ is the power at $\omega_0$, and $Q$ is the quality factor of oscillation. Following the reparameterization for the SHO PSD from the \celeritetwo documentation, we define
\begin{equation}
    \rho = \frac{2\pi}{\omega_0},
\end{equation}
\begin{equation}
    \tau = \frac{2 Q}{\omega_0},
\end{equation}
and
\begin{equation}
    \eta = \sqrt{S_0 \omega_0 Q},
\end{equation}
where $\rho$ is interpreted as the undamped fundamental period of the oscillator, $\tau$ is the characteristic timescale of the damping, and $\eta$ scales the amplitude of the GP (i.e., $\eta^2$ populates the diagonal of the GP covariance matrix). Rewriting Equation \ref{eqn:sho_kernel} in terms of $\rho$, $\tau$, and $\eta$, we have 
\begin{equation} \label{eqn:sho_kernel_RhoTauEta}
    S(\omega_f) = 8 \sqrt{2} \pi^{3/2} \frac{\eta^2}{\tau \rho^2}\Big[\big(\omega_f^2 - (\frac{2\pi}{\rho})^2\big)^2 + (\frac{2 \omega_f}{\tau})^2 \Big]^{-1}.
\end{equation}
The parameters and priors for this GP kernel are the same as used for the GP that flattens the light curve in the joint model (see \S\ref{sec:transit_modeling} and Table \ref{tab:joint_model}). We note that a lower bound of 1 d was placed on $\rho$ and $\tau$ to prevent the GP from overfitting the transits (see Figure \ref{fig:transit_gp_example}).

For \sysI, \sysIV, \sysV, and \sysVI, no 7$\sigma$ outliers were identified about this initial photometry-only model. \sysII, \sysIII, and \sysVII each had one 7$\sigma$ outlier that was removed, and \sysVIII had three. For the systems for which we identified these outliers, we repeated the initial transit model fitting with the outliers removed and found no remaining outliers about the fit. The maximum a posteriori (MAP) values from our initial models, save for $T_\mathrm{dur}$, were used as the starting values in the MAP optimization routine for the corresponding parameters in the joint models.

\subsection{Search for transit timing variations}
For completeness, we searched the \tess data for any signs of transit timing variations \citep[TTVs; e.g.,][]{haddenLithwick17}. We used the best-fitting transit times and orbital periods from the initial photometry-only transit model (above) as references for the expected transit times. We performed a MAP fit of the photometry that was analogous to the initial transit model, but now, for each planet, $\ln P$ and $T_\mathrm{c}$ were replaced with free parameters for the midpoint of each individual transit. We placed a Gaussian prior on each of the observed transit times centered at the expected time with a width of 1 d.

For each of the 11 transiting planets in our sample, we found that the maximum of the absolute difference between the observed and expected transit time ($O - C$) was $< 20$ min and the median of these maximum values was about 1 min across all planets. None of the $O - C$ time series show an obvious trend or sinusoidal variation. \sysIII b, \sysIV b, and \sysV b each had a maximum absolute value of $O - C$ between 10 and 20 min, and the scatter in $O - C$ for each of these three planets was about 8 min. These systems may warrant further investigation to determine whether the differences in the observed and expected transit times are significant. However, in the absence of a clear periodic TTV signal, we leave this work to future investigations. For the three multi-transiting planet systems in our sample (\sysI, \sysIII, and \sysVI), we note that none of the planet pairs have a near-integer period ratio, so TTVs may not be expected for these systems a priori. Given the lack of obvious evidence for TTVs in each system, we exclude them in our joint model.

\section{Search for RV trends and nontransiting companions} \label{sec:nt_search}
With all of the transits accounted for and the photometry cleaned, next we conducted a systematic search for long-term RV trends and the full orbits of nontransiting\footnote{We take ``nontransiting'' to mean that we did not observe a transit in the \tess photometry.} planetary signals in the RV time series. Long-term RV trends are indicative of massive, distant companions, which are more common for FGK hosts with close-in small planets (our sample) than for other stars \citep{zhu18, bryan19}. Our analysis identifies two systems with linear RV trends (\sysVI and \sysVIII) and one nontransiting, super-Jovian-mass planet on a moderately eccentric orbit (\sysVIII c).

\subsection{RV trends}
First, we attempted to determine which systems required a linear RV trend. We used the Akaike Information Criterion (AIC; \citealt{akaike74}) to choose between models with and without a linear trend. The AIC is defined as 
\begin{equation} \label{eqn:aic}
    \mathrm{AIC} = 2 k - 2 \ln \mathcal{\hat{L}},
\end{equation}
where $k$ is the number of free parameters in the model and $\mathcal{\hat{L}}$ is the maximum of the likelihood function with respect to the model parameters. In general, a lower AIC value is considered more favorable. Let $\Delta \mathrm{AIC}_i \equiv \mathrm{AIC}_i - \mathrm{AIC}_\mathrm{min}$, where AIC$_i$ is the AIC of the $i$th model under consideration and AIC$_\mathrm{min}$ is the lowest AIC value of all models considered. \cite{burnham04} provide the following guidelines in interpreting $\Delta$AIC values:
\begin{itemize}
    \item If $\Delta \mathrm{AIC}_i < 2$, the two models are nearly indistinguishable.
    \item If $2 < \Delta \mathrm{AIC}_i < 10$, the $i$th model is disfavored.
    \item If $\Delta \mathrm{AIC}_i > 10$, the $i$th model is essentially ruled out.
\end{itemize}
When two models had $\Delta \mathrm{AIC} < 4$, we chose the simpler model (e.g., even if including a linear RV trend reduces the AIC, if $\Delta \mathrm{AIC} < 4$, we adopted the model without a trend). There are only two systems that demand a linear RV trend, \sysVI and \sysVIII. For these systems we also attempted to include a quadratic term in addition to the linear trend, and while the AIC could not rule out models with curvature, there was no evidence to justify its inclusion.

\subsection{Nontransiting companions}
With linear RV trends either excluded or identified, we next used \rvsearch \citep{rosenthal21} to search for the full orbits of nontransiting planet candidates in the RV time series. \rvsearch employs an iterative Generalized Lomb-Scargle \citep[GLS;][]{lomb76, scargle82, zechmeister09} periodogram analysis to search for significant periodicity in the RV residuals. Significance is determined following the detection methodology of \cite{howard16}, where an empirical false alarm probability (FAP) threshold of 0.1\% is computed via the Bayesian Information Criterion (BIC; \citealt{schwarz78}). 

The BIC is defined as
\begin{equation} \label{eqn:bic}
    \mathrm{BIC} = k \ln n - 2 \ln \mathcal{\hat{L}},
\end{equation}
where $n$ is the size of the data and $k$ and $\mathcal{\hat{L}}$ are the same as in Equation \ref{eqn:aic}. While we could have also used the BIC to determine whether or not to include linear trends in our RV models, simulation studies suggest that for finite sample sizes, the BIC may be at risk of selecting very poor models \citep{burnham04, vrieze12}. In our analysis, the two comparison statistics typically agreed and could be used relatively interchangeably.

Before applying \rvsearch to our RV time series, we removed the signals of the transiting planets and, for \sysVI and \sysVIII, the linear RV trends. \rvsearch identified no signals above the 0.1\% FAP threshold in the RV residuals for all systems except \sysI, \sysV, and \sysVIII. However, for all but \sysVIII, it seems that the signals identified are related to the RV window function. We discuss our interpretation of the detections below.

In the case of \sysI, \rvsearch identified an eccentric signal ($e \approx 0.7$, $K \approx 4.5$ \mps) at $P = 122$ d. This is likely the second harmonic of the yearly observing alias (i.e., $122 \times 3 = 366$), and we do not interpret it as planetary in nature. Visually, it is clear that the \keckhires observations of \sysI can be roughly grouped into three observing seasons (see Figure \ref{fig:hip8152_phot_and_rvs}, right). This is probably contributing to the power in the RV window function around 365 d (Figure \ref{fig:hip8152_periodogram}). It should also be mentioned that $P = 122$ d is an alias commonly seen in archival \keckhires RV time series \citep{rosenthal21}. Furthermore, at periastron passage, a planet with $P = 122$ d and $e = 0.7$ would have a very close ($< 0.01$ AU) encounter with the orbit of \sysI c (\periodc = \periodIc d), suggesting that such an architecture is not stable. 

For \sysV, \rvsearch identified a moderately eccentric signal ($e \approx 0.3$, $K \approx 3.5$ \mps) at 45.6 d. However, like \sysI, there is significant power related to the yearly alias in the periodogram of \sysV's RV window function (in this case, at $2 \times 365$ d; Figure \ref{fig:hd135694_periodogram}) and the supposed period is likely a harmonic of this signal ($365.25 / 45.60 = 8.00$). Therefore, we also interpret the 45.6 d signal as an artifact of our RV sampling. If the $P = 45.6$ d signal truly is a planet, however, it would not cross orbits with \sysV b (\periodb = \periodVb d).

The moderately eccentric orbit of the nontransiting super-Jovian, \sysVIII c, near $P = 570$ d is visible in \sysVIII's RV time series (Figure \ref{fig:toi1736_phot_and_rvs}, right). For completeness, we conducted a blind search for the orbit of \sysVIII c after removing the transiting planet, \sysVIII b, and the system's linear RV trend. \rvsearch recovers the orbit of \sysVIII c with \periodc $= 573.6$ d, \transitTime $= 2272.4$ BTJD, $K_\mathrm{c} = 195$ \mps, $e_\mathrm{c} = 0.37$, and $\omega_\mathrm{c} = 162\degree$. Models of the RVs that either replaced the linear RV trend with the partial orbit of an even longer period giant planet or included a curvature term in addition to the linear trend were not preferred by the AIC. After removing the linear RV trend and the orbits of planets b and c, \rvsearch failed to identify any other signals above the 0.1\% FAP threshold in the RV residuals.

\section{Stellar activity considerations} \label{sec:stellar_activity_considerations}
Stellar activity mitigation is a key component of RV mass measurements for small planets, especially when the stellar rotation period or one of its harmonics is close to the period of the planet in question \citep[e.g.,][]{vanderburg16}. Most of the hosts in our sample show little Ca II H and K emission, implying that they are relatively inactive---this is in part why they were chosen for Doppler monitoring \citep{chontos22}. Using our \keckhires spectra, we measure \logrhk \citep{middelkoop82, noyes84} for each system and find a median value across all eight hosts of $-5.00$. For reference, over its magnetic cycle the Sun oscillates between \logrhk $= -5.05$ and $-4.84$ at the solar minimum and maximum, respectively \citep{meunier10}.

With each \keckhires spectrum we also measured \shk values, which trace Ca II H and K emission strength (\citealt{isaacson10}; Isaacson et al. in prep). While photometry can act as a proxy for stellar activity \citep[e.g.,][]{aigrain12, haywood14, grunblatt15}, if photometric and spectroscopic monitoring are not contemporaneous, the connection between the time-varying activity signal during the two sampling periods can be unclear \citep{kosiarek20}. Because the RV and \shk measurements are simultaneous, they offer a real-time view of the star's behavior and serve as a useful supplement to the \tess photometry.

\subsection{Correlated \shk values and RV residuals?}
As a first step toward understanding the connection between stellar activity and our RV measurements, we examined the correlation between the \keckhires \shk values and the \keckhires RVs after the planetary-attributed RV signals were removed. By ``planetary-attributed,'' we mean the RV signals from transiting planets and, for \sysVI and \sysVIII, linear RV trends that we assume are caused by distant giant companions. We attributed the strong linear trends in the RV time series for \sysVI and \sysVIII to distant giants rather than stellar activity because for these quiet stars (which have \logrhk = \logrhkVI and \logrhk = \logrhkVIII, respectively) we would expect the amplitude of the stellar activity signal to be on the order of a few \mps \citep[e.g.,][]{wright05, wright08}. However, the change in RV over the observing baselines for these systems is closer to 100 \mps and the corresponding change in \shk value is $< 0.02$, so it does not seem like stellar activity could be responsible for the large RV trends. For \sysVIII, we also removed the RV signal of \sysVIII c—a nontransiting, massive planet on a moderately eccentric orbit near $P = 570$ d—because it is clearly planetary.

\begin{figure*}
\gridline{\fig{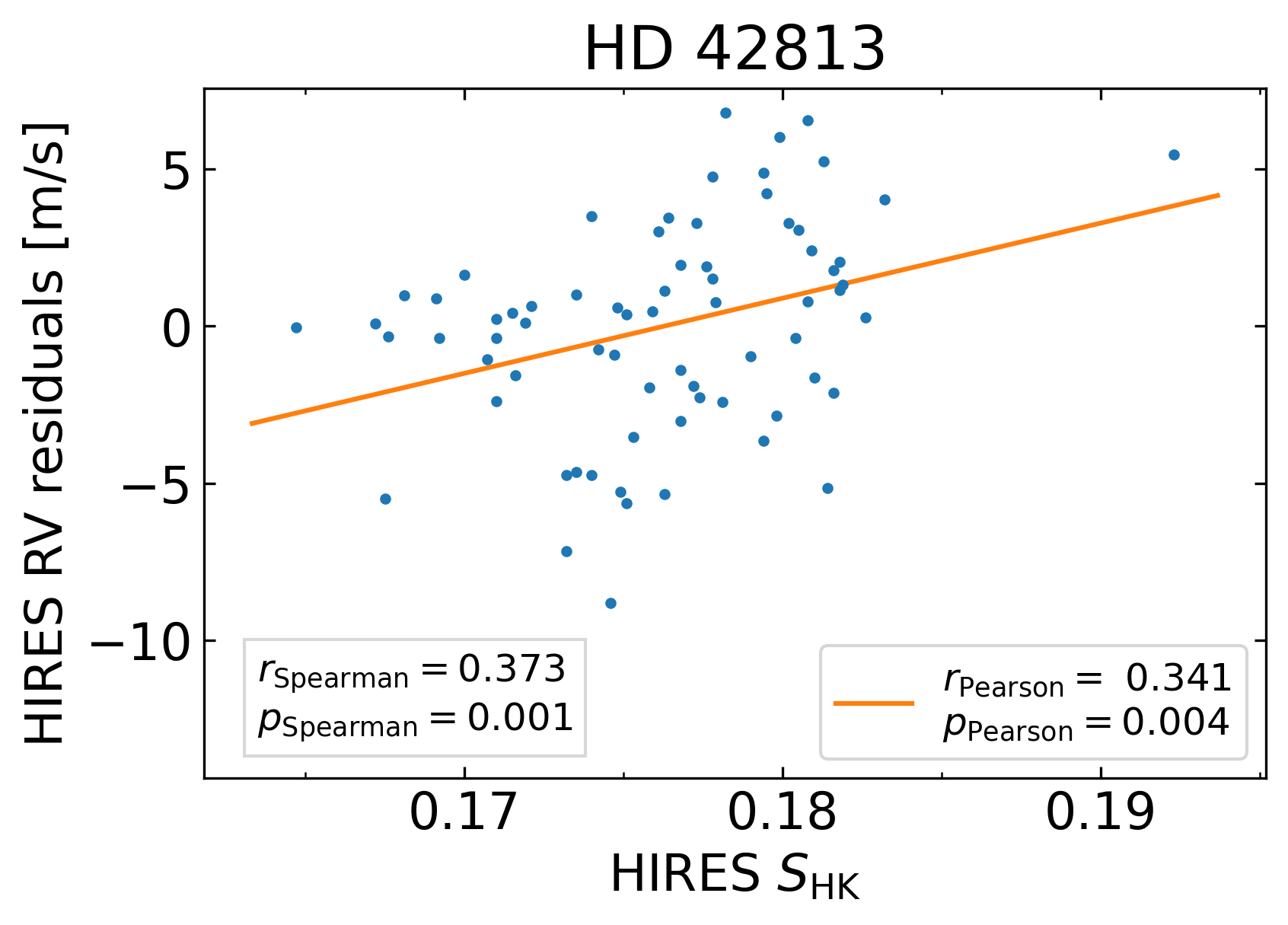}{0.5\textwidth}{}
          \fig{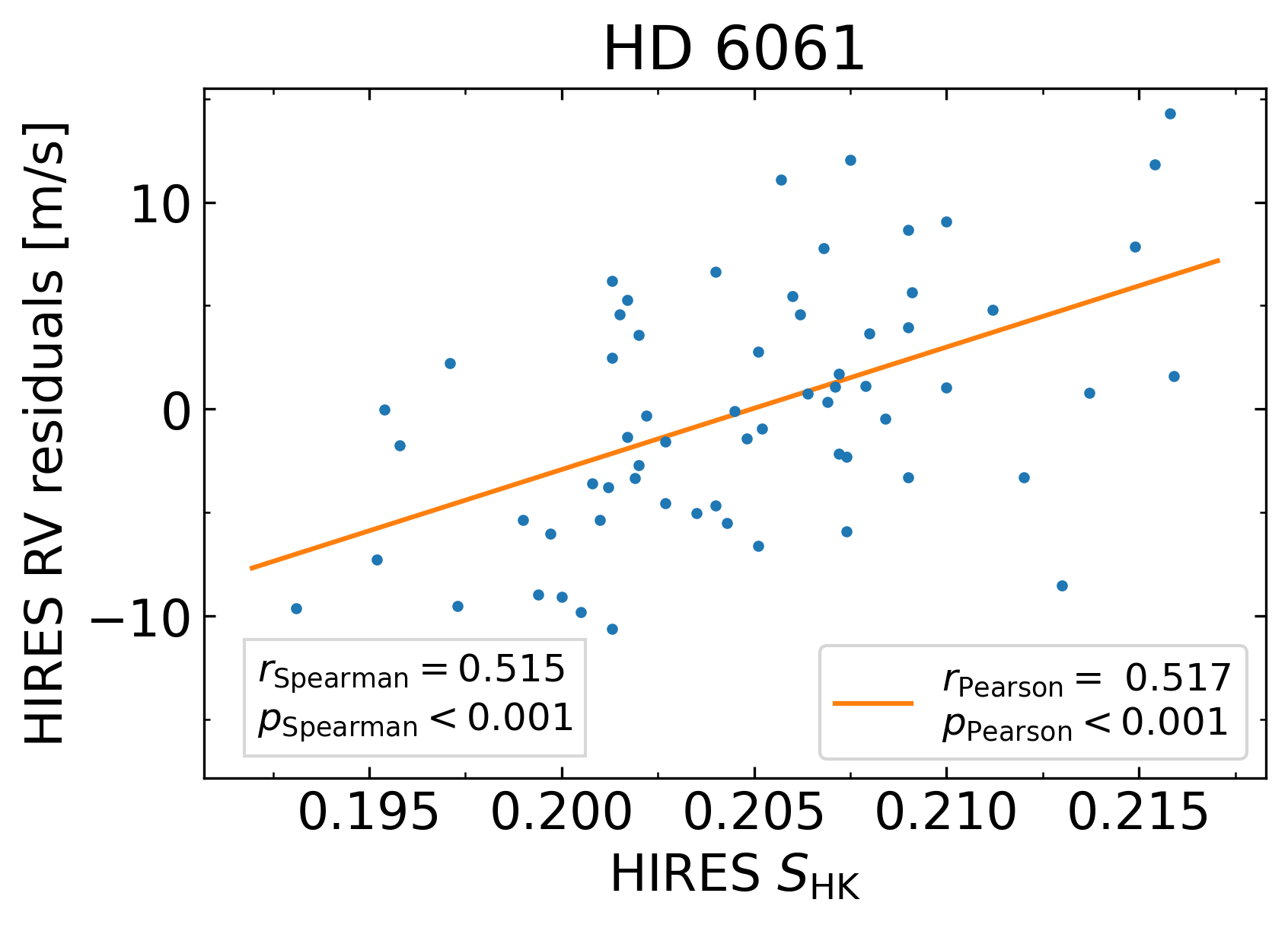}{0.5\textwidth}{}}
  \caption{\sysII (left) and \sysVII (right) are the only two systems where the $p$-value of either the Spearman or Pearson test (and in their cases, both) was $< 0.05$ for the \keckhires RV residuals and \shk values. In each panel the \keckhires RVs are shown in blue and a linear least-squares fit to the data is plotted in orange. We note that for \sysII, we removed the one observation with \shk $> 0.19$ and refit the data, but still find $p_\mathrm{Spearman}$ and $p_\mathrm{Pearson}$ are both $< 0.05$. The correlation between the RV residuals and the \shk values suggests that stellar activity might be manifesting itself in the RVs, but we emphasize that $p$-value testing can be unreliable (e.g., \citealt{colquhoun14}). This test is just one point of reference in our broader investigation into stellar activity contamination in the RVs.}\label{fig:svalue_correlation}
\end{figure*}

With planetary signals removed from the RV time series, we calculated both the Spearman rank-order correlation coefficient ($r_\mathrm{Spearman}$) and the Pearson correlation coefficient ($r_\mathrm{Pearson}$) for the \keckhires \shk values and the \keckhires RV residuals (e.g., \citealt{press92}). There were only two systems, \sysII and \sysVII, where the $p$-value for either the Spearman or Pearson test was $< 0.05$ (Figure \ref{fig:svalue_correlation}). These $p$-values may not be trustworthy given the relatively small sizes of the data sets ($N_\mathrm{HIRES} =$ \sysIIhiresjNbinnedRVs and \sysVIIhiresjNbinnedRVs for \sysII and \sysVII, respectively) and concerns regarding $p$-value testing in general (e.g., \citealt{colquhoun14}), but the apparent correlation between the RV residuals and \shk values in these systems spurred further investigation. According to Ca II H and K emission, \sysII is relatively inactive (\logrhk = \logrhkII), while \sysVII shows signs of moderate activity (\logrhk = \logrhkVII). For the other systems in our sample, while there appears to be no correlation between the RVs and the \keckhires \shk values, we still conducted a holistic examination of stellar activity.

\subsection{Periodogram analysis} \label{sec:periodogram_analysis}
Periodograms are a powerful tool for identifying periodic signals in time series data, though, as we saw with our search for nontransiting planets in \S\ref{sec:nt_search}, the physical interpretation of peaks in their power spectra should be treated with care. Caution should also be exercised when searching for signals across complementary data sets—\cite{kosiarek20} find that their periodogram and autocorrelation analyses of solar photometry correctly identify the solar rotation period less than half of the time. With these caveats in mind, we computed GLS periodograms to search for signs of periodicity related to stellar activity in the out-of-transit (OoT) PDCSAP \tess photometry and the \keckhires \shk values for each system. As we did for the periodograms of the RV residuals from \rvsearch, we also compared these to the periodogram of the RV window function to place purported signals in context with our imperfect time sampling. Periodograms for each system can be found in Appendix \ref{appendix:joint_models}. FAPs were calculated for these periodograms following \cite{baluev08}.

In general, we do not see an obvious stellar rotation period in the \tess PDCSAP (or SAP) photometry for any of our hosts. \sysVII is the only system whose PDCSAP light curve seems to exhibit rotational modulation by eye. The periodogram analysis from \cite{fetherolf22arxiv} finds that \sysVII's Sector 17 PDCSAP light curve is well fit by a single sinusoid with a period of $4.8 \pm 0.4$ d and an amplitude of about 0.2 ppt. This $P \approx 5$ d signal coincides with the strong peak we see in our own periodogram of the Sector 17 and 57 OoT PDCSAP photometry (Figure \ref{fig:hd6061_periodogram}). However, the activity-rotation relation from \cite{noyes84} suggests that the rotation period of this early-G dwarf should be closer to $P_\mathrm{rot} \approx 16$ d and the PDC algorithm is known to suppress stellar activity signals with $P \gtrsim 10$ d. A rotation period is not clear in \sysVII's SAP light curve, which seems to be heavily impacted by spacecraft and/or detector systematics (the light curve has sharp ramps and a low-frequency trend). In the end, perhaps the $P \approx 5$ d signal is a harmonic of the rotation period, or an artifact of the interplay between the PDC algorithm and the true astrophysical signal (if any).

Similar to the case for the \tess photometry, none of the systems exhibit a clear and obvious activity signal in the GLS periodogram of their \keckhires \shk values. For every system, the highest peak in the \shk periodogram is either the nightly alias or related to the yearly alias. We comment on other, seemingly inconsequential features of each system's \shk periodogram in \S\ref{sec:results}. 

\subsection{The case of \sysVII} \label{sec:sysVII_activity}
While most of the stars in our sample are relatively inactive (\logrhk $\lesssim -5.0$), \sysVII is the only host that would sit firmly among the ``active'' stars ($-5.0 <$ \logrhk $< -4.3$) in the activity-rotation analysis of \cite{mamajek08}. \sysVII is a G0 dwarf \citep{cannon93} with moderate Ca II H and K emission (\logrhk = \logrhkVII). The \sysVII \keckhires RV residuals and \shk measurements are strongly correlated, indicating that a stellar activity signal may be contaminating the RVs. What, then, is the star's rotation period and how does it compare to the orbital period of \sysVII b (\periodb $= 5.25$ d)?

As mentioned in \S\ref{sec:periodogram_analysis}, there appears to be some sort rotational modulation in \sysVII's \tess photometry (with $P \approx 5$ d), but its connection to the stellar rotation period is unclear. The activity-rotation relation from \cite{noyes84} suggests $P_\mathrm{rot} \approx 16$ d, so perhaps the signal in the \tess photometry is a harmonic of the true rotation period. Using \specMatchSynth, we find \vsini $=$ \vsiniVII km/s. After combining this with our stellar radius measurement (\rstar $=$ \rstarVII \rsun) and marginalizing over the inclination of the stellar spin axis, \sysVII's projected rotational velocity implies $P_\mathrm{rot} = 14^{+13}_{-9}$ d. While \sysVII's true rotation period remains uncertain, all of these clues suggest it is reasonable to expect that $P_\mathrm{rot}$ or its harmonics are in the neighborhood of the orbital period for planet b.

The GLS periodograms of the \sysVII observations do not point to a clear and obvious stellar rotation period, but they do appear to hint at unresolved signals. After accounting for instrumental offsets and removing the Keplerian signal of \sysVII b, there are several peaks in the GLS periodogram of the RV residuals. The highest peak is located at $P = 5.5$ d and rises above the 1\% FAP level. There are also peaks at the 10\% FAP level near $P = 7$, 10, and 16 d. In the GLS periodogram of the \keckhires \shk values, the highest peak between $P = 2$ d and $P = 100$ d (contributions from the window function dominate the power spectrum beyond this range) is a peak at $P = 13$ d that reaches the 10\% FAP level. If the stellar rotation period is somewhere between $P =$ 12--17 d, this would seem to agree with the \cite{noyes84} estimate, our \vsini measurement, and the peaks in the periodograms of the RV residuals and \shk values.

To summarize, for each system we explored the possibility of stellar activity contaminating the RV time series. We checked for a correlation between the \keckhires RV residuals and the \shk values. We also searched for signals in the GLS periodograms of the \tess photometry, \shk values, and RV residuals. \sysVII is the only system that seems to show an activity signal. Though the principal period of the activity signal is not entirely obvious, various estimates seem to suggest that the stellar rotation period is in the neighborhood of $P =$ 12--17 d. In \S\ref{sec:jointGP} we describe our formal approach for including a Gaussian process model of stellar activity in our joint model of the photometry, RVs, and \shk values.

\section{Joint photometry, radial velocity, and activity modeling} \label{sec:joint_model}
Here we describe our method for deriving planet properties. In short, we used a custom analysis pipeline based on the Python package \exoplanet \citep{exoplanet:foremanmackey18} to jointly model each system's photometry, radial velocities, and, if necessary, \keckhires \shk stellar activity indicators. We applied this framework homogeneously to each system in our sample. Our code and worked examples are publicly available online \citep{murphy23_7783386}.

We summarize our joint model of each system in Tables \ref{tab:joint_model} and \ref{tab:nt_shk_rv_gp_model}. All model parameters had relatively broad priors, save for the stellar mass and radius, whose informed Gaussian priors stemmed from our high-resolution spectroscopy and isochrone modeling (see \S\ref{sec:stellar}). The likelihood function of the joint model is the product of the likelihood of the transit model and the RV model and, if applicable, the \shk model, all of which assume Gaussian residuals.

\subsection{Transits} \label{sec:transit_modeling}
We parameterize the transit portion of the joint model in terms of $\ln P$, $T_\mathrm{c}$, $\ln R_\mathrm{p}/R_*$, $b$, and $\sqrt{e}\cos\omega$ and $\sqrt{e}\sin\omega$. As in our initial transit modeling, we use the quadratic limb darkening law from \cite{exoplanet:kipping13}. When modeling photometry alone, orbital eccentricity, argument of periastron, and impact parameter can be highly degenerate for low to moderate \snr transits \citep{petigura20}. This $e$-$\omega$-$b$ degeneracy can lead to multimodal MAP solutions and create regions of very high curvature on the posterior surface \citep[i.e., the dreaded ``funnel'' geometry known to plague hierarchical models;][]{neal03}. One of the main advantages that our joint model has over separate models of the photometry and RVs is that in most cases, the RVs are able to quickly rule out highly eccentric orbits for the transiting planets, thereby restricting the $e$-$\omega$-$b$ phase space and alleviating this degeneracy. 

We note that when \emph{both} the planet's transit and RV signals are low to moderate \snr, our joint model can still fall victim to the $e$-$\omega$-$b$ degeneracy because the RVs are not able to rule out cases of high $e$. For example, when fitting our joint model to \sysII, we found that a funnel would form at moderate impact parameter ($b \gtrsim 0.7$) and moderate eccentricity ($e \gtrsim 0.2$) because the planet's RV detection is not significant enough ($K/\sigma_K \approx 2.5$) to exclude orbits with large $e$ and small $K$. In this case, we fixed $e \equiv 0$, which removed the funnel and improved sampling reliability and performance. We encountered a similar situation for \sysIII. In any case, RV-only models of these systems show that the orbits are consistent with being circular (see \S\ref{sec:radvel}).

Assumptions of circular orbits, parameterizing with transit duration (which implies a stellar density), and importance sampling can be used to derive constraints on $e$ and $\omega$ when combined with a known stellar density from spectroscopy, for example. This strategy circumvents the $e$-$\omega$-$b$ degeneracy entirely \citep[e.g.,][]{macdougall21}. This is the approach we used for our initial photometric model (see \S\ref{sec:phot_cleaning_init_transits}) when removing outliers so as to avoid the $e$-$\omega$-$b$ degeneracy when fitting for the initial MAP solution. While the parameterization in \cite{macdougall21} offers a robust method of modeling the photometry alone in the presence of this degeneracy, we chose to fit a joint model for simplicity rather than fitting the photometry and RVs in series. As a sanity check, the posteriors of the transit parameters resulting from of our joint model were all 1$\sigma$ consistent with the corresponding MAP values we found from the photometry-only fit.

\begin{figure}
    \centering
    \includegraphics[width=\columnwidth]{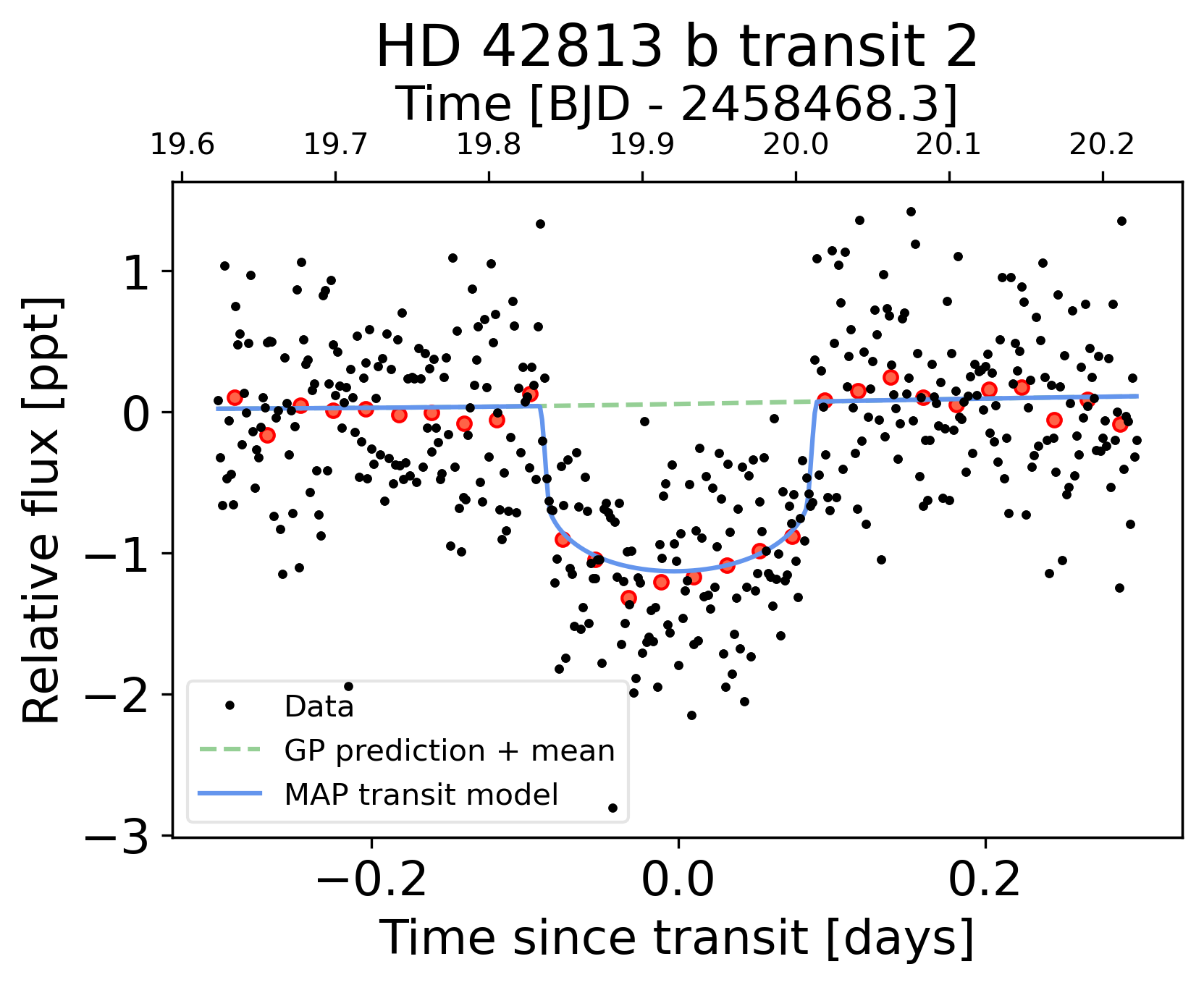}
    \caption{An example of our simultaneous transit and GP fitting for the second transit of \sysII b in Sector 6. The PDCSAP data are shown in black and data in 30-min bins are shown in red. The GP prediction across the transit (plus a small global offset fit to the data) is shown as the green dashed line, and the best-fitting transit model is shown as the blue line. We visually inspected each transit across all systems to ensure that the GP prediction did not absorb any of the transit signal.}
    \label{fig:transit_gp_example}
\end{figure}

As we did with our initial transit modeling (\S\ref{sec:phot_cleaning_init_transits}), we fit the transit model simultaneously with a GP using a kernel in the form of an overdamped SHO (Equation \ref{eqn:sho_kernel_RhoTauEta}) in order to flatten the light curve. To prevent the GP from absorbing part of the transit signal, we enforced that the GP's undamped period ($\rho$) and damping timescale ($\tau$) must both be $> 1$ d. For each system, we also visually inspected each transit to ensure that the GP's prediction was sufficiently smooth across the transit duration. Figure \ref{fig:transit_gp_example} illustrates the simultaneous transit and GP fitting for \sysII b's second transit in Sector 6. 
\subsection{Radial velocities} \label{sec:rv_modeling}
To describe the spectroscopic orbits of transiting planets, we used $\ln P$, $T_\mathrm{c}$, $\sqrt{e}\cos\omega$ and $\sqrt{e}\sin\omega$, and $\ln K$, where all but $\ln K$ were shared with the transit model. For each RV instrument we also included an offset ($\gamma$) and RV jitter term ($\sigma$), where the latter is added in quadrature with the pointwise RV measurement errors. As mentioned in \S\ref{sec:nt_search}, for each system we calculated the AIC of models that included or excluded a linear RV trend, $\dot{\gamma}$. \sysVI and \sysVIII are the only two systems where the AIC ruled out models without a trend. For these systems, we also tried adding a quadratic term to the background trend, but the AIC did not support including the curvature.

\sysVIII is the only system for which our adopted joint model includes a nontransiting planet. We treated the spectroscopic orbit of the nontransiting planet, \sysVIII c, in the same way as was done for transiting planets, save for the fact that we broadened the Gaussian priors on $\ln P$ and $T_\mathrm{c}$. The initial guesses for $\ln P$ and $T_\mathrm{c}$ were taken from our \rvsearch results for the system (see \S\ref{sec:nt_search}), and their priors were given a width of $\ln$ 50 d and 100 d, respectively. The parameters and priors for the RV signals of nontransiting companions in our joint model are found at the top of Table \ref{tab:nt_shk_rv_gp_model}.

\begin{deluxetable*}{lcccc}
\tablecaption{Joint model of the photometry and RVs \label{tab:joint_model}}
\tabletypesize{\footnotesize}
\tablehead{\colhead{Parameter} & \colhead{Symbol} & \colhead{Units} & \colhead{Prior} & \colhead{Notes}}
\startdata
\sidehead{\emph{Light curve parameters}}
Light curve mean offset & $\mu_\mathrm{phot}$ & ppt & $\mathcal{N}$(0, 10) & \\
Log photometric jitter & $\ln$ $\sigma_\mathrm{phot}$ & $\ln$ ppt & $\mathcal{N}$($\ln$ $s_\mathrm{phot}$, 2) & A\\
\sidehead{\emph{RV instrument parameters}}
Offset for RV instrument $i$ & $\gamma_i$ & \mps & $\mathcal{U}$[-250, 250] & \\ 
Log jitter for RV instrument $i$ & $\ln \sigma_{\mathrm{RV},\:i}$ & \mps & $\mathcal{N}$($\ln s_\mathrm{RV,\:i}$, 2)[$\ln 0.1$, $\ln 20$] & A \\
\sidehead{\emph{Stellar parameters}}
Limb-darkening parameter 1 & $q_1$ & & $\mathcal{U}$[0, 1] & B\\
Limb-darkening parameter 2 & $q_2$ & & $\mathcal{U}$[0, 1] & B\\
Stellar mass & \mstar & \msun & $\mathcal{N}$($M_*$, $\sigma_{M_*}$)[0, 3] & C \\
Stellar radius & \rstar & \rsun & $\mathcal{N}$($R_*$, $\sigma_{R_*}$)[0, 3] & C \\
\sidehead{\emph{Transiting planet parameters}}
Log orbital period & $\ln$ $P$ & $\ln$ d & $\mathcal{N}$($\ln$ $P_\mathrm{TOI}$, 1) & D \\
Time of inferior conjunction & $T_\mathrm{c}$ & d & $\mathcal{N}$($T_{\mathrm{c},\:\mathrm{TOI}}$, 1) & D \\
Log occultation fraction & $\ln$ $\frac{R_\mathrm{p}}{R_*}$ & & $\mathcal{N}$($\ln$ $\frac{R_\mathrm{p}}{R_*}_\mathrm{TOI}$, $\ln$ 10) & D \\
Impact parameter & $b$ & & $\mathcal{U}$[0, 1] & \\ 
$\sqrt{e}\cos({\omega})$ & $\xi_1$ & & $\mathcal{D}$($\xi_1$, $\xi_2$)[0, 1], VE($e | \bm{\theta}$) & E \\ 
$\sqrt{e}\sin({\omega})$ & $\xi_2$ & & $\mathcal{D}$($\xi_1$, $\xi_2$)[0, 1], VE($e | \bm{\theta}$) & E \\ 
Log RV semi-amplitude & $\ln$ $K$ & \mps & $\mathcal{N}$($\ln$ $s_\mathrm{RV}$, $\ln$ 50) & A \\
\sidehead{\emph{Light curve GP hyperparameters}}
Log GP amplitude & $\ln \eta_{\mathrm{phot}}$ & $\ln$ ppt & $\mathcal{N}$(0, 10) & F \\
Log GP undamped period & $\ln \rho_{\mathrm{phot}}$ & $\ln$ d & $\mathcal{N}$($\ln$ 10, $\ln$ 50)[$\ln$ 1, $\ln$ 200] & F \\
Log GP damping timescale & $\ln \tau_{\mathrm{phot}}$ & $\ln$ d & $\mathcal{N}$($\ln$ 10, $\ln$ 50)[$\ln$ 1, $\ln$ 200] & F \\
\enddata
\tablecomments{$\mathcal{N}$(X, Y) refers to a Gaussian distribution with mean X and standard deviation Y. $\mathcal{N}$(X, Y)[A, B] refers to a bounded Gaussian with mean X, standard deviation Y, and hard bounds at A and B. $\mathcal{U}$[X, Y] refers to a uniform distribution inclusive on the interval X and Y.\\
\textbf{A}: $\sigma_\mathrm{phot}$ is treated as a uniform pointwise flux measurement error. $s_\mathrm{phot}$ refers to the sample standard deviation of the PDCSAP light curve flux. $s_\mathrm{RV,\:i}$ refers to the same for the RVs of instrument $i$.\\
\textbf{B}: The parameterization $q_1 \equiv (u_1 + u_2)^2$ and $q_2 \equiv 0.5 u_1(u_1 + u_2)^{-1}$, where $u_1$ and $u_2$ are the usual quadratic limb darkening coefficients, follows the prescription by \cite{exoplanet:kipping13}. \\
\textbf{C}: The bounded Gaussian priors on stellar mass and radius have centers and widths corresponding to our derivation of the stellar parameters in \S\ref{sec:stellar}.\\
\textbf{D}: For some parameter, $x$, $x_\mathrm{TOI}$ refers to the value of that parameter as reported in the TOI catalog when accessed on 2022-Oct-04. The TOI catalog contains erroneous orbital properties for \sysI c and \sysVI c, but the correct orbital ephemerides are known from \cheops observations. \\
\textbf{E}: $\mathcal{D}$($\xi_1$, $\xi_2$)[0, 1] refers to a uniform distribution over the unit disk (i.e. $\sqrt{\xi_1^2 + \xi_2^2} \leq 1$). VE($e | \bm{\theta}$) refers to the mixture distribution from \cite{exoplanet:vaneylen19} which is used as a prior on $e$ and whose hyperparameters, $\bm{\theta}$, are fixed to the posterior medians from that work. \\
\textbf{F}: The hyperparameters of the GP used to flatten the light curve, which has a kernel whose power spectral density is in the form of a stochastic harmonic oscillator (SHO; see Equation \ref{eqn:sho_kernel}). $\rho_{\mathrm{phot}}$ and $\tau_{\mathrm{phot}}$, the undamped period and damping timescale of the SHO, respectively, are forced to be $> 1$ d to prevent the GP from overfitting low signal-to-noise transits. \\
}
\end{deluxetable*}

\begin{rotatetable*}\movetabledown=2cm 
\begin{deluxetable*}{lcccc}
\tablecaption{Additional parameters for nontransiting planets and joint GP modeling of the RVs and \keckhires \shk values \label{tab:nt_shk_rv_gp_model}}
\tabletypesize{\scriptsize}
\tablehead{\colhead{Parameter} & \colhead{Symbol} & \colhead{Units} & \colhead{Prior} & \colhead{Notes}}
\startdata
\sidehead{\emph{Nontransiting planet parameters}}
Log orbital period & $\ln$ $P$ & $\ln$ d & $\mathcal{N}$($\ln$ $P_\mathrm{NT}$, $\ln$ 50) & A \\
Time of inferior conjunction & $T_\mathrm{c}$ & d & $\mathcal{N}$($T_{\mathrm{c},\:\mathrm{NT}}$, 100) & A \\
$\sqrt{e}\cos({\omega})$ & $\xi_1$ & & $\mathcal{D}$($\xi_1$, $\xi_2$)[0, 1], VE($e | \bm{\theta}$) & B \\ 
$\sqrt{e}\sin({\omega})$ & $\xi_2$ & & $\mathcal{D}$($\xi_1$, $\xi_2$)[0, 1], VE($e | \bm{\theta}$) & B \\ 
Log RV semi-amplitude & $\ln$ $K$ & \mps & $\mathcal{N}$($\ln$ $s_\mathrm{RV}$, $\ln$ 50) & C \\
\sidehead{\emph{RV trends}}
Linear RV trend & $\dot{\gamma}$ & \mps/d & $\mathcal{N}$(0, 10) & \\
\sidehead{\emph{\shk instrument parameters}}
Offset for \keckhires \shk values & $\gamma_{S_\mathrm{HK},\:\mathrm{HIRES}}$ & & $\mathcal{U}$[0, 1] & \\ 
Log jitter for \keckhires \shk values & $\ln \sigma_{S_\mathrm{HK},\:\mathrm{HIRES}}$ & & $\mathcal{N}$($\ln s_{S_\mathrm{HK,\:HIRES}}$, 2) & C\\
\sidehead{\emph{Instrument-specific GP hyperparameters}}
GP amplitude of rotation term for \keckhires \shk values & $\eta_{\mathrm{rot},\:S_\mathrm{HK},\:\mathrm{HIRES}}$ & & Inv-$\Gamma$($\alpha=0.85$, $\beta=0.004$) & D\\
Log GP amplitude of exponential decay term for \keckhires \shk values & $\ln \eta_{\mathrm{dec},\:S_\mathrm{HK},\:\mathrm{HIRES}}$ & & $\mathcal{N}$(0, 10) & \\
Log GP amplitude of rotation term for RV instrument $i$ & $\ln \eta_{\mathrm{rot},\:\mathrm{RV},\:i}$ & $\ln$ \mps & $\mathcal{N}$(0, 10) & \\
Log GP amplitude of exponential decay term for RV instrument $i$ & $\ln \eta_{\mathrm{dec},\:\mathrm{RV},\:i}$ & $\ln$ \mps & $\mathcal{N}$(0, 10) & \\
\sidehead{\emph{RV and \shk shared hyperparameters for GP rotation term}}
Log GP rotation period & $\ln P_{\mathrm{rot}}$ & $\ln$ d & $\mathcal{N}$($\ln P_\mathrm{rot,\:guess}$, $\ln 1.5$)[$\ln 1$, $\ln 60$] & E \\
Log quality factor of secondary mode & $\ln Q_{0}$ &  & $\mathcal{N}$(0, 2) & \\
Log quality factor offset between primary and secondary modes & $\ln \delta Q$ &  & $\mathcal{N}$(0, 2) & \\
Fractional amplitude of secondary mode relative to primary mode & $f$ &  & $\mathcal{U}$[0.01, 1] & \\
\sidehead{\emph{RV and \shk shared hyperparameters for GP exponential decay term}}
Log undamped period of exponential decay term & $\ln \rho_{\mathrm{dec}}$ & $\ln$ d & $\mathcal{N}$($\ln$ 10, $\ln$ 50)[$\ln$ 1, $\ln$ 100] & \\
Quality factor of exponential decay term & $Q_{\mathrm{dec}}$ &  & $\equiv 1/\sqrt{2}$ & F \\
\enddata
\tablecomments{Notation in this table mirrors that in Table \ref{tab:joint_model}.\\
\textbf{A}: $P_\mathrm{NT}$ and $T_{\mathrm{c},\:\mathrm{NT}}$ refer to initial guesses for the values of the period and time of inferior conjunction, respectively, for a nontransiting planet's orbit. These were taken as the MAP values from our \rvsearch results. Each prior's width is such that we do not expect the initial guesses to bias the best-fitting nontransiting planet orbital parameters. \\
\textbf{B}: $\mathcal{D}$($\xi_1$, $\xi_2$)[0, 1] refers to a uniform distribution over the unit disk (i.e. $\sqrt{\xi_1^2 + \xi_2^2} \leq 1$). VE($e | \bm{\theta}$) refers to the mixture distribution from \cite{exoplanet:vaneylen19} which is used as a prior on $e$ and whose hyperparameters, $\bm{\theta}$, are fixed to the posterior medians from \cite{exoplanet:vaneylen19}. \\
\textbf{C}: $s_\mathrm{RV}$ refers to the sample standard deviation of the RV timeseries across all instruments. $s_{S_\mathrm{HK,\:HIRES}}$ is the sample standard deviation of the \keckhires \shk activity indices.\\
\textbf{D}: Inv-$\Gamma$ refers to the inverse Gamma distribution, the parameters of which have been chosen to define the tails of the distribution such that $p(x < 0.001$ $) < 0.01$ and $p(x > 1$ $) < 0.01$. This prior helps keep the amplitude of this GP component positive though with a lighter tail near zero as opposed to a Gamma distribution. \\
\textbf{E}: $P_\mathrm{rot,\:guess}$ is chosen using a periodogram analysis on a case-by-case basis. The default prescription is to assign the value to the period with the peak power in a GLS periodogram of the \tess photometry, though this can be superseded in the context of signals in the RV residuals and/or \keckhires \shk values. \\
\textbf{F}: Fixing $Q_{\mathrm{dec}} \equiv 1/\sqrt{2}$, gives this (overdamped) SHO the same power spectral density (PSD) as stellar granulation \citep{harvey85, kallinger14}. 
} 
\end{deluxetable*}
\end{rotatetable*}

\subsection{Gaussian process modeling of stellar activity} \label{sec:jointGP}
GPs are a popular tool for modeling correlated noise in RV data due to stellar activity (e.g., \citealt{robertson13, haywood14, grunblatt15, kosiarek20}). To further investigate contamination from stellar activity in our RV time series beyond our exploratory analysis in \S\ref{sec:stellar_activity_considerations}, we added a multidimensional GP to our joint model. We refer to this GP as multidimensional because it is fit to the RVs and \keckhires \shk values simultaneously. Each instrument (\apflevy RVs, \keckhires RVs, and \keckhires \shk) is assigned its own GP kernel which shares all hyperparameters with the other kernels, save for the GP amplitude (which we denote with $\eta$). In addition to the GP, the \keckhires \shk values are also fit with an offset and jitter term.

While \apflevy and \keckhires are both iodine-based RV instruments (i.e., they measure RVs in the same spectral region) we use different amplitude hyperparameters for their RV GP kernels. This is done as a catch-all to account for systematic differences in the manifestation of the stellar activity signal in their RV time series (e.g., perhaps those pertaining to their difference in spectral resolution, differences in long-term spectrograph stability, etc.). The practice of using independent GP amplitude hyperparameters for separate RV instruments that cover a similar spectral range is commonplace in the literature \citep[e.g.,][]{grunblatt15, kosiarek19, kosiarek21}.

The kernel of the multidimensional GP we used to model the stellar activity signal in the RVs and \keckhires \shk values is a mixture of three terms, each of which has a PSD in the form of an SHO (the kernel is the sum of \celeritetwo's \texttt{SHOTerm} and \texttt{RotationTerm}; \citealt{celerite2}). The first term is an overdamped oscillator, meant to describe exponentially-decaying behavior such as spot evolution, and is the same as the kernel of the GP used to flatten the light curve (Equation \ref{eqn:sho_kernel_RhoTauEta}). The only difference is that we fix the quality factor to $Q \equiv \frac{1}{\sqrt{2}}$ (which effectively sets the characteristic damping timescale, $\tau$), since this gives the SHO the same PSD as stellar granulation \citep{harvey85, kallinger14}. Plugging into and rearranging Equation \ref{eqn:sho_kernel_RhoTauEta}, for instrument $i$, we have
\begin{multline} \label{eqn:kernel_rv_shk_first_term}
    S_{\mathrm{dec},\:i}(\omega_f) = 16 \sqrt{2} \pi^{5/2} \frac{\eta^2_{\mathrm{dec},\:i}}{\rho^3}\Big[\big(\omega_f^2 - (\frac{2\pi}{\rho})^2\big)^2 \\ 
    + (\frac{2\sqrt{2} \pi \omega_f}{\rho})^2 \Big]^{-1}.
\end{multline}

The second and third terms of the kernel are underdamped SHOs, with fundamental frequencies corresponding to the stellar rotation period and its first harmonic, respectively. For instrument $i$, the PSDs of these terms can be written as
\begin{equation}
    S_{P_\mathrm{rot},\:i}(\omega_f) = \sqrt{\frac{2}{\pi}} \frac{S_{1,\:i}\: \omega_1^4}{(\omega_f^2 - \omega_1^2)^2 + \omega_1^2 \omega_f^2/Q_1^2}
\end{equation}
and 
\begin{equation}
    S_{P_\mathrm{rot}/2,\:i}(\omega_f) = \sqrt{\frac{2}{\pi}} \frac{S_{2,\:i}\: \omega_2^4}{(\omega_f^2 - \omega_2^2)^2 + \omega_2^2 \omega_f^2/Q_2^2}.
\end{equation}

The hyperparameters of $S_{P_\mathrm{rot},\:i}$ and $S_{P_\mathrm{rot}/2,\:i}$ are related via
\begin{align}
    Q_1 & = \frac{1}{2} + Q_0 + \delta Q \\ 
    \omega_1 & = \frac{4 \pi Q_1}{P_\mathrm{rot}\sqrt{4 Q_1^2 - 1}} \\
    S_{1,\:i} & = \frac{\eta^2_{\mathrm{rot},\:i}}{(1 + f) \omega_1 Q_1}
\end{align}
and
\begin{align}
    Q_2 & = \frac{1}{2} + Q_0 \\ 
    \omega_2 & = \frac{8 \pi Q_1}{P_\mathrm{rot}\sqrt{4 Q_1^2 - 1}} \\
    S_{2,\:i} & = \frac{f \eta^2_{\mathrm{rot},\:i}}{(1 + f) \omega_1 Q_1},
\end{align} 
where $\eta_{\mathrm{rot},\:i}$ is the amplitude of $S_{P_\mathrm{rot},\:i} + S_{P_\mathrm{rot}/2,\:i}$ relative to $S_{\mathrm{dec},\:i}$, $Q_0$ is the quality factor minus $1/2$ for the oscillator at $P_\mathrm{rot}/2$, $\delta Q$ is the difference between the quality factors of the oscillators at $P_\mathrm{rot}$ and $P_\mathrm{rot}/2$, $P_\mathrm{rot}$ is the primary period of variability (meant to represent the stellar rotation period), and $f$ is the fractional amplitude of the SHO at $P_\mathrm{rot}/2$ relative to the SHO at $P_\mathrm{rot}$.

Putting it all together, the PSD of the GP kernel for instrument $i$ is the sum of a term describing exponentially decaying behavior ($S_{\mathrm{dec},\:i}$) and a term describing periodic behavior ($S_{\mathrm{rot},\:i}$):
\begin{align}
    S_i(\omega_f) &= S_{\mathrm{dec},\:i}(\omega_f) + \big( S_{P_\mathrm{rot},\:i}(\omega_f) +  S_{P_\mathrm{rot}/2,\:i}(\omega_f)\big) \\
    &= S_{\mathrm{dec},\:i}(\omega_f) +  S_{\mathrm{rot},\:i}(\omega_f).
\end{align}

The GP parameters and priors are summarized in Table \ref{tab:nt_shk_rv_gp_model}. We experimented with variants of this kernel (e.g., removing the exponentially-decaying term, removing the first-harmonic term, removing the underdamped oscillators and adding a second overdamped oscillator), but found that this kernel was best at describing both the exponentially-decaying and periodic behavior.

\subsection{Posterior estimation} \label{sec:posterior_estimation}
We use No-U-Turn sampling \citep[NUTS;][]{nuts:hoffman14}, an adaptive form of Hamiltonian Monte Carlo \citep[HMC;][]{hmc:duane87, neal12} implemented with \exoplanet and \texttt{pymc3} \citep{pymc3}, to estimate the posterior distributions of the parameters in our joint model. HMC sampling uses the gradient of the posterior to help inform Markov transitions, enabling more efficient exploration of high-dimensional posterior surfaces than brute-force, guess-and-check methods like Metropolis-Hastings \citep{metropolis53, hastings70}.

For each system, a NUTS sampler ran 8 parallel chains with each chain taking at least 8000 ``tuning" steps before drawing 6000 samples. Samples drawn during the tuning period were discarded, similar to how various Markov Chain Monte Carlo (MCMC) methods discard burn-in samples. The chains were concatenated to produce a total of $N = 4.8 \times 10^4$ samples from the marginal posteriors of each model parameter. 

During the tuning stage the NUTS sampler optimizes hyperparameters, such as step size, to meet a targeted sample acceptance rate (in our case, 90\%) as it explores the posterior surface. This can help prevent the sampler from taking too large of a step while exploring a funnel on the posterior surface \citep{neal03}, where the gradient calculation can otherwise diverge and lead to biased inference \citep{betancourt13}. We found that the posterior geometries of the models of some systems were more prone to regions of high curvature than others, which prompted us to increase the number of tuning steps to prevent divergences (hence why each chain for each sampler took ``at least'' 8000 tuning steps).

In addition to being conscious of the number of tuning steps, for many joint model parameters that are strictly non-negative (e.g., planet orbital period, occultation fraction, RV semi-amplitude, etc.) we fit and explored the posterior of the natural logarithm of the parameter of interest rather than the parameter itself. We employed this parameterization because imposing hard bounds on the domains of model parameters can encourage the formation of a funnel. 

We assessed convergence of the HMC sampling through multiple diagnostic statistics. \cite{vehtari21} pointed out serious flaws with the standard Gelman-Rubin statistic ($\hat{R}$; \citealt{gelman92}), which is conventionally used to determine convergence for iterative stochastic algorithms like MCMC. Following their prescription, we instead assessed convergence by verifying a sufficiently small ($< 1.001$) \emph{rank-normalized} $\hat{R}$ for each model parameter. In brief, a rank-normalized $\hat{R}$ statistic is computed by calculating $\hat{R}$ on the normalized, rank-transformed chains of the parameter, rather than the values of the parameter itself. To ensure the chains could offer reliable confidence intervals, we also calculated the rank-normalized bulk and tail effective sample sizes from \cite{vehtari21} for each of the marginal posteriors (roughly, the effective sample sizes are the number of ``independent" samples obtained in the bulk and tails of the posterior). \cite{vehtari21} recommend that the effective sample size should be larger than 400 in both the bulk and the tails of the posterior. For every parameter we find the minimum between the bulk and tail effective sample sizes was comfortably larger than the recommended minimum threshold (typically we find $N_\mathrm{eff} \gtrsim 10^4$).

\section{Joint modeling results} \label{sec:results}
Here we describe the results of our stellar characterization, joint modeling, and posterior estimation for each system. In this section we include figures of the joint modeling results and periodograms for \sysI to inform their general format, but the rest can be found in Appendix \ref{appendix:joint_models}. Table \ref{tab:planet_prop_summary} contains a brief summary of the physical properties for all 12 planets. Tables of all measured and derived planet properties, as well as stellar properties, can be found for each system in Appendix \ref{appendix:joint_models}.
\begin{figure*}
\gridline{\fig{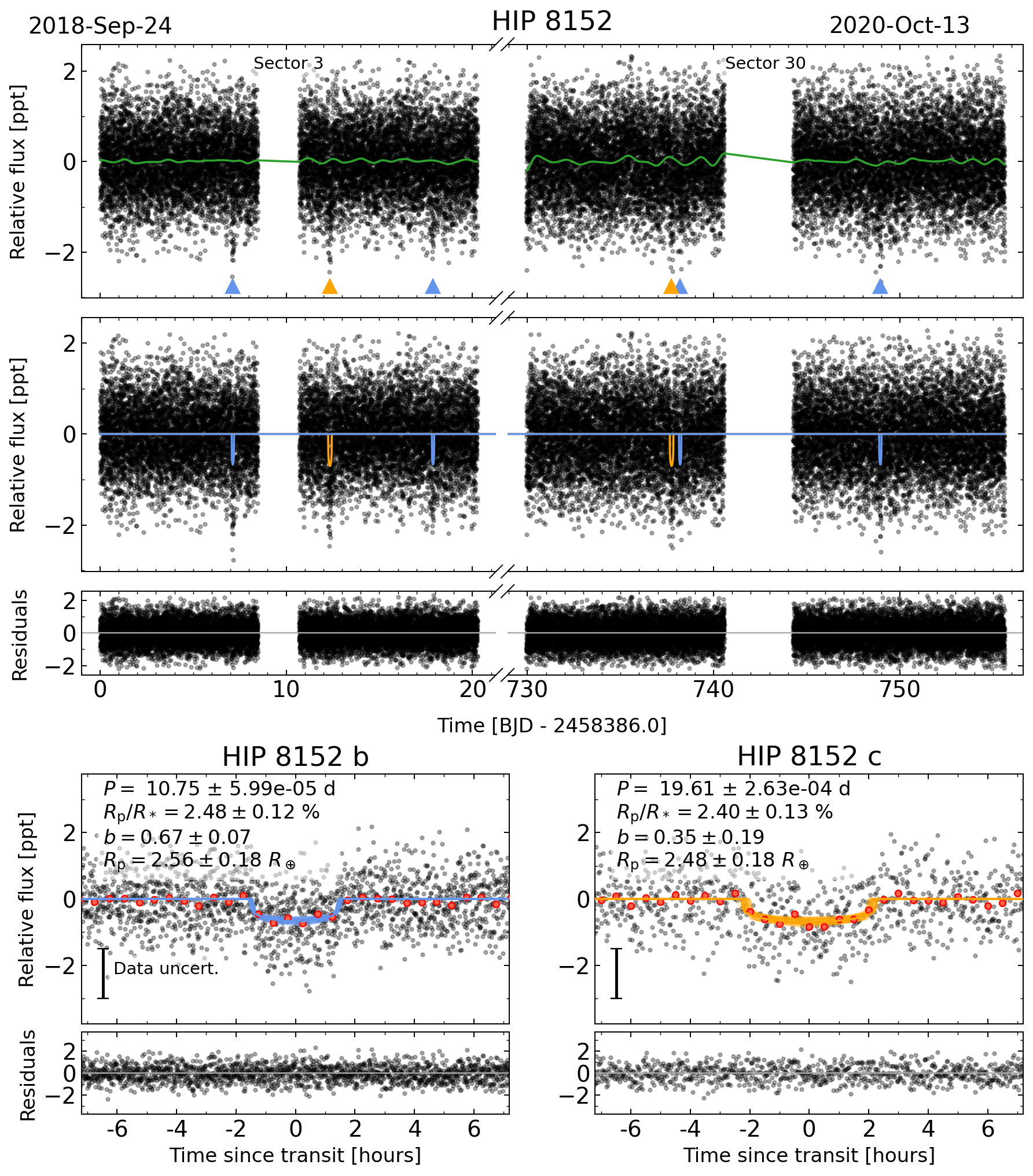}{0.5\textwidth}{}
          \fig{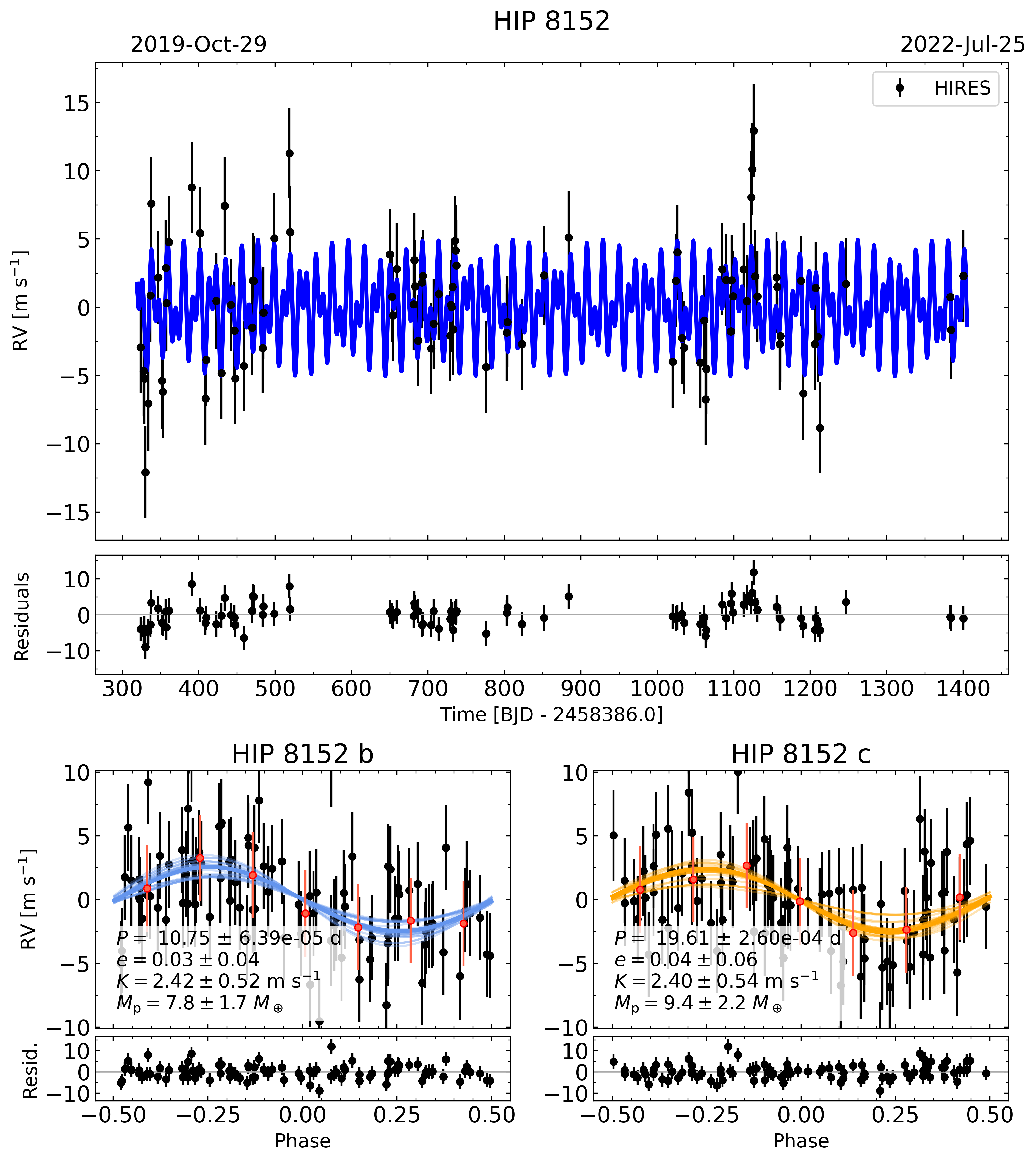}{0.5\textwidth}{}}
  \caption{Our joint modeling results for \sysI. Left: The photometric model. The top panel shows the PDCSAP light curve as the black points, with the GP used to flatten the light curve plotted as the green line. Triangles mark transits for planet b (blue) and c (orange). The middle panel shows the flattened light curve in black and the best-fitting transit models for planet b and c in blue and orange, respectively. Residuals are shown below. Phase-folded light curves and residuals are shown in the bottom panels for planets b (left) and c (right), with data in 30-min bins shown in red. The phase-folded best-fitting transit models for each planet are shown as the slightly thicker blue and orange lines, with 25 random posterior draws overplotted as the thinner lines. Right: The RV model. The top panel shows the RV time series with \keckhires data in black and the RV model in blue. Residuals are shown below. Phase-folded RV curves for planets b and c are shown in the bottom panels. Red points are data binned in 0.125 units of orbital phase. The phase-folded best-fitting RV models for each planet are shown as the slightly thicker blue and orange lines, with 25 random posterior draws overplotted as the thinner lines.}\label{fig:hip8152_phot_and_rvs}
\end{figure*}
\subsection{\sysI (\toiI)} \label{results:sysI}
\sysI is an inactive G dwarf for which we report the discovery of twin sub-Neptunes, \sysI b and \sysI c. Figure \ref{fig:hip8152_phot_and_rvs} summarizes our joint analysis and Table \ref{tab:hip8152_properties} summarizes the system properties.

As mentioned in \S\ref{sec:duo_transit_systems}, before \cheops follow-up, \sysI c constituted a duotransit scenario. \tess observed only two transits of \sysI c, one in Sector 3 and one in Sector 33, initially obfuscating the planet's true period. In 2022-May, a GLS periodogram analysis of the RVs with planet b removed showed a clear peak at $P = 19.6$ d. We built a joint model of the \tess photometry and \keckhires RVs with \monotools \citep{osborn22} which statistically ruled out all aliases other than the 19.6 d period. \cheops then confirmed the period of \sysI c by observing an additional transit in 2022-Aug (via private communication with the \cheops GTO team, point of contact H. Osborn).

\begin{figure*}
    \centering
    \includegraphics[width=\textwidth]{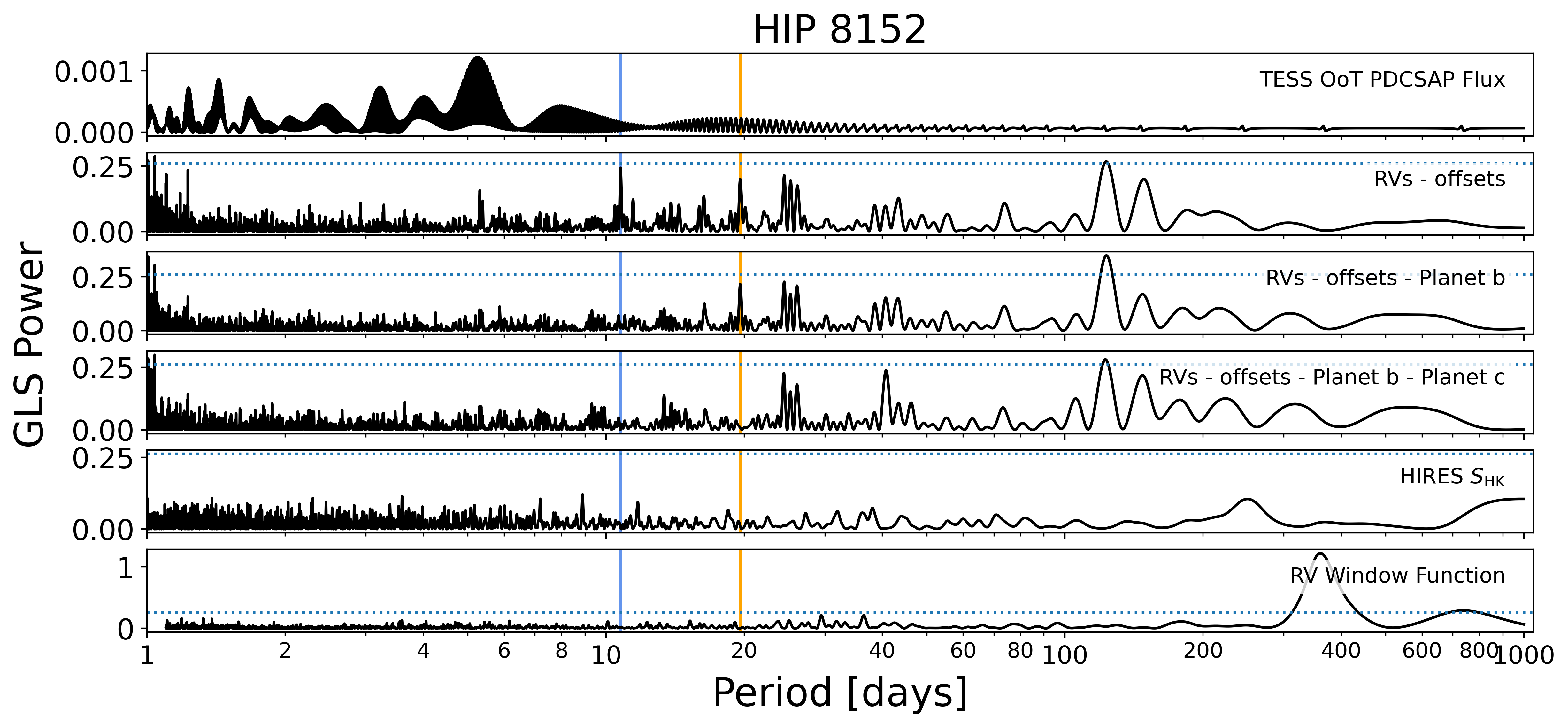}
    \caption{GLS periodograms for \sysI are shown in black. The vertical blue and orange lines mark the orbital periods of planet b and c, respectively. The dashed horizontal line indicates the 0.1\% FAP threshold \citep{baluev08}. We compute the GLS periodogram of the RV window function with a minimum period slightly longer than 1 d to avoid the strong nightly alias which otherwise skews the y-axis scale.}
    \label{fig:hip8152_periodogram}
\end{figure*}

We do not include a GP to model stellar activity in the RVs. \sysI is inactive according to Ca II H and K emission and the \keckhires RVs and \shk values are not correlated. Furthermore, there are no peaks rising above the 0.1\% FAP threshold in a GLS periodogram of the \keckhires \shk values. As discussed in \S\ref{sec:nt_search}, the only peak that rises above the 0.1\% FAP threshold in the RV residuals is near 120 d and is likely related to the RV window function. Figure \ref{fig:hip8152_periodogram} shows the GLS periodograms for the system.

\sysI b and c sit near the peak of the sub-Neptune distribution in the mass-radius plane and are on orbits slightly short of a 2:1 mean-motion resonance (MMR; \periodc/\periodb $= 1.8$). With nearly identical physical properties, these planets are attractive candidates for comparative studies in planet composition. Interestingly, \sysI b, though closer to the G dwarf host, is slightly less dense than \sysI c.
\subsection{\sysII (\toiII)} \label{results:sysII}
\sysII is an inactive, metal-rich K0V dwarf \citep{houk88}. We report the discovery of a transiting sub-Neptune in the system, \sysII b. Our adopted model assumes a circular orbit for \sysII b since posterior estimation for a model that included eccentricity was hindered by the planet's modest RV detection (see \S\ref{sec:transit_modeling}). In any case, a simpler, RV-only model that includes eccentricity finds that the planet's orbit is consistent with being circular (see \S\ref{sec:radvel}). Figure \ref{fig:hd42813_phot_and_rvs} summarizes our joint analysis and Table \ref{tab:hd42813_properties} summarizes the system properties. 

While the \keckhires RVs appear to be slightly correlated with the \keckhires \shk values (see Figure \ref{fig:svalue_correlation}, left), we decided to leave out a GP fit to the RVs and \shk values in our joint model. \sysII is seemingly inactive according to Ca II H and K emission to begin with. Furthermore, we do not find significant peaks in the GLS periodogram of the residuals of the Keplerian-only RV model. There are no peaks rising above the 0.1\% FAP threshold in the GLS periodogram of the \keckhires \shk values save for broad peaks that are consistent with the yearly alias and its harmonics. In addition, the GLS periodogram of the RV window function shows significant power at the yearly and lunar monthly aliases, making it difficult to disentangle a would-be activity signal in the RV residuals from artifacts of the RV time sampling (Figure \ref{fig:hd42813_periodogram}). We fit a joint model to the data that included a GP and recovered a mass measurement for \sysII b that was entirely consistent with the results of our Keplerian-only model. We therefore adopt a joint model of the \tess photometry and \keckhires RVs that does not include a GP fit to the RVs and \shk values.

With its seemingly low density, \sysII b is a very attractive candidate for atmospheric follow-up (TSM \tsmIIb at 98\% confidence). However, continued RV monitoring is first required to refine the planet's mass measurement. 

\subsection{\sysIII (\toiIII)} \label{results:sysIII}
As an F8 dwarf \citep{cannon93}, \sysIII has the earliest spectral type of all the hosts in our sample. We report the discovery of two transiting planets orbiting \sysIII: the sub-Neptune \sysIII b and the super-Earth \sysIII c. Similar to \sysII, we enforced circular orbits for planets b and c to avoid divergences during the HMC posterior estimation, though an RV-only model that included eccentricity found that the orbits were consistent with being circular (see \S\ref{sec:radvel}). Figure \ref{fig:hd25463_phot_and_rvs} summarizes our joint analysis and Table \ref{tab:hd25463_properties} summarizes the system properties.

\sysIII is seemingly inactive according to Ca II H and K emission (\logrhk $=$ \logrhkIII). However, we were prompted to explore a model that included a GP for activity mitigation because of the star's relatively rapid rotation. Using \specMatchSynth we find \vsini $=$ \vsiniIII km/s. When combining this with our stellar radius measurement and marginalizing over the inclination of the stellar spin axis, the star's projected rotation velocity implies $P_\mathrm{rot} = 4.3^{+1.7}_{-2.8}$ d. If this simplistic estimate is to be trusted, the stellar rotation period is very close to the orbital periods of the transiting planets (\periodb $= 7.0$ d and \periodc $= 3.0$ d). \cite{vanderburg16} highlight how stellar rotation can confuse the search for the Doppler signals of planets when $P_\mathrm{rot}$ and its first harmonic are in the neighborhood of the planets' orbital periods. Despite this concern, we do not find peaks above the 0.1\% FAP threshold in a GLS periodogram of the RVs after removing planets b and c. There are also no peaks that rise above the 0.1\% FAP threshold in a GLS periodogram of the \keckhires \shk values (see Figure \ref{fig:hd25463_periodogram}). Nevertheless, we added a GP to our joint model of the system with a Gaussian prior of $\mathcal{N}(4, 1.5)$ d on $P_\mathrm{rot}$, where 1.5 d is the Gaussian's standard deviation. The GP-enabled model finds best-fitting masses for planets b and c that are consistent with the posterior estimates of the non-GP model, but the HMC sampling had difficulty converging due to the large number of additional model parameters introduced by the GP kernel. We therefore adopt the non-GP model, whose HMC sampling does converge.

Given the brightness of the system ($V = 6.9$ mag, $J = 6.0$ mag), \sysIII is highly amenable to both ground- and space-based follow-up. With planets on opposite sides of the radius valley \citep{fulton17, vaneylen18}, \sysIII represents an opportunity for comparative studies in atmospheric mass loss. Though \sysIII is too bright for single object slitless spectroscopy (SOSS) with \jwst's Near Infrared Imager and Slitless Spectrograph (NIRISS; brightness limit of $J = 6.5$ mag), the system represents an attractive target for \hst. The mass measurement precision for the planets must be improved, however, before they are subjected to detailed atmospheric characterization \citep{batalha19}.

\subsection{\sysIV} \label{results:sysIV}
\sysIV is an inactive G dwarf for which we report the discovery of a hot transiting sub-Neptune, \sysIV b. Figure \ref{fig:toi669_phot_and_rvs} summarizes our joint analysis and Table \ref{tab:toi669_properties} summarizes the system properties.

We do not include a GP to model a stellar activity signal in the RVs and \shk values. \sysIV is inactive according to Ca II H and K emission, and the \keckhires RVs and \shk values are not correlated. Furthermore, there are no peaks rising above the 0.1\% FAP threshold in a GLS periodogram of the \keckhires \shk values or the \keckhires RV residuals (see Figure \ref{fig:toi669_periodogram}). 

A less significant peak near 9.6 d is visible in the periodogram of the RV residuals, but it is unclear whether the signal is planetary or related to the RV window function, which has significant power near 180 d and 25 d. If the $P = 9.6$ d signal is in fact a planet, assuming a circular orbit, $b > 1$ would imply an orbital inclination of \inclplanet $< 86.8\degree$. For reference, the orbit of \sysIV b has $i_\mathrm{b} =$ \inclIVb. A two-planet fit to the RVs using \radvel \citep{radvel} does not result in a significant detection of the $P = 9.6$ d signal (it finds $M_\mathrm{p} \sin i_\mathrm{p} = 5.0 \pm 2.6$ \mearth for a Keplerian at $P = 9.61 \pm 0.52$ d; the resulting mass of \sysIV b in this two-planet fit is consistent with our adopted joint model). The $\Delta$AIC between the one- and two-planet \radvel models is $< 1$, so there does not appear to be evidence for including the $P = 9.6$ d signal.

\sysIV is a relatively bright G dwarf ($J = 9.6$ mag) whose hot sub-Neptune (\teq $=$ \teqIVb K, \sincplanet $=$ \sincIVb \sincearth) lies just outside of the ``sub-Neptune desert" (planets with $2.2 < R_\mathrm{p} < 3.8$ \rearth and \sincplanet $> 650$ \sincearth; \citealt{lundkvist16}). \sysIV b's mass and radius measurements place it at the mode of the sub-Neptune mass-radius distribution.

\subsection{\sysV (\toiV)} \label{results:sysV}
\sysV is a K0 dwarf \citep{cannon93}. We report the discovery of a warm sub-Neptune in the system, \sysV b. Figure \ref{fig:hd135694_phot_and_rvs} summarizes our joint analysis and Table \ref{tab:hd135694_properties} summarizes the system properties. 

We do not include a GP to model stellar activity in the RV time series. \sysV's Ca II H and K emission (\logrhk $=$ \logrhkV) indicates that the star is relatively inactive and the \keckhires RV residuals and \shk values do not appear to be correlated. Although there are 14 sectors of \tess photometry available, a stellar rotation period is not readily apparent in either the PDCSAP or SAP light curve. There is a strong peak ($< 0.1\%$ FAP) in the RV residuals at 45.6 d (see Figure \ref{fig:hd135694_periodogram}), but we attribute this power to the RV window function given that 45.6 is a near-perfect divisor of 365.25. It is unclear why our mass measurement is so imprecise (about 2.7$\sigma$) when we have nearly 200 RV measurements between \apflevy and \keckhires. Below, we discuss scenarios that may be the cause of model misspecification.

After the $P = 45.6$ d peak, second-highest peak in the GLS periodogram of the RV residuals rises above the 10\% FAP threshold and is located at about $P = 32.5$ d. This period is not a clear harmonic of the yearly alias or the lunar monthly alias, but it is just about twice the period of \sysV b ($32.5 / 15.9 = 2.0$). The highest peak short of $P = 100$ d in the GLS periodogram of the \keckhires \shk values (beyond $P = 100$ d the GLS power is dominated by contributions from the RV window function) is at about $P = 31$ d and also rises above the 10\% FAP threshold. While the $P = 32.5$ d peak in the RV residuals is not overwhelmingly significant, it could represent either a nontransiting planet in a near 2:1 MMR with \sysV b, or the stellar rotation period. The latter explanation seems slightly more preferable given the $P \approx 31$ d peak in the GLS periodogram of the \keckhires \shk values. Furthermore, the activity-rotation relation from \cite{noyes84} suggests that \sysV has $P_\mathrm{rot} \approx 29$ d, making a rotation period of roughly 31--33 d seem reasonable for this K0 dwarf. 

For completeness, we added a GP component to our joint model following the methodology in \S\ref{sec:jointGP}. We placed a Gaussian prior on the GP rotation period at 32.5 d with a width of 1.5 d. The only other difference between this model and our adopted model of the photometry and RVs is that we forced \sysV b's orbit to be circular so as to prevent the HMC sampling from diverging (which it tended to do when allowing $e_\mathrm{b}$ and $\omega_\mathrm{b}$ to float). This GP-enabled model returned \mplanetb $= 6.1 \pm 2.1$ \mearth, in agreement with the results of our adopted model.

We also explored the idea that the signal near 32 d could be a nontransiting planet in a near 2:1 MMR with planet b. We fit a joint model where the GP in the model above was replaced with a nontransiting planet on a circular orbit. This model finds \mplanetb $= 6.6 \pm 2.1$ \mearth for \sysV b (which agrees with the results of our adopted, one-planet model) and $M_\mathrm{p} \sin i_\mathrm{p} = 10.6 \pm 2.6$ \mearth for the signal at $P = 32.5$ d. The AIC favors the one-planet plus GP model over the (adopted) one-planet model, which itself is favored over the two-planet model (all at the $\Delta$AIC $> 10$ level). It should be noted, however, that GPs can be susceptible to overfitting \citep[e.g.,][]{blunt23arxiv}, which can muddle the interpretability of Bayesian model comparison statistics. In this context, the AIC's preference for the GP-enabled model is not entirely surprising. Setting aside the AIC comparison, if there was a nontransiting planet at 32.5 d, assuming a circular orbit, $b > 1$ for the planet would imply \inclplanet $< 88.6\degree$. For reference, we find that planet b's orbit has $i_\mathrm{b} =$ \inclVb. Simulations similar to those conducted by \cite{lubin22} could be used to place a lower limit on the inclination of the potential nontransiting planet, but these are beyond the scope of this work. 

In summary, we cannot rule out the possibility that the $P = 32.5$ d signal represents either the stellar rotation period or a nontransiting planet. However, the signal's ambiguity, combined with the star's lack of Ca II H and K emission, encouraged us to adopt a model of the photometry and RVs that does not use a GP for stellar activity mitigation and does not include nontransiting planets. In any case, our experimentation with various models of the data reassures us that the mass measurement of planet b is seemingly insensitive to our choice of model. \sysV is a bright ($V = 9.1$ mag, $J = 7.9$ mag) K0 dwarf that adds another planet to the mode of the sub-Neptune mass-radius distribution. Continued Doppler monitoring is required to refine the planet's mass measurement precision.

\subsection{\sysVI (\toiVI)} \label{results:sysVI}
\subsubsection{Joint analysis from this work}\label{results:sysVI_this_work}
\sysVI is a G5 dwarf \citep{cannon93}. The system is host to two warm sub-Neptunes, \sysVI b and \sysVI c. We robustly detect a linear RV trend, indicating that there is also a distant, massive companion in the system. Figure \ref{fig:hip9618_phot_and_rvs} summarizes our joint analysis and Table \ref{tab:hip9618_properties} summarizes the system properties.

We exclude a GP fit to the RVs and \keckhires \shk values from our adopted model. \sysVI is nominally inactive based on its Ca II H and K emission levels (\logrhk $=$ \logrhkVI) and the \keckhires RV residuals and \shk values are not correlated. Stellar activity does not seem to be a concern for \sysVI according to our GLS periodograms (Figure \ref{fig:hip9618_periodogram}). There are no peaks that rise above the 0.1\% FAP threshold in the RV residuals of our joint model. In the GLS periodogram of the \keckhires \shk values, there is a peak just long of 30 d that rises above the 0.1\% FAP level, but it is unclear whether or not the power is related to the window function---the periodogram also shows significant power near 180 d and at $P > 700$ d. Even if the signal near $P = 30$ d in the \shk values is related to the stellar rotation period for this late G dwarf, the lack of power in the periodogram of the RV residuals indicates that activity is not greatly impacting the planet mass measurements.

The nature of the massive companion causing the linear RV trend is uncertain. We find that the companion must have $M \sin i_\mathrm{p} \gtrsim 4.7$ $M_\mathrm{Jup}$ and $a \gtrsim 5.0$ AU by making the following simplifying assumptions: (1) the RV trend is caused by a single companion, (2) the companion's orbit is circular, (3) the companion's orbital period is greater than four times our \apflevy and \keckhires RV baseline (about 1042 d), and (4) the RV semi-amplitude of the companion's orbit is greater than the $\Delta$RV caused by the trend over the baseline (about 62 \mps). A more detailed investigation is beyond the scope of this work---in \S\ref{results:sysVI_comparison} we summarize the results from \oXXIII, who conduct a thorough analysis of the trend that includes constraints from RVs, astrometry, and direct imaging. Continued RV monitoring is required to reveal the true nature of the distant companion.

\sysVI is perhaps the most exciting system in our sample for atmospheric follow-up with \jwst. As noted by \oXXIII, \sysVI is one of only five multi-transiting systems with $K_s < 8$ mag to host a planet with $P > 50$ d. Our mass constraints translate to TSM values of \tsmVIb and \tsmVIc for planet b and c, respectively (with the lower limit for planet c reflecting 98\% confidence). These values place both planets above the \cite{kempton18} TSM cutoff ($>84$) for follow-up of planets with $2.75 <$ \rplanet $< 4$ \rearth. Moreover, \sysVI b's TSM estimate places it in the top quartile of all planets in its radius range (top quartile cutoff of TSM $>146$). 

\begin{figure*}
    \centering
    \includegraphics[width=0.6\textwidth]{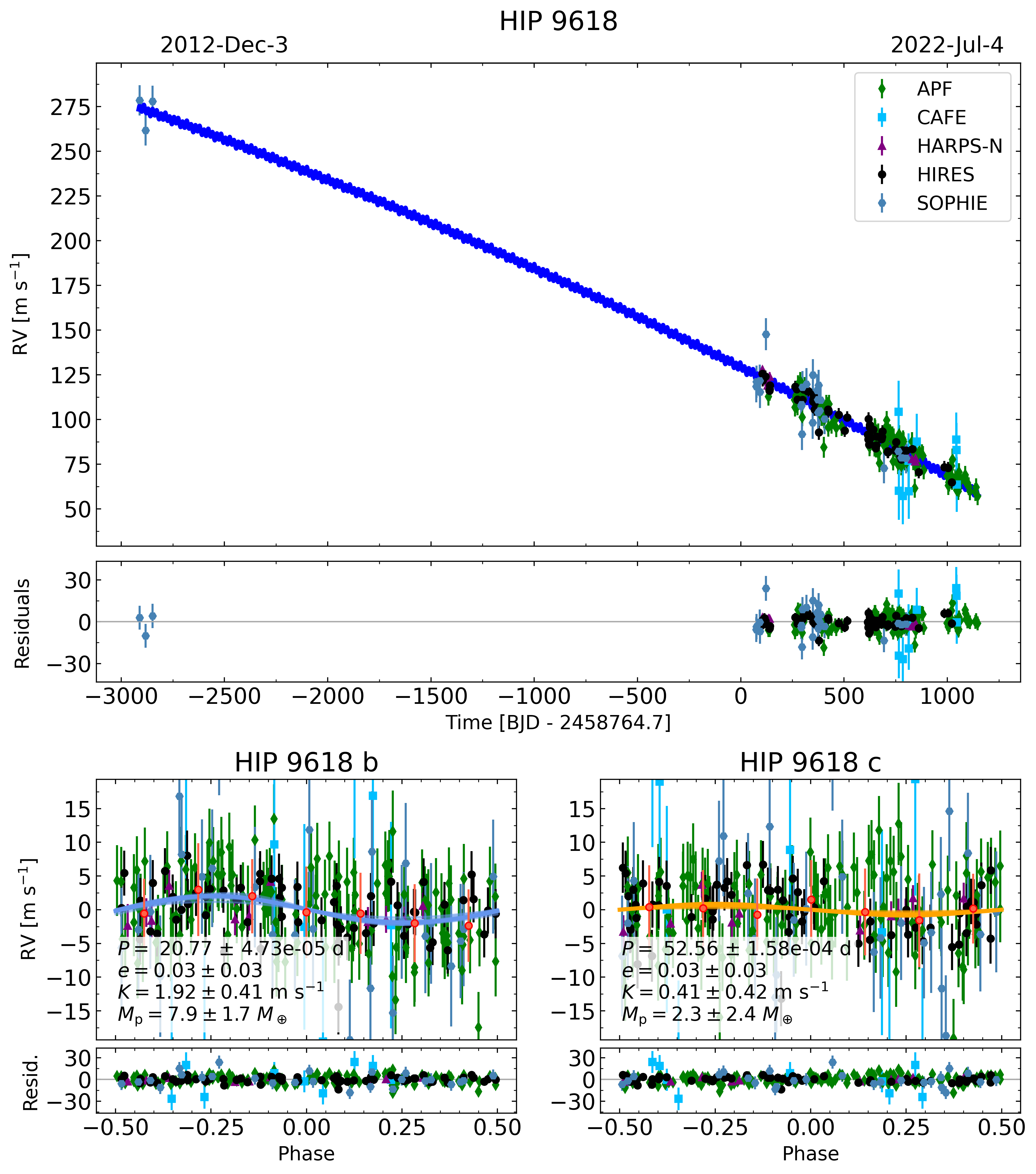}
    \caption{The RV portion of a joint model for \sysVI that includes the CAFE, HARPS-N, and SOPHIE RVs from \oXXIII in addition to the \apflevy and \keckhires data from this work. The figure description is the same as for Figure \ref{fig:hip8152_phot_and_rvs}. This RV model also includes a curvature term in addition to the linear trend, in order to mimic the adopted solution from \oXXIII. The results of this model are consistent with our joint model of the \apflevy and \keckhires RVs alone.}
    \label{fig:hip9618_rvs_all}
\end{figure*}

\subsubsection{Comparison with the \oXXIII results} \label{results:sysVI_comparison}
\oXXIII first reported the discovery and confirmation of \sysVI b and c using space-based photometry from \tess and \cheops in combination with a total of 49 RVs from CAFE, HARPS-N, and SOPHIE. The authors report masses of \mplanetb $= 10.0 \pm 3.1$ \mearth and \mplanetc $< 18$ \mearth (at 3$\sigma$ confidence). 

The only significant difference between the RV model presented in this work and the adopted model from \oXXIII is that, in addition to including a linear trend in their model of the RVs, \oXXIII also include a quadratic term ($\ddot{\gamma}$). \oXXIII find $\dot{\gamma} = -0.067 \pm 0.001$ m/s/d and $\ddot{\gamma} = -4.99 \times 10^{-6} \pm 6.4 \times 10^{-7}$ m/s/d$^2$. \oXXIII use \orvara \citep{orvara} to translate their reported RV trend and curvature, the lack of an astrometric detection with \hipparcos \citep{hipparcos} and \gaia, constraints from HRI, the lack of secondary lines in their high-resolution spectra of \sysVI, and the assumed stability of the inner transiting planet system into orbital separation and mass ratio posteriors for a distant, massive, single companion. \oXXIII suggest the companion is either a brown dwarf or low-mass M dwarf with $0.08^{+0.12}_{-0.05}$ \msun in an orbit at $26^{+19}_{-11}$ AU.

The curvature reported by \oXXIII is driven by just three SOPHIE RVs acquired between 2011-Oct and 2011-Dec---the next SOPHIE RV (which also happens to be the next RV from any of their three RV instruments) was taken eight years later in 2019-Dec. If we exclude the three SOPHIE RVs collected in 2011 from our RV analysis (either with our joint model or with \radvel), then the curvature detection disappears. \cite{sophie} quote SOPHIE's RV stability as being $\approx3$ \mps over several months. To explore the possibility that the purported curvature is in fact due to the instrument's RV zero-point drift over the eight-year ($\approx100$-month) gap between observations, we refit all of the available RVs in \radvel, included a linear RV trend (but no curvature), and treated the 2011 SOPHIE RVs as coming from their own instrument (i.e., we assigned them their own RV offset and jitter). This model finds a linear RV trend consistent with that of our adopted joint model of the \apflevy and \keckhires RVs. To enable this consistency, the 2011 SOPHIE RVs require about a 30 \mps offset. Our \apflevy and \keckhires RVs lack the baseline to independently confirm the RV curvature reported by \oXXIII. Additional, long-term monitoring is required to fully characterize the distant companion.

Finally, we added all of the RV data from \oXXIII to a joint model of the \tess photometry and our \apflevy and \keckhires RVs. For better comparison with \oXXIII, we include the 2011 SOPHIE RVs and RV curvature in addition to the linear trend. We include RV jitter terms and offsets for each instrument and fit the same jitter and offset term to both the pre- and post-2012 SOPHIE data. Our joint model of the combined data set finds \mplanetb $= 7.9 \pm 1.8$ \mearth and \mplanetc $< 8.0$ \mearth at 98\% confidence, which is consistent with our adopted joint model of the \apflevy and \keckhires RVs. The RV portion of this joint model can be seen in Figure \ref{fig:hip9618_rvs_all}. While we have discussed the results of modeling all of the available RV data for the sake of completeness, in the interest of homogeneity, the values found in Tables \ref{tab:planet_prop_summary} and \ref{tab:hip9618_properties} stem from our joint model that includes only the \apflevy and \keckhires RVs.

\subsection{\sysVII (\toiVII)} \label{results:sysVII}
\begin{figure}
    \centering
    \includegraphics[width=\columnwidth]{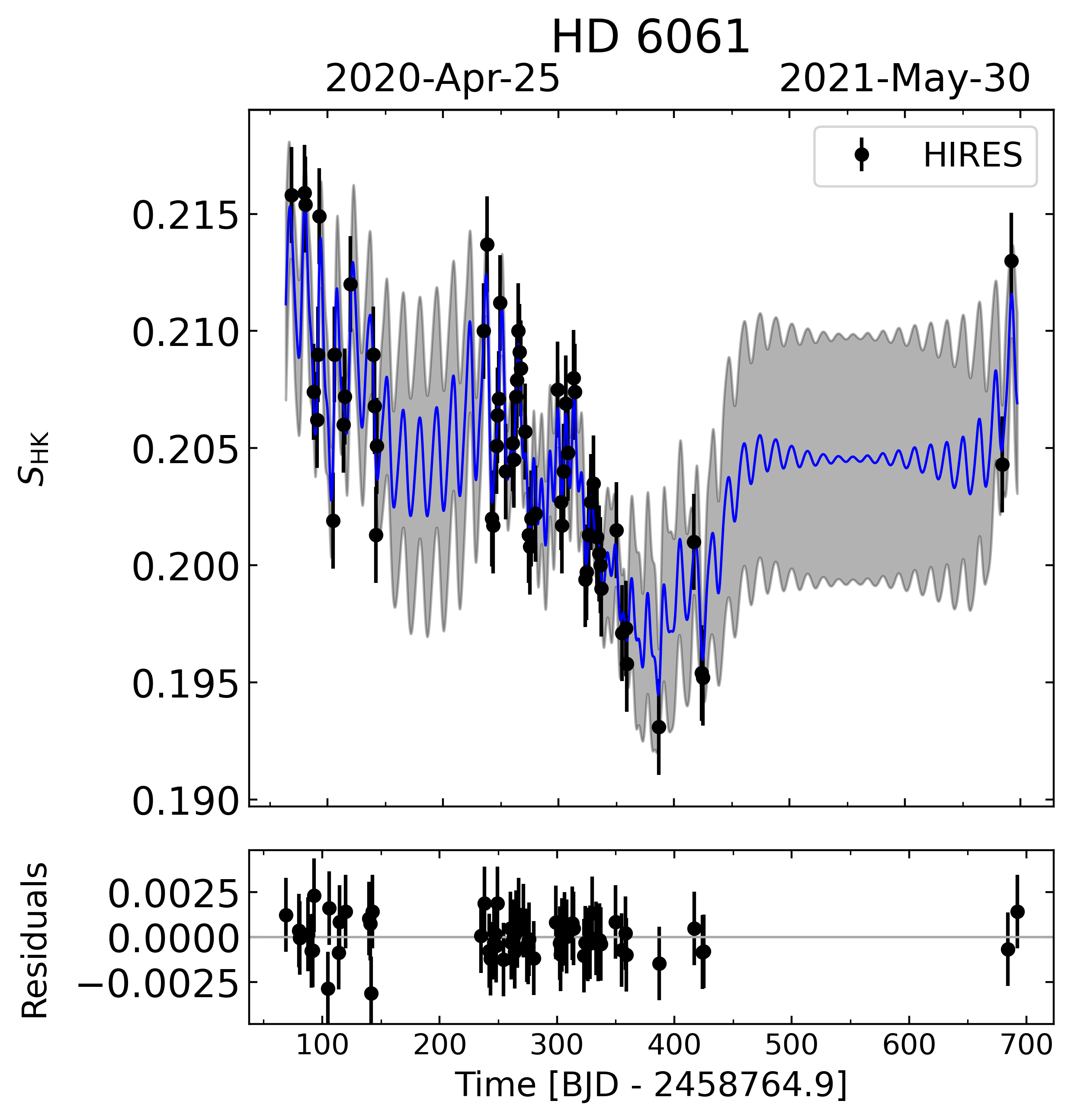}
    \caption{The \shk portion of the joint model for \sysVII, in which GPs with some shared hyperparameters are simultaneously fit to the RVs and the \keckhires \shk values. The GP kernel is described in \S\ref{sec:jointGP}. The best-fitting value of the GP period is $P_\mathrm{rot} = 13.9 \pm 1.5$ d. See Figure \ref{fig:hd6061_phot_and_rvs} for the photometry and RV portions of this joint model. The GP posterior prediction is shown as the blue line surrounded by a 1$\sigma$ error envelope. The \keckhires \shk values are shown as the black points with 1$\sigma$ errors, where the measurement error on each \shk value has been added in quadrature with the a jitter term that has been fit to the data. Residuals are shown in the bottom panel.}
    \label{fig:hd6061_svalues}
\end{figure}

\sysVII is a moderately active G0 dwarf \citep{cannon93} for which we find an M4/5V dwarf companion, \sysVIIcomp, that is almost certainly gravitationally bound (see \S\ref{sec:ao_sysVIIcomp}). We also report the discovery of the hot sub-Neptune, \sysVII b. Figure \ref{fig:hd6061_phot_and_rvs} summarizes our joint analysis and Table \ref{tab:hd6061_properties} summarizes the system properties. 

Following the methodology in \S\ref{sec:jointGP}, our adopted model includes a GP to address stellar activity. Since we include a GP in our model of the RVs, we assume a circular orbit for \sysVII b to reduce model complexity and improve the performance of the HMC sampling. For a model that included both the GP and orbital eccentricity for \sysVII b, the NUTS sampler suffered from divergences after tuning and failed to converge. For completeness, we fit a non-GP model of the system that included eccentricity and found that \sysVII b's orbit is consistent with being circular.

\sysVII's Ca II H and K emission suggests that it is moderately active (\logrhk $=$ \logrhkVII). We also find that the \keckhires RV residuals and \shk values are strongly correlated (Figure \ref{fig:svalue_correlation}, right). As discussed in \S\ref{sec:sysVII_activity}, various lines of inquiry suggest that the stellar rotation period is in the neighborhood of $P_\mathrm{rot} \approx$ 12--17 d. For our adopted, GP-enabled joint model, we placed a Gaussian prior on $P_\mathrm{rot}$ in the middle of this range, at $\mathcal{N}(14.5, 1.5)$ d. Figure \ref{fig:hd6061_svalues} shows the GP model of the \keckhires \shk values, which is fit simultaneously with the GP of the RVs. After removing the GP and the orbit of planet b, the GLS periodogram of the RV residuals contains no peaks rising above the 0.1\% FAP threshold (see Figure \ref{fig:hd6061_periodogram}). Our posterior estimation finds $P_\mathrm{rot} = 13.9 \pm 1.5$ d.

For completeness, we explored alternative models of the \sysVII observations to check for model overfitting, which can plague GP-based planet mass measurements \citep[e.g.,][]{blunt23arxiv}. A joint model of the photometry and RVs that did not include a GP and did not assume a circular orbit for \sysVII b finds \mplanetb $= 7.8 \pm 2.9$ \mearth and that $e_\mathrm{b}$ is consistent with zero. The mass measurement from this Keplerian-only model is nearly 1$\sigma$ consistent with the planet mass measurement from our adopted model. Perhaps the slightly lower mass measurement from the Keplerian-only model suggests that our adopted, GP-enabled model is overfitting slightly, but not egregiously so. In \S\ref{sec:radvelGP} we explore a model of the RVs that uses a different GP kernel and find a planet mass measurement that is consistent with our adopted model.

\sysVII is a bright ($V = 8.8$ mag, $J = 7.7$ mag) G dwarf with a close-in sub-Neptune planet that lands near the mode of the sub-Neptune mass-radius distribution. Continued RV monitoring is required to refine the planet mass measurement and better understand the stellar activity signal.

\subsection{\sysVIII} \label{results:sysVIII}
\sysVIII is a subgiant star for which we report the discovery of a transiting sub-Neptune, \sysVIII b, and a temperate super-Jovian-mass planet on a moderately eccentric orbit, \sysVIII c. We robustly detect a linear trend in the RVs, indicating that there is also a distant, massive companion in the system. Figure \ref{fig:toi1736_phot_and_rvs} summarizes our joint analysis and Table \ref{tab:toi1736_properties} summarizes the system properties.

We do not include a GP in our joint model. \sysVIII is seemingly inactive according to Ca II H and K emission (\logrhk $=$ \logrhkVIII) and the \keckhires RVs and \shk values do not appear to be correlated. Furthermore, there are no peaks in a GLS periodogram of the RV residuals that rise above the 0.1\% FAP threshold. We note a peak in the RV residuals near 55 d that rises above the 1\% FAP level, but it is unclear whether this signal is planetary, activity-related, or related to the RV window function.

The nature of the massive companion causing the linear RV trend is uncertain. Over our observing baseline of 909 d, the linear RV trend causes a $\Delta$RV of about 166 \mps. Using the same set of crude assumptions as we did for the case of \sysVI, we find that the companion must have $M \sin i_\mathrm{p} \gtrsim 12.9$ $M_\mathrm{Jup}$ and $a \gtrsim 4.7$ AU, tentatively suggesting that it is too massive to be a planet. Relaxing the assumption that the companion is on a circular orbit, with $a = 4.7$ AU it must have $e \lesssim 0.6$ so as not to cross the orbit of \sysVIII c, which implies a minimum mass limit of $M \sin i_\mathrm{p} \gtrsim 10.3$ $M_\mathrm{Jup}$. Dynamical simulations would better inform the allowed values for the orbital eccentricity of the massive companion to ensure stability, but these are beyond the scope of this work. In the end, continued RV monitoring is required to reveal the true nature of this companion.

\sysVIII is the only star of our eight systems that is slightly evolved, and the only system for which we detect the full orbit of a massive, presumably nontransiting planet. The system's architecture is intriguing: a transiting sub-Neptune interior to a temperate super-Jovian and a massive, potentially non-planetary companion. Given the system's architecture, evolutionary state, and precise physical properties, \sysVIII b (TSM $=$ \tsmVIIIb) represents an attractive opportunity for atmospheric observations with \jwst.

\begin{deluxetable*}{r c c c c c c}
\tablecaption{Summary of derived planet properties from our joint analysis \label{tab:planet_prop_summary}}
\tabletypesize{\normalsize}
\tablehead{
    \colhead{Planet name} & \colhead{$P$} & \colhead{\transitTime} & \colhead{$e$} & \colhead{\rplanet}  & \colhead{\mplanet}  & \colhead{\teq} \\ 
    \colhead{}            & \colhead{(d)} & \colhead{(BTJD)}       & \colhead{}    & \colhead{(\rearth)} & \colhead{(\mearth)} & \colhead{(K)}
}
\startdata
\sysI b    & \periodIb            & \tcBTJDIb    & \eccIb            & \rpIb    & \mpIb                         & \teqIb \\
\sysI c    & \periodIc            & \tcBTJDIc    & \eccIc            & \rpIc    & \mpIc                         & \teqIc \\
\sysII b   & \periodIIb           & \tcBTJDIIb   & \eccIIb           & \rpIIb   & \mpIIb                        & \teqIIb \\
\sysIII b  & \periodIIIb          & \tcBTJDIIIb  & \eccIIIb          & \rpIIIb  & \mpIIIb                       & \teqIIIb \\
\sysIII c  & \periodIIIc          & \tcBTJDIIIc  & \eccIIIc          & \rpIIIc  & \mpIIIc                       & \teqIIIc \\
\sysIV b   & \periodIVb           & \tcBTJDIVb   & \eccIVb           & \rpIVb   & \mpIVb                        & \teqIVb \\
\sysV b    & \periodVb            & \tcBTJDVb    & \eccVb            & \rpVb    & \mpVb                         & \teqVb \\
\sysVI b   & \periodVIb           & \tcBTJDVIb   & \eccVIb           & \rpVIb   & \mpVIb                        & \teqVIb \\
\sysVI c   & \periodVIc           & \tcBTJDVIc   & \eccVIc           & \rpVIc   & \mpVIc                        & \teqVIc \\
\sysVII b  & \periodVIIb          & \tcBTJDVIIb  & \eccVIIb          & \rpVIIb  & \mpVIIb                       & \teqVIIb \\
\sysVIII b & \periodVIIIb         & \tcBTJDVIIIb & \eccVIIIb         & \rpVIIIb & \mpVIIIb                      & \teqVIIIb \\
\sysVIII c & \nontransperiodVIIIc & \nodata      & \nontranseccVIIIc & \nodata  & \nontransmsiniVIIIc$^\dagger$ & \nontransteqVIIIc \\
\enddata
\tablecomments{A summary of the results of our joint modeling framework. The full results of our stellar characterization and joints analysis for each system can be found in Appendix \ref{appendix:joint_models}. $\mathrm{BTJD} = \mathrm{BJD} - 2457000$. Upper limits reflect 98\% confidence. \teq is calculated assuming zero Bond albedo and full day-night heat redistribution. $^\dagger$\sysVIII c is nontransiting, meaning that this value is in fact $M_\mathrm{p} \sin i_\mathrm{p}$.}
\end{deluxetable*}

\section{RV modeling with \radvel} \label{sec:radvel}
As alluded to in \S\ref{sec:results}, we also used the \radvel software package \citep{radvel} to measure the masses of the planets in our sample. We did this in order to compare the planet mass measurements from our custom joint modeling framework with a more established RV modeling tool (e.g., \citealt{rosenthal21, teske21}). For each system we used \radvel to model the RVs independent of the \tess photometry. We also used \radvel to experiment with a GP kernel that has fewer free parameters than the one described in \S\ref{sec:jointGP}. In general, we find all of the \radvel results are consistent with our adopted joint models.

\subsection{Keplerian-only modeling} \label{sec:radvelKeplerian}
First, we attempted to replicate the results of our joint models with Keplerian-only \radvel models of the RVs (i.e., no GPs). For each system, we used $P$, \transitTime, $K$, and $\sqrt{e}\cos\omega$ and $\sqrt{e}\sin\omega$ to describe the orbit of each planet. For transiting planets, $P$ and \transitTime were fixed to the posterior median values resulting from our adopted joint model. For the nontransiting planet \sysVIII c, we placed Gaussian priors on $P$ and \transitTime using the posteriors of our joint model. For each system we included an offset (with prior $\mathcal{U}$[$-250$, 250] \mps) and jitter term (with prior $\mathcal{U}$[0, 20] \mps) for each RV instrument. We also included a linear RV trend to see whether or not it was favored by the AIC. The only other prior we included was to force $e < 0.99$.

We performed a MAP fit to the data and conducted posterior estimation with \emcee \citep{emcee}. We followed the default prescriptions for burn-in criteria, number of walkers, number of steps, and convergence criteria from \cite{radvel}. We found that for each system, each planet's mass measurement was entirely consistent between our adopted joint model and our Keplerian-only \radvel model. The \radvel models show that the orbits of all planets (save for \sysVIII c) are consistent with being circular and for all but \sysVI and \sysVIII, a linear RV trend is not favored by the AIC.

\subsection{Gaussian process modeling} \label{sec:radvelGP}
When using GPs for regression, choosing a kernel can be somewhat subjective, so it is useful to compare models that use different kernels in order to ensure that the results are not biased. In the case of our joint model, the kernel we employ (see \S\ref{sec:jointGP}) is relatively complex compared to e.g., a squared exponential kernel or a Mat\'ern 3/2 kernel. The kernel introduces 11 free parameters for a system with both \apflevy and \keckhires data: six amplitude parameters ($\eta_{\mathrm{dec},\:i}$ and $\eta_{\mathrm{rot},\:i}$ for each instrument, $i =$ \apflevy RV, \keckhires RV, and \keckhires \shk), four shared hyperparameters to describe the rotation term ($Q_0$, $\delta Q$, $P_\mathrm{rot}$, and $f$), and one shared hyperparameter to describe the exponentially decaying term ($\rho_\mathrm{dec}$). The GP hyperparameters are summarized in Table \ref{tab:nt_shk_rv_gp_model}. As mentioned at the end of \S\ref{sec:jointGP}, we ultimately chose this kernel after experimenting with its variants. The authors of \exoplanet also suggest that it is a good kernel for modeling stellar activity.\footnote{\url{https://gallery.exoplanet.codes/tutorials/stellar-variability/}}

As a sanity check, we attempted to model the RVs in \radvel using a GP kernel with fewer hyperparameters. For instrument $i$, the kernel (sometimes referred to as the ``quasi-periodic'' kernel; e.g., \citealt{grunblatt15, kosiarek21}) quantifies covariance between data observed at times $t$ and $t'$ as
\begin{equation} \label{eqn:qp_kernel}
    k_i (t,t') = \eta_{1,\:i}^2 \ \mathrm{exp} \left[-\frac{(t-t')^2}{\eta_2^2}-\frac{\sin^2(\frac{\pi(t-t')}{\eta_3})}{2 \eta_4^2}\right].
\end{equation}
$\eta_{\text{1-4}}$ are the hyperparameters: $\eta_{1,\:i}$ represents the amplitude of the covariance for instrument $i$, $\eta_2$ is interpreted as the evolutionary timescale of active stellar regions, $\eta_3$ is interpreted as the stellar rotation period, and $\eta_4$ is the length scale of the covariance's periodicity. The hyperparameters are shared between instruments save for the amplitudes, $\eta_{1,\:i}$. To incorporate this GP into our \radvel models, we first trained the GP by fitting it to the \keckhires \shk values. The posteriors of $\eta_2$, $\eta_3$, and $\eta_4$ resulting from the training were then used as numerical priors for these hyperparameters when fitting the RVs. We also placed a uniform prior of $\mathcal{U}$[0, 20] \mps on $\eta_{1,\:i}$. This process is an in-series analog to our joint model's simultaneous fitting of the RVs and \keckhires \shk values.

For the GP training on the \shk values, we placed a uniform prior of $\mathcal{U}$[0, 1] on $\eta_{1,\:i}$ and broad Jeffreys priors \citep{jeffreys46} of $\mathcal{J}$[1, 500] d on $\eta_2$ and $\eta_3$. For $\eta_4$, we used the Gaussian prior $\mathcal{N}$(0.5, 0.05) per \cite{haywood18}. Training the GP on the \keckhires \shk values did not result in clear constraints on the hyperparameters for \sysIII, \sysV, and \sysVII. For \sysIII, we also tried fitting the \shk values using a prior of $\mathcal{N}$(4, 1.5) d on $\eta_3$ (like we did for $P_\mathrm{rot}$ when using the joint model's more complicated kernel). However, posterior estimation with \emcee failed to converge when adding the trained GP to the \radvel model of the RVs, which is likely a symptom of the lack of constraints on the other GP hyperparameters.

For \sysV, during the GP training we replaced the Jeffreys prior on $\eta_3$ with a relatively broad Gaussian, $\mathcal{N}$(32.5, 7.5) d, to hone in on the 32.5 d signal that we identified in the GLS periodograms of the RV residuals and the \keckhires \shk values. Adding the trained GP to the \radvel model of the RVs described in \S\ref{sec:radvelKeplerian}, we find \mplanetb $= 7.3 \pm 1.9$ \mearth for \sysV b. This mass is consistent with our joint model's result of \mplanetb $=$ \mpVb \mearth.

For \sysVII, the GP training on the \shk values resulted in a bimodal posterior for $\eta_3$ with peaks near 28 d and 14 d. Posterior estimation for a fit to the RVs using the trained GP did not converge due to walkers getting caught at one of the two $\eta_3$ peaks. As discussed in \S\ref{sec:sysVII_activity}, it seems as though $\eta_3 \approx 14$ d may be a better representation of the true rotation period for this G0 dwarf. As an experiment, we repeated the training but this time we replaced the Jeffreys prior on $\eta_3$ with $\mathcal{N}$(14, 1.5) d. This is the same prior we placed on $P_\mathrm{rot}$ in our adopted joint model (see \S\ref{results:sysVII}). Adding this trained GP to the RV model, we find \mplanetb $= 12.6 \pm 2.4$ \mearth for \sysVII b. This result is consistent with the mass measurement from our adopted, GP-enabled joint model of \mplanetb $=$ \mpVIIb \mearth. On the other hand, this GP-enabled \radvel model is in slight disagreement with the results of our Keplerian-only joint model of the data, which finds \mplanetb $= 7.8 \pm 2.9$ \mearth. Continued Doppler monitoring of this moderately active system should help cast light on the nature of the stellar activity signal. 

\begin{figure*}
    \centering
    \includegraphics[width=\textwidth]{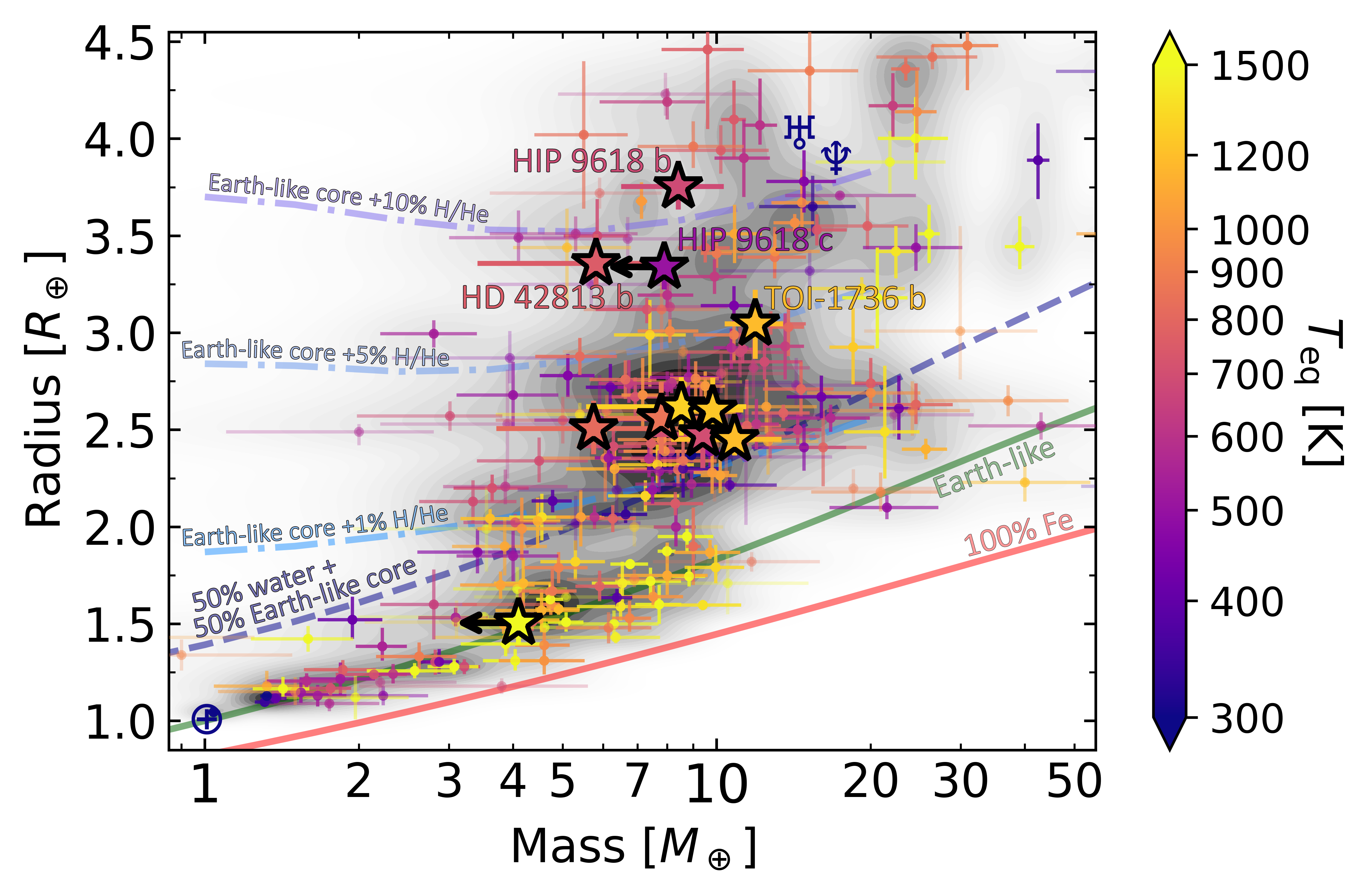}
    \caption{The mass-radius diagram for small planets. Data comes from the NASA Exoplanet Archive's planetary systems table, as accessed on 2022-Nov-17 \citep{NEA}. Planets with mass and radius measurements to better than 50\% and 15\% fractional precision, respectively, are shown as the circles with 1$\sigma$ error bars. The opacity of the points is proportional to mass measurement precision (i.e., a less precise mass measurement translates to a more transparent marker). Color corresponds to equilibrium temperature assuming zero Bond albedo and full day-night heat redistribution. Underlying contours come from Gaussian kernel density estimation (KDE) of the confirmed planets described above. The 11 transiting planets from this work are overplotted as the stars. Mass upper limits (98\% confidence) are plotted for \sysIII c (yellow star near the radius valley) and \sysVI c (labeled). A handful of composition curves are plotted for reference \citep{lopezforney14, zeng16, zeng19}. The curves of H/He envelopes atop Earth-like cores come from \cite{lopezforney14} and are chosen for a planet receiving 10$\times$ Earth's incident flux (i.e., $T_\mathrm{eq} \approx 500$ K) orbiting a 10 Gyr-old, solar-metallicity star. The 50\% water plus 50\% Earth-like composition curve from \cite{zeng19} is calculated for a fixed temperature of 700 K at 100 bar, which determines the planetary model's specific entropy. Note that familiar features of the planet radius distribution are now visible as two-dimensional features in the mass-radius plane. These include the radius valley \citep{fulton17, vaneylen18}, with center near 6 \mearth and 1.8 \rearth, and the radius cliff \citep[e.g.,][]{kite19}, as seen by in the steep drop off in the number of planets around 3 \rearth.}
    \label{fig:mr_diagram}
\end{figure*}

\section{Planet bulk composition} \label{sec:bulk_comp}
Here we contextualize the 11 transiting planets from this sample in the mass-radius diagram (Figure \ref{fig:mr_diagram}). We compare the planets' locations relative to models from \cite{lopezforney14} and \cite{zeng16, zeng19}, and interpolate over theoretical grids of composition to infer planet bulk properties \citep{piaulet21}. The planets fall into three categories: super-Earths (\sysIII c), typical sub-Neptunes (\sysI b and c, \sysIII b, \sysIV b, \sysV b, and \sysVII b), and puffy sub-Neptunes (\sysII b, \sysVI b and c, and \sysVIII b). 

\subsection{The mass-radius diagram}
In our sample, \sysIII c sits alone below the radius valley. While we do not measure a precise mass for the planet (\mplanet \mpIIIc \mearth at 98\% confidence), our upper limit implies that \sysIII c's core contains some fraction of volatiles or ices. Alternatively, the planet's core could be iron-poor. The planet is an attractive target for follow-up, though its small Doppler signal ($<1.5$ \mps) relative to the star's RV jitter ($\approx 7$ \mps) has frustrated our mass measurement efforts with \apflevy and \keckhires.

Six planets land on the mode of the sub-Neptune distribution near \modemass \mearth and \moderadius \rearth (clockwise from left in Figure \ref{fig:mr_diagram}: \sysV b, \sysI b, \sysIII b, \sysIV b, \sysVII b, and \sysI c). These planets have bulk densities that are roughly consistent with a 0.1-2\% H$_2$ envelope by mass sitting atop an Earth-like core. However, at low H$_2$ envelope mass fractions, the \cite{lopezforney14} models become degenerate with those invoking a water-rich bulk composition \citep[i.e., an Earth-like core with a small H$_2$ envelope becomes indistinguishable from a planet made of half ice and half rock;][]{aguichine21}. Indeed, such ``water worlds'' are predicted by formation theory \citep[e.g.,][]{raymond18}.

The last four planets (\sysII b, \sysVI b, \sysVI c, and \sysVIII c) sit just beyond the ``radius cliff''—the steep drop off in planet occurrence around 3 \rearth (e.g., \citealt{kite19})—and are all seemingly consistent with having a substantial ($> 2$\%) fraction of their mass in an H$_2$ envelope. Curiously, the densest of these planets, \sysVIII b, is also the only planet orbiting a subgiant star. The in-transit detection of He I absorption for \sysVIII b would provide evidence of ongoing photoevaporation \citep[e.g.,][]{zhang22}, which might have started when \sysVIII evolved off the main sequence.

\begin{deluxetable}{r|c|c|cc}
\tablecaption{\smint results \label{tab:smint}}
\tabletypesize{\normalsize}
\tablehead{
    \colhead{}            & \colhead{\LFXIV} & \colhead{\ZXVI}  & \multicolumn{2}{c}{\AXXI} \\ 
    \colhead{Planet name} & \colhead{\fenv}  & \colhead{\fhIIo} & \colhead{\fcore} & \colhead{\fhIIo}
}
\startdata
\sysIII c  & $<0.2$        & $60 \pm 30$ & $52 \pm 30$ & $< 15$      \\
\sysVII b  & $0.9 \pm 0.4$ & $60 \pm 24$ & $45 \pm 30$ & $29 \pm 11$ \\
\sysI c    & $1.5 \pm 0.8$ & $67 \pm 24$ & $43 \pm 30$ & $44 \pm 18$ \\
\sysV b    & $1.9 \pm 0.7$ & $83 \pm 16$ & $42 \pm 30$ & $57 \pm 15$ \\
\sysI b    & $1.9 \pm 0.9$ & $77 \pm 20$ & $43 \pm 31$ & $54 \pm 18$ \\
\sysIV b   & $1.4 \pm 0.7$ & $78 \pm 19$ & $42 \pm 30$ & $41 \pm 14$ \\
\sysIII b  & $1.6 \pm 0.7$ & $78 \pm 20$ & $44 \pm 31$ & $47 \pm 16$ \\
\sysVIII b & $3.3 \pm 0.9$ & \nodata     & \nodata     & \nodata     \\
\sysVI c   & $7.0 \pm 1.2$ & \nodata     & \nodata     & \nodata     \\
\sysII b   & $6.3 \pm 1.1$ & \nodata     & \nodata     & \nodata     \\
\sysVI b   & $9.2 \pm 1.2$ & \nodata     & \nodata     & \nodata     \\
\enddata
\tablecomments{Results from our interpolation on the grids of planet composition from \cite{lopezforney14}, \cite{zeng16}, and \cite{aguichine21}. All values are shown in percent. Planets appear in order of increasing radius. For the \LFXIV grid, \fenv is the fraction of the planet's mass contained in an \hhe-dominated, solar metallicity envelope, assuming a rocky core composition. For \ZXVI, \fhIIo is the planet's core \water mass fraction, assuming the planet is composed of \water ice and silicates. For \AXXI, \fcore is the fraction of the planet's refractory core that is iron, with the rest of the core being made up of silicates (e.g., \fcore $\approx 32\%$ for Earth). \fhIIo is the total mass fraction of the planet's \water content, which is contained in a supercritical fluid layer and a steam atmosphere. For \sysIII c, the upper limit on \fenv reflects 98\% confidence for the case where \sincplanet is fixed to 1000 \sincearth (the upper limit of the \LFXIV grid) in place of using the planet's actual instellation, \sincplanet $\approx 1400$ \sincearth. Similarly, for the \AXXI grid, the \fcore and \fhIIo values represent the case where \teq has been fixed to 1300 K (the upper limit of the grid) in place of \sysIII c's actual equilibrium temperature of \teq $\approx 1700$ K. The upper limit on \fhIIo reflects 98\% confidence. For the four puffy sub-Neptunes (\sysVIII c, \sysVI b, \sysII b, and \sysVI c) their large radii demand an \hhe envelope. Without one, both the \ZXVI and \AXXI \fhIIo values rail to 100\% and the planet mass is inflated such that it is inconsistent with the results of our joint photometry and RV analysis.}
\end{deluxetable}

\subsection{\smint analysis}
To make more quantitative statements about possible planet bulk properties, we used the Structure Model INTerpolator tool (\smint; \citealt{piaulet21}) to interpolate over the theoretical grids of planet composition from \cite{lopezforney14}, \cite{zeng16}, and \cite{aguichine21}. Hereafter, we refer to these works as \LFXIV, \ZXVI, and \AXXI, respectively.

The \LFXIV grid assumes a planet is composed of an \hhe-dominated, solar metallicity, envelope atop a rocky core. The planet is then thermally evolved over time according to the methods of \cite{lopez12}, but ignoring the influence of XUV- and EUV-driven photoevaporation. To interpolate over the \LFXIV grid, \smint takes inputs of planet mass, instellation flux, and system age and determines an \hhe envelope mass fraction (\fenv) that best matches the observed planet radius. For each of the transiting planets in our system, we placed Gaussian priors on planet mass, radius, and instellation flux according to our joint modeling results and a uniform prior on \fenv from 0.1\% to 20\%. Since age is typically difficult to infer for main sequence stars, we placed a uniform prior on the age of each system between 1 and 10 Gyr (including for the subgiant \sysVIII).

\begin{figure}
    \centering
    \includegraphics[width=\columnwidth]{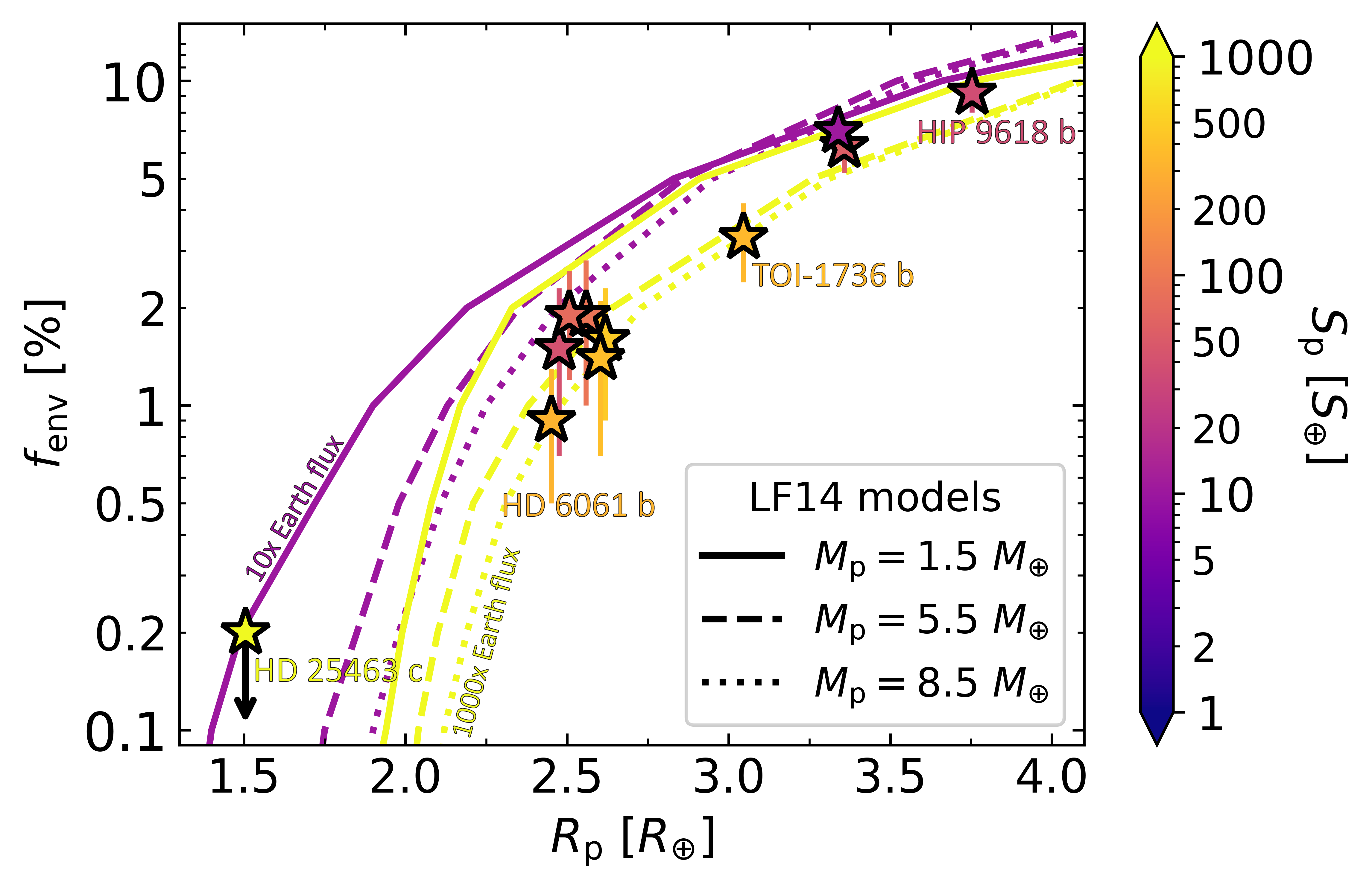}
    \caption{Inferred values of \fenv as a function of planet radius for the 11 transiting planets. \fenv values are estimated by applying \smint to the grid of thermal evolution from \LFXIV. Several slices of the \LFXIV grid are plotted for reference. Color corresponds to instellation flux in Earth units. An upper limit (which reflects 98\% confidence) is shown for \sysIII c, whose actual instellation (\sincplanet $\approx 1400$ \sincearth) surpasses the upper limit of the \LFXIV grid (\sincplanet $= 1000$ \sincearth).}
    \label{fig:rp_fenv}
\end{figure}

We explored the posteriors of planet mass, instellation flux, system age, and \fenv using \emcee. Each \emcee sampler used 50 chains with each chain taking at least 5000 steps. Chains continued sampling until they converged or the chains reached 10$^4$ steps. Convergence was determined by enforcing that each chain was at least 50$\times$ longer than the maximum autocorrelation time across all parameters \citep[$\tau_\mathrm{max}$;][]{goodman10} and that the maximum relative change in $\tau_\mathrm{max}$ between convergence checks (every 100 steps) was $<1\%$. After sampling was complete, the first 60\% of steps in each chain were discarded as burn-in and the remaining samples were concatenated. The inferred values of \fenv for each planet are summarized in Table \ref{tab:smint}. Figure \ref{fig:rp_fenv} plots the inferred \fenv values as a function of planet radius.

For \sysIII c, the planet is too highly irradiated (\sincplanet $\approx 1400$ \sincearth) for the \LFXIV grid (which has an upper limit of \sincplanet $= 1000$ \sincearth). To place an upper limit on \fenv for \sysIII c, we fixed \sincplanet $= 1000$ and placed a uniform prior on \mplanet between 1.5 \mearth and 4.1 \mearth, where the lower limit comes from the mass of a 1.5 \rearth planet lying on the 50\% water and 50\% rock isocomposition curve from \cite{zeng16}, and the upper limit comes from our Doppler observations. We use the 50\% water and 50\% rock isocomposition curve as a fiducial lower bound on the mass of \sysIII c because cosmic abundance measurements suggests that 50\% should be an upper limit on planet core water mass fractions. 

Recently, the idea that small planets may owe a substantial fraction of their mass to \water ice, liquid, and/or vapor---as opposed to strictly being composed of rock and \hhe---has found observational evidence \citep[e.g.,][]{zeng19, luque22} to support theories of ice-rich core formation \citep[e.g.,][]{raymond18}. To explore these so-called ``water world'' compositions, we also applied \smint to the grid from \ZXVI, which models planets as a mixture of liquid \water, high pressure \water ice, and silicates. Such a composition resembles that of the solar system's icy moons. To interpolate over the \ZXVI grid, \smint tunes planet mass and core \water mass fraction (\fhIIo) to best-fit the observed planet radius. We placed a uniform prior on \fhIIo between 0\% and 100\%. The posterior estimation was analogous to our procedure for the \LFXIV grid. \fhIIo estimates are summarized in Table \ref{tab:smint}.

One caveat of the \ZXVI model is that it does not necessarily apply to the short-period sub-Neptunes identified by \tess, since these planets are generally too highly irradiated to have all of their \water in the solid and liquid phases. More applicable are the models of \AXXI, in which Earth-like cores are surrounded by a supercritical \water fluid layer and a steam-dominated envelope. To interpolate on the \AXXI grid we fit the following free parameters to match the measured planet radius: the planet core mass fraction (\fcore; the fraction of the planet's refractory core that is iron, with the rest of the core being made up of silicates), planet water mass fraction (\fhIIo; which includes both the supercritical fluid and steam envelope components), irradiation temperature (for which we use \teq assuming zero Bond albedo and full day-night heat redistribution), and planet mass. We placed uniform priors on \fcore and \fhIIo between 0\% and 100\% and used informed Gaussian priors on \teq and \mplanet according to the results of our stellar characterization and joint photometry and RV analysis. The \emcee sampling then proceeded following our method for the \LFXIV and \ZXVI grids. \fcore and \fhIIo estimates are summarized in Table \ref{tab:smint}. 

Similar to the case of \sysIII c and the \LFXIV grid, the \AXXI grid has an upper limit of \teq $= 1300$ K, yet the planet has \teq $\approx 1700$ K. To estimate an upper limit on \fhIIo for \sysIII c, we fixed \teq to 1300 K and placed a uniform prior on \mplanetc between 1.5 \mearth and 4.1 \mearth (where we have again bounded \mplanet below by the 50\% water and 50\% rock isocomposition curve and above by our Doppler observations).

Our interpolation on the \ZXVI grid results in systematically higher \fhIIo values as compared to the \AXXI grid (typically 60--80\% versus 30--50\%). We interpret this as a symptom of the distinction made above, where the \AXXI model is better-suited to describe highly irradiated water worlds while the \ZXVI grid is more applicable to planets at low instellation flux. Since the \AXXI model can place water in an extended envelope, less overall water is needed to match a planet's radius. In the \ZXVI model, however, all of the water must go into the liquid and solid phases, which makes it difficult to replicate intermediate to low-density sub-Neptunes without large values for \fhIIo. Cosmic abundance measurements suggest that planetesimals forming beyond the snow line should be a 1:1 mixture of \water ice and rock. To this end, we note that \fhIIo $\gtrsim 50$\% is unphysical, so the results of our interpolation on the \ZXVI grid should be treated with care. For primordially icy cores, thermal processes such as radiogenic heating also work to reduce \fhIIo below 50\% \citep{grimm93, monteux18}.

Here we summarize the results of our bulk composition analysis. The super-Earth \sysIII c is too low-mass and too highly irradiated to host a volatile envelope. The \ZXVI grid suggests that the planet's core has a high water content, but given our comments above, the \ZXVI \fhIIo estimates should be treated with care at the high equilibrium temperatures of the planets in our sample. This is not to necessarily say that \sysIII c's core is not ice-rich, however. Indeed, the planet's core may contain some amount of volatiles given our mass upper limit (\mplanet \mpIIIc \mearth). Perhaps volatiles have been dissolved into the planet's core as a result of the interaction between a reactive iron core, silicate mantle, and a primordial envelope which has since been stripped away \citep{schlichting22}. Alternatively, \sysIII c may have a rocky core that is iron-poor---using our high-resolution, iodine-free \keckhires spectra, we find that \sysIII has [Fe/H] $=$ \fehIII dex.

For the planets at the mode of the sub-Neptune mass-radius distribution (\sysI b and c, \sysIII b, \sysIV b, \sysV b, and \sysVII b), \fenv and \fhIIo are degenerate. These planets can be reasonably explained either as an Earth-like core with a small ($1\% \lesssim$ \fenv $\lesssim 2\%$) \hhe envelope or as an irradiated water world consisting of an Earth-like core and roughly 30\% to 50\% of their mass in a supercritical water layer beneath a steam atmosphere. Again, while the \ZXVI grid suggests that these planets have \fhIIo $\gtrsim 60\%$, this is an overestimate of the water content. Transmission spectroscopy to measure the atmospheric H/O ratio may help break the degeneracy between these two compositions.

Finally, we find that the puffy sub-Neptunes (\sysII b, \sysVI b and c, and \sysVIII b) all demand a massive \hhe envelope ($3\% \lesssim$ \fenv $\lesssim 10\%$). Attempting to explain these planets using the \ZXVI and \AXXI models resulted in planet masses that were inconsistent with our joint photometry and RV analysis and water mass fractions that railed to 100\%. These planets could also have a slightly less massive \hhe envelope if their cores contain some fraction of \water ice, but they must host some sort of \hhe envelope regardless of their water content.

\section{Prospects for atmospheric characterization} \label{sec:atmo_char}
The sub-Neptune regime of the mass-radius plane is host to a confluence of theoretical models of bulk composition, making it difficult to infer the interiors of these planets from mass and radius measurements alone \citep{valencia07, adams08, zeng19, otegi20b}. Measurements of atmospheric metallicity, however, may be able to break these degeneracies and cast light on planet composition, which can, in turn, inform theories of formation and evolution \citep{rogersSeager10}. What are the prospects for characterizing the atmospheres of the transiting planets presented in this work? 

\begin{figure}
    \centering
    \includegraphics[width=\columnwidth]{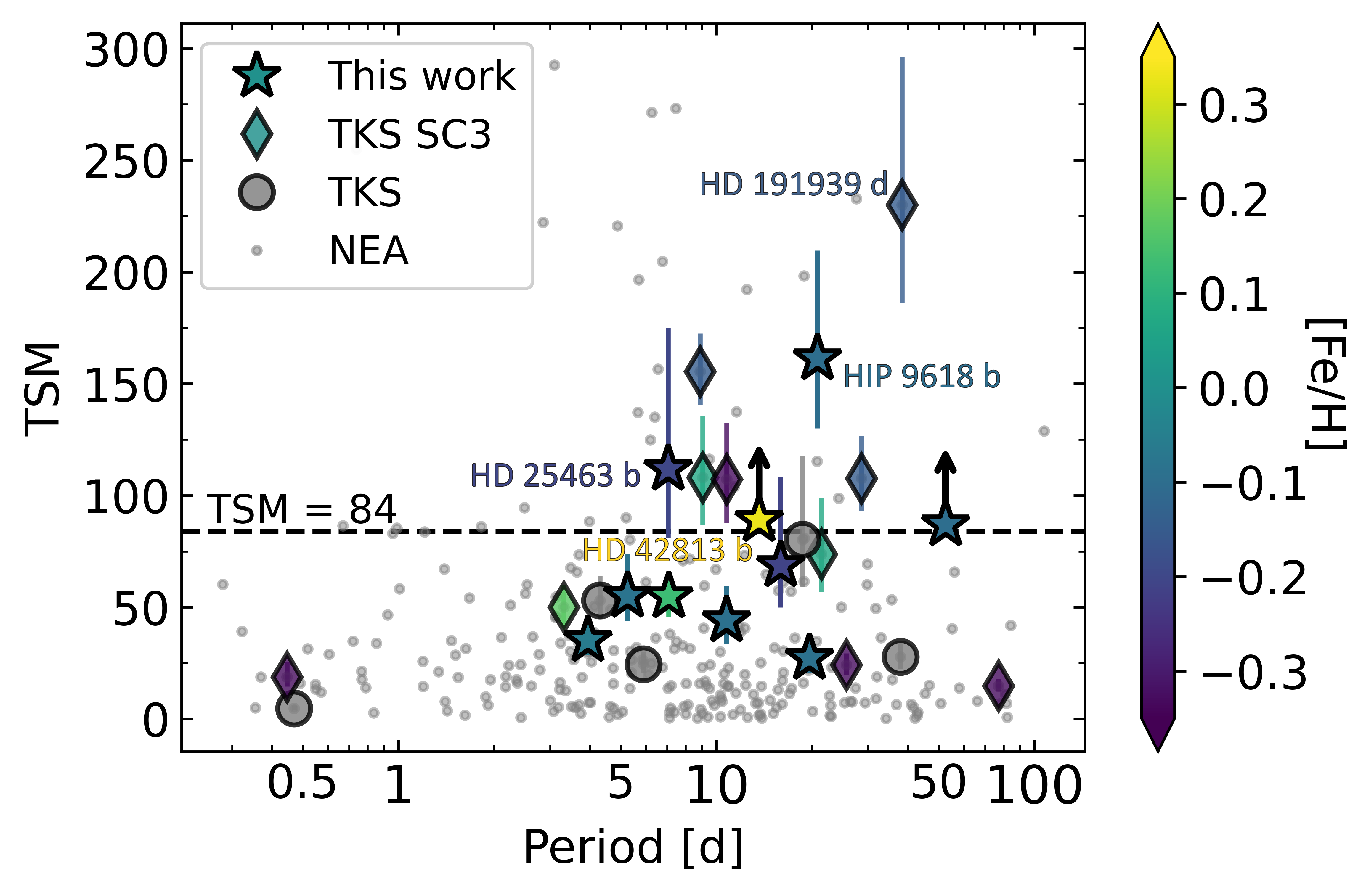}
    \caption{The same planets from Figure \ref{fig:mr_diagram} but now plotted in terms of orbital period versus TSM, a \jwst \snr proxy \citep{kempton18}. Planets from the NASA Exoplanet Archive (NEA) are shown as the gray dots. Confirmed planets from the TESS-Keck Survey (TKS) are shown as the gray circles. Planets from TKS Science Case 3 (SC3), TKS' follow-up of targets amenable to atmospheric characterization, are shown as the diamonds and stars where marker color represents host star metallicity, as measured using [Fe/H]. TKS SC3 planets from this work are the stars, where the lower limits for \sysII b and \sysVI c reflect 98\% confidence. \sysIII c is not shown because its mass constraint is too uninformative. The horizontal dashed line at TSM $= 84$ marks the TSM cutoff suggested by \cite{kempton18} for scheduling atmospheric observations of planets with $2.75 < R_\mathrm{p} < 4$ \rearth.}
    \label{fig:tsm_figure}
\end{figure}

\cite{kempton18} introduced the now widely used Transmission Spectroscopy Metric (TSM; see Equation \ref{eqn:tsm}) to quantify how amenable a planet might be to transit observations with \jwst. TSM is a proxy for the expected \snr of a 10 hr \jwst NIRISS-SOSS observing program assuming a cloud-free, solar-metallicity, H$_2$-dominated planet atmosphere. Figure \ref{fig:tsm_figure} shows the same planets from Figure \ref{fig:mr_diagram} but now plotted by TSM as a function of orbital period. Confirmed planets from the TKS SC3 program (the survey's planet atmospheres science theme) are plotted as the diamonds with transiting planets from this work being shown as the stars. Based on TSM value, some of the most exciting TKS SC3 systems for atmospheric characterization include the multi-transiting planet systems HD 191939 \citep{badenas-agusti20, lubin22, orell-miquel23} and \sysVI (\oXXIII; this work). 

As mentioned in \S\ref{sec:intro_targets_this_work}, while not every planet from this work has an extraordinarily high TSM value (e.g., \sysI b and c), we emphasize the results of \cite{batalha23}. The authors stress that the best samples for inferring population-level characteristics of small planets via transmission spectroscopy are not necessarily composed of the best individual targets. Furthermore, a high TSM value is not a guarantee for the detection of atmospheric molecular features given the seeming ubiquity of clouds and hazes (e.g., \citealt{gao21} and references therein). Multiple factors should be considered in addition to TSM value when selecting targets for atmospheric follow-up observations, including a planet's location in the mass-radius plane, host star properties, and system multiplicity, among others. To these ends, the transiting planets in this work represent a valuable addition to the sample of viable targets for space-based atmospheric observations. 

Detailed characterization of small planet transmission spectra requires a precise planet mass measurement in order to break the degeneracy between planet surface gravity and atmospheric mean molecular weight. \cite{batalha19} demonstrate that with a $5\sigma$ planet mass, uncertainty in the atmospheric characterization process is dominated by the quality of the transmission spectra. On the other hand, a $2\sigma$ mass is still useful for atmospheric characterization, but the dominant source of uncertainty remains the degeneracy between surface gravity and mean molecular weight. $5\sigma$ precision is still lacking for several planets presented in this work. We encourage future Doppler surveys to continue to monitor these targets in order to improve their mass measurements. It is unclear why we did not reach higher precision for some of the planet mass measurements given our large number of RVs (at least 60 \keckhires RVs for each target), but the possible culprits may include unmitigated stellar activity (e.g., perhaps for \sysVII b) and/or inadequate RV measurement precision (e.g., in the case of the low-mass planet, \sysIII c).

Planets are often selected for Doppler follow-up and subsequent atmospheric characterization because of their novelty. This selection bias disfavors targets which are in fact the most common products of planet formation in our Galaxy. The mode of the sub-Neptune mass-radius distribution can now be clearly identified around \modemass \mearth and \moderadius \rearth. With mass and radius held fixed, how might changes in instellation, host star metallicity, and/or system multiplicity affect the (atmospheric) composition of different planets on the mode? Much of the \jwst Cycle 1 exoplanet transit observations are dedicated to hot, giant planets and small planets orbiting cool stars, yet few are earmarked for planets sitting on the mode of the sub-Neptune mass-radius distribution (HD 15337 c is the only such planet in Cycle 1). Six planets from this work all land on the mode (\sysI b and c, \sysIII b, \sysIV b, \sysV b, and \sysVII b). These planets have similar masses and radii, but span more than an order of magnitude in instellation flux. Furthermore, the five host stars are all similar in mass and \teff. Precisely characterized planets sitting on the mode of the sub-Neptune mass-radius distribution offer a unique opportunity for the inter-system comparison of atmospheric composition by way of their commonality.

Regarding planet multiplicity, three of the eight systems in this work, \sysI, \sysIII, and \sysVI, host multiple transiting planets. Multiplanet systems are testbeds for theories of planet formation and evolution. Systems with multiple transiting planets are even more valuable, as they enable the intra-system comparison of atmospheric composition. \sysI b and c are nearly identical in physical properties save for instellation flux, offering a rare opportunity to compare the atmospheric composition of two planets while freezing out all other nuisance parameters (e.g., planet bulk density, stellar properties, system age, etc.). HD 63935 b and c, also TKS SC3 planets, represent a similar case of precisely-characterized twin sub-Neptunes orbiting a G dwarf \citep{scarsdale21}. Like \sysI and HD 63935, \sysVI is G dwarf with multiple transiting sub-Neptunes but at lower instellation flux.

\section{Conclusion} \label{sec:conclusion}
In this paper, we used nearly two years' worth of Doppler monitoring to report mass measurements for 11 planets transiting eight bright host stars. We also reported the discovery and confirmation of a super-Jovian-mass planet on a moderately eccentric orbit. Four systems, \sysI, \sysIII, \sysVI, and \sysVIII, host multiple planets, with the first three systems hosting multiple transiting planets. Two systems, \sysVI and \sysVIII, also exhibit long-term RV trends, indicative of distant, massive companions. In addition to these planet confirmations, we also report what is likely a gravitationally-bound mid-M dwarf companion (\sysVIIcomp) to \sysVII. The two stars have a sky-projected separation of 200 AU.

Planet properties were derived in a homogeneous manner using a joint photometry and RV modeling framework, and careful consideration was given to mitigating signs of stellar activity in the spectroscopic data. We contextualized these planets in the mass-radius diagram and examined their prospects for future atmospheric characterization. We highlight \sysVI b as a very attractive target for atmospheric characterization with \jwst. A summary of derived planet properties can be found in Table \ref{tab:planet_prop_summary}. A full list of system parameters, including stellar properties, can be found in Tables \ref{tab:hip8152_properties} through \ref{tab:toi1736_properties}.

Thanks in large part to the wealth of new discoveries from \tess, long-term Doppler follow-up continues to enrich the mass-radius diagram. As we enter the era of \jwst, measurements of atmospheric metallicity will hopefully disambiguate the interior compositions of these planets and inform our understanding of the physics of their formation. 

\vspace{0.5cm}
\begin{center}
ACKNOWLEDGMENTS
\end{center}

The authors wish to recognize and acknowledge the very significant cultural role and reverence that Maunakea has always had within the indigenous Hawaiian community. We are most fortunate to have the opportunity to conduct observations from this sacred mountain, which is now colonized land.

The authors thank the anonymous referee for their helpful comments which improved the manuscript. J.M.A.M. is supported by the National Science Foundation (NSF) Graduate Research Fellowship Program (GRFP) under Grant No. DGE-1842400. A.B. and R.A.R. are supported by the NSF GRFP under Grant No. DGE-1745301. I.J.M.C. acknowledges support from the NSF through Grant No. AST-1824644. J.M.A.M. acknowledges the LSSTC Data Science Fellowship Program, which is funded by LSSTC, NSF Cybertraining Grant No. 1829740, the Brinson Foundation, and the Moore Foundation; his participation in the program has benefited this work. N.M.B. acknowledges support from NASA’S Interdisciplinary Consortia for Astrobiology Research (NNH19ZDA001N-ICAR) under award number 19-ICAR19\_2-0041. J.V.Z. acknowledges support from the Future Investigators in NASA Earth and Space Science and Technology (FINESST) grant 80NSSC22K1606. T.F. acknowledges support from the University of California President's Postdoctoral Fellowship Program. P.D. acknowledges support from a 51 Pegasi b Postdoctoral Fellowship from the Heising-Simons Foundation. E.A.P. acknowledges support from the Alfred P. Sloan Foundation. D.H. acknowledges support from the Alfred P. Sloan Foundation and the National Aeronautics and Space Administration (80NSSC21K0652). C.D.D. acknowledges the support of the Hellman Family Faculty Fund, the Alfred P. Sloan Foundation, the David \& Lucile Packard Foundation, and the National Aeronautics and Space Administration via the \tess Guest Investigator Program (80NSSC18K1583).

J.M.A.M. thanks (in alphabetical order) Artyom Aguichine, Aarynn Carter, Hugh Osborn, James Rogers, and Hilke Schlichting for insightful discussions and their openness to collaboration.

This work used \texttt{Expanse} at the San Diego Supercomputer Center through allocation PHY220015 from the Advanced Cyberinfrastructure Coordination Ecosystem: Services \& Support (ACCESS) program, which is supported by NSF grants 2138259, 2138286, 2138307, 2137603, and 2138296.

We acknowledge the use of public \tess data from pipelines at the \tess Science Office and at the \tess Science Processing Operations Center. Resources supporting this work were provided by the NASA High-End Computing (HEC) Program through the NASA Advanced Supercomputing (NAS) Division at Ames Research Center for the production of the SPOC data products.

We thank Ken and Gloria Levy, who supported the construction of the Levy Spectrometer on the Automated Planet Finder. We thank the University of California and Google for supporting Lick Observatory, and the UCO staff for their dedicated work scheduling and operating the telescopes of Lick Observatory.

This work presents results from the European Space Agency (ESA) space mission \gaia. \gaia data are being processed by the \gaia Data Processing and Analysis Consortium (DPAC). Funding for the DPAC is provided by national institutions, in particular the institutions participating in the \gaia MultiLateral Agreement (MLA).

This publication makes use of data products from the Two Micron All Sky Survey, which is a joint project of the University of Massachusetts and the Infrared Processing and Analysis Center/California Institute of Technology, funded by the National Aeronautics and Space Administration and the NSF.

\facilities{APF, Hale (PHARO), Keck:I (HIRES), Keck:II (NIRC2), TESS.}

\software{\texttt{astropy} \citep{exoplanet:astropy13, exoplanet:astropy18}, \celeritetwo \citep{celerite2}, \exoplanet \citep{exoplanet:exoplanet}, \isoclassify \citep{huber17, berger20}, \texttt{matplotlib} \citep{matplotlib}, \texttt{numpy} \citep{numpy}, \texttt{pandas} \citep{pandas}, \texttt{pymc3} \citep{exoplanet:pymc3}, Python 3 \citep{python3}, \radvel \citep{radvel}, \sklearn \citep{sklearn}, \texttt{scipy} \citep{scipy}, \specMatchEmp \citep{yee17}, \specMatchSynth \citep{specmatchsynth}, \texttt{starry} \citep{starry}, \texttt{tessla} \citep{murphy23_7783386}, \texttt{theano} \citep{exoplanet:theano}.}

\bibliography{main}{}

\begin{thebibliography}{}
\expandafter\ifx\csname natexlab\endcsname\relax\def\natexlab#1{#1}\fi
\providecommand{\url}[1]{\href{#1}{#1}}
\providecommand{\dodoi}[1]{doi:~\href{http://doi.org/#1}{\nolinkurl{#1}}}
\providecommand{\doeprint}[1]{\href{http://ascl.net/#1}{\nolinkurl{http://ascl.net/#1}}}
\providecommand{\doarXiv}[1]{\href{https://arxiv.org/abs/#1}{\nolinkurl{https://arxiv.org/abs/#1}}}

\bibitem[{{Aceituno} {et~al.}(2013){Aceituno}, {S{\'a}nchez}, {Grupp}, {Lillo},
  {Hern{\'a}n-Obispo}, {Benitez}, {Montoya}, {Thiele}, {Pedraz}, {Barrado},
  {Dreizler}, \& {Bean}}]{cafe}
{Aceituno}, J., {S{\'a}nchez}, S.~F., {Grupp}, F., {et~al.} 2013, \aap, 552,
  A31, \dodoi{10.1051/0004-6361/201220361}

\bibitem[{{Adams} {et~al.}(2008){Adams}, {Seager}, \&
  {Elkins-Tanton}}]{adams08}
{Adams}, E.~R., {Seager}, S., \& {Elkins-Tanton}, L. 2008, \apj, 673, 1160,
  \dodoi{10.1086/524925}

\bibitem[{{Aguichine} {et~al.}(2021){Aguichine}, {Mousis}, {Deleuil}, \&
  {Marcq}}]{aguichine21}
{Aguichine}, A., {Mousis}, O., {Deleuil}, M., \& {Marcq}, E. 2021, \apj, 914,
  84, \dodoi{10.3847/1538-4357/abfa99}

\bibitem[{{Aigrain} {et~al.}(2012){Aigrain}, {Pont}, \& {Zucker}}]{aigrain12}
{Aigrain}, S., {Pont}, F., \& {Zucker}, S. 2012, \mnras, 419, 3147,
  \dodoi{10.1111/j.1365-2966.2011.19960.x}

\bibitem[{{Akaike}(1974)}]{akaike74}
{Akaike}, H. 1974, IEEE Transactions on Automatic Control, 19, 716

\bibitem[{{Akana Murphy}(2023)}]{murphy23_7783386}
{Akana Murphy}, J.~M. 2023, {tessla: Joint modeling of TESS photometry and
  radial velocities}, 1.0,  Zenodo, \dodoi{10.5281/zenodo.7783386}

\bibitem[{{Astropy Collaboration} {et~al.}(2013){Astropy Collaboration},
  {Robitaille}, {Tollerud}, {Greenfield}, {Droettboom}, {Bray}, {Aldcroft},
  {Davis}, {Ginsburg}, {Price-Whelan}, {Kerzendorf}, {Conley}, {Crighton},
  {Barbary}, {Muna}, {Ferguson}, {Grollier}, {Parikh}, {Nair}, {Unther},
  {Deil}, {Woillez}, {Conseil}, {Kramer}, {Turner}, {Singer}, {Fox}, {Weaver},
  {Zabalza}, {Edwards}, {Azalee Bostroem}, {Burke}, {Casey}, {Crawford},
  {Dencheva}, {Ely}, {Jenness}, {Labrie}, {Lim}, {Pierfederici}, {Pontzen},
  {Ptak}, {Refsdal}, {Servillat}, \& {Streicher}}]{exoplanet:astropy13}
{Astropy Collaboration}, {Robitaille}, T.~P., {Tollerud}, E.~J., {et~al.} 2013,
  \aap, 558, A33, \dodoi{10.1051/0004-6361/201322068}

\bibitem[{{Astropy Collaboration} {et~al.}(2018){Astropy Collaboration},
  {Price-Whelan}, {Sip{\H o}cz}, {G{\"u}nther}, {Lim}, {Crawford}, {Conseil},
  {Shupe}, {Craig}, {Dencheva}, {Ginsburg}, {VanderPlas}, {Bradley},
  {P{\'e}rez-Su{\'a}rez}, {de Val-Borro}, {Aldcroft}, {Cruz}, {Robitaille},
  {Tollerud}, {Ardelean}, {Babej}, {Bach}, {Bachetti}, {Bakanov}, {Bamford},
  {Barentsen}, {Barmby}, {Baumbach}, {Berry}, {Biscani}, {Boquien}, {Bostroem},
  {Bouma}, {Brammer}, {Bray}, {Breytenbach}, {Buddelmeijer}, {Burke},
  {Calderone}, {Cano Rodr{\'{\i}}guez}, {Cara}, {Cardoso}, {Cheedella},
  {Copin}, {Corrales}, {Crichton}, {D'Avella}, {Deil}, {Depagne}, {Dietrich},
  {Donath}, {Droettboom}, {Earl}, {Erben}, {Fabbro}, {Ferreira}, {Finethy},
  {Fox}, {Garrison}, {Gibbons}, {Goldstein}, {Gommers}, {Greco}, {Greenfield},
  {Groener}, {Grollier}, {Hagen}, {Hirst}, {Homeier}, {Horton}, {Hosseinzadeh},
  {Hu}, {Hunkeler}, {Ivezi{\'c}}, {Jain}, {Jenness}, {Kanarek}, {Kendrew},
  {Kern}, {Kerzendorf}, {Khvalko}, {King}, {Kirkby}, {Kulkarni}, {Kumar},
  {Lee}, {Lenz}, {Littlefair}, {Ma}, {Macleod}, {Mastropietro}, {McCully},
  {Montagnac}, {Morris}, {Mueller}, {Mumford}, {Muna}, {Murphy}, {Nelson},
  {Nguyen}, {Ninan}, {N{\"o}the}, {Ogaz}, {Oh}, {Parejko}, {Parley}, {Pascual},
  {Patil}, {Patil}, {Plunkett}, {Prochaska}, {Rastogi}, {Reddy Janga},
  {Sabater}, {Sakurikar}, {Seifert}, {Sherbert}, {Sherwood-Taylor}, {Shih},
  {Sick}, {Silbiger}, {Singanamalla}, {Singer}, {Sladen}, {Sooley},
  {Sornarajah}, {Streicher}, {Teuben}, {Thomas}, {Tremblay}, {Turner},
  {Terr{\'o}n}, {van Kerkwijk}, {de la Vega}, {Watkins}, {Weaver}, {Whitmore},
  {Woillez}, {Zabalza}, \& {Astropy Contributors}}]{exoplanet:astropy18}
{Astropy Collaboration}, {Price-Whelan}, A.~M., {Sip{\H o}cz}, B.~M., {et~al.}
  2018, \aj, 156, 123, \dodoi{10.3847/1538-3881/aabc4f}

\bibitem[{{Badenas-Agusti} {et~al.}(2020){Badenas-Agusti}, {G{\"u}nther},
  {Daylan}, {Mikal-Evans}, {Vanderburg}, {Huang}, {Matthews}, {Rackham},
  {Bieryla}, {Stassun}, {Kane}, {Shporer}, {Fulton}, {Hill}, {Nowak}, {Ribas},
  {Pall{\'e}}, {Jenkins}, {Latham}, {Seager}, {Ricker}, {Vanderspek}, {Winn},
  {Abril-Pla}, {Collins}, {Serra}, {Niraula}, {Rustamkulov}, {Barclay},
  {Crossfield}, {Howell}, {Ciardi}, {Gonzales}, {Schlieder}, {Caldwell},
  {Fausnaugh}, {McDermott}, {Paegert}, {Pepper}, {Rose}, \&
  {Twicken}}]{badenas-agusti20}
{Badenas-Agusti}, M., {G{\"u}nther}, M.~N., {Daylan}, T., {et~al.} 2020, \aj,
  160, 113, \dodoi{10.3847/1538-3881/aba0b5}

\bibitem[{{Baluev}(2008)}]{baluev08}
{Baluev}, R.~V. 2008, \mnras, 385, 1279,
  \dodoi{10.1111/j.1365-2966.2008.12689.x}

\bibitem[{{Batalha} {et~al.}(2019){Batalha}, {Lewis}, {Fortney}, {Batalha},
  {Kempton}, {Lewis}, \& {Line}}]{batalha19}
{Batalha}, N.~E., {Lewis}, T., {Fortney}, J.~J., {et~al.} 2019, \apjl, 885,
  L25, \dodoi{10.3847/2041-8213/ab4909}

\bibitem[{{Batalha} {et~al.}(2023){Batalha}, {Wolfgang}, {Teske}, {Alam},
  {Alderson}, {Batalha}, {L{\'o}pez-Morales}, \& {Wakeford}}]{batalha23}
{Batalha}, N.~E., {Wolfgang}, A., {Teske}, J., {et~al.} 2023, \aj, 165, 14,
  \dodoi{10.3847/1538-3881/ac9f45}

\bibitem[{{Batalha} {et~al.}(2013){Batalha}, {Rowe}, {Bryson}, {Barclay},
  {Burke}, {Caldwell}, {Christiansen}, {Mullally}, {Thompson}, {Brown},
  {Dupree}, {Fabrycky}, {Ford}, {Fortney}, {Gilliland}, {Isaacson}, {Latham},
  {Marcy}, {Quinn}, {Ragozzine}, {Shporer}, {Borucki}, {Ciardi}, {Gautier},
  {Haas}, {Jenkins}, {Koch}, {Lissauer}, {Rapin}, {Basri}, {Boss}, {Buchhave},
  {Carter}, {Charbonneau}, {Christensen-Dalsgaard}, {Clarke}, {Cochran},
  {Demory}, {Desert}, {Devore}, {Doyle}, {Esquerdo}, {Everett}, {Fressin},
  {Geary}, {Girouard}, {Gould}, {Hall}, {Holman}, {Howard}, {Howell},
  {Ibrahim}, {Kinemuchi}, {Kjeldsen}, {Klaus}, {Li}, {Lucas}, {Meibom},
  {Morris}, {Pr{\v{s}}a}, {Quintana}, {Sanderfer}, {Sasselov}, {Seader},
  {Smith}, {Steffen}, {Still}, {Stumpe}, {Tarter}, {Tenenbaum}, {Torres},
  {Twicken}, {Uddin}, {Van Cleve}, {Walkowicz}, \& {Welsh}}]{batalha13}
{Batalha}, N.~M., {Rowe}, J.~F., {Bryson}, S.~T., {et~al.} 2013, \apjs, 204,
  24, \dodoi{10.1088/0067-0049/204/2/24}

\bibitem[{{Benz} {et~al.}(2021){Benz}, {Broeg}, {Fortier}, {Rando}, {Beck},
  {Beck}, {Queloz}, {Ehrenreich}, {Maxted}, {Isaak}, {Billot}, {Alibert},
  {Alonso}, {Ant{\'o}nio}, {Asquier}, {Bandy}, {B{\'a}rczy}, {Barrado},
  {Barros}, {Baumjohann}, {Bekkelien}, {Bergomi}, {Biondi}, {Bonfils},
  {Borsato}, {Brandeker}, {Busch}, {Cabrera}, {Cessa}, {Charnoz}, {Chazelas},
  {Collier Cameron}, {Corral Van Damme}, {Cortes}, {Davies}, {Deleuil},
  {Deline}, {Delrez}, {Demangeon}, {Demory}, {Erikson}, {Farinato}, {Fossati},
  {Fridlund}, {Futyan}, {Gandolfi}, {Garcia Munoz}, {Gillon}, {Guterman},
  {Gutierrez}, {Hasiba}, {Heng}, {Hernandez}, {Hoyer}, {Kiss}, {Kovacs},
  {Kuntzer}, {Laskar}, {Lecavelier des Etangs}, {Lendl}, {L{\'o}pez}, {Lora},
  {Lovis}, {L{\"u}ftinger}, {Magrin}, {Malvasio}, {Marafatto}, {Michaelis}, {de
  Miguel}, {Modrego}, {Munari}, {Nascimbeni}, {Olofsson}, {Ottacher},
  {Ottensamer}, {Pagano}, {Palacios}, {Pall{\'e}}, {Peter}, {Piazza}, {Piotto},
  {Pizarro}, {Pollaco}, {Ragazzoni}, {Ratti}, {Rauer}, {Ribas}, {Rieder},
  {Rohlfs}, {Safa}, {Salatti}, {Santos}, {Scandariato}, {S{\'e}gransan},
  {Simon}, {Smith}, {Sordet}, {Sousa}, {Steller}, {Szab{\'o}}, {Szoke},
  {Thomas}, {Tschentscher}, {Udry}, {Van Grootel}, {Viotto}, {Walter},
  {Walton}, {Wildi}, \& {Wolter}}]{cheops}
{Benz}, W., {Broeg}, C., {Fortier}, A., {et~al.} 2021, Experimental Astronomy,
  51, 109, \dodoi{10.1007/s10686-020-09679-4}

\bibitem[{{Berger} {et~al.}(2020){Berger}, {Huber}, {van Saders}, {Gaidos},
  {Tayar}, \& {Kraus}}]{berger20}
{Berger}, T.~A., {Huber}, D., {van Saders}, J.~L., {et~al.} 2020, \aj, 159,
  280, \dodoi{10.3847/1538-3881/159/6/280}

\bibitem[{{Betancourt} \& {Girolami}(2013)}]{betancourt13}
{Betancourt}, M.~J., \& {Girolami}, M. 2013, arXiv e-prints, arXiv:1312.0906.
\newblock \doarXiv{1312.0906}

\bibitem[{{Blunt} {et~al.}(2023){Blunt}, {Carvalho}, {David}, {Beichman},
  {Zink}, {Gaidos}, {Behmard}, {Bouma}, {Cody}, {Dai}, {Foreman-Mackey},
  {Grunblatt}, {Howard}, {Kosiarek}, {Knutson}, {Rubenzahl}, {Beard},
  {Chontos}, {Giacalone}, {Hirano}, {Johnson}, {Lubin}, {Akana Murphy},
  {Petigura}, {Van Zandt}, \& {Weiss}}]{blunt23arxiv}
{Blunt}, S., {Carvalho}, A., {David}, T.~J., {et~al.} 2023, arXiv e-prints,
  arXiv:2306.08145, \dodoi{10.48550/arXiv.2306.08145}

\bibitem[{Borucki {et~al.}(2010)Borucki, Koch, Basri, Batalha, Brown, Caldwell,
  Caldwell, Christensen-Dalsgaard, Cochran, DeVore, Dunham, Dupree, Gautier,
  Geary, Gilliland, Gould, Howell, Jenkins, Kondo, Latham, Marcy, Meibom,
  Kjeldsen, Lissauer, Monet, Morrison, Sasselov, Tarter, Boss, Brownlee, Owen,
  Buzasi, Charbonneau, Doyle, Fortney, Ford, Holman, Seager, Steffen, Welsh,
  Rowe, Anderson, Buchhave, Ciardi, Walkowicz, Sherry, Horch, Isaacson,
  Everett, Fischer, Torres, Johnson, Endl, MacQueen, Bryson, Dotson, Haas,
  Kolodziejczak, Van~Cleve, Chandrasekaran, Twicken, Quintana, Clarke, Allen,
  Li, Wu, Tenenbaum, Verner, Bruhweiler, Barnes, \& Prsa}]{borucki10}
Borucki, W.~J., Koch, D., Basri, G., {et~al.} 2010, Science, 327, 977,
  \dodoi{10.1126/science.1185402}

\bibitem[{{Borucki} {et~al.}(2012){Borucki}, {Koch}, {Batalha}, {Bryson},
  {Rowe}, {Fressin}, {Torres}, {Caldwell}, {Christensen-Dalsgaard}, {Cochran},
  {DeVore}, {Gautier}, {Geary}, {Gilliland}, {Gould}, {Howell}, {Jenkins},
  {Latham}, {Lissauer}, {Marcy}, {Sasselov}, {Boss}, {Charbonneau}, {Ciardi},
  {Kaltenegger}, {Doyle}, {Dupree}, {Ford}, {Fortney}, {Holman}, {Steffen},
  {Mullally}, {Still}, {Tarter}, {Ballard}, {Buchhave}, {Carter},
  {Christiansen}, {Demory}, {D{\'e}sert}, {Dressing}, {Endl}, {Fabrycky},
  {Fischer}, {Haas}, {Henze}, {Horch}, {Howard}, {Isaacson}, {Kjeldsen},
  {Johnson}, {Klaus}, {Kolodziejczak}, {Barclay}, {Li}, {Meibom}, {Prsa},
  {Quinn}, {Quintana}, {Robertson}, {Sherry}, {Shporer}, {Tenenbaum},
  {Thompson}, {Twicken}, {Van Cleve}, {Welsh}, {Basu}, {Chaplin}, {Miglio},
  {Kawaler}, {Arentoft}, {Stello}, {Metcalfe}, {Verner}, {Karoff}, {Lundkvist},
  {Lund}, {Handberg}, {Elsworth}, {Hekker}, {Huber}, {Bedding}, \&
  {Rapin}}]{borucki12}
{Borucki}, W.~J., {Koch}, D.~G., {Batalha}, N., {et~al.} 2012, \apj, 745, 120,
  \dodoi{10.1088/0004-637X/745/2/120}

\bibitem[{{Brandt} {et~al.}(2021){Brandt}, {Dupuy}, {Li}, {Brandt}, {Zeng},
  {Michalik}, {Bardalez Gagliuffi}, \& {Raposo-Pulido}}]{orvara}
{Brandt}, T.~D., {Dupuy}, T.~J., {Li}, Y., {et~al.} 2021, \aj, 162, 186,
  \dodoi{10.3847/1538-3881/ac042e}

\bibitem[{{Bryan} {et~al.}(2019){Bryan}, {Knutson}, {Lee}, {Fulton}, {Batygin},
  {Ngo}, \& {Meshkat}}]{bryan19}
{Bryan}, M.~L., {Knutson}, H.~A., {Lee}, E.~J., {et~al.} 2019, \aj, 157, 52,
  \dodoi{10.3847/1538-3881/aaf57f}

\bibitem[{{Bryson} {et~al.}(2021){Bryson}, {Kunimoto}, {Kopparapu}, {Coughlin},
  {Borucki}, {Koch}, {Aguirre}, {Allen}, {Barentsen}, {Batalha}, {Berger},
  {Boss}, {Buchhave}, {Burke}, {Caldwell}, {Campbell}, {Catanzarite},
  {Chandrasekaran}, {Chaplin}, {Christiansen}, {Christensen-Dalsgaard},
  {Ciardi}, {Clarke}, {Cochran}, {Dotson}, {Doyle}, {Duarte}, {Dunham},
  {Dupree}, {Endl}, {Fanson}, {Ford}, {Fujieh}, {Gautier}, {Geary},
  {Gilliland}, {Girouard}, {Gould}, {Haas}, {Henze}, {Holman}, {Howard},
  {Howell}, {Huber}, {Hunter}, {Jenkins}, {Kjeldsen}, {Kolodziejczak},
  {Larson}, {Latham}, {Li}, {Mathur}, {Meibom}, {Middour}, {Morris}, {Morton},
  {Mullally}, {Mullally}, {Pletcher}, {Prsa}, {Quinn}, {Quintana}, {Ragozzine},
  {Ramirez}, {Sanderfer}, {Sasselov}, {Seader}, {Shabram}, {Shporer}, {Smith},
  {Steffen}, {Still}, {Torres}, {Troeltzsch}, {Twicken}, {Uddin}, {Van Cleve},
  {Voss}, {Weiss}, {Welsh}, {Wohler}, \& {Zamudio}}]{bryson21}
{Bryson}, S., {Kunimoto}, M., {Kopparapu}, R.~K., {et~al.} 2021, \aj, 161, 36,
  \dodoi{10.3847/1538-3881/abc418}

\bibitem[{Burnham \& Anderson(2004)}]{burnham04}
Burnham, K.~P., \& Anderson, D.~R. 2004, Sociological Methods \& Research, 33,
  261, \dodoi{10.1177/0049124104268644}

\bibitem[{Butler {et~al.}(1996)Butler, Marcy, Williams, McCarthy, Dosanjh, \&
  Vogt}]{butler96}
Butler, R.~P., Marcy, G.~W., Williams, E., {et~al.} 1996, Publications of the
  Astronomical Society of the Pacific, 108, 500, \dodoi{10.1086/133755}

\bibitem[{{Cannon} \& {Pickering}(1993)}]{cannon93}
{Cannon}, A.~J., \& {Pickering}, E.~C. 1993, VizieR Online Data Catalog,
  III/135A

\bibitem[{{Chen} \& {Kipping}(2017)}]{chenKipping17}
{Chen}, J., \& {Kipping}, D. 2017, \apj, 834, 17,
  \dodoi{10.3847/1538-4357/834/1/17}

\bibitem[{{Choi} {et~al.}(2016){Choi}, {Dotter}, {Conroy}, {Cantiello},
  {Paxton}, \& {Johnson}}]{choi16}
{Choi}, J., {Dotter}, A., {Conroy}, C., {et~al.} 2016, \apj, 823, 102,
  \dodoi{10.3847/0004-637X/823/2/102}

\bibitem[{{Chontos} {et~al.}(2022){Chontos}, {Murphy}, {MacDougall},
  {Fetherolf}, {Van Zandt}, {Rubenzahl}, {Beard}, {Huber}, {Batalha},
  {Crossfield}, {Dressing}, {Fulton}, {Howard}, {Isaacson}, {Kane}, {Petigura},
  {Robertson}, {Roy}, {Weiss}, {Behmard}, {Dai}, {Dalba}, {Giacalone}, {Hill},
  {Lubin}, {Mayo}, {Mo{\v{c}}nik}, {Polanski}, {Rosenthal}, {Scarsdale},
  {Turtelboom}, {Ricker}, {Vanderspek}, {Latham}, {Seager}, {Winn}, {Jenkins},
  {Quinn}, {Guerrero}, {Collins}, {Ciardi}, {Shporer}, {Goeke}, {Levine},
  {Ting}, {Bieryla}, {Collins}, {Kielkopf}, {Barkaoui}, {Benni},
  {Esparza-Borges}, {Conti}, {Hooton}, {Kagetani}, {Laloum}, {Marino},
  {Massey}, {Murgas}, {Papini}, {Schwarz}, {Srdoc}, {Stockdale}, {Wang},
  {Wittrock}, \& {Zou}}]{chontos22}
{Chontos}, A., {Murphy}, J. M.~A., {MacDougall}, M.~G., {et~al.} 2022, \aj,
  163, 297, \dodoi{10.3847/1538-3881/ac6266}

\bibitem[{{Ciardi} {et~al.}(2018){Ciardi}, {Crossfield}, {Feinstein},
  {Schlieder}, {Petigura}, {David}, {Bristow}, {Patel}, {Arnold}, {Benneke},
  {Christiansen}, {Dressing}, {Fulton}, {Howard}, {Isaacson}, {Sinukoff}, \&
  {Thackeray}}]{ciardi18}
{Ciardi}, D.~R., {Crossfield}, I. J.~M., {Feinstein}, A.~D., {et~al.} 2018,
  \aj, 155, 10, \dodoi{10.3847/1538-3881/aa9921}

\bibitem[{{Colquhoun}(2014)}]{colquhoun14}
{Colquhoun}, D. 2014, Royal Society Open Science, 1, 140216,
  \dodoi{10.1098/rsos.140216}

\bibitem[{{Cosentino} {et~al.}(2012){Cosentino}, {Lovis}, {Pepe}, {Collier
  Cameron}, {Latham}, {Molinari}, {Udry}, {Bezawada}, {Black}, {Born},
  {Buchschacher}, {Charbonneau}, {Figueira}, {Fleury}, {Galli}, {Gallie},
  {Gao}, {Ghedina}, {Gonzalez}, {Gonzalez}, {Guerra}, {Henry}, {Horne},
  {Hughes}, {Kelly}, {Lodi}, {Lunney}, {Maire}, {Mayor}, {Micela}, {Ordway},
  {Peacock}, {Phillips}, {Piotto}, {Pollacco}, {Queloz}, {Rice}, {Riverol},
  {Riverol}, {San Juan}, {Sasselov}, {Segransan}, {Sozzetti}, {Sosnowska},
  {Stobie}, {Szentgyorgyi}, {Vick}, \& {Weber}}]{harpsn}
{Cosentino}, R., {Lovis}, C., {Pepe}, F., {et~al.} 2012, in Society of
  Photo-Optical Instrumentation Engineers (SPIE) Conference Series, Vol. 8446,
  Ground-based and Airborne Instrumentation for Astronomy IV, ed. I.~S.
  {McLean}, S.~K. {Ramsay}, \& H.~{Takami}, 84461V, \dodoi{10.1117/12.925738}

\bibitem[{{Dai} {et~al.}(2020){Dai}, {Roy}, {Fulton}, {Robertson}, {Hirsch},
  {Isaacson}, {Albrecht}, {Mann}, {Kristiansen}, {Batalha}, {Beard}, {Behmard},
  {Chontos}, {Crossfield}, {Dalba}, {Dressing}, {Giacalone}, {Hill}, {Howard},
  {Huber}, {Kane}, {Kosiarek}, {Lubin}, {Mayo}, {Mocnik}, {Akana Murphy},
  {Petigura}, {Rosenthal}, {Rubenzahl}, {Scarsdale}, {Weiss}, {Van Zandt},
  {Ricker}, {Vanderspek}, {Latham}, {Seager}, {Winn}, {Jenkins}, {Caldwell},
  {Charbonneau}, {Daylan}, {G{\"u}nther}, {Morgan}, {Quinn}, {Rose}, \&
  {Smith}}]{dai20}
{Dai}, F., {Roy}, A., {Fulton}, B., {et~al.} 2020, \aj, 160, 193,
  \dodoi{10.3847/1538-3881/abb3bd}

\bibitem[{{Dalba} {et~al.}(2022){Dalba}, {Kane}, {Dragomir}, {Villanueva},
  {Collins}, {Jacobs}, {LaCourse}, {Gagliano}, {Kristiansen}, {Omohundro},
  {Schwengeler}, {Terentev}, {Vanderburg}, {Fulton}, {Isaacson}, {Van Zandt},
  {Howard}, {Thorngren}, {Howell}, {Batalha}, {Chontos}, {Crossfield},
  {Dressing}, {Huber}, {Petigura}, {Robertson}, {Roy}, {Weiss}, {Behmard},
  {Beard}, {Brinkman}, {Giacalone}, {Hill}, {Lubin}, {Mayo}, {Mo{\v{c}}nik},
  {Akana Murphy}, {Polanski}, {Rice}, {Rosenthal}, {Rubenzahl}, {Scarsdale},
  {Turtelboom}, {Tyler}, {Benni}, {Boyce}, {Esposito}, {Girardin}, {Laloum},
  {Lewin}, {Mann}, {Marchis}, {Schwarz}, {Srdoc}, {Steuer}, {Sivarani}, {Unni},
  {Eisner}, {Fetherolf}, {Li}, {Yao}, {Pepper}, {Ricker}, {Vanderspek},
  {Latham}, {Seager}, {Winn}, {Jenkins}, {Burke}, {Eastman}, {Lund},
  {Rodriguez}, {Rowden}, {Ting}, \& {Villase{\~n}or}}]{dalba22}
{Dalba}, P.~A., {Kane}, S.~R., {Dragomir}, D., {et~al.} 2022, \aj, 163, 61,
  \dodoi{10.3847/1538-3881/ac415b}

\bibitem[{{Dekany} {et~al.}(2013){Dekany}, {Roberts}, {Burruss}, {Bouchez},
  {Truong}, {Baranec}, {Guiwits}, {Hale}, {Angione}, {Trinh}, {Zolkower},
  {Shelton}, {Palmer}, {Henning}, {Croner}, {Troy}, {McKenna}, {Tesch},
  {Hildebrandt}, \& {Milburn}}]{dekany13}
{Dekany}, R., {Roberts}, J., {Burruss}, R., {et~al.} 2013, \apj, 776, 130,
  \dodoi{10.1088/0004-637X/776/2/130}

\bibitem[{{Dressing} \& {Charbonneau}(2013)}]{dressing13}
{Dressing}, C.~D., \& {Charbonneau}, D. 2013, \apj, 767, 95,
  \dodoi{10.1088/0004-637X/767/1/95}

\bibitem[{{Drimmel} {et~al.}(2003){Drimmel}, {Cabrera-Lavers}, \&
  {L{\'o}pez-Corredoira}}]{drimmel03}
{Drimmel}, R., {Cabrera-Lavers}, A., \& {L{\'o}pez-Corredoira}, M. 2003, \aap,
  409, 205, \dodoi{10.1051/0004-6361:20031070}

\bibitem[{{Duane} {et~al.}(1987){Duane}, {Kennedy}, {Pendleton}, \&
  {Roweth}}]{hmc:duane87}
{Duane}, S., {Kennedy}, A.~D., {Pendleton}, B.~J., \& {Roweth}, D. 1987,
  Physics Letters B, 195, 216, \dodoi{10.1016/0370-2693(87)91197-X}

\bibitem[{{Eastman} {et~al.}(2010){Eastman}, {Siverd}, \& {Gaudi}}]{eastman10}
{Eastman}, J., {Siverd}, R., \& {Gaudi}, B.~S. 2010, \pasp, 122, 935,
  \dodoi{10.1086/655938}

\bibitem[{{Fetherolf} {et~al.}(2022){Fetherolf}, {Pepper}, {Simpson}, {Kane},
  {Mocnik}, {Antoci}, {Huber}, {Jenkins}, {Stassun}, {Twicken}, {Vanderspek},
  \& {Winn}}]{fetherolf22arxiv}
{Fetherolf}, T., {Pepper}, J., {Simpson}, E., {et~al.} 2022, arXiv e-prints,
  arXiv:2208.11721.
\newblock \doarXiv{2208.11721}

\bibitem[{{Foreman-Mackey}(2018{\natexlab{a}})}]{celerite2}
{Foreman-Mackey}, D. 2018{\natexlab{a}}, Research Notes of the American
  Astronomical Society, 2, 31, \dodoi{10.3847/2515-5172/aaaf6c}

\bibitem[{{Foreman-Mackey}(2018{\natexlab{b}})}]{exoplanet:foremanmackey18}
---. 2018{\natexlab{b}}, Research Notes of the American Astronomical Society,
  2, 31, \dodoi{10.3847/2515-5172/aaaf6c}

\bibitem[{{Foreman-Mackey} {et~al.}(2013){Foreman-Mackey}, {Hogg}, {Lang}, \&
  {Goodman}}]{emcee}
{Foreman-Mackey}, D., {Hogg}, D.~W., {Lang}, D., \& {Goodman}, J. 2013, \pasp,
  125, 306, \dodoi{10.1086/670067}

\bibitem[{Foreman-Mackey {et~al.}(2020)Foreman-Mackey, Luger, Czekala, Agol,
  Price-Whelan, \& Barclay}]{exoplanet:exoplanet}
Foreman-Mackey, D., Luger, R., Czekala, I., {et~al.} 2020,
  exoplanet-dev/exoplanet v0.3.2, \dodoi{10.5281/zenodo.1998447}

\bibitem[{{Fressin} {et~al.}(2013){Fressin}, {Torres}, {Charbonneau}, {Bryson},
  {Christiansen}, {Dressing}, {Jenkins}, {Walkowicz}, \& {Batalha}}]{fressin13}
{Fressin}, F., {Torres}, G., {Charbonneau}, D., {et~al.} 2013, \apj, 766, 81,
  \dodoi{10.1088/0004-637X/766/2/81}

\bibitem[{{Fulton} {et~al.}(2018){Fulton}, {Petigura}, {Blunt}, \&
  {Sinukoff}}]{radvel}
{Fulton}, B.~J., {Petigura}, E.~A., {Blunt}, S., \& {Sinukoff}, E. 2018, \pasp,
  130, 044504, \dodoi{10.1088/1538-3873/aaaaa8}

\bibitem[{{Fulton} {et~al.}(2017){Fulton}, {Petigura}, {Howard}, {Isaacson},
  {Marcy}, {Cargile}, {Hebb}, {Weiss}, {Johnson}, {Morton}, {Sinukoff},
  {Crossfield}, \& {Hirsch}}]{fulton17}
{Fulton}, B.~J., {Petigura}, E.~A., {Howard}, A.~W., {et~al.} 2017, \aj, 154,
  109, \dodoi{10.3847/1538-3881/aa80eb}

\bibitem[{{Furlan} {et~al.}(2017){Furlan}, {Ciardi}, {Everett}, {Saylors},
  {Teske}, {Horch}, {Howell}, {van Belle}, {Hirsch}, {Gautier}, {Adams},
  {Barrado}, {Cartier}, {Dressing}, {Dupree}, {Gilliland}, {Lillo-Box},
  {Lucas}, \& {Wang}}]{furlan17}
{Furlan}, E., {Ciardi}, D.~R., {Everett}, M.~E., {et~al.} 2017, \aj, 153, 71,
  \dodoi{10.3847/1538-3881/153/2/71}

\bibitem[{{Gaia Collaboration} {et~al.}(2016){Gaia Collaboration}, {Prusti},
  {de Bruijne}, {Brown}, {Vallenari}, {Babusiaux}, {Bailer-Jones}, {Bastian},
  {Biermann}, {Evans}, {Eyer}, {Jansen}, {Jordi}, {Klioner}, {Lammers},
  {Lindegren}, {Luri}, {Mignard}, {Milligan}, {Panem}, {Poinsignon},
  {Pourbaix}, {Randich}, {Sarri}, {Sartoretti}, {Siddiqui}, {Soubiran},
  {Valette}, {van Leeuwen}, {Walton}, {Aerts}, {Arenou}, {Cropper}, {Drimmel},
  {H{\o}g}, {Katz}, {Lattanzi}, {O'Mullane}, {Grebel}, {Holland}, {Huc},
  {Passot}, {Bramante}, {Cacciari}, {Casta{\~n}eda}, {Chaoul}, {Cheek}, {De
  Angeli}, {Fabricius}, {Guerra}, {Hern{\'a}ndez}, {Jean-Antoine-Piccolo},
  {Masana}, {Messineo}, {Mowlavi}, {Nienartowicz}, {Ord{\'o}{\~n}ez-Blanco},
  {Panuzzo}, {Portell}, {Richards}, {Riello}, {Seabroke}, {Tanga},
  {Th{\'e}venin}, {Torra}, {Els}, {Gracia-Abril}, {Comoretto},
  {Garcia-Reinaldos}, {Lock}, {Mercier}, {Altmann}, {Andrae}, {Astraatmadja},
  {Bellas-Velidis}, {Benson}, {Berthier}, {Blomme}, {Busso}, {Carry},
  {Cellino}, {Clementini}, {Cowell}, {Creevey}, {Cuypers}, {Davidson}, {De
  Ridder}, {de Torres}, {Delchambre}, {Dell'Oro}, {Ducourant}, {Fr{\'e}mat},
  {Garc{\'\i}a-Torres}, {Gosset}, {Halbwachs}, {Hambly}, {Harrison}, {Hauser},
  {Hestroffer}, {Hodgkin}, {Huckle}, {Hutton}, {Jasniewicz}, {Jordan},
  {Kontizas}, {Korn}, {Lanzafame}, {Manteiga}, {Moitinho}, {Muinonen},
  {Osinde}, {Pancino}, {Pauwels}, {Petit}, {Recio-Blanco}, {Robin}, {Sarro},
  {Siopis}, {Smith}, {Smith}, {Sozzetti}, {Thuillot}, {van Reeven}, {Viala},
  {Abbas}, {Abreu Aramburu}, {Accart}, {Aguado}, {Allan}, {Allasia},
  {Altavilla}, {{\'A}lvarez}, {Alves}, {Anderson}, {Andrei}, {Anglada Varela},
  {Antiche}, {Antoja}, {Ant{\'o}n}, {Arcay}, {Atzei}, {Ayache}, {Bach},
  {Baker}, {Balaguer-N{\'u}{\~n}ez}, {Barache}, {Barata}, {Barbier}, {Barblan},
  {Baroni}, {Barrado y Navascu{\'e}s}, {Barros}, {Barstow}, {Becciani},
  {Bellazzini}, {Bellei}, {Bello Garc{\'\i}a}, {Belokurov}, {Bendjoya},
  {Berihuete}, {Bianchi}, {Bienaym{\'e}}, {Billebaud}, {Blagorodnova},
  {Blanco-Cuaresma}, {Boch}, {Bombrun}, {Borrachero}, {Bouquillon}, {Bourda},
  {Bouy}, {Bragaglia}, {Breddels}, {Brouillet}, {Br{\"u}semeister},
  {Bucciarelli}, {Budnik}, {Burgess}, {Burgon}, {Burlacu}, {Busonero}, {Buzzi},
  {Caffau}, {Cambras}, {Campbell}, {Cancelliere}, {Cantat-Gaudin}, {Carlucci},
  {Carrasco}, {Castellani}, {Charlot}, {Charnas}, {Charvet}, {Chassat},
  {Chiavassa}, {Clotet}, {Cocozza}, {Collins}, {Collins}, {Costigan}, {Crifo},
  {Cross}, {Crosta}, {Crowley}, {Dafonte}, {Damerdji}, {Dapergolas}, {David},
  {David}, {De Cat}, {de Felice}, {de Laverny}, {De Luise}, {De March}, {de
  Martino}, {de Souza}, {Debosscher}, {del Pozo}, {Delbo}, {Delgado},
  {Delgado}, {di Marco}, {Di Matteo}, {Diakite}, {Distefano}, {Dolding}, {Dos
  Anjos}, {Drazinos}, {Dur{\'a}n}, {Dzigan}, {Ecale}, {Edvardsson}, {Enke},
  {Erdmann}, {Escolar}, {Espina}, {Evans}, {Eynard Bontemps}, {Fabre},
  {Fabrizio}, {Faigler}, {Falc{\~a}o}, {Farr{\`a}s Casas}, {Faye}, {Federici},
  {Fedorets}, {Fern{\'a}ndez-Hern{\'a}ndez}, {Fernique}, {Fienga}, {Figueras},
  {Filippi}, {Findeisen}, {Fonti}, {Fouesneau}, {Fraile}, {Fraser}, {Fuchs},
  {Furnell}, {Gai}, {Galleti}, {Galluccio}, {Garabato}, {Garc{\'\i}a-Sedano},
  {Gar{\'e}}, {Garofalo}, {Garralda}, {Gavras}, {Gerssen}, {Geyer}, {Gilmore},
  {Girona}, {Giuffrida}, {Gomes}, {Gonz{\'a}lez-Marcos},
  {Gonz{\'a}lez-N{\'u}{\~n}ez}, {Gonz{\'a}lez-Vidal}, {Granvik}, {Guerrier},
  {Guillout}, {Guiraud}, {G{\'u}rpide}, {Guti{\'e}rrez-S{\'a}nchez}, {Guy},
  {Haigron}, {Hatzidimitriou}, {Haywood}, {Heiter}, {Helmi}, {Hobbs},
  {Hofmann}, {Holl}, {Holland }, {Hunt}, {Hypki}, {Icardi}, {Irwin}, {Jevardat
  de Fombelle}, {Jofr{\'e}}, {Jonker}, {Jorissen}, {Julbe}, {Karampelas},
  {Kochoska}, {Kohley}, {Kolenberg}, {Kontizas}, {Koposov}, {Kordopatis},
  {Koubsky}, {Kowalczyk}, {Krone-Martins}, {Kudryashova}, {Kull}, {Bachchan},
  {Lacoste-Seris}, {Lanza}, {Lavigne}, {Le Poncin-Lafitte}, {Lebreton},
  {Lebzelter}, {Leccia}, {Leclerc}, {Lecoeur-Taibi}, {Lemaitre}, {Lenhardt},
  {Leroux}, {Liao}, {Licata}, {Lindstr{\o}m}, {Lister}, {Livanou}, {Lobel},
  {L{\"o}ffler}, {L{\'o}pez}, {Lopez-Lozano}, {Lorenz}, {Loureiro},
  {MacDonald}, {Magalh{\~a}es Fernandes}, {Managau}, {Mann}, {Mantelet},
  {Marchal}, {Marchant}, {Marconi}, {Marie}, {Marinoni}, {Marrese},
  {Marschalk{\'o}}, {Marshall}, {Mart{\'\i}n-Fleitas}, {Martino}, {Mary},
  {Matijevi{\v{c}}}, {Mazeh}, {McMillan}, {Messina}, {Mestre}, {Michalik},
  {Millar}, {Miranda}, {Molina}, {Molinaro}, {Molinaro}, {Moln{\'a}r},
  {Moniez}, {Montegriffo}, {Monteiro}, {Mor}, {Mora}, {Morbidelli}, {Morel},
  {Morgenthaler}, {Morley}, {Morris}, {Mulone}, {Muraveva}, {Musella},
  {Narbonne}, {Nelemans}, {Nicastro}, {Noval}, {Ord{\'e}novic},
  {Ordieres-Mer{\'e}}, {Osborne}, {Pagani}, {Pagano}, {Pailler}, {Palacin},
  {Palaversa}, {Parsons}, {Paulsen}, {Pecoraro}, {Pedrosa}, {Pentik{\"a}inen},
  {Pereira}, {Pichon}, {Piersimoni}, {Pineau}, {Plachy}, {Plum}, {Poujoulet},
  {Pr{\v{s}}a}, {Pulone}, {Ragaini}, {Rago}, {Rambaux}, {Ramos-Lerate},
  {Ranalli}, {Rauw}, {Read}, {Regibo}, {Renk}, {Reyl{\'e}}, {Ribeiro},
  {Rimoldini}, {Ripepi}, {Riva}, {Rixon}, {Roelens}, {Romero-G{\'o}mez},
  {Rowell}, {Royer}, {Rudolph}, {Ruiz-Dern}, {Sadowski}, {Sagrist{\`a}
  Sell{\'e}s}, {Sahlmann}, {Salgado}, {Salguero}, {Sarasso}, {Savietto},
  {Schnorhk}, {Schultheis}, {Sciacca}, {Segol}, {Segovia}, {Segransan},
  {Serpell}, {Shih}, {Smareglia}, {Smart}, {Smith}, {Solano}, {Solitro},
  {Sordo}, {Soria Nieto}, {Souchay}, {Spagna}, {Spoto}, {Stampa}, {Steele},
  {Steidelm{\"u}ller}, {Stephenson}, {Stoev}, {Suess}, {S{\"u}veges}, {Surdej},
  {Szabados}, {Szegedi-Elek}, {Tapiador}, {Taris}, {Tauran}, {Taylor},
  {Teixeira}, {Terrett}, {Tingley}, {Trager}, {Turon}, {Ulla}, {Utrilla},
  {Valentini}, {van Elteren}, {Van Hemelryck}, {van Leeuwen}, {Varadi},
  {Vecchiato}, {Veljanoski}, {Via}, {Vicente}, {Vogt}, {Voss}, {Votruba},
  {Voutsinas}, {Walmsley}, {Weiler}, {Weingrill}, {Werner}, {Wevers},
  {Whitehead}, {Wyrzykowski}, {Yoldas}, {{\v{Z}}erjal}, {Zucker}, {Zurbach},
  {Zwitter}, {Alecu}, {Allen}, {Allende Prieto}, {Amorim},
  {Anglada-Escud{\'e}}, {Arsenijevic}, {Azaz}, {Balm}, {Beck}, {Bernstein},
  {Bigot}, {Bijaoui}, {Blasco}, {Bonfigli}, {Bono}, {Boudreault}, {Bressan},
  {Brown}, {Brunet}, {Bunclark}, {Buonanno}, {Butkevich}, {Carret}, {Carrion},
  {Chemin}, {Ch{\'e}reau}, {Corcione}, {Darmigny}, {de Boer}, {de Teodoro}, {de
  Zeeuw}, {Delle Luche}, {Domingues}, {Dubath}, {Fodor}, {Fr{\'e}zouls},
  {Fries}, {Fustes}, {Fyfe}, {Gallardo}, {Gallegos}, {Gardiol}, {Gebran},
  {Gomboc}, {G{\'o}mez}, {Grux}, {Gueguen}, {Heyrovsky}, {Hoar}, {Iannicola},
  {Isasi Parache}, {Janotto}, {Joliet}, {Jonckheere}, {Keil}, {Kim},
  {Klagyivik}, {Klar}, {Knude}, {Kochukhov}, {Kolka}, {Kos}, {Kutka}, {Lainey},
  {LeBouquin}, {Liu}, {Loreggia}, {Makarov}, {Marseille}, {Martayan},
  {Martinez-Rubi}, {Massart}, {Meynadier}, {Mignot}, {Munari}, {Nguyen},
  {Nordlander}, {Ocvirk}, {O'Flaherty}, {Olias Sanz}, {Ortiz}, {Osorio},
  {Oszkiewicz}, {Ouzounis}, {Palmer}, {Park}, {Pasquato}, {Peltzer}, {Peralta},
  {P{\'e}turaud}, {Pieniluoma}, {Pigozzi}, {Poels}, {Prat}, {Prod'homme},
  {Raison}, {Rebordao}, {Risquez}, {Rocca-Volmerange}, {Rosen}, {Ruiz-Fuertes},
  {Russo}, {Sembay}, {Serraller Vizcaino}, {Short}, {Siebert}, {Silva},
  {Sinachopoulos}, {Slezak}, {Soffel}, {Sosnowska}, {Strai{\v{z}}ys}, {ter
  Linden}, {Terrell}, {Theil}, {Tiede}, {Troisi}, {Tsalmantza}, {Tur},
  {Vaccari}, {Vachier}, {Valles}, {Van Hamme}, {Veltz}, {Virtanen}, {Wallut},
  {Wichmann}, {Wilkinson}, {Ziaeepour}, \& {Zschocke}}]{gaia}
{Gaia Collaboration}, {Prusti}, T., {de Bruijne}, J.~H.~J., {et~al.} 2016,
  \aap, 595, A1, \dodoi{10.1051/0004-6361/201629272}

\bibitem[{{Gaia Collaboration} {et~al.}(2018){Gaia Collaboration}, {Brown},
  {Vallenari}, {Prusti}, {de Bruijne}, {Babusiaux}, {Bailer-Jones}, {Biermann},
  {Evans}, {Eyer}, {Jansen}, {Jordi}, {Klioner}, {Lammers}, {Lindegren},
  {Luri}, {Mignard}, {Panem}, {Pourbaix}, {Randich}, {Sartoretti}, {Siddiqui},
  {Soubiran}, {van Leeuwen}, {Walton}, {Arenou}, {Bastian}, {Cropper},
  {Drimmel}, {Katz}, {Lattanzi}, {Bakker}, {Cacciari}, {Casta{\~n}eda},
  {Chaoul}, {Cheek}, {De Angeli}, {Fabricius}, {Guerra}, {Holl}, {Masana},
  {Messineo}, {Mowlavi}, {Nienartowicz}, {Panuzzo}, {Portell}, {Riello},
  {Seabroke}, {Tanga}, {Th{\'e}venin}, {Gracia-Abril}, {Comoretto},
  {Garcia-Reinaldos}, {Teyssier}, {Altmann}, {Andrae}, {Audard},
  {Bellas-Velidis}, {Benson}, {Berthier}, {Blomme}, {Burgess}, {Busso},
  {Carry}, {Cellino}, {Clementini}, {Clotet}, {Creevey}, {Davidson}, {De
  Ridder}, {Delchambre}, {Dell'Oro}, {Ducourant},
  {Fern{\'a}ndez-Hern{\'a}ndez}, {Fouesneau}, {Fr{\'e}mat}, {Galluccio},
  {Garc{\'\i}a-Torres}, {Gonz{\'a}lez-N{\'u}{\~n}ez}, {Gonz{\'a}lez-Vidal},
  {Gosset}, {Guy}, {Halbwachs}, {Hambly}, {Harrison}, {Hern{\'a}ndez},
  {Hestroffer}, {Hodgkin}, {Hutton}, {Jasniewicz}, {Jean-Antoine-Piccolo},
  {Jordan}, {Korn}, {Krone-Martins}, {Lanzafame}, {Lebzelter}, {L{\"o}ffler},
  {Manteiga}, {Marrese}, {Mart{\'\i}n-Fleitas}, {Moitinho}, {Mora}, {Muinonen},
  {Osinde}, {Pancino}, {Pauwels}, {Petit}, {Recio-Blanco}, {Richards},
  {Rimoldini}, {Robin}, {Sarro}, {Siopis}, {Smith}, {Sozzetti}, {S{\"u}veges},
  {Torra}, {van Reeven}, {Abbas}, {Abreu Aramburu}, {Accart}, {Aerts},
  {Altavilla}, {{\'A}lvarez}, {Alvarez}, {Alves}, {Anderson}, {Andrei},
  {Anglada Varela}, {Antiche}, {Antoja}, {Arcay}, {Astraatmadja}, {Bach},
  {Baker}, {Balaguer-N{\'u}{\~n}ez}, {Balm}, {Barache}, {Barata}, {Barbato},
  {Barblan}, {Barklem}, {Barrado}, {Barros}, {Barstow}, {Bartholom{\'e}
  Mu{\~n}oz}, {Bassilana}, {Becciani}, {Bellazzini}, {Berihuete}, {Bertone},
  {Bianchi}, {Bienaym{\'e}}, {Blanco-Cuaresma}, {Boch}, {Boeche}, {Bombrun},
  {Borrachero}, {Bossini}, {Bouquillon}, {Bourda}, {Bragaglia}, {Bramante},
  {Breddels}, {Bressan}, {Brouillet}, {Br{\"u}semeister}, {Brugaletta},
  {Bucciarelli}, {Burlacu}, {Busonero}, {Butkevich}, {Buzzi}, {Caffau},
  {Cancelliere}, {Cannizzaro}, {Cantat-Gaudin}, {Carballo}, {Carlucci},
  {Carrasco}, {Casamiquela}, {Castellani}, {Castro-Ginard}, {Charlot},
  {Chemin}, {Chiavassa}, {Cocozza}, {Costigan}, {Cowell}, {Crifo}, {Crosta},
  {Crowley}, {Cuypers}, {Dafonte}, {Damerdji}, {Dapergolas}, {David}, {David},
  {de Laverny}, {De Luise}, {De March}, {de Martino}, {de Souza}, {de Torres},
  {Debosscher}, {del Pozo}, {Delbo}, {Delgado}, {Delgado}, {Di Matteo},
  {Diakite}, {Diener}, {Distefano}, {Dolding}, {Drazinos}, {Dur{\'a}n},
  {Edvardsson}, {Enke}, {Eriksson}, {Esquej}, {Eynard Bontemps}, {Fabre},
  {Fabrizio}, {Faigler}, {Falc{\~a}o}, {Farr{\`a}s Casas}, {Federici},
  {Fedorets}, {Fernique}, {Figueras}, {Filippi}, {Findeisen}, {Fonti},
  {Fraile}, {Fraser}, {Fr{\'e}zouls}, {Gai}, {Galleti}, {Garabato},
  {Garc{\'\i}a-Sedano}, {Garofalo}, {Garralda}, {Gavel}, {Gavras}, {Gerssen},
  {Geyer}, {Giacobbe}, {Gilmore}, {Girona}, {Giuffrida}, {Glass}, {Gomes},
  {Granvik}, {Gueguen}, {Guerrier}, {Guiraud}, {Guti{\'e}rrez-S{\'a}nchez},
  {Haigron}, {Hatzidimitriou}, {Hauser}, {Haywood}, {Heiter}, {Helmi}, {Heu},
  {Hilger}, {Hobbs}, {Hofmann}, {Holland}, {Huckle}, {Hypki}, {Icardi},
  {Jan{\ss}en}, {Jevardat de Fombelle}, {Jonker}, {Juh{\'a}sz}, {Julbe},
  {Karampelas}, {Kewley}, {Klar}, {Kochoska}, {Kohley}, {Kolenberg},
  {Kontizas}, {Kontizas}, {Koposov}, {Kordopatis}, {Kostrzewa-Rutkowska},
  {Koubsky}, {Lambert}, {Lanza}, {Lasne}, {Lavigne}, {Le Fustec}, {Le
  Poncin-Lafitte}, {Lebreton}, {Leccia}, {Leclerc}, {Lecoeur-Taibi},
  {Lenhardt}, {Leroux}, {Liao}, {Licata}, {Lindstr{\o}m}, {Lister}, {Livanou},
  {Lobel}, {L{\'o}pez}, {Managau}, {Mann}, {Mantelet}, {Marchal}, {Marchant},
  {Marconi}, {Marinoni}, {Marschalk{\'o}}, {Marshall}, {Martino}, {Marton},
  {Mary}, {Massari}, {Matijevi{\v{c}}}, {Mazeh}, {McMillan}, {Messina},
  {Michalik}, {Millar}, {Molina}, {Molinaro}, {Moln{\'a}r}, {Montegriffo},
  {Mor}, {Morbidelli}, {Morel}, {Morris}, {Mulone}, {Muraveva}, {Musella},
  {Nelemans}, {Nicastro}, {Noval}, {O'Mullane}, {Ord{\'e}novic},
  {Ord{\'o}{\~n}ez-Blanco}, {Osborne}, {Pagani}, {Pagano}, {Pailler},
  {Palacin}, {Palaversa}, {Panahi}, {Pawlak}, {Piersimoni}, {Pineau}, {Plachy},
  {Plum}, {Poggio}, {Poujoulet}, {Pr{\v{s}}a}, {Pulone}, {Racero}, {Ragaini},
  {Rambaux}, {Ramos-Lerate}, {Regibo}, {Reyl{\'e}}, {Riclet}, {Ripepi}, {Riva},
  {Rivard}, {Rixon}, {Roegiers}, {Roelens}, {Romero-G{\'o}mez}, {Rowell},
  {Royer}, {Ruiz-Dern}, {Sadowski}, {Sagrist{\`a} Sell{\'e}s}, {Sahlmann},
  {Salgado}, {Salguero}, {Sanna}, {Santana-Ros}, {Sarasso}, {Savietto},
  {Schultheis}, {Sciacca}, {Segol}, {Segovia}, {S{\'e}gransan}, {Shih},
  {Siltala}, {Silva}, {Smart}, {Smith}, {Solano}, {Solitro}, {Sordo}, {Soria
  Nieto}, {Souchay}, {Spagna}, {Spoto}, {Stampa}, {Steele},
  {Steidelm{\"u}ller}, {Stephenson}, {Stoev}, {Suess}, {Surdej}, {Szabados},
  {Szegedi-Elek}, {Tapiador}, {Taris}, {Tauran}, {Taylor}, {Teixeira},
  {Terrett}, {Teyssand ier}, {Thuillot}, {Titarenko}, {Torra Clotet}, {Turon},
  {Ulla}, {Utrilla}, {Uzzi}, {Vaillant}, {Valentini}, {Valette}, {van Elteren},
  {Van Hemelryck}, {van Leeuwen}, {Vaschetto}, {Vecchiato}, {Veljanoski},
  {Viala}, {Vicente}, {Vogt}, {von Essen}, {Voss}, {Votruba}, {Voutsinas},
  {Walmsley}, {Weiler}, {Wertz}, {Wevers}, {Wyrzykowski}, {Yoldas},
  {{\v{Z}}erjal}, {Ziaeepour}, {Zorec}, {Zschocke}, {Zucker}, {Zurbach}, \&
  {Zwitter}}]{gaiadr2}
{Gaia Collaboration}, {Brown}, A.~G.~A., {Vallenari}, A., {et~al.} 2018, \aap,
  616, A1, \dodoi{10.1051/0004-6361/201833051}

\bibitem[{{Gaia Collaboration} {et~al.}(2022){Gaia Collaboration}, {Vallenari},
  {Brown}, {Prusti}, {de Bruijne}, {Arenou}, {Babusiaux}, {Biermann},
  {Creevey}, {Ducourant}, {Evans}, {Eyer}, {Guerra}, {Hutton}, {Jordi},
  {Klioner}, {Lammers}, {Lindegren}, {Luri}, {Mignard}, {Panem}, {Pourbaix},
  {Randich}, {Sartoretti}, {Soubiran}, {Tanga}, {Walton}, {Bailer-Jones},
  {Bastian}, {Drimmel}, {Jansen}, {Katz}, {Lattanzi}, {van Leeuwen}, {Bakker},
  {Cacciari}, {Casta{\~n}eda}, {De Angeli}, {Fabricius}, {Fouesneau},
  {Fr{\'e}mat}, {Galluccio}, {Guerrier}, {Heiter}, {Masana}, {Messineo},
  {Mowlavi}, {Nicolas}, {Nienartowicz}, {Pailler}, {Panuzzo}, {Riclet}, {Roux},
  {Seabroke}, {Sordo{\o}rcit}, {Th{\'e}venin}, {Gracia-Abril}, {Portell},
  {Teyssier}, {Altmann}, {Andrae}, {Audard}, {Bellas-Velidis}, {Benson},
  {Berthier}, {Blomme}, {Burgess}, {Busonero}, {Busso}, {C{\'a}novas}, {Carry},
  {Cellino}, {Cheek}, {Clementini}, {Damerdji}, {Davidson}, {de Teodoro},
  {Nu{\~n}ez Campos}, {Delchambre}, {Dell'Oro}, {Esquej},
  {Fern{\'a}ndez-Hern{\'a}ndez}, {Fraile}, {Garabato}, {Garc{\'\i}a-Lario},
  {Gosset}, {Haigron}, {Halbwachs}, {Hambly}, {Harrison}, {Hern{\'a}ndez},
  {Hestroffer}, {Hodgkin}, {Holl}, {Jan{\ss}en}, {Jevardat de Fombelle},
  {Jordan}, {Krone-Martins}, {Lanzafame}, {L{\"o}ffler}, {Marchal}, {Marrese},
  {Moitinho}, {Muinonen}, {Osborne}, {Pancino}, {Pauwels}, {Recio-Blanco},
  {Reyl{\'e}}, {Riello}, {Rimoldini}, {Roegiers}, {Rybizki}, {Sarro}, {Siopis},
  {Smith}, {Sozzetti}, {Utrilla}, {van Leeuwen}, {Abbas}, {{\'A}brah{\'a}m},
  {Abreu Aramburu}, {Aerts}, {Aguado}, {Ajaj}, {Aldea-Montero}, {Altavilla},
  {{\'A}lvarez}, {Alves}, {Anders}, {Anderson}, {Anglada Varela}, {Antoja},
  {Baines}, {Baker}, {Balaguer-N{\'u}{\~n}ez}, {Balbinot}, {Balog}, {Barache},
  {Barbato}, {Barros}, {Barstow}, {Bartolom{\'e}}, {Bassilana}, {Bauchet},
  {Becciani}, {Bellazzini}, {Berihuete}, {Bernet}, {Bertone}, {Bianchi},
  {Binnenfeld}, {Blanco-Cuaresma}, {Blazere}, {Boch}, {Bombrun}, {Bossini},
  {Bouquillon}, {Bragaglia}, {Bramante}, {Breedt}, {Bressan}, {Brouillet},
  {Brugaletta}, {Bucciarelli}, {Burlacu}, {Butkevich}, {Buzzi}, {Caffau},
  {Cancelliere}, {Cantat-Gaudin}, {Carballo}, {Carlucci}, {Carnerero},
  {Carrasco}, {Casamiquela}, {Castellani}, {Castro-Ginard}, {Chaoul},
  {Charlot}, {Chemin}, {Chiaramida}, {Chiavassa}, {Chornay}, {Comoretto},
  {Contursi}, {Cooper}, {Cornez}, {Cowell}, {Crifo}, {Cropper}, {Crosta},
  {Crowley}, {Dafonte}, {Dapergolas}, {David}, {David}, {de Laverny}, {De
  Luise}, {De March}, {De Ridder}, {de Souza}, {de Torres}, {del Peloso}, {del
  Pozo}, {Delbo}, {Delgado}, {Delisle}, {Demouchy}, {Dharmawardena}, {Di
  Matteo}, {Diakite}, {Diener}, {Distefano}, {Dolding}, {Edvardsson}, {Enke},
  {Fabre}, {Fabrizio}, {Faigler}, {Fedorets}, {Fernique}, {Fienga}, {Figueras},
  {Fournier}, {Fouron}, {Fragkoudi}, {Gai}, {Garcia-Gutierrez},
  {Garcia-Reinaldos}, {Garc{\'\i}a-Torres}, {Garofalo}, {Gavel}, {Gavras},
  {Gerlach}, {Geyer}, {Giacobbe}, {Gilmore}, {Girona}, {Giuffrida}, {Gomel},
  {Gomez}, {Gonz{\'a}lez-N{\'u}{\~n}ez}, {Gonz{\'a}lez-Santamar{\'\i}a},
  {Gonz{\'a}lez-Vidal}, {Granvik}, {Guillout}, {Guiraud},
  {Guti{\'e}rrez-S{\'a}nchez}, {Guy}, {Hatzidimitriou}, {Hauser}, {Haywood},
  {Helmer}, {Helmi}, {Sarmiento}, {Hidalgo}, {Hilger}, {H{\l}adczuk}, {Hobbs},
  {Holland}, {Huckle}, {Jardine}, {Jasniewicz}, {Jean-Antoine Piccolo},
  {Jim{\'e}nez-Arranz}, {Jorissen}, {Juaristi Campillo}, {Julbe}, {Karbevska},
  {Kervella}, {Khanna}, {Kontizas}, {Kordopatis}, {Korn}, {K{\'o}sp{\'a}l},
  {Kostrzewa-Rutkowska}, {Kruszy{\'n}ska}, {Kun}, {Laizeau}, {Lambert},
  {Lanza}, {Lasne}, {Le Campion}, {Lebreton}, {Lebzelter}, {Leccia}, {Leclerc},
  {Lecoeur-Taibi}, {Liao}, {Licata}, {Lindstr{\o}m}, {Lister}, {Livanou},
  {Lobel}, {Lorca}, {Loup}, {Madrero Pardo}, {Magdaleno Romeo}, {Managau},
  {Mann}, {Manteiga}, {Marchant}, {Marconi}, {Marcos}, {Marcos Santos},
  {Mar{\'\i}n Pina}, {Marinoni}, {Marocco}, {Marshall}, {Polo},
  {Mart{\'\i}n-Fleitas}, {Marton}, {Mary}, {Masip}, {Massari},
  {Mastrobuono-Battisti}, {Mazeh}, {McMillan}, {Messina}, {Michalik}, {Millar},
  {Mints}, {Molina}, {Molinaro}, {Moln{\'a}r}, {Monari}, {Mongui{\'o}},
  {Montegriffo}, {Montero}, {Mor}, {Mora}, {Morbidelli}, {Morel}, {Morris},
  {Muraveva}, {Murphy}, {Musella}, {Nagy}, {Noval}, {Oca{\~n}a}, {Ogden},
  {Ordenovic}, {Osinde}, {Pagani}, {Pagano}, {Palaversa}, {Palicio},
  {Pallas-Quintela}, {Panahi}, {Payne-Wardenaar}, {Pe{\~n}alosa Esteller},
  {Penttil{\"a}}, {Pichon}, {Piersimoni}, {Pineau}, {Plachy}, {Plum}, {Poggio},
  {Pr{\v{s}}a}, {Pulone}, {Racero}, {Ragaini}, {Rainer}, {Raiteri}, {Rambaux},
  {Ramos}, {Ramos-Lerate}, {Re Fiorentin}, {Regibo}, {Richards}, {Rios Diaz},
  {Ripepi}, {Riva}, {Rix}, {Rixon}, {Robichon}, {Robin}, {Robin}, {Roelens},
  {Rogues}, {Rohrbasser}, {Romero-G{\'o}mez}, {Rowell}, {Royer}, {Ruz Mieres},
  {Rybicki}, {Sadowski}, {S{\'a}ez N{\'u}{\~n}ez}, {Sagrist{\`a} Sell{\'e}s},
  {Sahlmann}, {Salguero}, {Samaras}, {Sanchez Gimenez}, {Sanna},
  {Santove{\~n}a}, {Sarasso}, {Schultheis}, {Sciacca}, {Segol}, {Segovia},
  {S{\'e}gransan}, {Semeux}, {Shahaf}, {Siddiqui}, {Siebert}, {Siltala},
  {Silvelo}, {Slezak}, {Slezak}, {Smart}, {Snaith}, {Solano}, {Solitro},
  {Souami}, {Souchay}, {Spagna}, {Spina}, {Spoto}, {Steele},
  {Steidelm{\"u}ller}, {Stephenson}, {S{\"u}veges}, {Surdej}, {Szabados},
  {Szegedi-Elek}, {Taris}, {Taylo}, {Teixeira}, {Tolomei}, {Tonello}, {Torra},
  {Torra}, {Torralba Elipe}, {Trabucchi}, {Tsounis}, {Turon}, {Ulla}, {Unger},
  {Vaillant}, {van Dillen}, {van Reeven}, {Vanel}, {Vecchiato}, {Viala},
  {Vicente}, {Voutsinas}, {Weiler}, {Wevers}, {Wyrzykowski}, {Yoldas}, {Yvard},
  {Zhao}, {Zorec}, {Zucker}, \& {Zwitter}}]{gaiadr3arxiv}
{Gaia Collaboration}, {Vallenari}, A., {Brown}, A.~G.~A., {et~al.} 2022, arXiv
  e-prints, arXiv:2208.00211.
\newblock \doarXiv{2208.00211}

\bibitem[{{Gao} {et~al.}(2021){Gao}, {Wakeford}, {Moran}, \&
  {Parmentier}}]{gao21}
{Gao}, P., {Wakeford}, H.~R., {Moran}, S.~E., \& {Parmentier}, V. 2021, Journal
  of Geophysical Research (Planets), 126, e06655, \dodoi{10.1029/2020JE006655}

\bibitem[{{Gelman} \& {Rubin}(1992)}]{gelman92}
{Gelman}, A., \& {Rubin}, D.~B. 1992, Statistical Science, 7, 457,
  \dodoi{10.1214/ss/1177011136}

\bibitem[{{Giacalone} {et~al.}(2021){Giacalone}, {Dressing}, {Jensen},
  {Collins}, {Ricker}, {Vanderspek}, {Seager}, {Winn}, {Jenkins}, {Barclay},
  {Barkaoui}, {Cadieux}, {Charbonneau}, {Collins}, {Conti}, {Doyon}, {Evans},
  {Ghachoui}, {Gillon}, {Guerrero}, {Hart}, {Jehin}, {Kielkopf}, {McLean},
  {Murgas}, {Palle}, {Parviainen}, {Pozuelos}, {Relles}, {Shporer}, {Socia},
  {Stockdale}, {Tan}, {Torres}, {Twicken}, {Waalkes}, \& {Waite}}]{giacalone21}
{Giacalone}, S., {Dressing}, C.~D., {Jensen}, E. L.~N., {et~al.} 2021, \aj,
  161, 24, \dodoi{10.3847/1538-3881/abc6af}

\bibitem[{{Goodman} \& {Weare}(2010)}]{goodman10}
{Goodman}, J., \& {Weare}, J. 2010, Communications in Applied Mathematics and
  Computational Science, 5, 65, \dodoi{10.2140/camcos.2010.5.65}

\bibitem[{{Green} {et~al.}(2019){Green}, {Schlafly}, {Zucker}, {Speagle}, \&
  {Finkbeiner}}]{green19}
{Green}, G.~M., {Schlafly}, E., {Zucker}, C., {Speagle}, J.~S., \&
  {Finkbeiner}, D. 2019, \apj, 887, 93, \dodoi{10.3847/1538-4357/ab5362}

\bibitem[{{Grimm} \& {McSween}(1993)}]{grimm93}
{Grimm}, R.~E., \& {McSween}, H.~Y. 1993, Science, 259, 653

\bibitem[{{Grunblatt} {et~al.}(2015){Grunblatt}, {Howard}, \&
  {Haywood}}]{grunblatt15}
{Grunblatt}, S.~K., {Howard}, A.~W., \& {Haywood}, R.~D. 2015, \apj, 808, 127,
  \dodoi{10.1088/0004-637X/808/2/127}

\bibitem[{{Guerrero} {et~al.}(2021){Guerrero}, {Seager}, {Huang}, {Vanderburg},
  {Garcia Soto}, {Mireles}, {Hesse}, {Fong}, {Glidden}, {Shporer}, {Latham},
  {Collins}, {Quinn}, {Burt}, {Dragomir}, {Crossfield}, {Vanderspek},
  {Fausnaugh}, {Burke}, {Ricker}, {Daylan}, {Essack}, {G{\"u}nther}, {Osborn},
  {Pepper}, {Rowden}, {Sha}, {Villanueva}, {Yahalomi}, {Yu}, {Ballard},
  {Batalha}, {Berardo}, {Chontos}, {Dittmann}, {Esquerdo}, {Mikal-Evans},
  {Jayaraman}, {Krishnamurthy}, {Louie}, {Mehrle}, {Niraula}, {Rackham},
  {Rodriguez}, {Rowden}, {Sousa-Silva}, {Watanabe}, {Wong}, {Zhan},
  {Zivanovic}, {Christiansen}, {Ciardi}, {Swain}, {Lund}, {Mullally},
  {Fleming}, {Rodriguez}, {Boyd}, {Quintana}, {Barclay}, {Col{\'o}n},
  {Rinehart}, {Schlieder}, {Clampin}, {Jenkins}, {Twicken}, {Caldwell},
  {Coughlin}, {Henze}, {Lissauer}, {Morris}, {Rose}, {Smith}, {Tenenbaum},
  {Ting}, {Wohler}, {Bakos}, {Bean}, {Berta-Thompson}, {Bieryla}, {Bouma},
  {Buchhave}, {Butler}, {Charbonneau}, {Doty}, {Ge}, {Holman}, {Howard},
  {Kaltenegger}, {Kane}, {Kjeldsen}, {Kreidberg}, {Lin}, {Minsky}, {Narita},
  {Paegert}, {P{\'a}l}, {Palle}, {Sasselov}, {Spencer}, {Sozzetti}, {Stassun},
  {Torres}, {Udry}, \& {Winn}}]{guerrero21}
{Guerrero}, N.~M., {Seager}, S., {Huang}, C.~X., {et~al.} 2021, \apjs, 254, 39,
  \dodoi{10.3847/1538-4365/abefe1}

\bibitem[{{Hadden} \& {Lithwick}(2017)}]{haddenLithwick17}
{Hadden}, S., \& {Lithwick}, Y. 2017, \aj, 154, 5,
  \dodoi{10.3847/1538-3881/aa71ef}

\bibitem[{Harris {et~al.}(2020)Harris, Millman, van~der Walt, Gommers,
  Virtanen, Cournapeau, Wieser, Taylor, Berg, Smith, Kern, Picus, Hoyer, van
  Kerkwijk, Brett, Haldane, del R{'{\i}}o, Wiebe, Peterson,
  G{'{e}}rard-Marchant, Sheppard, Reddy, Weckesser, Abbasi, Gohlke, \&
  Oliphant}]{numpy}
Harris, C.~R., Millman, K.~J., van~der Walt, S.~J., {et~al.} 2020, Nature, 585,
  357, \dodoi{10.1038/s41586-020-2649-2}

\bibitem[{{Harvey}(1985)}]{harvey85}
{Harvey}, J. 1985, in ESA Special Publication, Vol. 235, Future Missions in
  Solar, Heliospheric \& Space Plasma Physics, ed. E.~{Rolfe} \& B.~{Battrick},
  199

\bibitem[{{Hastings}(1970)}]{hastings70}
{Hastings}, W.~K. 1970, Biometrika, 57, 97, \dodoi{10.1093/biomet/57.1.97}

\bibitem[{{Hayward} {et~al.}(2001){Hayward}, {Brandl}, {Pirger}, {Blacken},
  {Gull}, {Schoenwald}, \& {Houck}}]{pharo}
{Hayward}, T.~L., {Brandl}, B., {Pirger}, B., {et~al.} 2001, \pasp, 113, 105,
  \dodoi{10.1086/317969}

\bibitem[{Haywood {et~al.}(2014)Haywood, Collier~Cameron, Queloz, Barros,
  Deleuil, Fares, Gillon, Lanza, Lovis, Moutou, Pepe, Pollacco, Santerne,
  Ségransan, \& Unruh}]{haywood14}
Haywood, R.~D., Collier~Cameron, A., Queloz, D., {et~al.} 2014, Monthly Notices
  of the Royal Astronomical Society, 443, 2517, \dodoi{10.1093/mnras/stu1320}

\bibitem[{{Haywood} {et~al.}(2018){Haywood}, {Vanderburg}, {Mortier}, {Giles},
  {L{\'o}pez-Morales}, {Lopez}, {Malavolta}, {Charbonneau}, {Collier Cameron},
  {Coughlin}, {Dressing}, {Nava}, {Latham}, {Dumusque}, {Lovis}, {Molinari},
  {Pepe}, {Sozzetti}, {Udry}, {Bouchy}, {Johnson}, {Mayor}, {Micela},
  {Phillips}, {Piotto}, {Rice}, {Sasselov}, {S{\'e}gransan}, {Watson}, {Affer},
  {Bonomo}, {Buchhave}, {Ciardi}, {Fiorenzano}, \& {Harutyunyan}}]{haywood18}
{Haywood}, R.~D., {Vanderburg}, A., {Mortier}, A., {et~al.} 2018, \aj, 155,
  203, \dodoi{10.3847/1538-3881/aab8f3}

\bibitem[{Hoffman \& Gelman(2014)}]{nuts:hoffman14}
Hoffman, M., \& Gelman, A. 2014, J. Mach. Learn. Res., 15, 1593

\bibitem[{{Houk} \& {Smith-Moore}(1988)}]{houk88}
{Houk}, N., \& {Smith-Moore}, M. 1988, {Michigan Catalogue of Two-dimensional
  Spectral Types for the HD Stars. Volume 4, Declinations -26{\textdegree}.0 to
  -12{\textdegree}.0.}, Vol.~4 (Univ. of Michigan)

\bibitem[{{Howard} \& {Fulton}(2016)}]{howard16}
{Howard}, A.~W., \& {Fulton}, B.~J. 2016, \pasp, 128, 114401,
  \dodoi{10.1088/1538-3873/128/969/114401}

\bibitem[{Howard {et~al.}(2010)Howard, Johnson, Marcy, Fischer, Wright, Bernat,
  Henry, Peek, Isaacson, Apps, Endl, Cochran, Valenti, Anderson, \&
  Piskunov}]{howard10}
Howard, A.~W., Johnson, J.~A., Marcy, G.~W., {et~al.} 2010, The Astrophysical
  Journal, 721, 1467, \dodoi{10.1088/0004-637x/721/2/1467}

\bibitem[{{Howard} {et~al.}(2012){Howard}, {Marcy}, {Bryson}, {Jenkins},
  {Rowe}, {Batalha}, {Borucki}, {Koch}, {Dunham}, {Gautier}, {Van Cleve},
  {Cochran}, {Latham}, {Lissauer}, {Torres}, {Brown}, {Gilliland}, {Buchhave},
  {Caldwell}, {Christensen-Dalsgaard}, {Ciardi}, {Fressin}, {Haas}, {Howell},
  {Kjeldsen}, {Seager}, {Rogers}, {Sasselov}, {Steffen}, {Basri},
  {Charbonneau}, {Christiansen}, {Clarke}, {Dupree}, {Fabrycky}, {Fischer},
  {Ford}, {Fortney}, {Tarter}, {Girouard}, {Holman}, {Johnson}, {Klaus},
  {Machalek}, {Moorhead}, {Morehead}, {Ragozzine}, {Tenenbaum}, {Twicken},
  {Quinn}, {Isaacson}, {Shporer}, {Lucas}, {Walkowicz}, {Welsh}, {Boss},
  {Devore}, {Gould}, {Smith}, {Morris}, {Prsa}, {Morton}, {Still}, {Thompson},
  {Mullally}, {Endl}, \& {MacQueen}}]{howard12}
{Howard}, A.~W., {Marcy}, G.~W., {Bryson}, S.~T., {et~al.} 2012, \apjs, 201,
  15, \dodoi{10.1088/0067-0049/201/2/15}

\bibitem[{{Huber} {et~al.}(2017){Huber}, {Zinn}, {Bojsen-Hansen},
  {Pinsonneault}, {Sahlholdt}, {Serenelli}, {Silva Aguirre}, {Stassun},
  {Stello}, {Tayar}, {Bastien}, {Bedding}, {Buchhave}, {Chaplin}, {Davies},
  {Garc{\'\i}a}, {Latham}, {Mathur}, {Mosser}, \& {Sharma}}]{huber17}
{Huber}, D., {Zinn}, J., {Bojsen-Hansen}, M., {et~al.} 2017, \apj, 844, 102,
  \dodoi{10.3847/1538-4357/aa75ca}

\bibitem[{Hunter(2007)}]{matplotlib}
Hunter, J.~D. 2007, Computing in Science \& Engineering, 9, 90,
  \dodoi{10.1109/MCSE.2007.55}

\bibitem[{{Isaacson} \& {Fischer}(2010)}]{isaacson10}
{Isaacson}, H., \& {Fischer}, D. 2010, \apj, 725, 875,
  \dodoi{10.1088/0004-637X/725/1/875}

\bibitem[{{Jeffreys}(1946)}]{jeffreys46}
{Jeffreys}, H. 1946, Proceedings of the Royal Society of London Series A, 186,
  453, \dodoi{10.1098/rspa.1946.0056}

\bibitem[{{Jenkins}(2002)}]{jenkins02}
{Jenkins}, J.~M. 2002, \apj, 575, 493, \dodoi{10.1086/341136}

\bibitem[{{Jenkins} {et~al.}(2020){Jenkins}, {Tenenbaum}, {Seader}, {Burke},
  {McCauliff}, {Smith}, {Twicken}, \& {Chandrasekaran}}]{jenkins20}
{Jenkins}, J.~M., {Tenenbaum}, P., {Seader}, S., {et~al.} 2020, {Kepler Data
  Processing Handbook: Transiting Planet Search}, Kepler Science Document
  KSCI-19081-003, id. 9. Edited by Jon M. Jenkins.

\bibitem[{{Jenkins} {et~al.}(2010){Jenkins}, {Chandrasekaran}, {McCauliff},
  {Caldwell}, {Tenenbaum}, {Li}, {Klaus}, {Cote}, \& {Middour}}]{jenkins10}
{Jenkins}, J.~M., {Chandrasekaran}, H., {McCauliff}, S.~D., {et~al.} 2010, in
  Society of Photo-Optical Instrumentation Engineers (SPIE) Conference Series,
  Vol. 7740, Software and Cyberinfrastructure for Astronomy, ed. N.~M.
  {Radziwill} \& A.~{Bridger}, 77400D, \dodoi{10.1117/12.856764}

\bibitem[{{Jenkins} {et~al.}(2016){Jenkins}, {Twicken}, {McCauliff},
  {Campbell}, {Sanderfer}, {Lung}, {Mansouri-Samani}, {Girouard}, {Tenenbaum},
  {Klaus}, {Smith}, {Caldwell}, {Chacon}, {Henze}, {Heiges}, {Latham},
  {Morgan}, {Swade}, {Rinehart}, \& {Vanderspek}}]{jenkins16}
{Jenkins}, J.~M., {Twicken}, J.~D., {McCauliff}, S., {et~al.} 2016, in Society
  of Photo-Optical Instrumentation Engineers (SPIE) Conference Series, Vol.
  9913, Software and Cyberinfrastructure for Astronomy IV, ed. G.~{Chiozzi} \&
  J.~C. {Guzman}, 99133E, \dodoi{10.1117/12.2233418}

\bibitem[{{Kallinger} {et~al.}(2014){Kallinger}, {De Ridder}, {Hekker},
  {Mathur}, {Mosser}, {Gruberbauer}, {Garc{\'\i}a}, {Karoff}, \&
  {Ballot}}]{kallinger14}
{Kallinger}, T., {De Ridder}, J., {Hekker}, S., {et~al.} 2014, \aap, 570, A41,
  \dodoi{10.1051/0004-6361/201424313}

\bibitem[{{Kempton} {et~al.}(2018){Kempton}, {Bean}, {Louie}, {Deming}, {Koll},
  {Mansfield}, {Christiansen}, {L{\'o}pez-Morales}, {Swain}, {Zellem},
  {Ballard}, {Barclay}, {Barstow}, {Batalha}, {Beatty}, {Berta-Thompson},
  {Birkby}, {Buchhave}, {Charbonneau}, {Cowan}, {Crossfield}, {de Val-Borro},
  {Doyon}, {Dragomir}, {Gaidos}, {Heng}, {Hu}, {Kane}, {Kreidberg}, {Mallonn},
  {Morley}, {Narita}, {Nascimbeni}, {Pall{\'e}}, {Quintana}, {Rauscher},
  {Seager}, {Shkolnik}, {Sing}, {Sozzetti}, {Stassun}, {Valenti}, \& {von
  Essen}}]{kempton18}
{Kempton}, E. M.~R., {Bean}, J.~L., {Louie}, D.~R., {et~al.} 2018, \pasp, 130,
  114401, \dodoi{10.1088/1538-3873/aadf6f}

\bibitem[{{Kipping}(2013)}]{exoplanet:kipping13}
{Kipping}, D.~M. 2013, \mnras, 435, 2152, \dodoi{10.1093/mnras/stt1435}

\bibitem[{{Kite} {et~al.}(2019){Kite}, {Fegley}, {Schaefer}, \&
  {Ford}}]{kite19}
{Kite}, E.~S., {Fegley}, Bruce, J., {Schaefer}, L., \& {Ford}, E.~B. 2019,
  \apjl, 887, L33, \dodoi{10.3847/2041-8213/ab59d9}

\bibitem[{{Kite} {et~al.}(2020){Kite}, {Fegley}, {Schaefer}, \&
  {Ford}}]{kite20}
---. 2020, \apj, 891, 111, \dodoi{10.3847/1538-4357/ab6ffb}

\bibitem[{{Kosiarek} \& {Crossfield}(2020)}]{kosiarek20}
{Kosiarek}, M.~R., \& {Crossfield}, I. J.~M. 2020, \aj, 159, 271,
  \dodoi{10.3847/1538-3881/ab8d3a}

\bibitem[{{Kosiarek} {et~al.}(2019){Kosiarek}, {Crossfield},
  {Hardegree-Ullman}, {Livingston}, {Benneke}, {Henry}, {Howard}, {Berardo},
  {Blunt}, {Fulton}, {Hirsch}, {Howard}, {Isaacson}, {Petigura}, {Sinukoff},
  {Weiss}, {Bonfils}, {Dressing}, {Knutson}, {Schlieder}, {Werner}, {Gorjian},
  {Krick}, {Morales}, {Astudillo-Defru}, {Almenara}, {Delfosse}, {Forveille},
  {Lovis}, {Mayor}, {Murgas}, {Pepe}, {Santos}, {Udry}, {Corbett}, {Fors},
  {Law}, {Ratzloff}, \& {del Ser}}]{kosiarek19}
{Kosiarek}, M.~R., {Crossfield}, I. J.~M., {Hardegree-Ullman}, K.~K., {et~al.}
  2019, \aj, 157, 97, \dodoi{10.3847/1538-3881/aaf79c}

\bibitem[{{Kosiarek} {et~al.}(2021){Kosiarek}, {Berardo}, {Crossfield},
  {Laguna}, {Piaulet}, {Akana Murphy}, {Howell}, {Henry}, {Isaacson}, {Fulton},
  {Weiss}, {Petigura}, {Behmard}, {Hirsch}, {Teske}, {Burt}, {Mills},
  {Chontos}, {Mo{\v{c}}nik}, {Howard}, {Werner}, {Livingston}, {Krick},
  {Beichman}, {Gorjian}, {Kreidberg}, {Morley}, {Christiansen}, {Morales},
  {Scott}, {Crane}, {Wang}, {Shectman}, {Rosenthal}, {Grunblatt}, {Rubenzahl},
  {Dalba}, {Giacalone}, {Villanueva}, {Liu}, {Dai}, {Hill}, {Rice}, {Kane}, \&
  {Mayo}}]{kosiarek21}
{Kosiarek}, M.~R., {Berardo}, D.~A., {Crossfield}, I. J.~M., {et~al.} 2021,
  \aj, 161, 47, \dodoi{10.3847/1538-3881/abca39}

\bibitem[{{Kov{\'a}cs} {et~al.}(2002){Kov{\'a}cs}, {Zucker}, \&
  {Mazeh}}]{kovacs02}
{Kov{\'a}cs}, G., {Zucker}, S., \& {Mazeh}, T. 2002, \aap, 391, 369,
  \dodoi{10.1051/0004-6361:20020802}

\bibitem[{{Li} {et~al.}(2019){Li}, {Tenenbaum}, {Twicken}, {Burke}, {Jenkins},
  {Quintana}, {Rowe}, \& {Seader}}]{li19}
{Li}, J., {Tenenbaum}, P., {Twicken}, J.~D., {et~al.} 2019, \pasp, 131, 024506,
  \dodoi{10.1088/1538-3873/aaf44d}

\bibitem[{{Lightkurve Collaboration} {et~al.}(2018){Lightkurve Collaboration},
  {Cardoso}, {Hedges}, {Gully-Santiago}, {Saunders}, {Cody}, {Barclay}, {Hall},
  {Sagear}, {Turtelboom}, {Zhang}, {Tzanidakis}, {Mighell}, {Coughlin}, {Bell},
  {Berta-Thompson}, {Williams}, {Dotson}, \& {Barentsen}}]{lightkurve}
{Lightkurve Collaboration}, {Cardoso}, J.~V.~d.~M., {Hedges}, C., {et~al.}
  2018, {Lightkurve: Kepler and TESS time series analysis in Python},
  Astrophysics Source Code Library.
\newblock \doeprint{1812.013}

\bibitem[{{Lindegren} {et~al.}(1997){Lindegren}, {Mignard}, {S{\"o}derhjelm},
  {Badiali}, {Bernstein}, {Lampens}, {Pannunzio}, {Arenou}, {Bernacca},
  {Falin}, {Froeschl{\'e}}, {Kovalevsky}, {Martin}, {Perryman}, \&
  {Wielen}}]{hipparcos}
{Lindegren}, L., {Mignard}, F., {S{\"o}derhjelm}, S., {et~al.} 1997, \aap, 323,
  L53

\bibitem[{{Lissauer} {et~al.}(2012){Lissauer}, {Marcy}, {Rowe}, {Bryson},
  {Adams}, {Buchhave}, {Ciardi}, {Cochran}, {Fabrycky}, {Ford}, {Fressin},
  {Geary}, {Gilliland}, {Holman}, {Howell}, {Jenkins}, {Kinemuchi}, {Koch},
  {Morehead}, {Ragozzine}, {Seader}, {Tanenbaum}, {Torres}, \&
  {Twicken}}]{lissauer12}
{Lissauer}, J.~J., {Marcy}, G.~W., {Rowe}, J.~F., {et~al.} 2012, \apj, 750,
  112, \dodoi{10.1088/0004-637X/750/2/112}

\bibitem[{{Lomb}(1976)}]{lomb76}
{Lomb}, N.~R. 1976, \apss, 39, 447, \dodoi{10.1007/BF00648343}

\bibitem[{{Lopez} \& {Fortney}(2014)}]{lopezforney14}
{Lopez}, E.~D., \& {Fortney}, J.~J. 2014, \apj, 792, 1,
  \dodoi{10.1088/0004-637X/792/1/1}

\bibitem[{{Lopez} {et~al.}(2012){Lopez}, {Fortney}, \& {Miller}}]{lopez12}
{Lopez}, E.~D., {Fortney}, J.~J., \& {Miller}, N. 2012, \apj, 761, 59,
  \dodoi{10.1088/0004-637X/761/1/59}

\bibitem[{{Lubin} {et~al.}(2022){Lubin}, {Van Zandt}, {Holcomb}, {Weiss},
  {Petigura}, {Robertson}, {Akana Murphy}, {Scarsdale}, {Batygin}, {Polanski},
  {Batalha}, {Crossfield}, {Dressing}, {Fulton}, {Howard}, {Huber}, {Isaacson},
  {Kane}, {Roy}, {Beard}, {Blunt}, {Chontos}, {Dai}, {Dalba}, {Gary},
  {Giacalone}, {Hill}, {Mayo}, {Mo{\v{c}}nik}, {Kosiarek}, {Rice}, {Rubenzahl},
  {Latham}, {Seager}, {Winn}, \& {Gary}}]{lubin22}
{Lubin}, J., {Van Zandt}, J., {Holcomb}, R., {et~al.} 2022, \aj, 163, 101,
  \dodoi{10.3847/1538-3881/ac3d38}

\bibitem[{{Luger} {et~al.}(2019){Luger}, {Agol}, {Foreman-Mackey}, {Fleming},
  {Lustig-Yaeger}, \& {Deitrick}}]{starry}
{Luger}, R., {Agol}, E., {Foreman-Mackey}, D., {et~al.} 2019, \aj, 157, 64,
  \dodoi{10.3847/1538-3881/aae8e5}

\bibitem[{{Lundkvist} {et~al.}(2016){Lundkvist}, {Kjeldsen}, {Albrecht},
  {Davies}, {Basu}, {Huber}, {Justesen}, {Karoff}, {Silva Aguirre}, {van
  Eylen}, {Vang}, {Arentoft}, {Barclay}, {Bedding}, {Campante}, {Chaplin},
  {Christensen-Dalsgaard}, {Elsworth}, {Gilliland}, {Handberg}, {Hekker},
  {Kawaler}, {Lund}, {Metcalfe}, {Miglio}, {Rowe}, {Stello}, {Tingley}, \&
  {White}}]{lundkvist16}
{Lundkvist}, M.~S., {Kjeldsen}, H., {Albrecht}, S., {et~al.} 2016, Nature
  Communications, 7, 11201, \dodoi{10.1038/ncomms11201}

\bibitem[{{Luque} \& {Pall{\'e}}(2022)}]{luque22}
{Luque}, R., \& {Pall{\'e}}, E. 2022, Science, 377, 1211,
  \dodoi{10.1126/science.abl7164}

\bibitem[{{MacDougall} {et~al.}(2021){MacDougall}, {Petigura}, {Angelo},
  {Lubin}, {Batalha}, {Beard}, {Behmard}, {Blunt}, {Brinkman}, {Chontos},
  {Crossfield}, {Dai}, {Dalba}, {Dressing}, {Fulton}, {Giacalone}, {Hill},
  {Howard}, {Huber}, {Isaacson}, {Kane}, {Mayo}, {Mo{\v{c}}nik}, {Akana
  Murphy}, {Polanski}, {Rice}, {Robertson}, {Rosenthal}, {Roy}, {Rubenzahl},
  {Scarsdale}, {Turtelboom}, {Zandt}, {Weiss}, {Matthews}, {Jenkins}, {Latham},
  {Ricker}, {Seager}, {Vanderspek}, {Winn}, {Brasseur}, {Doty}, {Fausnaugh},
  {Guerrero}, {Henze}, {Lund}, \& {Shporer}}]{macdougall21}
{MacDougall}, M.~G., {Petigura}, E.~A., {Angelo}, I., {et~al.} 2021, \aj, 162,
  265, \dodoi{10.3847/1538-3881/ac295e}

\bibitem[{{Madhusudhan}(2019)}]{madhusudhan19}
{Madhusudhan}, N. 2019, \araa, 57, 617,
  \dodoi{10.1146/annurev-astro-081817-051846}

\bibitem[{{Mamajek} \& {Hillenbrand}(2008)}]{mamajek08}
{Mamajek}, E.~E., \& {Hillenbrand}, L.~A. 2008, \apj, 687, 1264,
  \dodoi{10.1086/591785}

\bibitem[{{Marshall} {et~al.}(2006){Marshall}, {Robin}, {Reyl{\'e}},
  {Schultheis}, \& {Picaud}}]{marshall06}
{Marshall}, D.~J., {Robin}, A.~C., {Reyl{\'e}}, C., {Schultheis}, M., \&
  {Picaud}, S. 2006, \aap, 453, 635, \dodoi{10.1051/0004-6361:20053842}

\bibitem[{{Metropolis} {et~al.}(1953){Metropolis}, {Rosenbluth}, {Rosenbluth},
  {Teller}, \& {Teller}}]{metropolis53}
{Metropolis}, N., {Rosenbluth}, A.~W., {Rosenbluth}, M.~N., {Teller}, A.~H., \&
  {Teller}, E. 1953, \jcp, 21, 1087, \dodoi{10.1063/1.1699114}

\bibitem[{{Meunier} {et~al.}(2010){Meunier}, {Desort}, \&
  {Lagrange}}]{meunier10}
{Meunier}, N., {Desort}, M., \& {Lagrange}, A.~M. 2010, \aap, 512, A39,
  \dodoi{10.1051/0004-6361/200913551}

\bibitem[{{Middelkoop}(1982)}]{middelkoop82}
{Middelkoop}, F. 1982, \aap, 107, 31

\bibitem[{{Monteux} {et~al.}(2018){Monteux}, {Golabek}, {Rubie}, {Tobie}, \&
  {Young}}]{monteux18}
{Monteux}, J., {Golabek}, G.~J., {Rubie}, D.~C., {Tobie}, G., \& {Young}, E.~D.
  2018, \ssr, 214, 39, \dodoi{10.1007/s11214-018-0473-x}

\bibitem[{{Morris} {et~al.}(2020){Morris}, {Twicken}, {Smith}, {Clarke},
  {Jenkins}, {Bryson}, {Girouard}, \& {Klaus}}]{morris20}
{Morris}, R.~L., {Twicken}, J.~D., {Smith}, J.~C., {et~al.} 2020, {Kepler Data
  Processing Handbook: Photometric Analysis}, Kepler Science Document
  KSCI-19081-003, id. 6. Edited by Jon M. Jenkins.

\bibitem[{{Morton} {et~al.}(2016){Morton}, {Bryson}, {Coughlin}, {Rowe},
  {Ravichandran}, {Petigura}, {Haas}, \& {Batalha}}]{morton16}
{Morton}, T.~D., {Bryson}, S.~T., {Coughlin}, J.~L., {et~al.} 2016, \apj, 822,
  86, \dodoi{10.3847/0004-637X/822/2/86}

\bibitem[{{NASA Exoplanet Archive}(2022)}]{NEA}
{NASA Exoplanet Archive}. 2022, Planetary Systems, Version: 2022-11-17,
  NExScI-Caltech/IPAC, \dodoi{10.26133/NEA12}

\bibitem[{Neal(2003)}]{neal03}
Neal, R.~M. 2003, The Annals of Statistics, 31, 705–767,
  \dodoi{10.1214/aos/1056562461}

\bibitem[{{Neal}(2012)}]{neal12}
{Neal}, R.~M. 2012, arXiv e-prints, arXiv:1206.1901.
\newblock \doarXiv{1206.1901}

\bibitem[{{Noyes} {et~al.}(1984){Noyes}, {Hartmann}, {Baliunas}, {Duncan}, \&
  {Vaughan}}]{noyes84}
{Noyes}, R.~W., {Hartmann}, L.~W., {Baliunas}, S.~L., {Duncan}, D.~K., \&
  {Vaughan}, A.~H. 1984, \apj, 279, 763, \dodoi{10.1086/161945}

\bibitem[{{Orell-Miquel} {et~al.}(2023){Orell-Miquel}, {Nowak}, {Murgas},
  {Palle}, {Morello}, {Luque}, {Badenas-Agusti}, {Ribas}, {Lafarga},
  {Espinoza}, {Morales}, {Zechmeister}, {Alqasim}, {Cochran}, {Gandolfi},
  {Goffo}, {Kab{\'a}th}, {Korth}, {Lam}, {Livingston}, {Muresan}, {Persson}, \&
  {Van Eylen}}]{orell-miquel23}
{Orell-Miquel}, J., {Nowak}, G., {Murgas}, F., {et~al.} 2023, \aap, 669, A40,
  \dodoi{10.1051/0004-6361/202244120}

\bibitem[{{Osborn} {et~al.}(2022){Osborn}, {Bonfanti}, {Gandolfi}, {Hedges},
  {Leleu}, {Fortier}, {Futyan}, {Gutermann}, {Maxted}, {Borsato}, {Collins},
  {Gomes da Silva}, {G{\'o}mez Maqueo Chew}, {Hooton}, {Lendl}, {Parviainen},
  {Salmon}, {Schanche}, {Serrano}, {Sousa}, {Tuson}, {Ulmer-Moll}, {Van
  Grootel}, {Wells}, {Wilson}, {Alibert}, {Alonso}, {Anglada}, {Asquier},
  {Barrado y Navascues}, {Baumjohann}, {Beck}, {Benz}, {Biondi}, {Bonfils},
  {Bouchy}, {Brandeker}, {Broeg}, {B{\'a}rczy}, {Barros}, {Cabrera}, {Charnoz},
  {Collier Cameron}, {Csizmadia}, {Davies}, {Deleuil}, {Delrez}, {Demory},
  {Ehrenreich}, {Erikson}, {Fossati}, {Fridlund}, {Gillon}, {G{\"o}mez-Munoz},
  {G{\"u}del}, {Heng}, {Hoyer}, {Isaak}, {Kiss}, {Laskar}, {Lecavelier des
  Etangs}, {Lovis}, {Magrin}, {Malavolta}, {McCormac}, {Nascimbeni},
  {Olofsson}, {Ottensamer}, {Pagano}, {Pall{\'e}}, {Peter}, {Piazza}, {Piotto},
  {Pollacco}, {Queloz}, {Ragazzoni}, {Rando}, {Rauer}, {Reimers}, {Ribas},
  {Demangeon}, {Smith}, {Sabin}, {Santos}, {Scandariato}, {Schroffenegger},
  {Schwarz}, {Shporer}, {Simon}, {Steller}, {Szab{\'o}}, {S{\'e}gransan},
  {Thomas}, {Udry}, {Walter}, \& {Walton}}]{osborn22}
{Osborn}, H.~P., {Bonfanti}, A., {Gandolfi}, D., {et~al.} 2022, \aap, 664,
  A156, \dodoi{10.1051/0004-6361/202243065}

\bibitem[{{Osborn} {et~al.}(2023){Osborn}, {Nowak}, {H{\'e}brard}, {Masseron},
  {Lillo-Box}, {Pall{\'e}}, {Bekkelien}, {Flor{\'e}n}, {Guterman}, {Simon},
  {Adibekyan}, {Bieryla}, {Borsato}, {Brandeker}, {Ciardi}, {Collier Cameron},
  {Collins}, {Egger}, {Gandolfi}, {Hooton}, {Latham}, {Lendl}, {Matthews},
  {Tuson}, {Ulmer-Moll}, {Vanderburg}, {Wilson}, {Ziegler}, {Alibert},
  {Alonso}, {Anglada}, {Arnold}, {Asquier}, {Barrado y Navascues},
  {Baumjohann}, {Beck}, {Belinski}, {Benz}, {Biondi}, {Boisse}, {Bonfils},
  {Broeg}, {Buchhave}, {B{\'a}rczy}, {Barros}, {Cabrera}, {Cardona Guillen},
  {Carleo}, {Castro-Gonz{\'a}lez}, {Charnoz}, {Christiansen}, {Cortes-Zuleta},
  {Csizmadia}, {Dalal}, {Davies}, {Deleuil}, {Delfosse}, {Delrez}, {Demory},
  {Dunlavey}, {Ehrenreich}, {Erikson}, {Fernandes}, {Fortier}, {Forveille},
  {Fossati}, {Fridlund}, {Gillon}, {Goeke}, {Goliguzova}, {Gonzales},
  {G{\"u}nther}, {G{\"u}del}, {Heidari}, {Henze}, {Howell}, {Hoyer}, {Immanuel
  Frey}, {Isaak}, {Jenkins}, {Kiefer}, {Kiss}, {Korth}, {Maxted}, {Laskar},
  {Lecavelier des Etangs}, {Lovis}, {Lund}, {Luque}, {Magrin}, {Almenara},
  {Martioli}, {Mecina}, {Medina}, {Moldovan}, {Morales-Calder{\'o}n},
  {Morello}, {Moutou}, {Murgas}, {Jensen}, {Nascimbeni}, {Olofsson},
  {Ottensamer}, {Pagano}, {Peter}, {Piotto}, {Pollacco}, {Queloz}, {Ragazzoni},
  {Rando}, {Rauer}, {Ribas}, {Ricker}, {Demangeon}, {Smith}, {Santos},
  {Scandariato}, {Seager}, {Sousa}, {Steller}, {Szab{\'o}}, {S{\'e}gransan},
  {Thomas}, {Udry}, {Ulmer}, {Van Grootel}, {Vanderspek}, {Walton}, \&
  {Winn}}]{osborn23arxiv}
{Osborn}, H.~P., {Nowak}, G., {H{\'e}brard}, G., {et~al.} 2023, arXiv e-prints,
  arXiv:2306.04450, \dodoi{10.48550/arXiv.2306.04450}

\bibitem[{{Otegi} {et~al.}(2020){Otegi}, {Dorn}, {Helled}, {Bouchy},
  {Haldemann}, \& {Alibert}}]{otegi20b}
{Otegi}, J.~F., {Dorn}, C., {Helled}, R., {et~al.} 2020, \aap, 640, A135,
  \dodoi{10.1051/0004-6361/202038006}

\bibitem[{pandas~development team(2020)}]{pandas}
pandas~development team, T. 2020, pandas-dev/pandas: Pandas, latest,  Zenodo,
  \dodoi{10.5281/zenodo.3509134}

\bibitem[{Pedregosa {et~al.}(2011)Pedregosa, Varoquaux, Gramfort, Michel,
  Thirion, Grisel, Blondel, Prettenhofer, Weiss, Dubourg, Vanderplas, Passos,
  Cournapeau, Brucher, Perrot, \& Duchesnay}]{sklearn}
Pedregosa, F., Varoquaux, G., Gramfort, A., {et~al.} 2011, Journal of Machine
  Learning Research, 12, 2825

\bibitem[{{Perruchot} {et~al.}(2008){Perruchot}, {Kohler}, {Bouchy}, {Richaud},
  {Richaud}, {Moreaux}, {Merzougui}, {Sottile}, {Hill}, {Knispel}, {Regal},
  {Meunier}, {Ilovaisky}, {Le Coroller}, {Gillet}, {Schmitt}, {Pepe}, {Fleury},
  {Sosnowska}, {Vors}, {M{\'e}gevand}, {Blanc}, {Carol}, {Point}, {Laloge}, \&
  {Brunel}}]{sophie}
{Perruchot}, S., {Kohler}, D., {Bouchy}, F., {et~al.} 2008, in Society of
  Photo-Optical Instrumentation Engineers (SPIE) Conference Series, Vol. 7014,
  Ground-based and Airborne Instrumentation for Astronomy II, ed. I.~S.
  {McLean} \& M.~M. {Casali}, 70140J, \dodoi{10.1117/12.787379}

\bibitem[{{Petigura}(2020)}]{petigura20}
{Petigura}, E.~A. 2020, \aj, 160, 89, \dodoi{10.3847/1538-3881/ab9fff}

\bibitem[{{Petigura} {et~al.}(2013){Petigura}, {Howard}, \&
  {Marcy}}]{petigura13}
{Petigura}, E.~A., {Howard}, A.~W., \& {Marcy}, G.~W. 2013, Proceedings of the
  National Academy of Science, 110, 19273, \dodoi{10.1073/pnas.1319909110}

\bibitem[{Petigura {et~al.}(2017)Petigura, Howard, Marcy, Johnson, Isaacson,
  Cargile, Hebb, Fulton, Weiss, Morton, Winn, Rogers, Sinukoff, Hirsch, \&
  Crossfield}]{specmatchsynth}
Petigura, E.~A., Howard, A.~W., Marcy, G.~W., {et~al.} 2017, The Astronomical
  Journal, 154, 107, \dodoi{10.3847/1538-3881/aa80de}

\bibitem[{{Piaulet} {et~al.}(2021){Piaulet}, {Benneke}, {Rubenzahl}, {Howard},
  {Lee}, {Thorngren}, {Angus}, {Peterson}, {Schlieder}, {Werner}, {Kreidberg},
  {Jaouni}, {Crossfield}, {Ciardi}, {Petigura}, {Livingston}, {Dressing},
  {Fulton}, {Beichman}, {Christiansen}, {Gorjian}, {Hardegree-Ullman}, {Krick},
  \& {Sinukoff}}]{piaulet21}
{Piaulet}, C., {Benneke}, B., {Rubenzahl}, R.~A., {et~al.} 2021, \aj, 161, 70,
  \dodoi{10.3847/1538-3881/abcd3c}

\bibitem[{{Plavchan} {et~al.}(2015){Plavchan}, {Latham}, {Gaudi}, {Crepp},
  {Dumusque}, {Furesz}, {Vanderburg}, {Blake}, {Fischer}, {Prato}, {White},
  {Makarov}, {Marcy}, {Stapelfeldt}, {Haywood}, {Collier-Cameron},
  {Quirrenbach}, {Mahadevan}, {Anglada}, \& {Muirhead}}]{plavchan15}
{Plavchan}, P., {Latham}, D., {Gaudi}, S., {et~al.} 2015, arXiv e-prints,
  arXiv:1503.01770.
\newblock \doarXiv{1503.01770}

\bibitem[{{Press} {et~al.}(1992){Press}, {Teukolsky}, {Vetterling}, \&
  {Flannery}}]{press92}
{Press}, W.~H., {Teukolsky}, S.~A., {Vetterling}, W.~T., \& {Flannery}, B.~P.
  1992, {Numerical recipes in C. The art of scientific computing} (Cambridge:
  University Press)

\bibitem[{{Rasmussen} \& {Williams}(2006)}]{rasmussen06}
{Rasmussen}, C.~E., \& {Williams}, C. K.~I. 2006, {Gaussian Processes for
  Machine Learning} (MIT Press)

\bibitem[{{Raymond} {et~al.}(2018){Raymond}, {Boulet}, {Izidoro}, {Esteves}, \&
  {Bitsch}}]{raymond18}
{Raymond}, S.~N., {Boulet}, T., {Izidoro}, A., {Esteves}, L., \& {Bitsch}, B.
  2018, \mnras, 479, L81, \dodoi{10.1093/mnrasl/sly100}

\bibitem[{Ricker {et~al.}(2014)Ricker, Winn, Vanderspek, Latham, Bakos, Bean,
  Berta-Thompson, Brown, Buchhave, Butler, Butler, Chaplin, Charbonneau,
  Christensen-Dalsgaard, Clampin, Deming, Doty, Lee, Dressing, Dunham, Endl,
  Fressin, Ge, Henning, Holman, Howard, Ida, Jenkins, Jernigan, Johnson,
  Kaltenegger, Kawai, Kjeldsen, Laughlin, Levine, Lin, Lissauer, MacQueen,
  Marcy, McCullough, Morton, Narita, Paegert, Palle, Pepe, Pepper, Quirrenbach,
  Rinehart, Sasselov, Sato, Seager, Sozzetti, Stassun, Sullivan, Szentgyorgyi,
  Torres, Udry, \& Villasenor}]{ricker14}
Ricker, G.~R., Winn, J.~N., Vanderspek, R., {et~al.} 2014, Journal of
  Astronomical Telescopes, Instruments, and Systems, 1, 1 ,
  \dodoi{10.1117/1.JATIS.1.1.014003}

\bibitem[{{Robertson} {et~al.}(2013){Robertson}, {Endl}, {Cochran}, \&
  {Dodson-Robinson}}]{robertson13}
{Robertson}, P., {Endl}, M., {Cochran}, W.~D., \& {Dodson-Robinson}, S.~E.
  2013, \apj, 764, 3, \dodoi{10.1088/0004-637X/764/1/3}

\bibitem[{{Rogers} \& {Seager}(2010)}]{rogersSeager10}
{Rogers}, L.~A., \& {Seager}, S. 2010, \apj, 712, 974,
  \dodoi{10.1088/0004-637X/712/2/974}

\bibitem[{{Rosenthal} {et~al.}(2021){Rosenthal}, {Fulton}, {Hirsch},
  {Isaacson}, {Howard}, {Dedrick}, {Sherstyuk}, {Blunt}, {Petigura}, {Knutson},
  {Behmard}, {Chontos}, {Crepp}, {Crossfield}, {Dalba}, {Fischer}, {Henry},
  {Kane}, {Kosiarek}, {Marcy}, {Rubenzahl}, {Weiss}, \& {Wright}}]{rosenthal21}
{Rosenthal}, L.~J., {Fulton}, B.~J., {Hirsch}, L.~A., {et~al.} 2021, \apjs,
  255, 8, \dodoi{10.3847/1538-4365/abe23c}

\bibitem[{Salvatier {et~al.}(2016{\natexlab{a}})Salvatier, Wiecki, \&
  Fonnesbeck}]{pymc3}
Salvatier, J., Wiecki, T.~V., \& Fonnesbeck, C. 2016{\natexlab{a}}, PeerJ
  Computer Science, 2, \dodoi{10.7717/peerj-cs.55}

\bibitem[{Salvatier {et~al.}(2016{\natexlab{b}})Salvatier, Wiecki, \&
  Fonnesbeck}]{exoplanet:pymc3}
---. 2016{\natexlab{b}}, PeerJ Computer Science, 2, e55

\bibitem[{{Savitzky} \& {Golay}(1964)}]{savitzky64}
{Savitzky}, A., \& {Golay}, M.~J.~E. 1964, Analytical Chemistry, 36, 1627

\bibitem[{{Scargle}(1982)}]{scargle82}
{Scargle}, J.~D. 1982, \apj, 263, 835, \dodoi{10.1086/160554}

\bibitem[{{Scarsdale} {et~al.}(2021){Scarsdale}, {Murphy}, {Batalha},
  {Crossfield}, {Dressing}, {Fulton}, {Howard}, {Huber}, {Isaacson}, {Kane},
  {Petigura}, {Robertson}, {Roy}, {Weiss}, {Beard}, {Behmard}, {Chontos},
  {Christiansen}, {Ciardi}, {Claytor}, {Collins}, {Collins}, {Dai}, {Dalba},
  {Dragomir}, {Fetherolf}, {Fukui}, {Giacalone}, {Gonzales}, {Hill}, {Hirsch},
  {Jensen}, {Kosiarek}, {de Leon}, {Lubin}, {Lund}, {Luque}, {Mayo},
  {Mo{\v{c}}nik}, {Mori}, {Narita}, {Nowak}, {Pall{\'e}}, {Rabus}, {Rosenthal},
  {Rubenzahl}, {Schlieder}, {Shporer}, {Stassun}, {Twicken}, {Wang},
  {Yahalomi}, {Jenkins}, {Latham}, {Ricker}, {Seager}, {Vanderspek}, \&
  {Winn}}]{scarsdale21}
{Scarsdale}, N., {Murphy}, J. M.~A., {Batalha}, N.~M., {et~al.} 2021, \aj, 162,
  215, \dodoi{10.3847/1538-3881/ac18cb}

\bibitem[{{Schlafly} \& {Finkbeiner}(2011)}]{schlafly11}
{Schlafly}, E.~F., \& {Finkbeiner}, D.~P. 2011, \apj, 737, 103,
  \dodoi{10.1088/0004-637X/737/2/103}

\bibitem[{{Schlichting} \& {Young}(2022)}]{schlichting22}
{Schlichting}, H.~E., \& {Young}, E.~D. 2022, \psj, 3, 127,
  \dodoi{10.3847/PSJ/ac68e6}

\bibitem[{{Schlieder} {et~al.}(2021){Schlieder}, {Gonzales}, {Ciardi}, {Patel},
  {Crossfield}, {Crepp}, {Dressing}, {Barclay}, \& {Howard}}]{schlieder21}
{Schlieder}, J.~E., {Gonzales}, E.~J., {Ciardi}, D.~R., {et~al.} 2021,
  Frontiers in Astronomy and Space Sciences, 8, 63,
  \dodoi{10.3389/fspas.2021.628396}

\bibitem[{{Schwarz}(1978)}]{schwarz78}
{Schwarz}, G. 1978, Annals of Statistics, 6, 461

\bibitem[{{Skrutskie} {et~al.}(2006){Skrutskie}, {Cutri}, {Stiening},
  {Weinberg}, {Schneider}, {Carpenter}, {Beichman}, {Capps}, {Chester},
  {Elias}, {Huchra}, {Liebert}, {Lonsdale}, {Monet}, {Price}, {Seitzer},
  {Jarrett}, {Kirkpatrick}, {Gizis}, {Howard}, {Evans}, {Fowler}, {Fullmer},
  {Hurt}, {Light}, {Kopan}, {Marsh}, {McCallon}, {Tam}, {Van Dyk}, \&
  {Wheelock}}]{2mass}
{Skrutskie}, M.~F., {Cutri}, R.~M., {Stiening}, R., {et~al.} 2006, \aj, 131,
  1163, \dodoi{10.1086/498708}

\bibitem[{Smith {et~al.}(2012)Smith, Stumpe, Cleve, Jenkins, Barclay, Fanelli,
  Girouard, Kolodziejczak, McCauliff, Morris, \& Twicken}]{smith12}
Smith, J.~C., Stumpe, M.~C., Cleve, J. E.~V., {et~al.} 2012, Publications of
  the Astronomical Society of the Pacific, 124, 1000, \dodoi{10.1086/667697}

\bibitem[{{Stassun} {et~al.}(2019){Stassun}, {Oelkers}, {Paegert}, {Torres},
  {Pepper}, {De Lee}, {Collins}, {Latham}, {Muirhead}, {Chittidi},
  {Rojas-Ayala}, {Fleming}, {Rose}, {Tenenbaum}, {Ting}, {Kane}, {Barclay},
  {Bean}, {Brassuer}, {Charbonneau}, {Ge}, {Lissauer}, {Mann}, {McLean},
  {Mullally}, {Narita}, {Plavchan}, {Ricker}, {Sasselov}, {Seager}, {Sharma},
  {Shiao}, {Sozzetti}, {Stello}, {Vanderspek}, {Wallace}, \& {Winn}}]{tic19}
{Stassun}, K.~G., {Oelkers}, R.~J., {Paegert}, M., {et~al.} 2019, \aj, 158,
  138, \dodoi{10.3847/1538-3881/ab3467}

\bibitem[{{Stumpe} {et~al.}(2014){Stumpe}, {Smith}, {Catanzarite}, {Van Cleve},
  {Jenkins}, {Twicken}, \& {Girouard}}]{stumpe14}
{Stumpe}, M.~C., {Smith}, J.~C., {Catanzarite}, J.~H., {et~al.} 2014, \pasp,
  126, 100, \dodoi{10.1086/674989}

\bibitem[{Stumpe {et~al.}(2012)Stumpe, Smith, Cleve, Twicken, Barclay, Fanelli,
  Girouard, Jenkins, Kolodziejczak, McCauliff, \& Morris}]{stumpe12}
Stumpe, M.~C., Smith, J.~C., Cleve, J. E.~V., {et~al.} 2012, Publications of
  the Astronomical Society of the Pacific, 124, 985, \dodoi{10.1086/667698}

\bibitem[{{Tayar} {et~al.}(2022){Tayar}, {Claytor}, {Huber}, \& {van
  Saders}}]{tayar22}
{Tayar}, J., {Claytor}, Z.~R., {Huber}, D., \& {van Saders}, J. 2022, \apj,
  927, 31, \dodoi{10.3847/1538-4357/ac4bbc}

\bibitem[{{Teske} {et~al.}(2021){Teske}, {Wang}, {Wolfgang}, {Gan},
  {Plotnykov}, {Armstrong}, {Butler}, {Cale}, {Crane}, {Howard}, {Jensen},
  {Law}, {Shectman}, {Plavchan}, {Valencia}, {Vanderburg}, {Ricker},
  {Vanderspek}, {Latham}, {Seager}, {Winn}, {Jenkins}, {Adibekyan}, {Barrado},
  {Barros}, {Benkhaldoun}, {Brown}, {Bryant}, {Burt}, {Caldwell},
  {Charbonneau}, {Cloutier}, {Collins}, {Collins}, {Colon}, {Conti},
  {Demangeon}, {Eastman}, {Elmufti}, {Feng}, {Flowers}, {Guerrero},
  {Hojjatpanah}, {Irwin}, {Isopi}, {Lillo-Box}, {Mallia}, {Massey}, {Mori},
  {Mullally}, {Narita}, {Nishiumi}, {Osborn}, {Paegert}, {de Leon}, {Quinn},
  {Reefe}, {Schwarz}, {Shporer}, {Soubkiou}, {Sousa}, {Stockdale}, {Str{\o}m},
  {Tan}, {Tang}, {Tenenbaum}, {Wheatley}, {Wittrock}, {Yahalomi}, \&
  {Zohrabi}}]{teske21}
{Teske}, J., {Wang}, S.~X., {Wolfgang}, A., {et~al.} 2021, \apjs, 256, 33,
  \dodoi{10.3847/1538-4365/ac0f0a}

\bibitem[{{Theano Development Team}(2016)}]{exoplanet:theano}
{Theano Development Team}. 2016, arXiv e-prints, abs/1605.02688.
\newblock \url{http://arxiv.org/abs/1605.02688}

\bibitem[{{Twicken} {et~al.}(2010){Twicken}, {Clarke}, {Bryson}, {Tenenbaum},
  {Wu}, {Jenkins}, {Girouard}, \& {Klaus}}]{twicken10}
{Twicken}, J.~D., {Clarke}, B.~D., {Bryson}, S.~T., {et~al.} 2010, in Society
  of Photo-Optical Instrumentation Engineers (SPIE) Conference Series, Vol.
  7740, Software and Cyberinfrastructure for Astronomy, ed. N.~M. {Radziwill}
  \& A.~{Bridger}, 774023, \dodoi{10.1117/12.856790}

\bibitem[{{Twicken} {et~al.}(2018){Twicken}, {Catanzarite}, {Clarke},
  {Girouard}, {Jenkins}, {Klaus}, {Li}, {McCauliff}, {Seader}, {Tenenbaum},
  {Wohler}, {Bryson}, {Burke}, {Caldwell}, {Haas}, {Henze}, \&
  {Sanderfer}}]{twicken18}
{Twicken}, J.~D., {Catanzarite}, J.~H., {Clarke}, B.~D., {et~al.} 2018, \pasp,
  130, 064502, \dodoi{10.1088/1538-3873/aab694}

\bibitem[{{Valencia} {et~al.}(2007){Valencia}, {Sasselov}, \&
  {O'Connell}}]{valencia07}
{Valencia}, D., {Sasselov}, D.~D., \& {O'Connell}, R.~J. 2007, \apj, 665, 1413,
  \dodoi{10.1086/519554}

\bibitem[{{Van Eylen} {et~al.}(2018){Van Eylen}, {Agentoft}, {Lundkvist},
  {Kjeldsen}, {Owen}, {Fulton}, {Petigura}, \& {Snellen}}]{vaneylen18}
{Van Eylen}, V., {Agentoft}, C., {Lundkvist}, M.~S., {et~al.} 2018, \mnras,
  479, 4786, \dodoi{10.1093/mnras/sty1783}

\bibitem[{{Van Eylen} {et~al.}(2019){Van Eylen}, {Albrecht}, {Huang},
  {MacDonald}, {Dawson}, {Cai}, {Foreman-Mackey}, {Lundkvist}, {Silva Aguirre},
  {Snellen}, \& {Winn}}]{exoplanet:vaneylen19}
{Van Eylen}, V., {Albrecht}, S., {Huang}, X., {et~al.} 2019, \aj, 157, 61,
  \dodoi{10.3847/1538-3881/aaf22f}

\bibitem[{Van~Rossum \& Drake(2009)}]{python3}
Van~Rossum, G., \& Drake, F.~L. 2009, Python 3 Reference Manual (Scotts Valley,
  CA: CreateSpace)

\bibitem[{{Vanderburg} {et~al.}(2016){Vanderburg}, {Plavchan}, {Johnson},
  {Ciardi}, {Swift}, \& {Kane}}]{vanderburg16}
{Vanderburg}, A., {Plavchan}, P., {Johnson}, J.~A., {et~al.} 2016, \mnras, 459,
  3565, \dodoi{10.1093/mnras/stw863}

\bibitem[{{Vanderburg} {et~al.}(2019){Vanderburg}, {Huang}, {Rodriguez},
  {Becker}, {Ricker}, {Vanderspek}, {Latham}, {Seager}, {Winn}, {Jenkins},
  {Addison}, {Bieryla}, {Brice{\~n}o}, {Bowler}, {Brown}, {Burke}, {Burt},
  {Caldwell}, {Clark}, {Crossfield}, {Dittmann}, {Dynes}, {Fulton}, {Guerrero},
  {Harbeck}, {Horner}, {Kane}, {Kielkopf}, {Kraus}, {Kreidberg}, {Law}, {Mann},
  {Mengel}, {Morton}, {Okumura}, {Pearce}, {Plavchan}, {Quinn}, {Rabus},
  {Rose}, {Rowden}, {Shporer}, {Siverd}, {Smith}, {Stassun}, {Tinney},
  {Wittenmyer}, {Wright}, {Zhang}, {Zhou}, \& {Ziegler}}]{vanderburg19}
{Vanderburg}, A., {Huang}, C.~X., {Rodriguez}, J.~E., {et~al.} 2019, \apjl,
  881, L19, \dodoi{10.3847/2041-8213/ab322d}

\bibitem[{{Vehtari} {et~al.}(2021){Vehtari}, {Gelman}, {Simpson}, {Carpenter},
  \& {B{\"u}rkner}}]{vehtari21}
{Vehtari}, A., {Gelman}, A., {Simpson}, D., {Carpenter}, B., \& {B{\"u}rkner},
  P.-C. 2021, Bayesian Analysis, 16, 667, \dodoi{10.1214/20-BA1221}

\bibitem[{Virtanen {et~al.}(2020)Virtanen, Gommers, Oliphant, Haberland, Reddy,
  Cournapeau, Burovski, Peterson, {Weckesser}, {Bright}, {van der Walt},
  {Brett}, {Wilson}, {Jarrod Millman}, {Mayorov}, {Nelson}, {Jones}, {Kern},
  {Larson}, {Carey}, {Polat}, {Feng}, {Moore}, {Vand erPlas}, {Laxalde},
  {Perktold}, {Cimrman}, {Henriksen}, {Quintero}, {Harris}, {Archibald},
  {Ribeiro}, {Pedregosa}, {van Mulbregt}, \& {Contributors}}]{scipy}
Virtanen, P., Gommers, R., Oliphant, T.~E., {et~al.} 2020, Nature Methods

\bibitem[{{Vogt} {et~al.}(1994){Vogt}, {Allen}, {Bigelow}, {Bresee}, {Brown},
  {Cantrall}, {Conrad}, {Couture}, {Delaney}, {Epps}, {Hilyard}, {Hilyard},
  {Horn}, {Jern}, {Kanto}, {Keane}, {Kibrick}, {Lewis}, {Osborne},
  {Pardeilhan}, {Pfister}, {Ricketts}, {Robinson}, {Stover}, {Tucker}, {Ward},
  \& {Wei}}]{vogt94}
{Vogt}, S.~S., {Allen}, S.~L., {Bigelow}, B.~C., {et~al.} 1994, in Society of
  Photo-Optical Instrumentation Engineers (SPIE) Conference Series, Vol. 2198,
  Instrumentation in Astronomy VIII, ed. D.~L. {Crawford} \& E.~R. {Craine},
  362, \dodoi{10.1117/12.176725}

\bibitem[{{Vogt} {et~al.}(2014){Vogt}, {Radovan}, {Kibrick}, {Butler},
  {Alcott}, {Allen}, {Arriagada}, {Bolte}, {Burt}, {Cabak}, {Chloros},
  {Cowley}, {Deich}, {Dupraw}, {Earthman}, {Epps}, {Faber}, {Fischer}, {Gates},
  {Hilyard}, {Holden}, {Johnston}, {Keiser}, {Kanto}, {Katsuki}, {Laiterman},
  {Lanclos}, {Laughlin}, {Lewis}, {Lockwood}, {Lynam}, {Marcy}, {McLean},
  {Miller}, {Misch}, {Peck}, {Pfister}, {Phillips}, {Rivera}, {Sandford},
  {Saylor}, {Stover}, {Thompson}, {Walp}, {Ward}, {Wareham}, {Wei}, \&
  {Wright}}]{vogt14}
{Vogt}, S.~S., {Radovan}, M., {Kibrick}, R., {et~al.} 2014, \pasp, 126, 359,
  \dodoi{10.1086/676120}

\bibitem[{Vrieze(2012)}]{vrieze12}
Vrieze, S. 2012, Psychological methods, 17 2, 228

\bibitem[{{Wizinowich} {et~al.}(2000){Wizinowich}, {Acton}, {Shelton},
  {Stomski}, {Gathright}, {Ho}, {Lupton}, {Tsubota}, {Lai}, {Max}, {Brase},
  {An}, {Avicola}, {Olivier}, {Gavel}, {Macintosh}, {Ghez}, \&
  {Larkin}}]{nirc2}
{Wizinowich}, P., {Acton}, D.~S., {Shelton}, C., {et~al.} 2000, \pasp, 112,
  315, \dodoi{10.1086/316543}

\bibitem[{{Wright}(2005)}]{wright05}
{Wright}, J.~T. 2005, \pasp, 117, 657, \dodoi{10.1086/430369}

\bibitem[{{Wright} {et~al.}(2008){Wright}, {Marcy}, {Butler}, {Vogt}, {Henry},
  {Isaacson}, \& {Howard}}]{wright08}
{Wright}, J.~T., {Marcy}, G.~W., {Butler}, R.~P., {et~al.} 2008, \apjl, 683,
  L63, \dodoi{10.1086/587461}

\bibitem[{Yee {et~al.}(2017)Yee, Petigura, \& von Braun}]{yee17}
Yee, S.~W., Petigura, E.~A., \& von Braun, K. 2017, The Astrophysical Journal,
  836, 77, \dodoi{10.3847/1538-4357/836/1/77}

\bibitem[{{Zechmeister} \& {K{\"u}rster}(2009)}]{zechmeister09}
{Zechmeister}, M., \& {K{\"u}rster}, M. 2009, \aap, 496, 577,
  \dodoi{10.1051/0004-6361:200811296}

\bibitem[{{Zeng} {et~al.}(2016){Zeng}, {Sasselov}, \& {Jacobsen}}]{zeng16}
{Zeng}, L., {Sasselov}, D.~D., \& {Jacobsen}, S.~B. 2016, \apj, 819, 127,
  \dodoi{10.3847/0004-637X/819/2/127}

\bibitem[{{Zeng} {et~al.}(2019){Zeng}, {Jacobsen}, {Sasselov}, {Petaev},
  {Vanderburg}, {Lopez-Morales}, {Perez-Mercader}, {Mattsson}, {Li}, {Heising},
  {Bonomo}, {Damasso}, {Berger}, {Cao}, {Levi}, \& {Wordsworth}}]{zeng19}
{Zeng}, L., {Jacobsen}, S.~B., {Sasselov}, D.~D., {et~al.} 2019, Proceedings of
  the National Academy of Science, 116, 9723, \dodoi{10.1073/pnas.1812905116}

\bibitem[{{Zhang} {et~al.}(2022){Zhang}, {Knutson}, {Wang}, {Dai}, \&
  {Barrag{\'a}n}}]{zhang22}
{Zhang}, M., {Knutson}, H.~A., {Wang}, L., {Dai}, F., \& {Barrag{\'a}n}, O.
  2022, \aj, 163, 67, \dodoi{10.3847/1538-3881/ac3fa7}

\bibitem[{{Zhu} {et~al.}(2018){Zhu}, {Petrovich}, {Wu}, {Dong}, \&
  {Xie}}]{zhu18}
{Zhu}, W., {Petrovich}, C., {Wu}, Y., {Dong}, S., \& {Xie}, J. 2018, \apj, 860,
  101, \dodoi{10.3847/1538-4357/aac6d5}

\end{thebibliography}
\bibliographystyle{aasjournal}
\appendix

\section{Joint modeling results} \label{appendix:joint_models}
\begin{deluxetable*}{lcccc}
\tablecaption{\sysI system parameters \label{tab:hip8152_properties}}
\tabletypesize{\scriptsize}
\startdata
\tablehead{
    \vspace{0.01cm} \\ 
    \multicolumn{5}{c}{\textbf{Stellar Parameters}} \\
    \colhead{Parameter} & \colhead{Symbol} & \colhead{Units} & \colhead{Value} & \colhead{Provenance}
}\\
\sidehead{\emph{Identifying information}}
TOI ID & & & 266 & Guerrero \\
TIC ID & & & 164767175 & Guerrero \\
R.A. & & deg (J2000) & $26.20$ & \gaiadrthree \\
decl. & & deg (J2000) & $-18.40$ & \gaiadrthree \\
Parallax & $\pi$ & mas & $9.83 \pm 0.01$ & \gaiadrthree \\
Johnson V-band apparent magnitude & $V$ & mag & $10.07 \pm 0.03$ & TIC \\
J-band apparent magnitude & $J$ & mag & $8.85 \pm 0.03$ & \twomass \\
K$_s$-band apparent magnitude & $K_s$ & mag & $8.45 \pm 0.02$ & \twomass \\
\sidehead{\emph{Spectroscopy}}
Effective temperature & \teff & K & \teffI &  \specMatchEmp \\
Metallicity & \feh & dex & \fehI & \specMatchEmp \\
Ca II H \& K emission & \logrhk & & \logrhkI & Isaacson \\
\sidehead{\emph{Isochrone modeling}}
Mass & \mstar & \msun & \mstarI  & \isoclassify \\
Radius & \rstar & \rsun & \rstarI & \isoclassify \\
Age & & Gyr & \ageI & \isoclassify \\
\sidehead{\emph{Transit modeling}}
Limb-darkening parameter 1 & $u_1$ & & \uOneI & Joint model \\
Limb-darkening parameter 2 & $u_2$ & & \uTwoI & Joint model \\
\vspace{0.01cm} \\ 
\multicolumn{5}{c}{\textbf{Planet Parameters}} \\
\colhead{Parameter} & \colhead{Symbol} & \colhead{Units} & \colhead{\sysI b value} & \colhead{\sysI c value} \\
\hline
\sidehead{\emph{Measured quantities}}
Orbital period & $P$ & d & \periodIb & \periodIc \\ 
Time of inferior conjunction & \transitTime & BTJD & \tcBTJDIb & \tcBTJDIc \\
Occultation fraction & $R_\mathrm{p}/R_*$ & & \rorIb & \rorIc \\
Impact parameter & $b$ & & \bIb & \bIc \\
Orbital eccentricity & $e$ & & \eccIb & \eccIc \\
Argument of periastron & $\omega$ & deg & \omegafoldeddegIb & \omegafoldeddegIc \\
RV semi-amplitude & $K$ & \mps & \KIb & \KIc \\
\sidehead{\emph{Derived quantities}}
Orbital separation & $a/R_*$ & & \aorIb & \aorIc \\
Orbital semimajor axis & $a$ & AU & \aIb & \aIc \\
Radius & \rplanet & \rearth & \rpIb & \rpIc \\
Mass & \mplanet & \mearth & \mpIb & \mpIc \\
Bulk density & $\rho$ & \gcc & \rhoIb & \rhoIc \\ 
Equilibrium temperature & \teq & K & \teqIb & \teqIc \\ 
Instellation flux & \sincplanet & \sincearth & \sincIb & \sincIc \\
Transit duration & \tdur & hr & \durhrIb & \durhrIc \\
TSM & & & \tsmIb & \tsmIc \\
\vspace{0.01cm} \\ 
\multicolumn{5}{c}{\textbf{Additional Parameters}} \\
\colhead{Parameter} & \colhead{Symbol} & \colhead{Units} & \colhead{Value} & \colhead{} \\
\hline
\tess photometric offset & $\mu_\mathrm{TESS}$ & ppt & \meanfluxI & \\
\tess photometric jitter & $\sigma_\mathrm{TESS}$ & ppt & \sigmaphotI & \\
\keckhires RV offset & $\gamma_\mathrm{HIRES}$ & \mps & \gammarvHIRESI & \\
\keckhires RV jitter & $\sigma_\mathrm{HIRES}$ & \mps & \sigmarvHIRESI & \\
\enddata
\tablecomments{Errors on stellar mass and radius have been inflated according to \cite{tayar22}. Upper limits reflect 98\% confidence. \teq is calculated assuming zero Bond albedo and full day-night heat redistribution. The Transmission Spectroscopy Metric (TSM) from \cite{kempton18} is a \jwst \snr proxy. References for the provenance values in the order in which they appear in the table: ``Guerrero'' refers to the \tess Primary Mission TOI Catalog \citep{guerrero21}, TIC (TESS Input Catalog; \citealt{tic19}), \gaiadrthree (\citealt{gaia, gaiadr3arxiv}), \twomass \citep{2mass}, \specMatchEmp \citep{yee17}, ``Isaacson'' refers to the methods described in \cite{isaacson10}, \isoclassify (\citealt{huber17}; \citealt{berger20}).}
\end{deluxetable*}

\begin{figure*}
\gridline{\fig{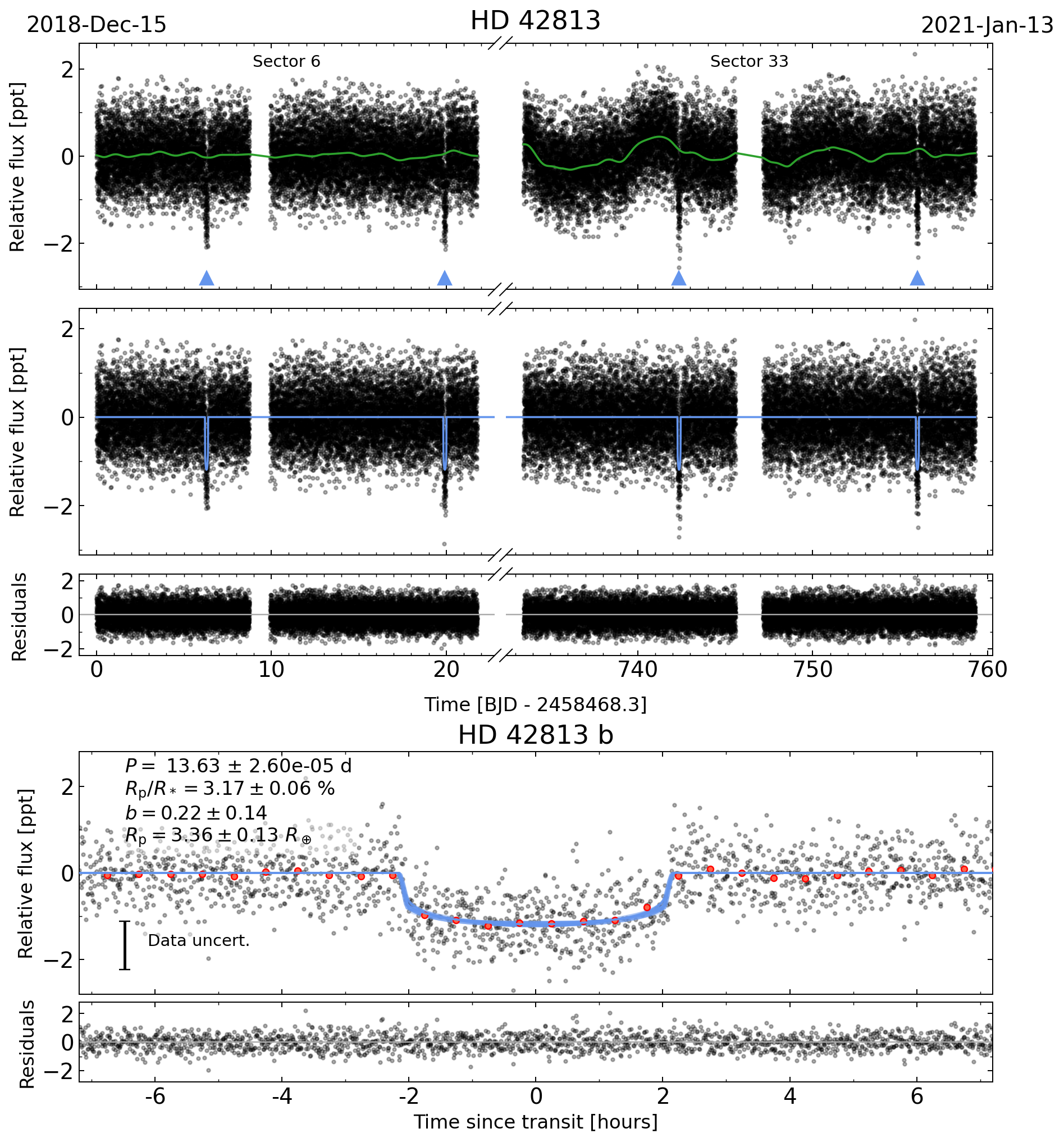}{0.5\textwidth}{}
          \fig{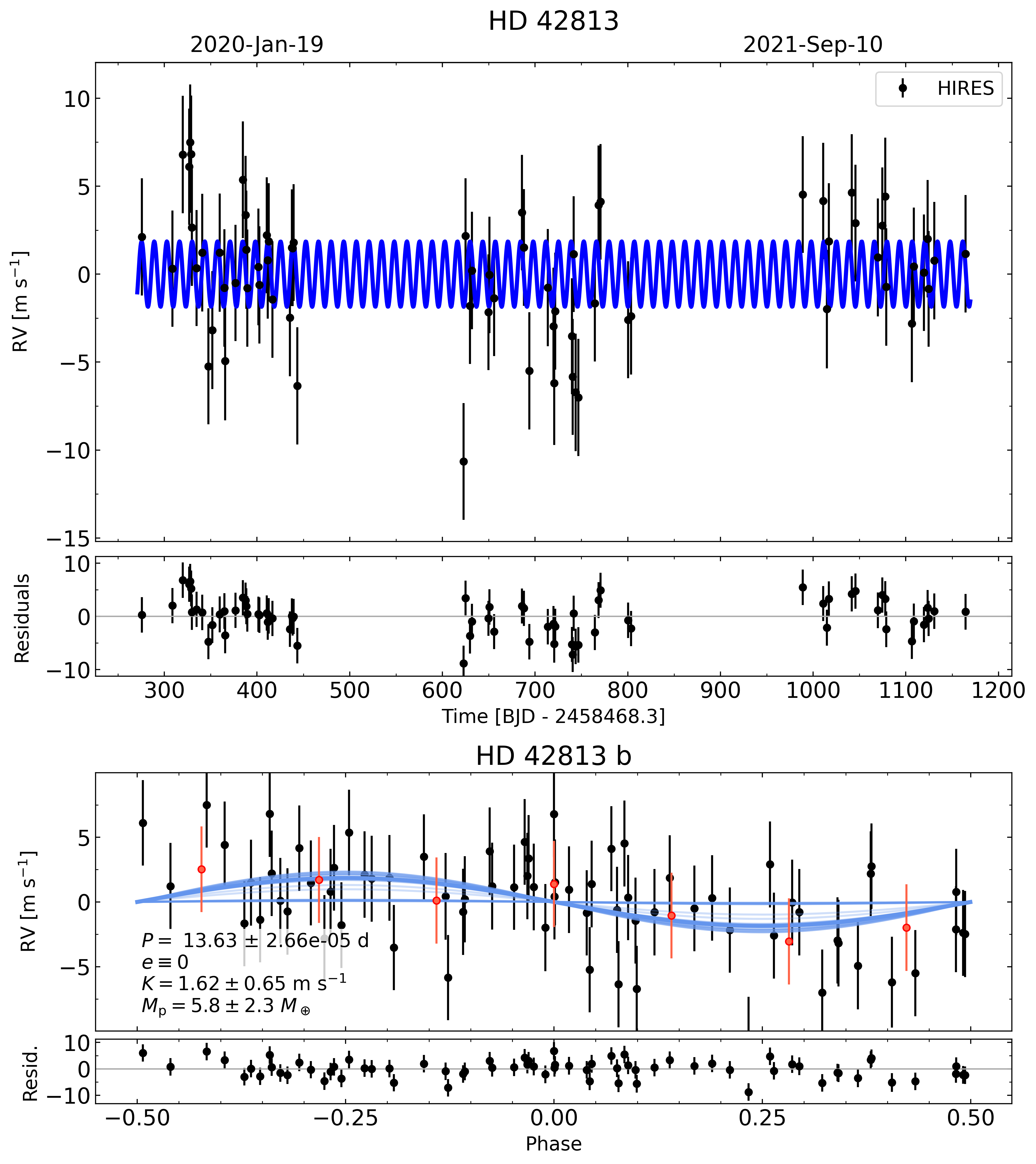}{0.5\textwidth}{}}
  \caption{Our joint modeling results for \sysII. The figure description is the same as for Figure \ref{fig:hip8152_phot_and_rvs}.}\label{fig:hd42813_phot_and_rvs}
\end{figure*}

\begin{figure*}
    \centering
    \includegraphics[width=\textwidth]{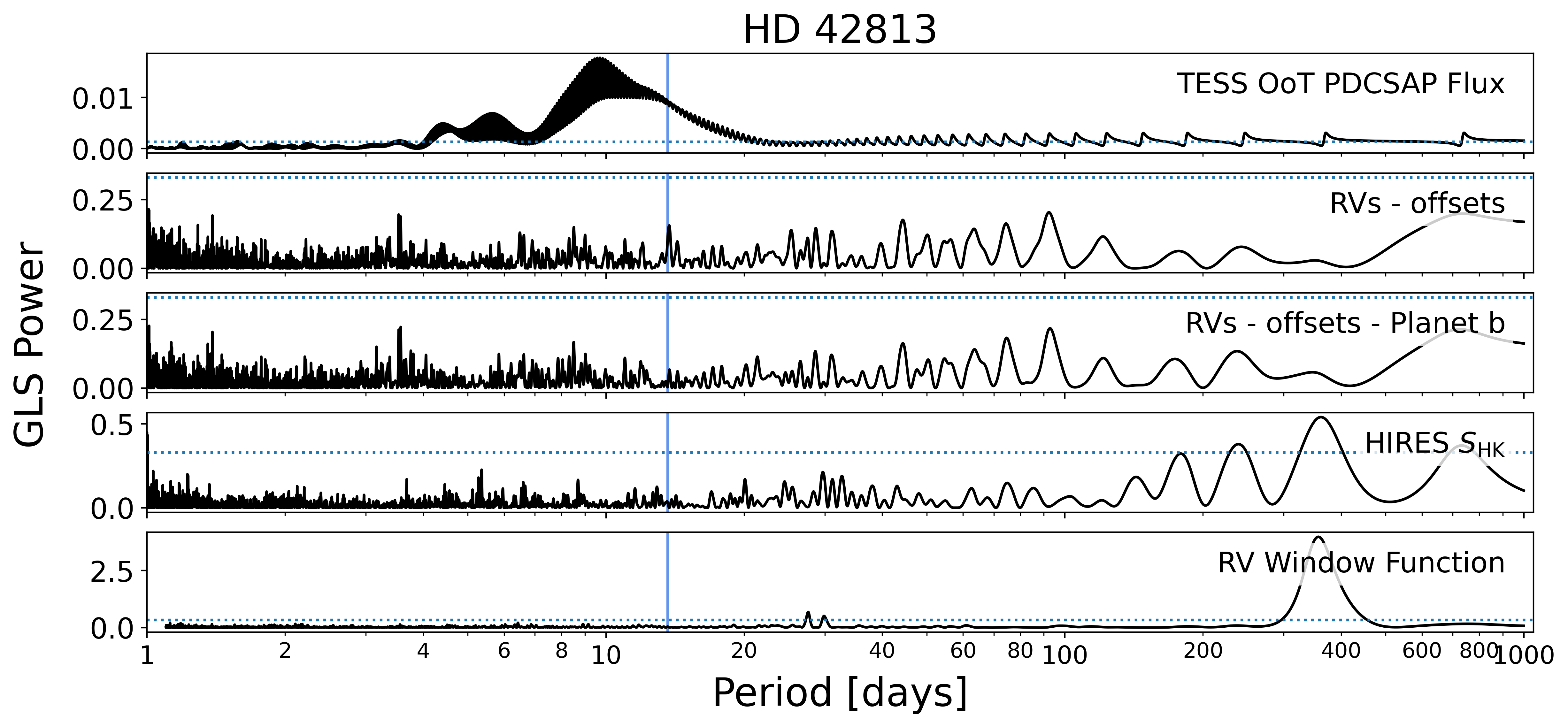}
    \caption{GLS periodograms for \sysII. The figure description is the same as for Figure \ref{fig:hip8152_periodogram}.}
    \label{fig:hd42813_periodogram}
\end{figure*}

\begin{deluxetable*}{lcccc}
\tablecaption{\sysII system parameters \label{tab:hd42813_properties}}
\tabletypesize{\scriptsize}
\startdata
\tablehead{
    \vspace{0.01cm} \\ 
    \multicolumn{5}{c}{\textbf{Stellar Parameters}} \\
    \colhead{Parameter} & \colhead{Symbol} & \colhead{Units} & \colhead{Value} & \colhead{Provenance}
}\\
\sidehead{\emph{Identifying information}}
TOI ID & & & 469 & Guerrero \\
TIC ID & & & 33692729 & Guerrero \\
R.A. & & deg (J2000) & $93.05$ & \gaiadrthree \\
decl. & & deg (J2000) & $-14.64$ & \gaiadrthree \\
Parallax & $\pi$ & mas & $14.71 \pm 0.02$ & \gaiadrthree \\
Johnson V-band apparent magnitude & $V$ & mag & $9.49 \pm 0.03$ & TIC \\
J-band apparent magnitude & $J$ & mag & $8.06 \pm 0.03$ & \twomass \\
K$_s$-band apparent magnitude & $K_s$ & mag & $7.59 \pm 0.02$ & \twomass \\
\sidehead{\emph{Spectroscopy}}
Effective temperature & \teff & K & \teffII & \specMatchEmp \\
Metallicity & \feh & dex & \fehII & \specMatchEmp \\
Ca II H \& K emission & \logrhk & & \logrhkII & Isaacson \\
\sidehead{\emph{Isochrone modeling}}
Mass & \mstar & \msun & \mstarII  & \isoclassify \\
Radius & \rstar & \rsun & \rstarII & \isoclassify \\
Age & & Gyr & \ageII & \isoclassify \\
\sidehead{\emph{Transit modeling}}
Limb-darkening parameter 1 & $u_1$ & & \uOneII & Joint model \\
Limb-darkening parameter 2 & $u_2$ & & \uTwoII & Joint model \\
\vspace{0.01cm} \\ 
\multicolumn{5}{c}{\textbf{Planet Parameters}} \\
\colhead{Parameter} & \colhead{Symbol} & \colhead{Units} & \colhead{\sysII b value} \\
\hline
\sidehead{\emph{Measured quantities}}
Orbital period & $P$ & d & \periodIIb & \\ 
Time of inferior conjunction & \transitTime & BTJD & \tcBTJDIIb & \\
Occultation fraction & $R_\mathrm{p}/R_*$ & & \rorIIb & \\
Impact parameter & $b$ & & \bIIb & \\
Orbital eccentricity & $e$ & & \eccIIb & \\
Argument of periastron & $\omega$ & deg & \omegafoldeddegIIb & \\
RV semi-amplitude & $K$ & \mps & \KIIb & \\
\sidehead{\emph{Derived quantities}}
Orbital separation & $a/R_*$ & & \aorIIb & \\
Orbital semimajor axis & $a$ & AU & \aIIb & \\
Radius & \rplanet & \rearth & \rpIIb & \\
Mass & \mplanet & \mearth & \mpIIb & \\
Bulk density & $\rho$ & \gcc & \rhoIIb & \\ 
Equilibrium temperature & \teq & K & \teqIIb & \\ 
Instellation flux & \sincplanet & \sincearth & \sincIIb & \\
Transit duration & \tdur & hr & \durhrIIb & \\
TSM & & & \tsmIIb & \\
\vspace{0.01cm} \\ 
\multicolumn{5}{c}{\textbf{Additional Parameters}} \\
\colhead{Parameter} & \colhead{Symbol} & \colhead{Units} & \colhead{Value} & \colhead{} \\
\hline
\tess photometric offset & $\mu_\mathrm{TESS}$ & ppt & \meanfluxII & \\
\tess photometric jitter & $\sigma_\mathrm{TESS}$ & ppt & \sigmaphotII & \\
\keckhires RV offset & $\gamma_\mathrm{HIRES}$ & \mps & \gammarvHIRESII & \\
\keckhires RV jitter & $\sigma_\mathrm{HIRES}$ & \mps & \sigmarvHIRESII & \\
\enddata
\tablecomments{Table notes are the same as found at the bottom of Table \ref{tab:hip8152_properties}. The lower limit on TSM reflects 98\% confidence.}
\end{deluxetable*}

\begin{figure*} 
\gridline{\fig{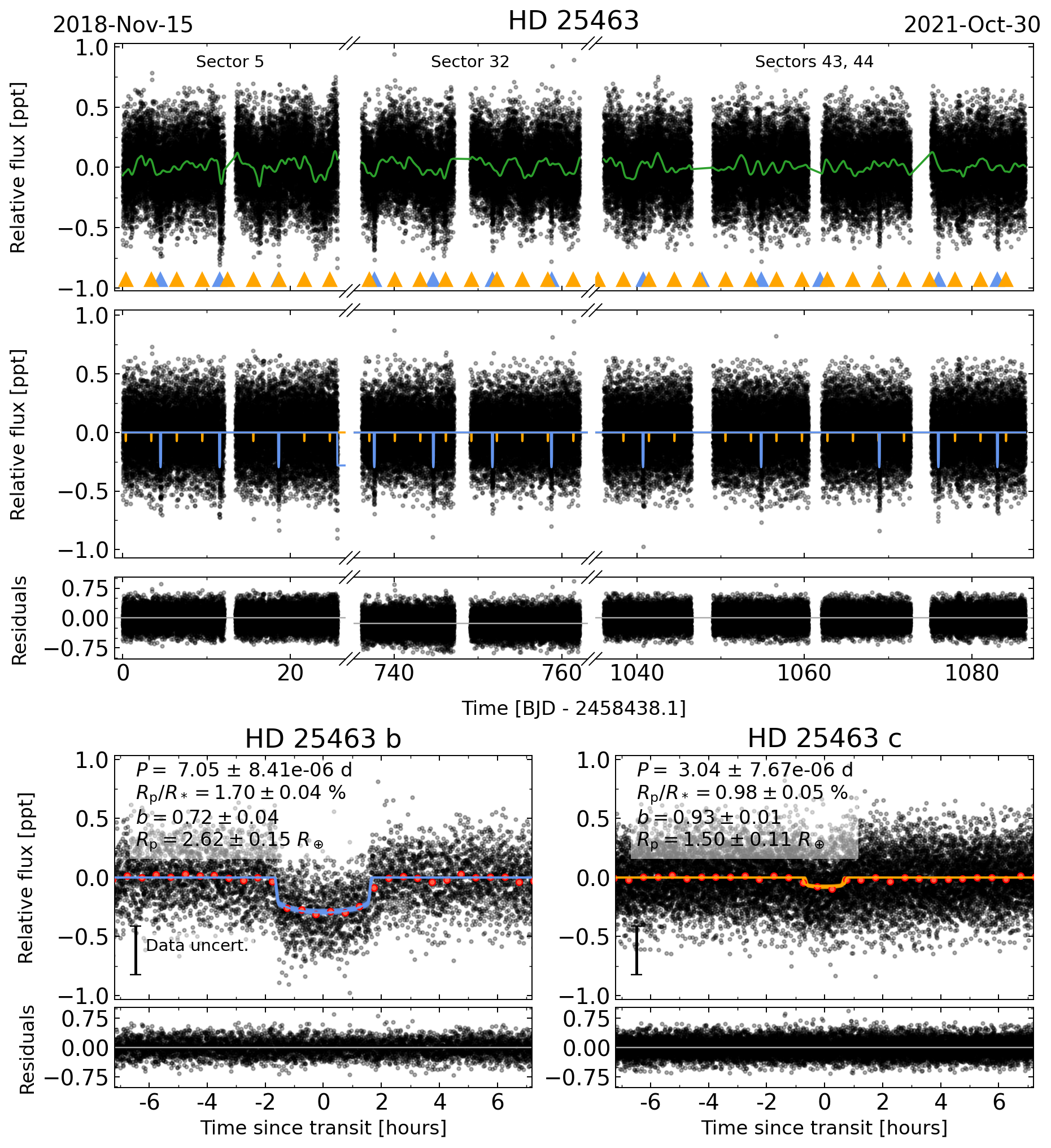}{0.5\textwidth}{}
          \fig{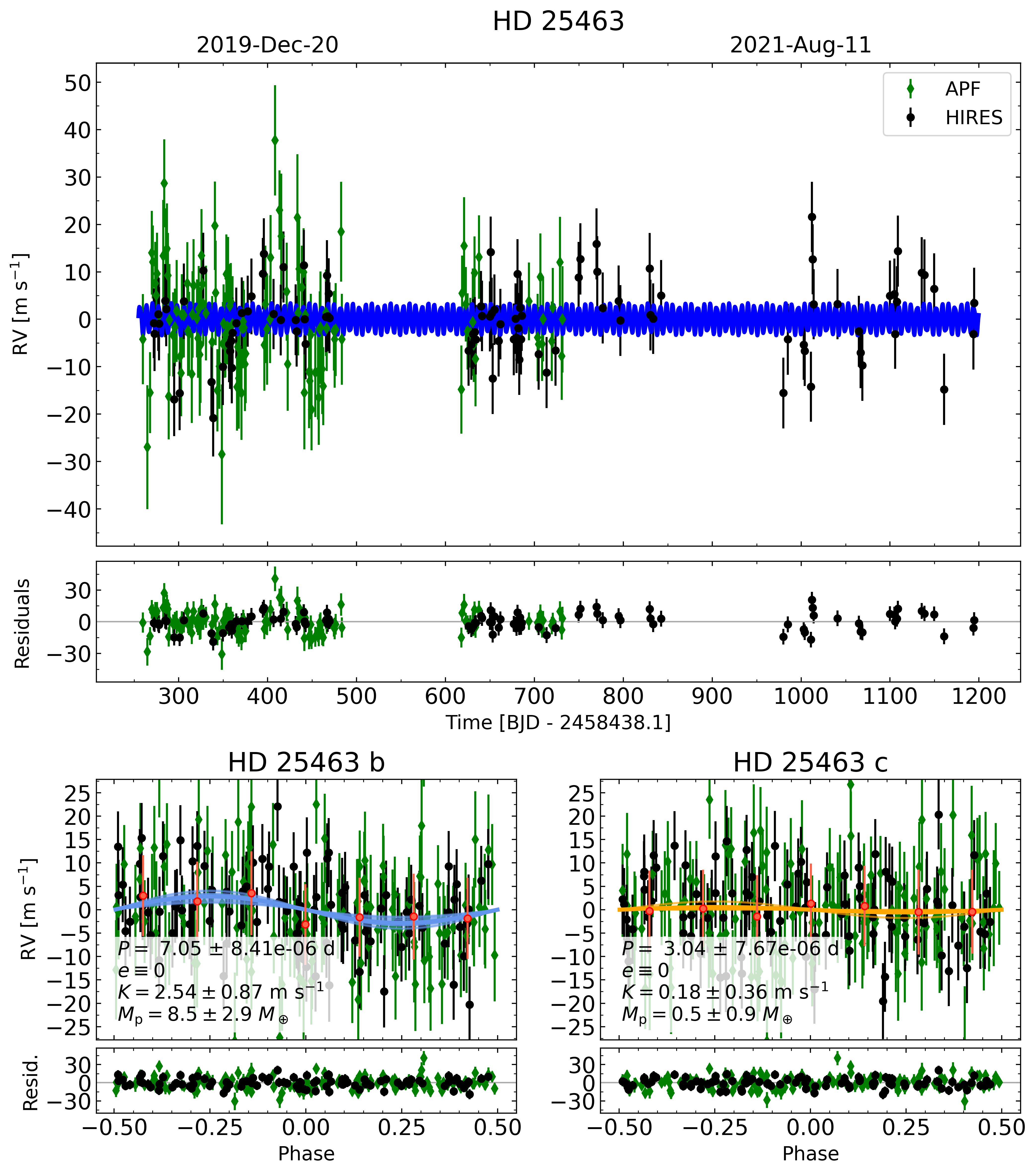}{0.5\textwidth}{}}
  \caption{Our joint modeling results for \sysIII. The figure description is the same as for Figure \ref{fig:hip8152_phot_and_rvs}.}\label{fig:hd25463_phot_and_rvs}
\end{figure*}

\begin{figure*}
    \centering
    \includegraphics[width=\textwidth]{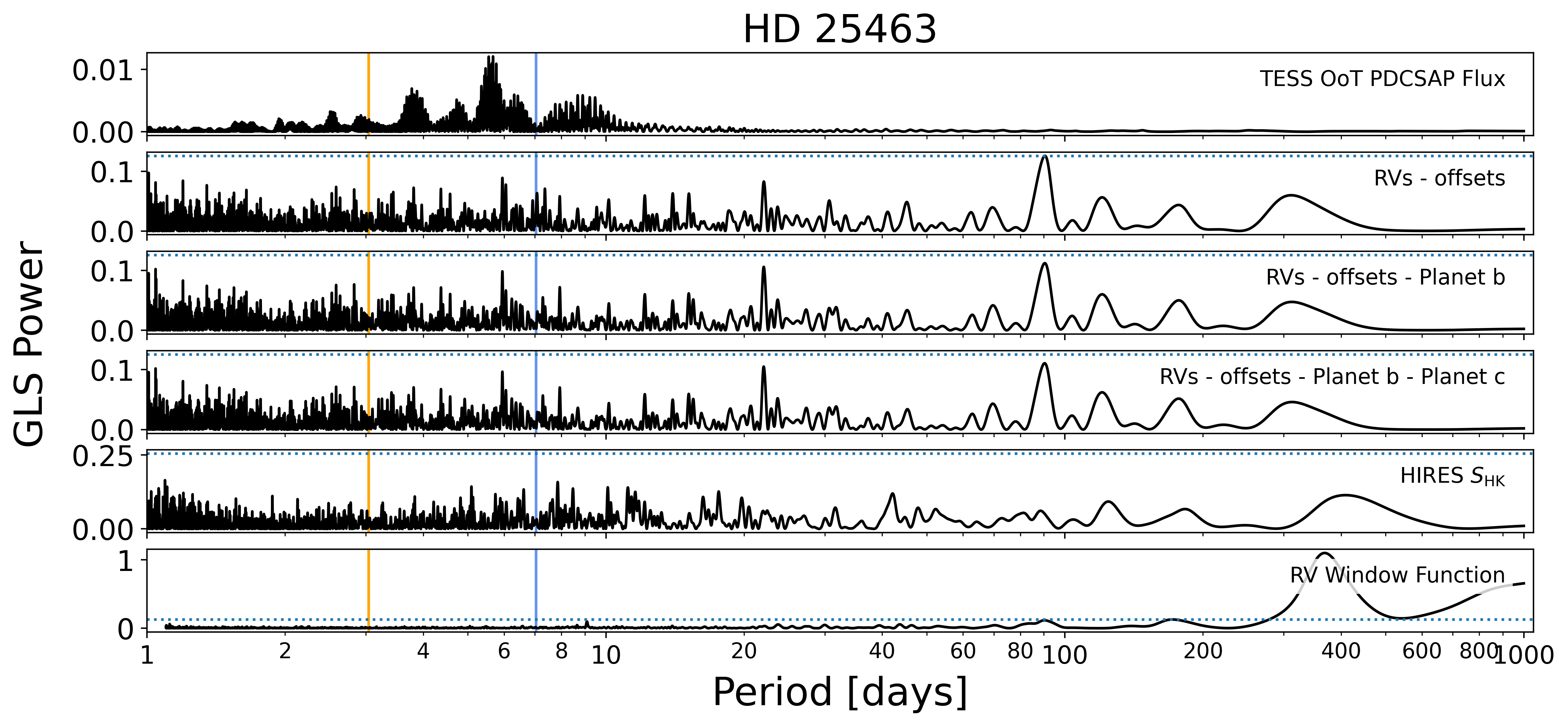}
    \caption{GLS periodograms for \sysIII. The figure description is the same as for Figure \ref{fig:hip8152_periodogram}.}
    \label{fig:hd25463_periodogram}
\end{figure*}

\begin{deluxetable*}{lcccc}
\tablecaption{\sysIII system parameters \label{tab:hd25463_properties}}
\tabletypesize{\scriptsize}
\startdata
\tablehead{
    \vspace{0.01cm} \\ 
    \multicolumn{5}{c}{\textbf{Stellar Parameters}} \\
    \colhead{Parameter} & \colhead{Symbol} & \colhead{Units} & \colhead{Value} & \colhead{Provenance}
}\\
\sidehead{\emph{Identifying information}}
TOI ID & & & 554 & Guerrero \\
TIC ID & & & 407966340 & Guerrero \\
R.A. & & deg (J2000) & $60.74$ & \gaiadrthree \\
decl. & & deg (J2000) & $9.20$ & \gaiadrthree \\
Parallax & $\pi$ & mas & $22.13 \pm 0.02$ & \gaiadrthree \\
Johnson V-band apparent magnitude & $V$ & mag & $6.91 \pm 0.02$ & TIC \\
J-band apparent magnitude & $J$ & mag & $5.95 \pm 0.02$ & \twomass \\
K$_s$-band apparent magnitude & $K_s$ & mag & $5.71 \pm 0.02$ & \twomass \\
\sidehead{\emph{Spectroscopy}}
Effective temperature & \teff & K & \teffIII &  \specMatchEmp \\
Metallicity & \feh & dex & \fehIII & \specMatchEmp \\
Ca II H \& K emission & \logrhk & & \logrhkIII & Isaacson \\
\sidehead{\emph{Isochrone modeling}}
Mass & \mstar & \msun & \mstarIII  & \isoclassify \\
Radius & \rstar & \rsun & \rstarIII & \isoclassify \\
Age & & Gyr & \ageIII & \isoclassify \\
\sidehead{\emph{Transit modeling}}
Limb-darkening parameter 1 & $u_1$ & & \uOneIII & Joint model \\
Limb-darkening parameter 2 & $u_2$ & & \uTwoIII & Joint model \\
\vspace{0.01cm} \\ 
\multicolumn{5}{c}{\textbf{Planet Parameters}} \\
\colhead{Parameter} & \colhead{Symbol} & \colhead{Units} & \colhead{\sysIII b value} & \colhead{\sysIII c value} \\
\hline
\sidehead{\emph{Measured quantities}}
Orbital period & $P$ & d & \periodIIIb & \periodIIIc \\ 
Time of inferior conjunction & \transitTime & BTJD & \tcBTJDIIIb & \tcBTJDIIIc \\
Occultation fraction & $R_\mathrm{p}/R_*$ & & \rorIIIb & \rorIIIc \\
Impact parameter & $b$ & & \bIIIb & \bIIIc \\
Orbital eccentricity & $e$ & & \eccIIIb & \eccIIIc \\
Argument of periastron & $\omega$ & deg & \omegafoldeddegIIIb & \omegafoldeddegIIIc \\
RV semi-amplitude & $K$ & \mps & \KIIIb & \KIIIc \\
\sidehead{\emph{Derived quantities}}
Orbital separation & $a/R_*$ & & \aorIIIb & \aorIIIc \\
Orbital semimajor axis & $a$ & AU & \aIIIb & \aIIIc \\
Radius & \rplanet & \rearth & \rpIIIb & \rpIIIc \\
Mass & \mplanet & \mearth & \mpIIIb & \mpIIIc \\
Bulk density & $\rho$ & \gcc & \rhoIIIb & \rhoIIIc \\ 
Equilibrium temperature & \teq & K & \teqIIIb & \teqIIIc \\ 
Instellation flux & \sincplanet & \sincearth & \sincIIIb & \sincIIIc \\
Transit duration & \tdur & hr & \durhrIIIb & \durhrIIIc \\
TSM & & & \tsmIIIb & \nodata \\
\vspace{0.01cm} \\ 
\multicolumn{5}{c}{\textbf{Additional Parameters}} \\
\colhead{Parameter} & \colhead{Symbol} & \colhead{Units} & \colhead{Value} & \colhead{} \\
\hline
\tess photometric offset & $\mu_\mathrm{TESS}$ & ppt & \meanfluxIII & \\
\tess photometric jitter & $\sigma_\mathrm{TESS}$ & ppt & \sigmaphotIII & \\
\keckhires RV offset & $\gamma_\mathrm{HIRES}$ & \mps & \gammarvHIRESIII & \\
\keckhires RV jitter & $\sigma_\mathrm{HIRES}$ & \mps & \sigmarvHIRESIII & \\
\apflevy RV offset & $\gamma_\mathrm{APF}$ & \mps & \gammarvAPFIII & \\
\apflevy RV jitter & $\sigma_\mathrm{APF}$ & \mps & \sigmarvAPFIII & \\
\enddata
\tablecomments{Table notes are the same as found at the bottom of Table \ref{tab:hip8152_properties}. The upper limits on the RV semi-amplitude, mass, and bulk density of planet c reflect 98\% confidence. We choose not to report a TSM value for planet c because its mass is too unconstrained. Planet properties come from the joint model shown in Figure \ref{fig:hd25463_phot_and_rvs}, which uses the \igrand reduction method to compute RVs from the \apflevy spectra.}
\end{deluxetable*}

\begin{figure*}
\gridline{\fig{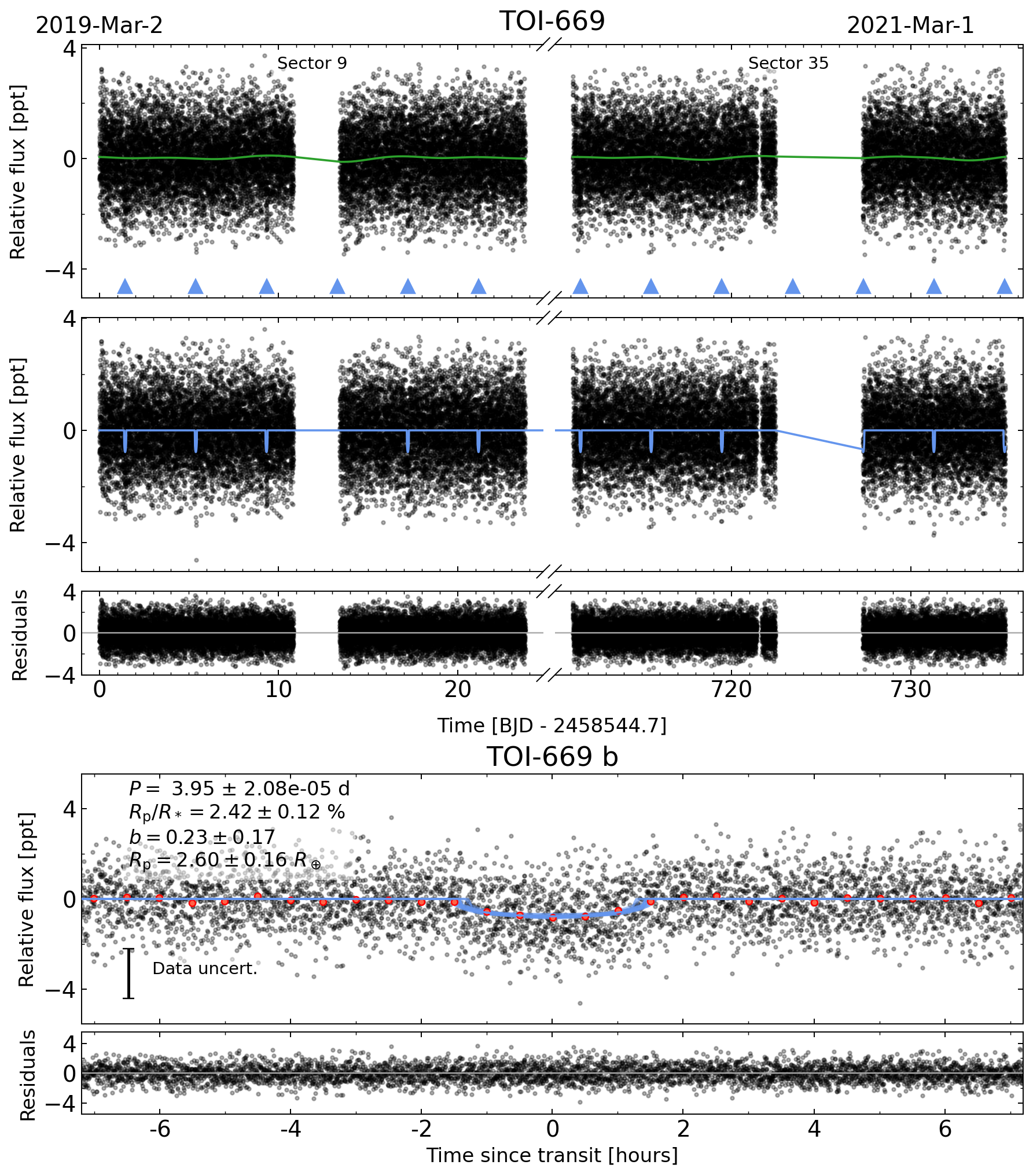}{0.5\textwidth}{}
          \fig{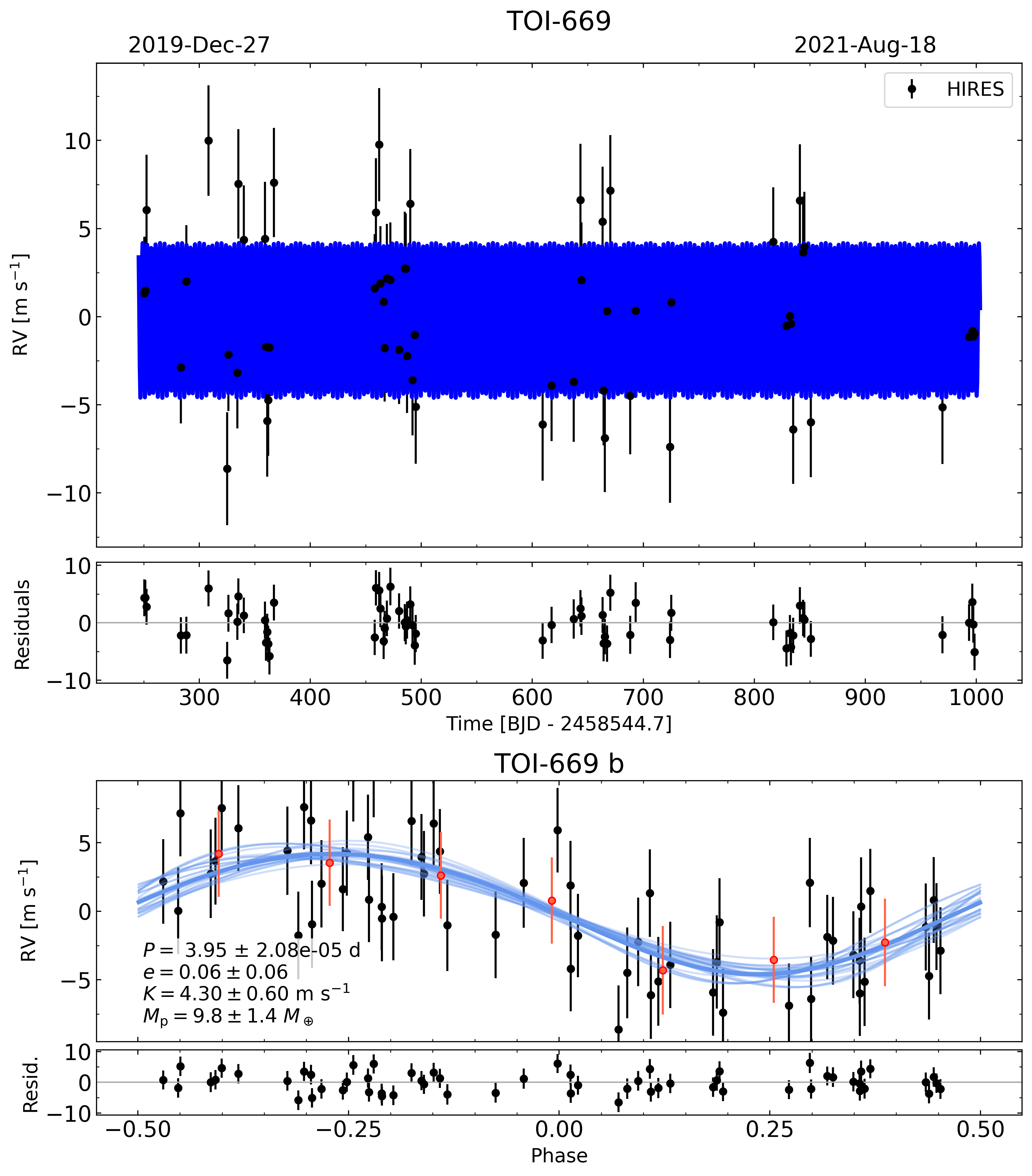}{0.5\textwidth}{}}
  \caption{Our joint modeling results for \sysIV. The figure description is the same as for Figure \ref{fig:hip8152_phot_and_rvs}.}\label{fig:toi669_phot_and_rvs}
\end{figure*}

\begin{figure*}
    \centering
    \includegraphics[width=\textwidth]{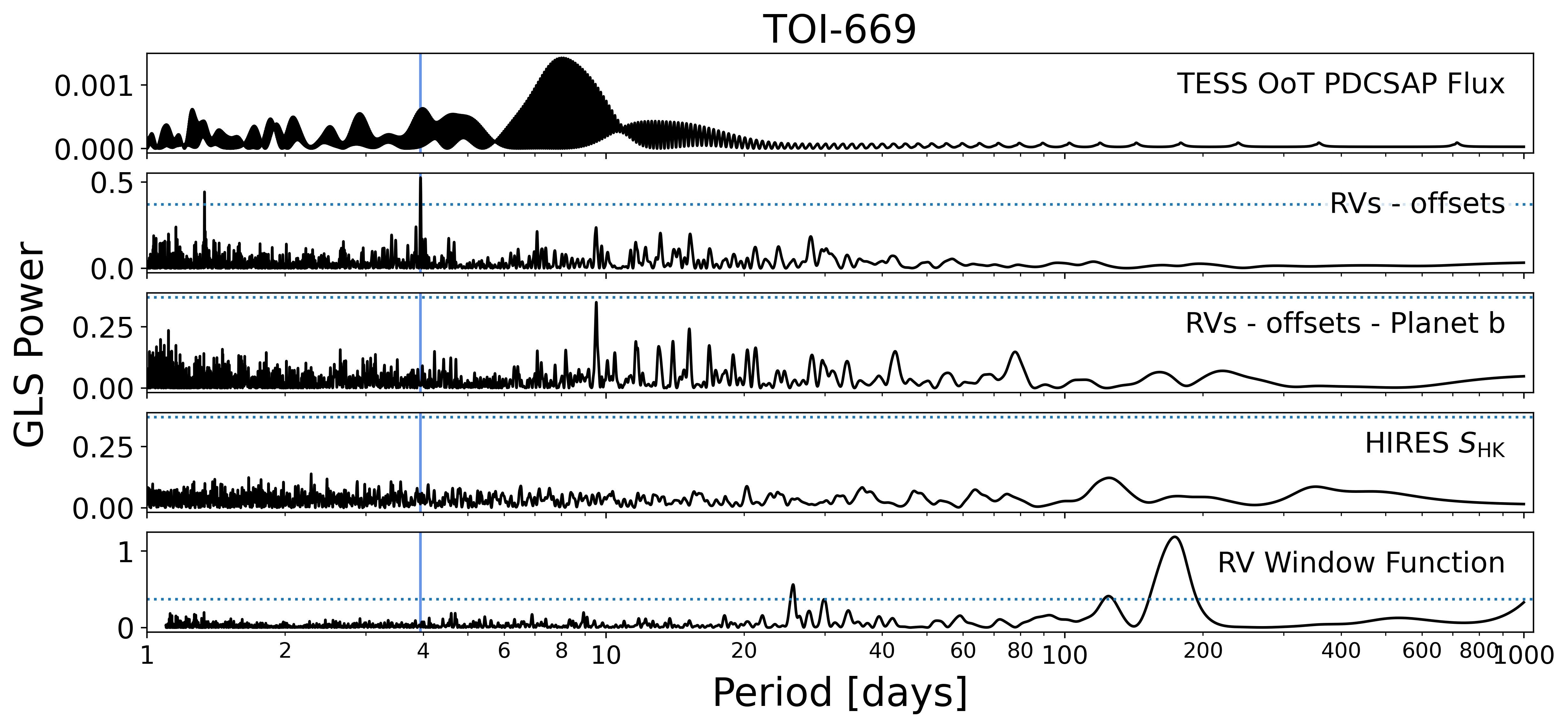}
    \caption{GLS periodograms for \sysIV. The figure description is the same as for Figure \ref{fig:hip8152_periodogram}.}
    \label{fig:toi669_periodogram}
\end{figure*}

\begin{deluxetable*}{lcccc}
\tablecaption{\sysIV system parameters \label{tab:toi669_properties}}
\tabletypesize{\scriptsize}
\startdata
\tablehead{
    \vspace{0.01cm} \\ 
    \multicolumn{5}{c}{\textbf{Stellar Parameters}} \\
    \colhead{Parameter} & \colhead{Symbol} & \colhead{Units} & \colhead{Value} & \colhead{Provenance}
}\\
\sidehead{\emph{Identifying information}}
TOI ID & & & 669 & Guerrero \\
TIC ID & & & 124573851 & Guerrero \\
R.A. & & deg (J2000) & $158.90$ & \gaiadrthree \\
decl. & & deg (J2000) & $-5.18$ & \gaiadrthree \\
Parallax & $\pi$ & mas & $7.01 \pm 0.01$ & \gaiadrthree \\
Johnson V-band apparent magnitude & $V$ & mag & $10.61 \pm 0.01$ & TIC \\
J-band apparent magnitude & $J$ & mag & $9.56 \pm 0.02$ & \twomass \\
K$_s$-band apparent magnitude & $K_s$ & mag & $9.13 \pm 0.02$ & \twomass \\
\sidehead{\emph{Spectroscopy}}
Effective temperature & \teff & K & \teffIV & \specMatchEmp \\
Metallicity & \feh & dex & \fehIV & \specMatchEmp \\
Ca II H \& K emission & \logrhk & & \logrhkIV & Isaacson \\
\sidehead{\emph{Isochrone modeling}}
Mass & \mstar & \msun & \mstarIV  & \isoclassify \\
Radius & \rstar & \rsun & \rstarIV & \isoclassify \\
Age & & Gyr & \ageIV & \isoclassify \\
\sidehead{\emph{Transit modeling}}
Limb-darkening parameter 1 & $u_1$ & & \uOneIV & Joint model \\
Limb-darkening parameter 2 & $u_2$ & & \uTwoIV & Joint model \\
\vspace{0.01cm} \\ 
\multicolumn{5}{c}{\textbf{Planet Parameters}} \\
\colhead{Parameter} & \colhead{Symbol} & \colhead{Units} & \colhead{\sysIV b value} \\
\hline
\sidehead{\emph{Measured quantities}}
Orbital period & $P$ & d & \periodIVb & \\ 
Time of inferior conjunction & \transitTime & BTJD & \tcBTJDIVb & \\
Occultation fraction & $R_\mathrm{p}/R_*$ & & \rorIVb & \\
Impact parameter & $b$ & & \bIVb & \\
Orbital eccentricity & $e$ & & \eccIVb & \\
Argument of periastron & $\omega$ & deg & \omegafoldeddegIVb & \\
RV semi-amplitude & $K$ & \mps & \KIVb & \\
\sidehead{\emph{Derived quantities}}
Orbital separation & $a/R_*$ & & \aorIVb & \\
Orbital semimajor axis & $a$ & AU & \aIVb & \\
Radius & \rplanet & \rearth & \rpIVb & \\
Mass & \mplanet & \mearth & \mpIVb & \\
Bulk density & $\rho$ & \gcc & \rhoIVb & \\ 
Equilibrium temperature & \teq & K & \teqIVb & \\ 
Instellation flux & \sincplanet & \sincearth & \sincIVb & \\
Transit duration & \tdur & hr & \durhrIVb & \\
TSM & & & \tsmIVb & \\
\vspace{0.01cm} \\ 
\multicolumn{5}{c}{\textbf{Additional Parameters}} \\
\colhead{Parameter} & \colhead{Symbol} & \colhead{Units} & \colhead{Value} & \colhead{} \\
\hline
\tess photometric offset & $\mu_\mathrm{TESS}$ & ppt & \meanfluxIV & \\
\tess photometric jitter & $\sigma_\mathrm{TESS}$ & ppt & \sigmaphotIV & \\
\keckhires RV offset & $\gamma_\mathrm{HIRES}$ & \mps & \gammarvHIRESIV & \\
\keckhires RV jitter & $\sigma_\mathrm{HIRES}$ & \mps & \sigmarvHIRESIV & \\
\enddata
\tablecomments{Table notes are the same as found at the bottom of Table \ref{tab:hip8152_properties}.}
\end{deluxetable*}

\begin{figure*}
\gridline{\fig{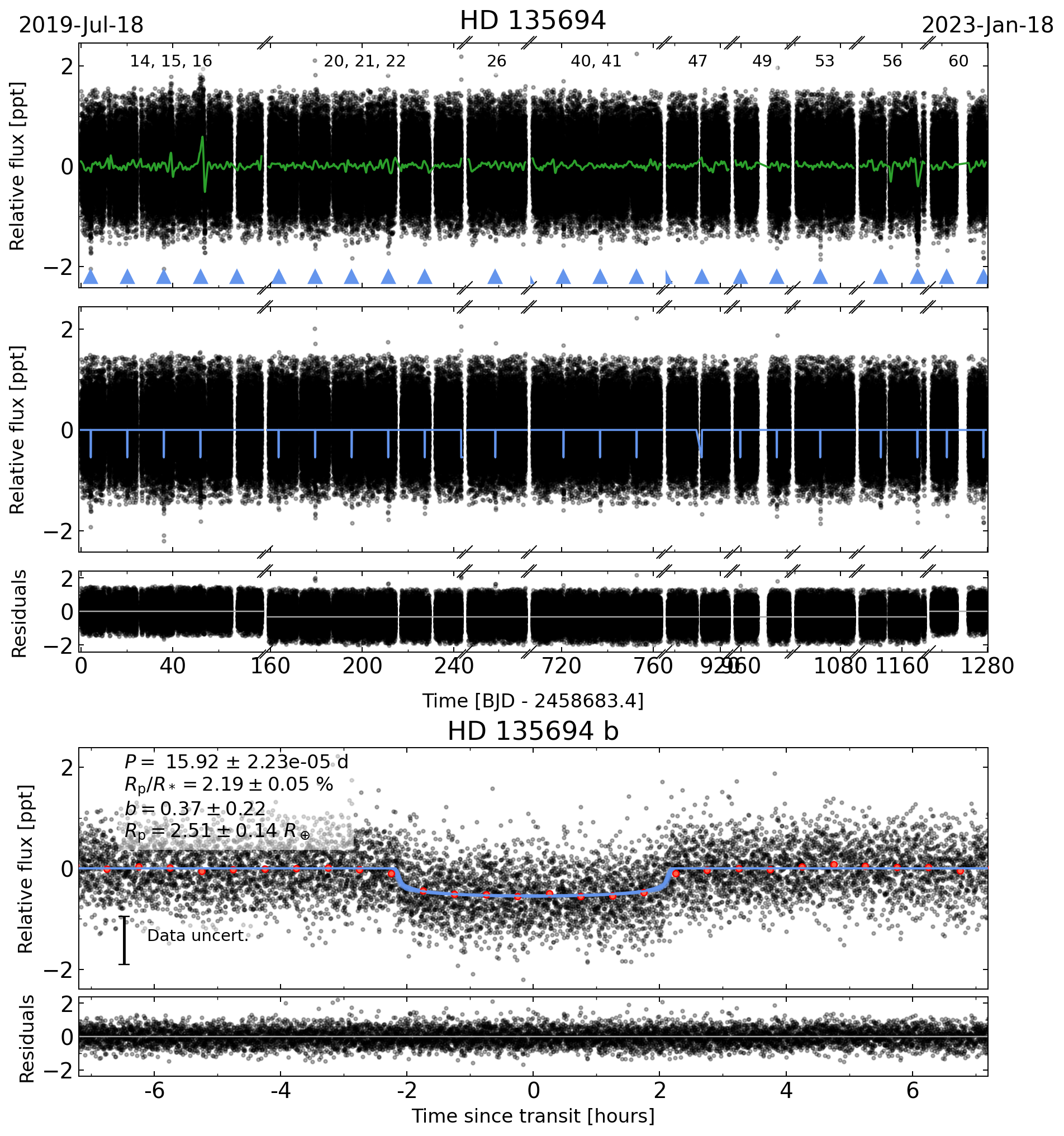}{0.5\textwidth}{}
          \fig{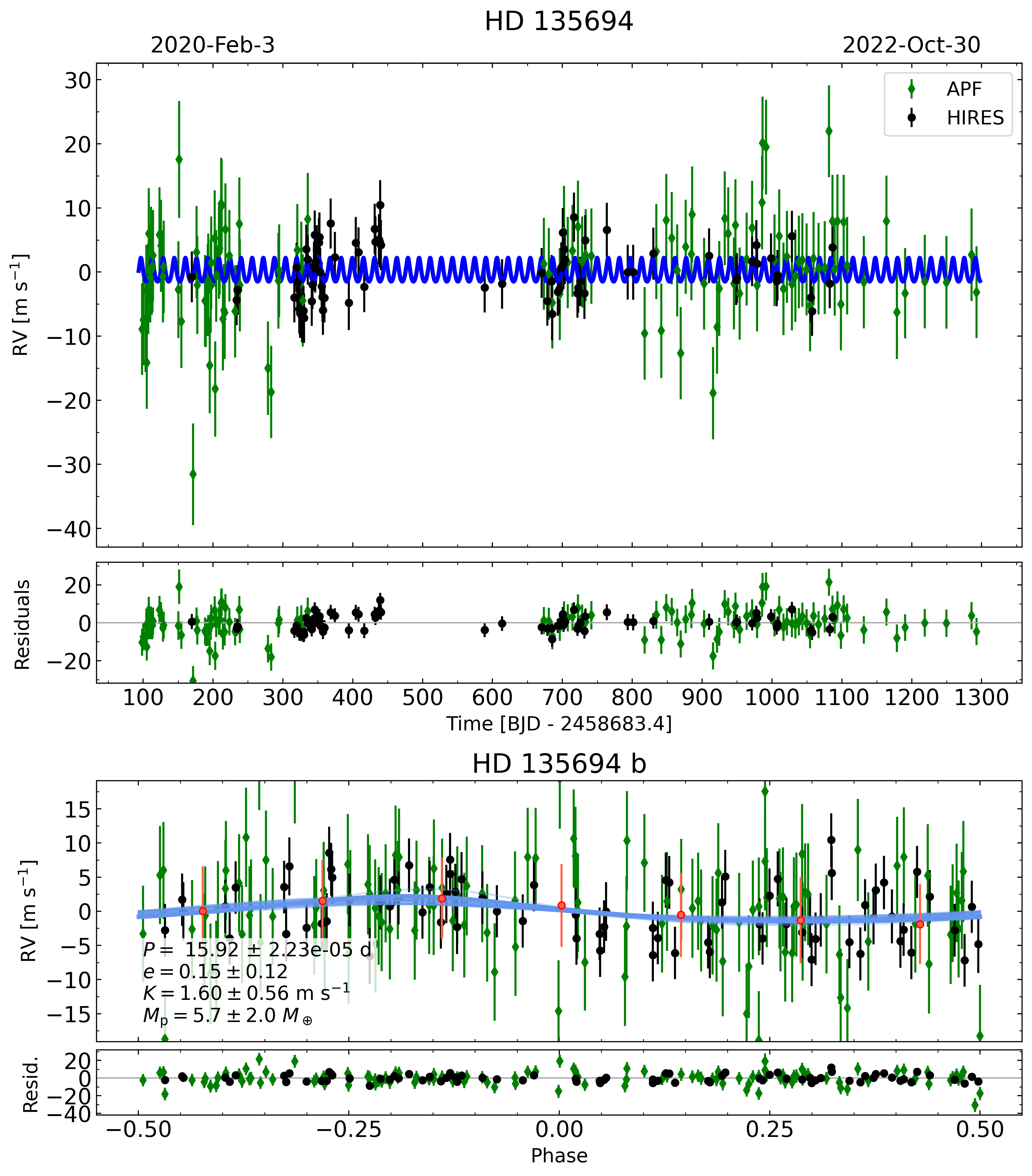}{0.5\textwidth}{}}
  \caption{Our joint modeling results for \sysV. The figure description is the same as for Figure \ref{fig:hip8152_phot_and_rvs}.}\label{fig:hd135694_phot_and_rvs}
\end{figure*}

\begin{figure*}
    \centering
    \includegraphics[width=\textwidth]{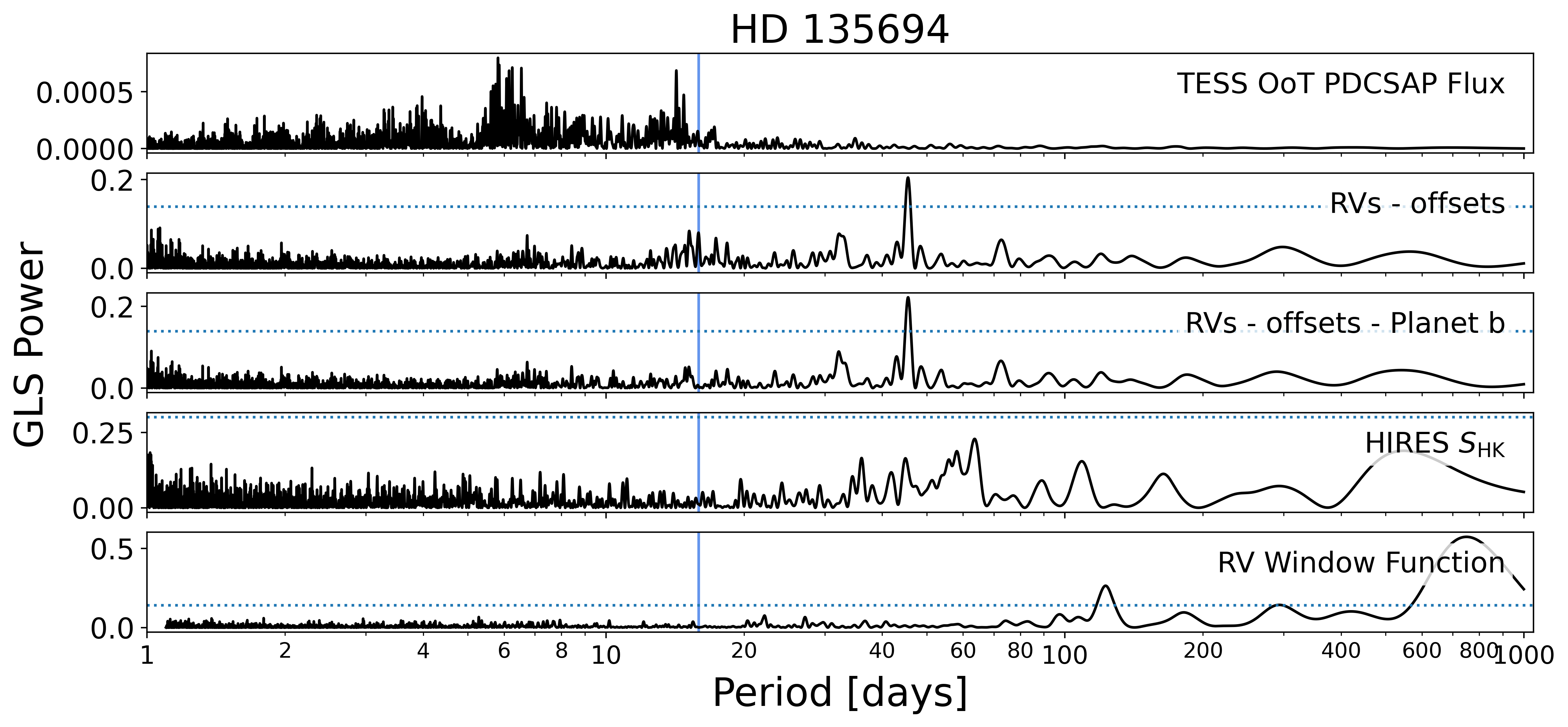}
    \caption{GLS periodograms for \sysV. The figure description is the same as for Figure \ref{fig:hip8152_periodogram}.}
    \label{fig:hd135694_periodogram}
\end{figure*}

\begin{deluxetable*}{lcccc}
\tablecaption{\sysV system parameters \label{tab:hd135694_properties}}
\tabletypesize{\scriptsize}
\startdata
\tablehead{
    \vspace{0.01cm} \\ 
    \multicolumn{5}{c}{\textbf{Stellar Parameters}} \\
    \colhead{Parameter} & \colhead{Symbol} & \colhead{Units} & \colhead{Value} & \colhead{Provenance}
}\\
\sidehead{\emph{Identifying information}}
TOI ID & & & 1247 & Guerrero \\
TIC ID & & & 232540264 & Guerrero \\
R.A. & & deg (J2000) & $227.87$ & \gaiadrthree \\
decl. & & deg (J2000) & $71.84$ & \gaiadrthree \\
Parallax & $\pi$ & mas & $13.63 \pm 0.01$ & \gaiadrthree \\
Johnson V-band apparent magnitude & $V$ & mag & $9.08 \pm 0.03$ & TIC \\
J-band apparent magnitude & $J$ & mag & $7.87 \pm 0.02$ & \twomass \\
K$_s$-band apparent magnitude & $K_s$ & mag & $7.50 \pm 0.02$ & \twomass \\
\sidehead{\emph{Spectroscopy}}
Effective temperature & \teff & K & \teffV & \specMatchEmp \\
Metallicity & \feh & dex & \fehV & \specMatchEmp \\
Ca II H \& K emission & \logrhk & & \logrhkV & Isaacson \\
\sidehead{\emph{Isochrone modeling}}
Mass & \mstar & \msun & \mstarV  & \isoclassify \\
Radius & \rstar & \rsun & \rstarV & \isoclassify \\
Age & & Gyr & \ageV & \isoclassify \\
\sidehead{\emph{Transit modeling}}
Limb-darkening parameter 1 & $u_1$ & & \uOneV & Joint model \\
Limb-darkening parameter 2 & $u_2$ & & \uTwoV & Joint model \\
\vspace{0.01cm} \\ 
\multicolumn{5}{c}{\textbf{Planet Parameters}} \\
\colhead{Parameter} & \colhead{Symbol} & \colhead{Units} & \colhead{\sysV b value} \\
\hline
\sidehead{\emph{Measured quantities}}
Orbital period & $P$ & d & \periodVb & \\ 
Time of inferior conjunction & \transitTime & BTJD & \tcBTJDVb & \\
Occultation fraction & $R_\mathrm{p}/R_*$ & & \rorVb & \\
Impact parameter & $b$ & & \bVb & \\
Orbital eccentricity & $e$ & & \eccVb & \\
Argument of periastron & $\omega$ & deg & \omegafoldeddegVb & \\
RV semi-amplitude & $K$ & \mps & \KVb & \\
\sidehead{\emph{Derived quantities}}
Orbital separation & $a/R_*$ & & \aorVb & \\
Orbital semimajor axis & $a$ & AU & \aVb & \\
Radius & \rplanet & \rearth & \rpVb & \\
Mass & \mplanet & \mearth & \mpVb & \\
Bulk density & $\rho$ & \gcc & \rhoVb & \\ 
Equilibrium temperature & \teq & K & \teqVb & \\ 
Instellation flux & \sincplanet & \sincearth & \sincVb & \\
Transit duration & \tdur & hr & \durhrVb & \\
TSM & & & \tsmVb & \\
\vspace{0.01cm} \\ 
\multicolumn{5}{c}{\textbf{Additional Parameters}} \\
\colhead{Parameter} & \colhead{Symbol} & \colhead{Units} & \colhead{Value} & \colhead{} \\
\hline
\tess photometric offset & $\mu_\mathrm{TESS}$ & ppt & \meanfluxV & \\
\tess photometric jitter & $\sigma_\mathrm{TESS}$ & ppt & \sigmaphotV & \\
\keckhires RV offset & $\gamma_\mathrm{HIRES}$ & \mps & \gammarvHIRESV & \\
\keckhires RV jitter & $\sigma_\mathrm{HIRES}$ & \mps & \sigmarvHIRESV & \\
\apflevy RV offset & $\gamma_\mathrm{APF}$ & \mps & \gammarvAPFV & \\
\apflevy RV jitter & $\sigma_\mathrm{APF}$ & \mps & \sigmarvAPFV & \\
\enddata
\tablecomments{Table notes are the same as found at the bottom of Table \ref{tab:hip8152_properties}.}
\end{deluxetable*}

\begin{figure*}
\gridline{\fig{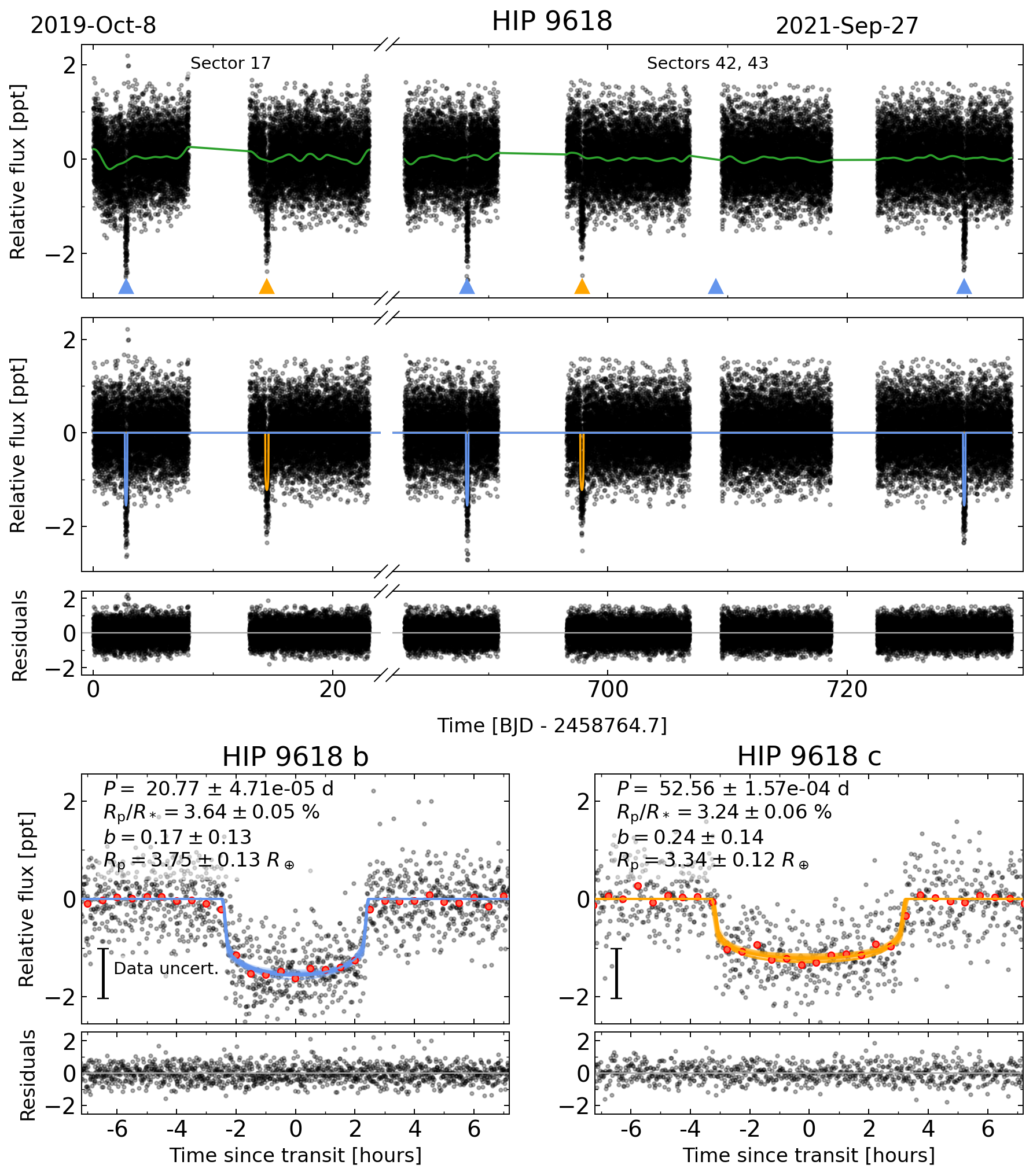}{0.5\textwidth}{}
          \fig{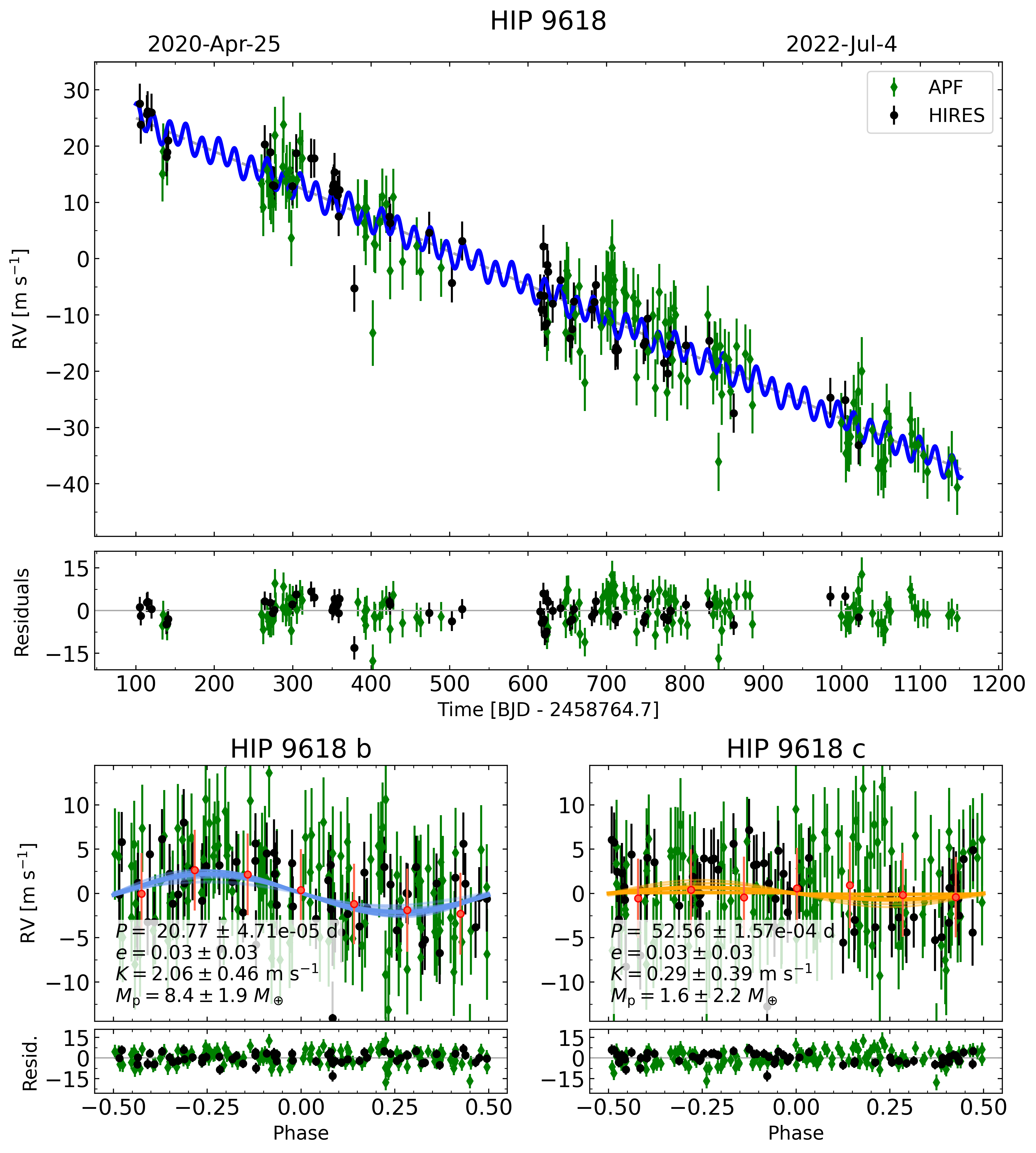}{0.5\textwidth}{}}
  \caption{Our joint modeling results for \sysVI. The figure description is the same as for Figure \ref{fig:hip8152_phot_and_rvs}.}\label{fig:hip9618_phot_and_rvs}
\end{figure*}

\begin{figure*}
    \centering
    \includegraphics[width=\textwidth]{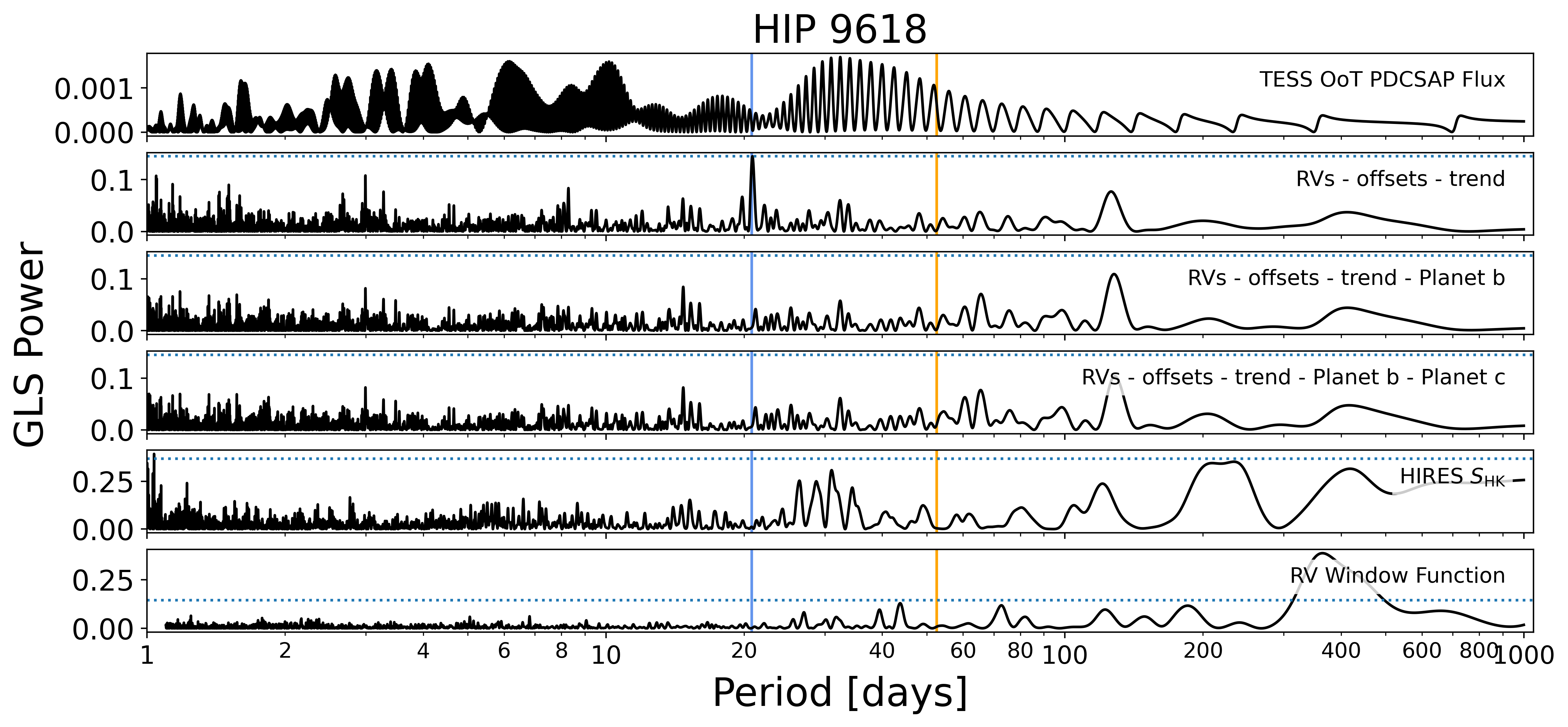}
    \caption{GLS periodograms for \sysVI. The figure description is the same as for Figure \ref{fig:hip8152_periodogram}.}
    \label{fig:hip9618_periodogram}
\end{figure*}

\begin{deluxetable*}{lcccc}
\tablecaption{\sysVI system parameters \label{tab:hip9618_properties}}
\tabletypesize{\scriptsize}
\startdata
\tablehead{
    \vspace{0.01cm} \\ 
    \multicolumn{5}{c}{\textbf{Stellar Parameters}} \\
    \colhead{Parameter} & \colhead{Symbol} & \colhead{Units} & \colhead{Value} & \colhead{Provenance}
}\\
\sidehead{\emph{Identifying information}}
TOI ID & & & 1471 & Guerrero \\
TIC ID & & & 306263608 & Guerrero \\
R.A. & & deg (J2000) & $30.91$ & \gaiadrthree \\
decl. & & deg (J2000) & $21.28$ & \gaiadrthree \\
Parallax & $\pi$ & mas & $14.86 \pm 0.02$ & \gaiadrthree \\
Johnson V-band apparent magnitude & $V$ & mag & $9.20 \pm 0.03$ & TIC \\
J-band apparent magnitude & $J$ & mag & $7.92 \pm 0.03$ & \twomass \\
K$_s$-band apparent magnitude & $K_s$ & mag & $7.56 \pm 0.02$ & \twomass \\
\sidehead{\emph{Spectroscopy}}
Effective temperature & \teff & K & \teffVI &  \specMatchEmp \\
Metallicity & \feh & dex & \fehVI & \specMatchEmp \\
Ca II H \& K emission & \logrhk & & \logrhkVI & Isaacson \\
\sidehead{\emph{Isochrone modeling}}
Mass & \mstar & \msun & \mstarVI  & \isoclassify \\
Radius & \rstar & \rsun & \rstarVI & \isoclassify \\
Age & & Gyr & \ageVI & \isoclassify \\
\sidehead{\emph{Transit modeling}}
Limb-darkening parameter 1 & $u_1$ & & \uOneVI & Joint model \\
Limb-darkening parameter 2 & $u_2$ & & \uTwoVI & Joint model \\
\vspace{0.01cm} \\ 
\multicolumn{5}{c}{\textbf{Planet Parameters}} \\
\colhead{Parameter} & \colhead{Symbol} & \colhead{Units} & \colhead{\sysVI b value} & \colhead{\sysVI c value} \\
\hline
\sidehead{\emph{Measured quantities}}
Orbital period & $P$ & d & \periodVIb & \periodVIc \\ 
Time of inferior conjunction & \transitTime & BTJD & \tcBTJDVIb & \tcBTJDVIc \\
Occultation fraction & $R_\mathrm{p}/R_*$ & & \rorVIb & \rorVIc \\
Impact parameter & $b$ & & \bVIb & \bVIc \\
Orbital eccentricity & $e$ & & \eccVIb & \eccVIc \\
Argument of periastron & $\omega$ & deg & \omegafoldeddegVIb & \omegafoldeddegVIc \\
RV semi-amplitude & $K$ & \mps & \KVIb & \KVIc \\
\sidehead{\emph{Derived quantities}}
Orbital separation & $a/R_*$ & & \aorVIb & \aorVIc \\
Orbital semimajor axis & $a$ & AU & \aVIb & \aVIc \\
Radius & \rplanet & \rearth & \rpVIb & \rpVIc \\
Mass & \mplanet & \mearth & \mpVIb & \mpVIc \\
Bulk density & $\rho$ & \gcc & \rhoVIb & \rhoVIc \\ 
Equilibrium temperature & \teq & K & \teqVIb & \teqVIc \\ 
Instellation flux & \sincplanet & \sincearth & \sincVIb & \sincVIc \\
Transit duration & \tdur & hr & \durhrVIb & \durhrVIc \\
TSM & & & \tsmVIb & \tsmVIc \\
\vspace{0.01cm} \\ 
\multicolumn{5}{c}{\textbf{Additional Parameters}} \\
\colhead{Parameter} & \colhead{Symbol} & \colhead{Units} & \colhead{Value} & \colhead{} \\
\hline
\tess photometric offset & $\mu_\mathrm{TESS}$ & ppt & \meanfluxVI & \\
\tess photometric jitter & $\sigma_\mathrm{TESS}$ & ppt & \sigmaphotVI & \\
\keckhires RV offset & $\gamma_\mathrm{HIRES}$ & \mps & \gammarvHIRESVI & \\
\keckhires RV jitter & $\sigma_\mathrm{HIRES}$ & \mps & \sigmarvHIRESVI & \\
\apflevy RV offset & $\gamma_\mathrm{APF}$ & \mps & \gammarvAPFVI & \\
\apflevy RV jitter & $\sigma_\mathrm{APF}$ & \mps & \sigmarvAPFVI & \\
Linear RV trend & $\dot{\gamma}$ & \mps/d & \trendrvslopeVI & \\
\enddata
\tablecomments{Table notes are the same as found at the bottom of Table \ref{tab:hip8152_properties}. All limits reflect 98\% confidence.}
\end{deluxetable*}

\begin{figure*}
\gridline{\fig{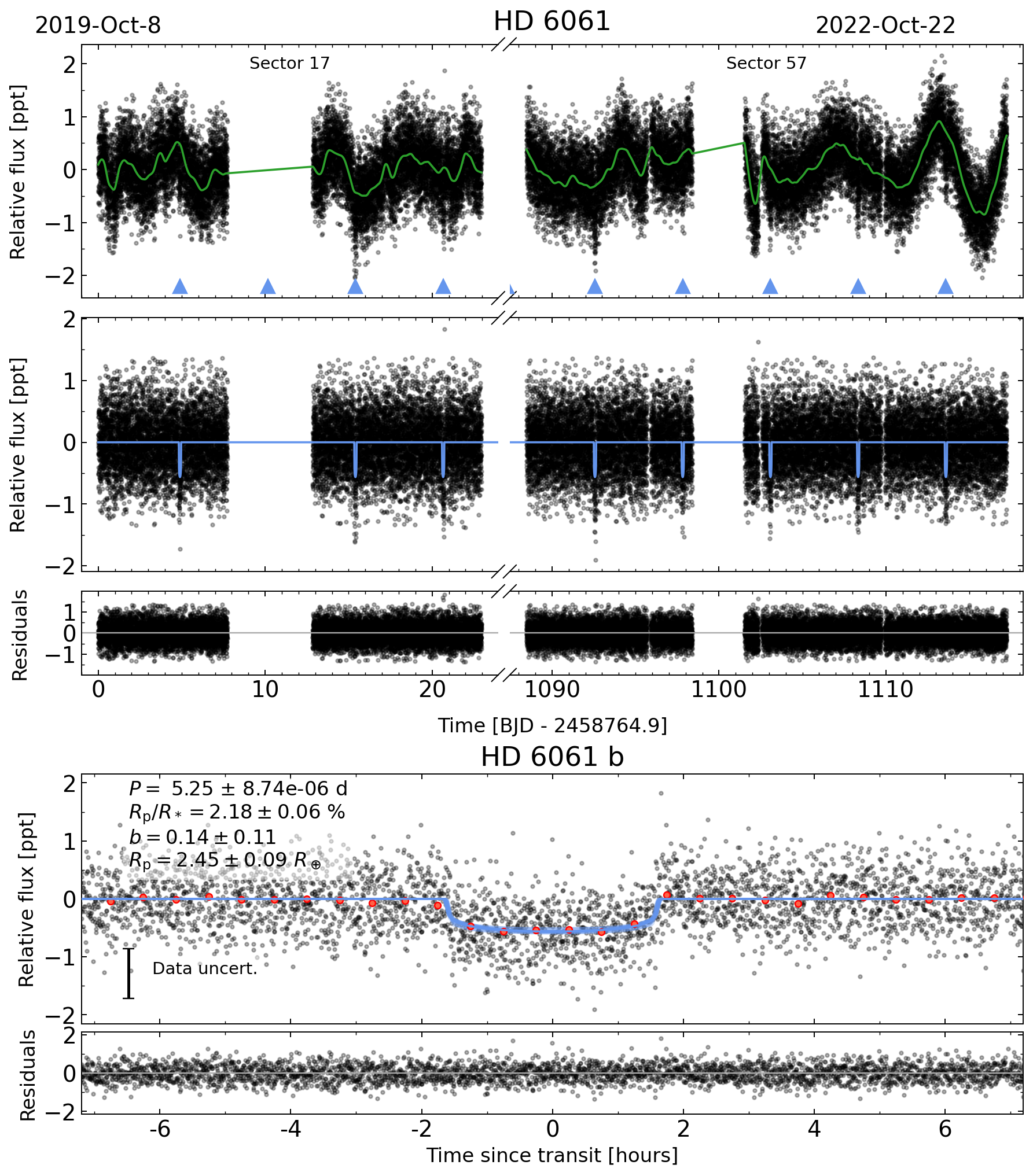}{0.5\textwidth}{}
          \fig{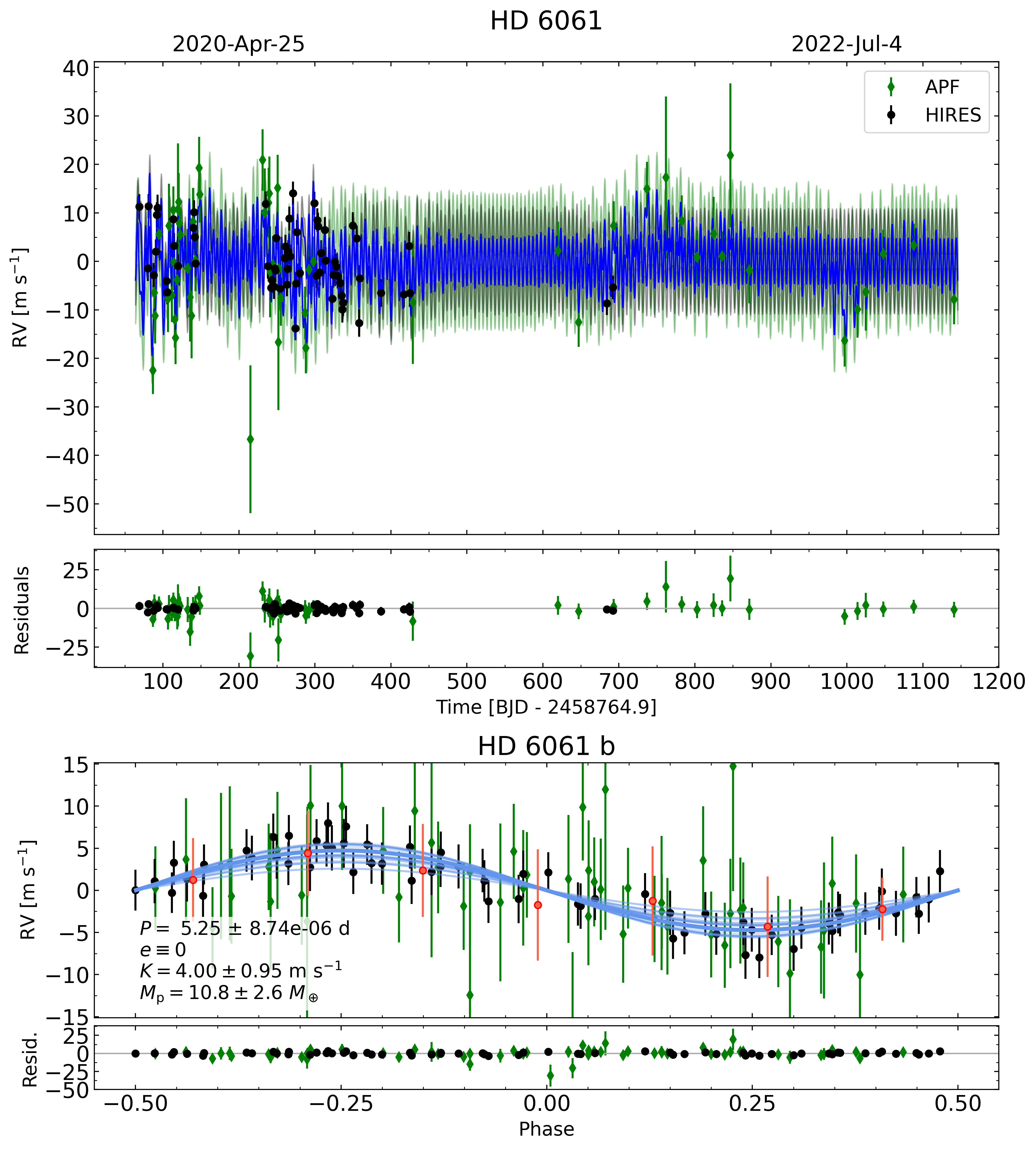}{0.5\textwidth}{}}
  \caption{Our joint modeling results for \sysVII. The figure description is the same as for Figure \ref{fig:hip8152_phot_and_rvs}, save for the RV model (right), which also shows the GP model of stellar activity. The combined Keplerian plus GP predictions for \keckhires and \apflevy are shown in blue, with 1$\sigma$ error envelopes in gray and green, respectively.}\label{fig:hd6061_phot_and_rvs}
\end{figure*}

\begin{figure*}
    \centering
    \includegraphics[width=\textwidth]{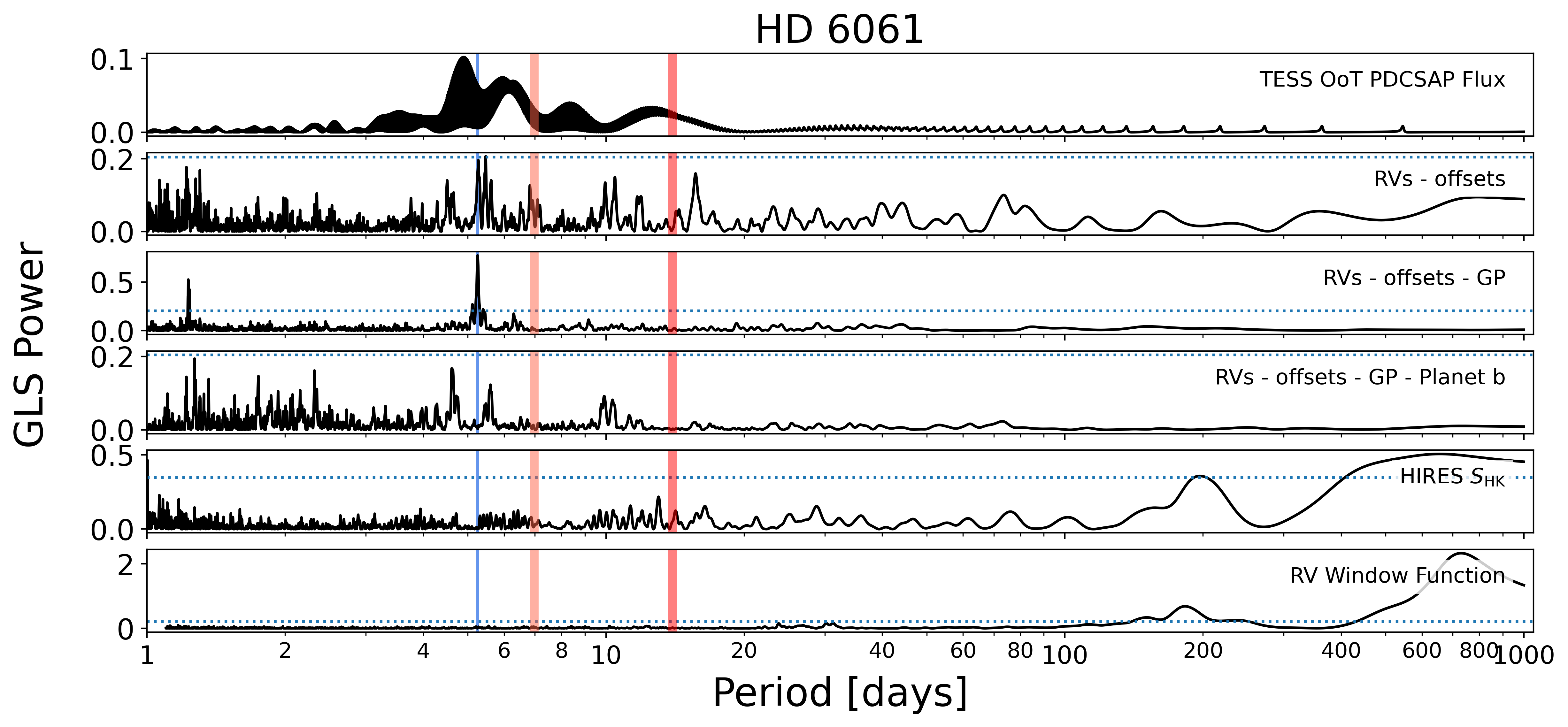}
    \caption{GLS periodograms for \sysVII. The figure description is the same as for Figure \ref{fig:hip8152_periodogram}. The period of the GP used to model stellar activity in the RVs is marked with the red vertical line and its first harmonic is marked with the light red line.}
    \label{fig:hd6061_periodogram}
\end{figure*}

\begin{deluxetable*}{lcccc}
\tablecaption{\sysVII system parameters \label{tab:hd6061_properties}}
\tabletypesize{\scriptsize}
\startdata
\tablehead{
    \vspace{0.01cm} \\ 
    \multicolumn{5}{c}{\textbf{Stellar Parameters}} \\
    \colhead{Parameter} & \colhead{Symbol} & \colhead{Units} & \colhead{Value} & \colhead{Provenance}
}\\
\sidehead{\emph{Identifying information}}
TOI ID & & & 1473 & Guerrero \\
TIC ID & & & 352413427 & Guerrero \\
R.A. & & deg (J2000) & $15.60$ & \gaiadrthree \\
decl. & & deg (J2000) & $37.19$ & \gaiadrthree \\
Parallax & $\pi$ & mas & $14.77 \pm 0.02$ & \gaiadrthree \\
Johnson V-band apparent magnitude & $V$ & mag & $8.84 \pm 0.01$ & TIC \\
J-band apparent magnitude & $J$ & mag & $7.72 \pm 0.03$ & AO \\
K$_s$-band apparent magnitude & $K_s$ & mag & $7.41 \pm 0.02$ & AO \\
\sidehead{\emph{Spectroscopy}}
Effective temperature & \teff & K & \teffVII & \specMatchEmp \\
Metallicity & \feh & dex & \fehVII & \specMatchEmp \\
Ca II H \& K emission & \logrhk & & \logrhkVII & Isaacson \\
\sidehead{\emph{Isochrone modeling}}
Mass & \mstar & \msun & \mstarVII  & \isoclassify \\
Radius & \rstar & \rsun & \rstarVII & \isoclassify \\
Age & & Gyr & \ageVII & \isoclassify \\
\sidehead{\emph{Transit modeling}}
Limb-darkening parameter 1 & $u_1$ & & \uOneVII & Joint model \\
Limb-darkening parameter 2 & $u_2$ & & \uTwoVII & Joint model \\
\vspace{0.01cm} \\ 
\multicolumn{5}{c}{\textbf{Planet Parameters}} \\
\colhead{Parameter} & \colhead{Symbol} & \colhead{Units} & \colhead{\sysVII b value} \\
\hline
\sidehead{\emph{Measured quantities}}
Orbital period & $P$ & d & \periodVIIb & \\ 
Time of inferior conjunction & \transitTime & BTJD & \tcBTJDVIIb & \\
Occultation fraction & $R_\mathrm{p}/R_*$ & & \rorVIIb & \\
Impact parameter & $b$ & & \bVIIb & \\
Orbital eccentricity & $e$ & & \eccVIIb & \\
Argument of periastron & $\omega$ & deg & \omegafoldeddegVIIb & \\
RV semi-amplitude & $K$ & \mps & \KVIIb & \\
\sidehead{\emph{Derived quantities}}
Orbital separation & $a/R_*$ & & \aorVIIb & \\
Orbital semimajor axis & $a$ & AU & \aVIIb & \\
Radius & \rplanet & \rearth & \rpVIIb & \\
Mass & \mplanet & \mearth & \mpVIIb & \\
Bulk density & $\rho$ & \gcc & \rhoVIIb & \\ 
Equilibrium temperature & \teq & K & \teqVIIb & \\ 
Instellation flux & \sincplanet & \sincearth & \sincVIIb & \\
Transit duration & \tdur & hr & \durhrVIIb & \\
TSM & & & \tsmVIIb & \\
\vspace{0.01cm} \\ 
\multicolumn{5}{c}{\textbf{Additional Parameters}} \\
\colhead{Parameter} & \colhead{Symbol} & \colhead{Units} & \colhead{Value} & \colhead{} \\
\hline
\tess photometric offset & $\mu_\mathrm{TESS}$ & ppt & \meanfluxVII & \\
\tess photometric jitter & $\sigma_\mathrm{TESS}$ & ppt & \sigmaphotVII & \\
\keckhires RV offset & $\gamma_\mathrm{HIRES}$ & \mps & \gammarvHIRESVII & \\
\keckhires RV jitter & $\sigma_\mathrm{HIRES}$ & \mps & \sigmarvHIRESVII & \\
\apflevy RV offset & $\gamma_\mathrm{APF}$ & \mps & \gammarvAPFVII & \\
\apflevy RV jitter & $\sigma_\mathrm{APF}$ & \mps & \sigmarvAPFVII & \\
\enddata
\tablecomments{Table notes are the same as found at the bottom of Table \ref{tab:hip8152_properties}. For \sysVII's $J$ and $K_s$ apparent magnitudes, the ``AO'' provenance denotes that these values have been deblended using our AO imaging observations to account for the M4/5V companion, \sysVIIcomp.}
\end{deluxetable*}

\begin{figure*}
\gridline{\fig{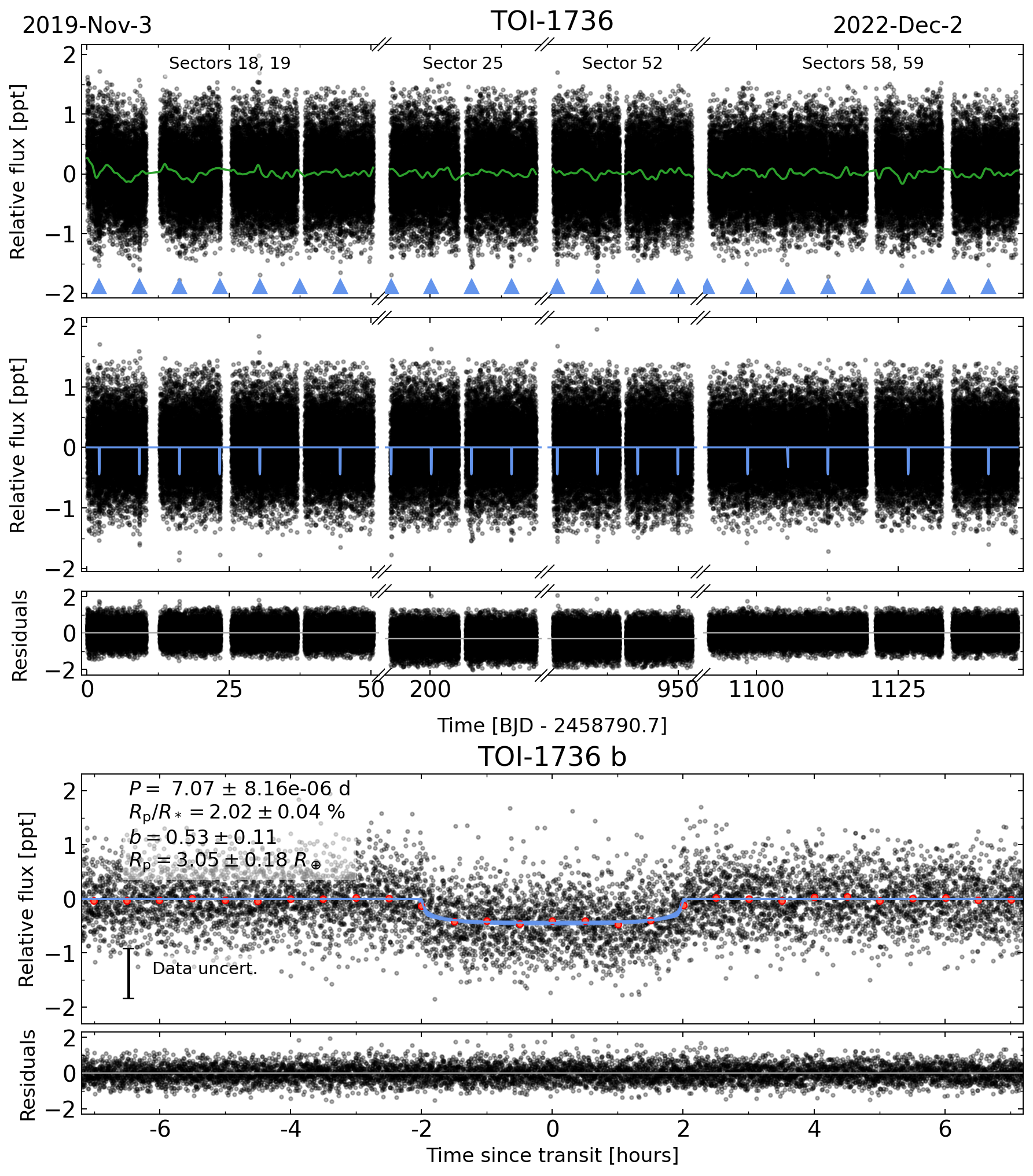}{0.5\textwidth}{}
          \fig{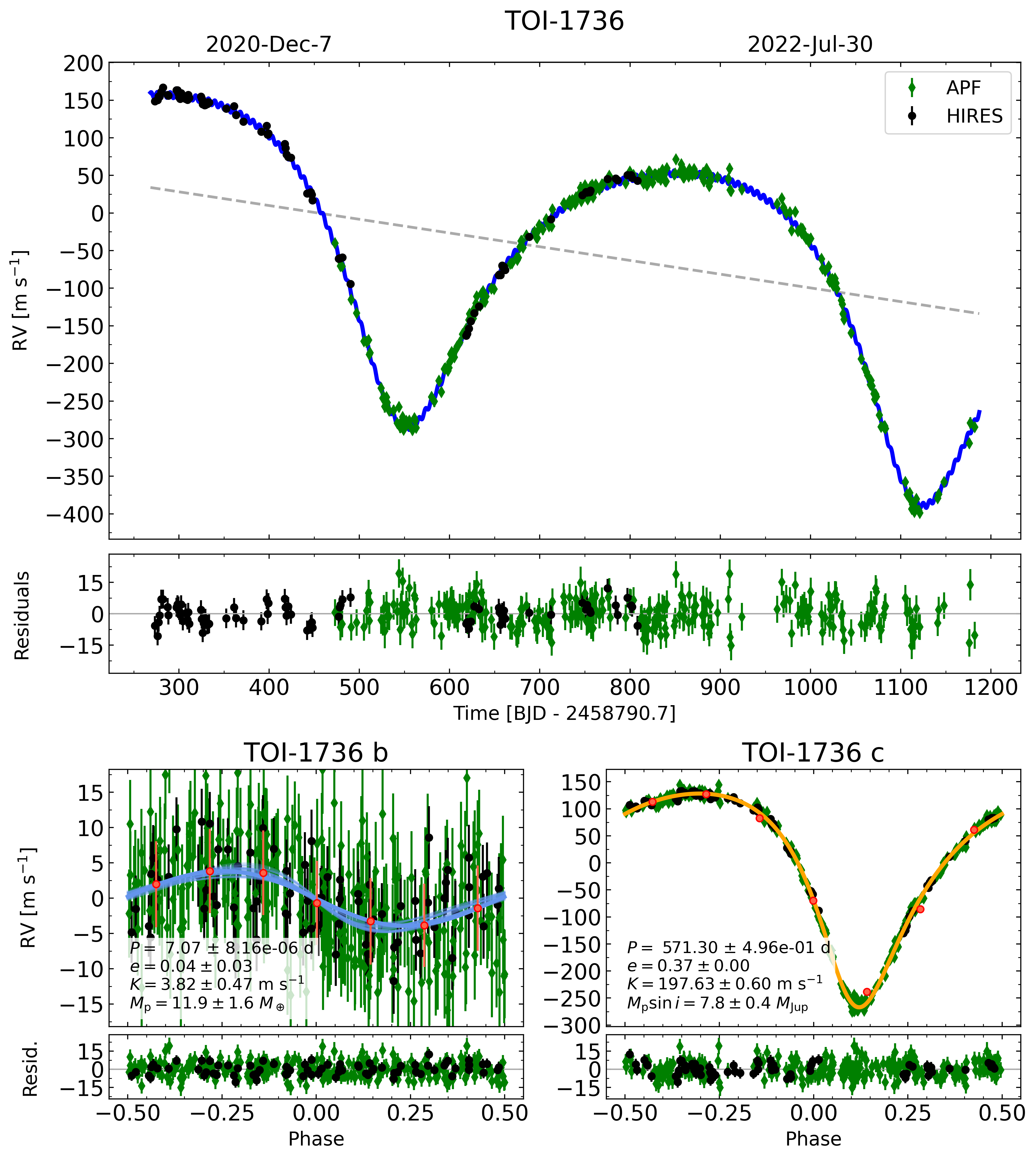}{0.5\textwidth}{}}
  \caption{Our joint modeling results for \sysVIII. The figure description is the same as for Figure \ref{fig:hip8152_phot_and_rvs}. The dashed gray line in the top panel of the RV figure represents the linear RV trend detected in the system.}\label{fig:toi1736_phot_and_rvs}
\end{figure*}

\begin{figure*}
    \centering
    \includegraphics[width=\textwidth]{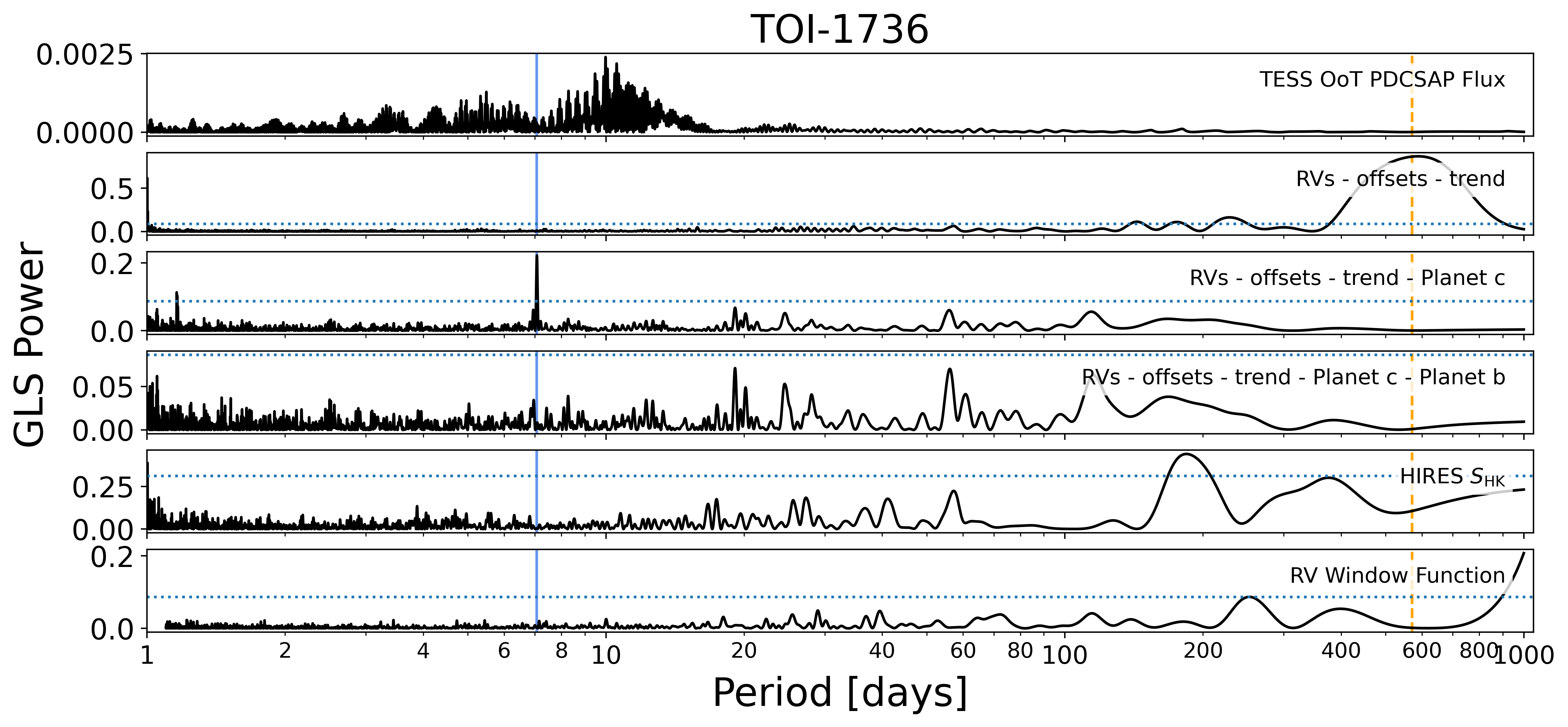}
    \caption{GLS periodograms for \sysVIII. The figure description is the same as for Figure \ref{fig:hip8152_periodogram}. The dashed vertical line for planet c indicates that it is nontransiting.}
    \label{fig:toi1736_periodogram}
\end{figure*}

\begin{deluxetable*}{lcccc}
\tablecaption{\sysVIII system parameters \label{tab:toi1736_properties}}
\tabletypesize{\scriptsize}
\startdata
\tablehead{
    \vspace{0.01cm} \\ 
    \multicolumn{5}{c}{\textbf{Stellar Parameters}} \\
    \colhead{Parameter} & \colhead{Symbol} & \colhead{Units} & \colhead{Value} & \colhead{Provenance}
}\\
\sidehead{\emph{Identifying information}}
TOI ID & & & 1736 & Guerrero \\
TIC ID & & & 408618999 & Guerrero \\
R.A. & & deg (J2000) & $43.43$ & \gaiadrthree \\
decl. & & deg (J2000) & $69.10$ & \gaiadrthree \\
Parallax & $\pi$ & mas & $11.34 \pm 0.02$ & \gaiadrthree \\
Johnson V-band apparent magnitude & $V$ & mag & $8.953 \pm 0.002$ & TIC \\
J-band apparent magnitude & $J$ & mag & $7.69 \pm 0.02$ & \twomass \\
K$_s$-band apparent magnitude & $K_s$ & mag & $7.28 \pm 0.02$ & \twomass \\
\sidehead{\emph{Spectroscopy}}
Effective temperature & \teff & K & \teffVIII &  \specMatchEmp \\
Metallicity & \feh & dex & \fehVIII & \specMatchEmp \\
Ca II H \& K emission & \logrhk & & \logrhkVIII & Isaacson \\
\sidehead{\emph{Isochrone modeling}}
Mass & \mstar & \msun & \mstarVIII  & \isoclassify \\
Radius & \rstar & \rsun & \rstarVIII & \isoclassify \\
Age & & Gyr & \ageVIII & \isoclassify \\
\sidehead{\emph{Transit modeling}}
Limb-darkening parameter 1 & $u_1$ & & \uOneVIII & Joint model \\
Limb-darkening parameter 2 & $u_2$ & & \uTwoVIII & Joint model \\
\vspace{0.01cm} \\ 
\multicolumn{5}{c}{\textbf{Planet Parameters}} \\
\colhead{Parameter} & \colhead{Symbol} & \colhead{Units} & \colhead{\sysVIII b value} & \colhead{\sysVIII c value} \\
\hline
\sidehead{\emph{Measured quantities}}
Orbital period & $P$ & d & \periodVIIIb & \nontransperiodVIIIc \\ 
Time of inferior conjunction & \transitTime & BTJD & \tcBTJDVIIIb & \nontranstcBTJDVIIIc \\
Occultation fraction & $R_\mathrm{p}/R_*$ & & \rorVIIIb & \nodata \\
Impact parameter & $b$ & & \bVIIIb & \nodata \\
Orbital eccentricity & $e$ & & \eccVIIIb & \nontranseccVIIIc \\
Argument of periastron & $\omega$ & deg & \omegafoldeddegVIIIb & \nontransomegafoldeddegVIIIc \\
RV semi-amplitude & $K$ & \mps & \KVIIIb & \nontransKVIIIc \\
\sidehead{\emph{Derived quantities}}
Orbital separation & $a/R_*$ & & \aorVIIIb & \nontransaorVIIIc \\
Orbital semimajor axis & $a$ & AU & \aVIIIb & \nontransaVIIIc \\
Radius & \rplanet & \rearth & \rpVIIIb & \nodata \\
Minimum mass & \mplanet $\sin i$ & \mearth & \msiniVIIIb & \nontransmsiniVIIIc \\
Minimum mass & \mplanet $\sin i$ & $M_\mathrm{Jup}$ & \msinijupVIIIb  & \nontransmsinijupVIIIc \\
Mass & \mplanet & \mearth & \mpVIIIb & \nodata \\
Bulk density & $\rho$ & \gcc & \rhoVIIIb & \nodata \\ 
Equilibrium temperature & \teq & K & \teqVIIIb & \nontransteqVIIIc \\ 
Instellation flux & \sincplanet & \sincearth & \sincVIIIb & \nontranssincVIIIc \\
Transit duration & \tdur & hr & \durhrVIIIb & \nodata \\
TSM & & & \tsmVIIIb & \nodata \\
\vspace{0.01cm} \\ 
\multicolumn{5}{c}{\textbf{Additional Parameters}} \\
\colhead{Parameter} & \colhead{Symbol} & \colhead{Units} & \colhead{Value} & \colhead{} \\
\hline
\tess photometric offset & $\mu_\mathrm{TESS}$ & ppt & \meanfluxVIII & \\
\tess photometric jitter & $\sigma_\mathrm{TESS}$ & ppt & \sigmaphotVIII & \\
\keckhires RV offset & $\gamma_\mathrm{HIRES}$ & \mps & \gammarvHIRESVIII & \\
\keckhires RV jitter & $\sigma_\mathrm{HIRES}$ & \mps & \sigmarvHIRESVIII & \\
\apflevy RV offset & $\gamma_\mathrm{APF}$ & \mps & \gammarvAPFVIII & \\
\apflevy RV jitter & $\sigma_\mathrm{APF}$ & \mps & \sigmarvAPFVIII & \\
Linear RV trend & $\dot{\gamma}$ & \mps/d & \trendrvslopeVIII & \\
\enddata
\tablecomments{Table notes are the same as found at the bottom of Table \ref{tab:hip8152_properties}.}
\end{deluxetable*}

\section{Default \sysIII \apflevy Radial Velocity Measurements}\label{appendix:apf_grand}
\begin{figure}
    \centering
    \includegraphics[width=\columnwidth]{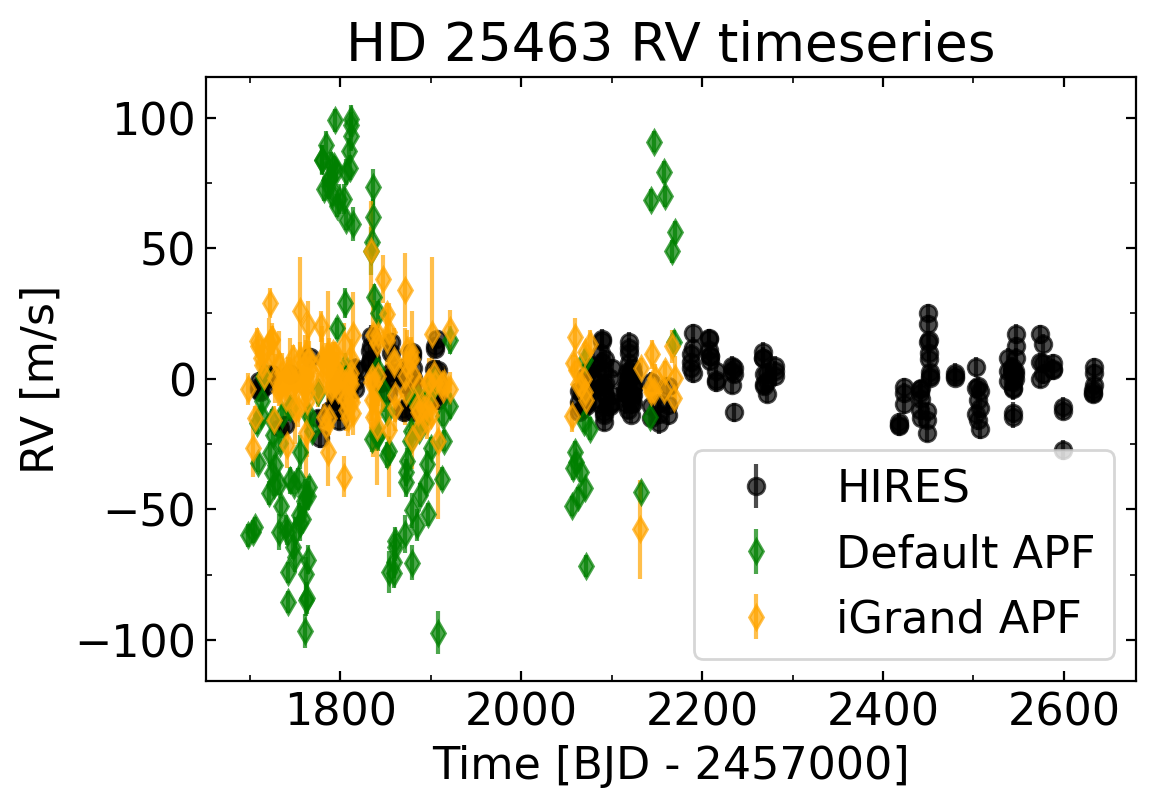}
    \caption{The \sysIII RVs. \keckhires RVs are shown as the black circles, the default pipeline's \apflevy RVs are shown as the green diamonds, and the \igrand pipeline's \apflevy RVs are shown as the orange diamonds. The default \apflevy RVs show scatter that is seemingly inconsistent with the astrophysical jitter when compared to contemporaneous \keckhires RVs. The \igrand reduction method seems to mitigate these systematics.}
    \label{fig:hd25463_rv_timeseries}
\end{figure}

\begin{figure*}
    \centering
    \includegraphics[width=0.6\textwidth]{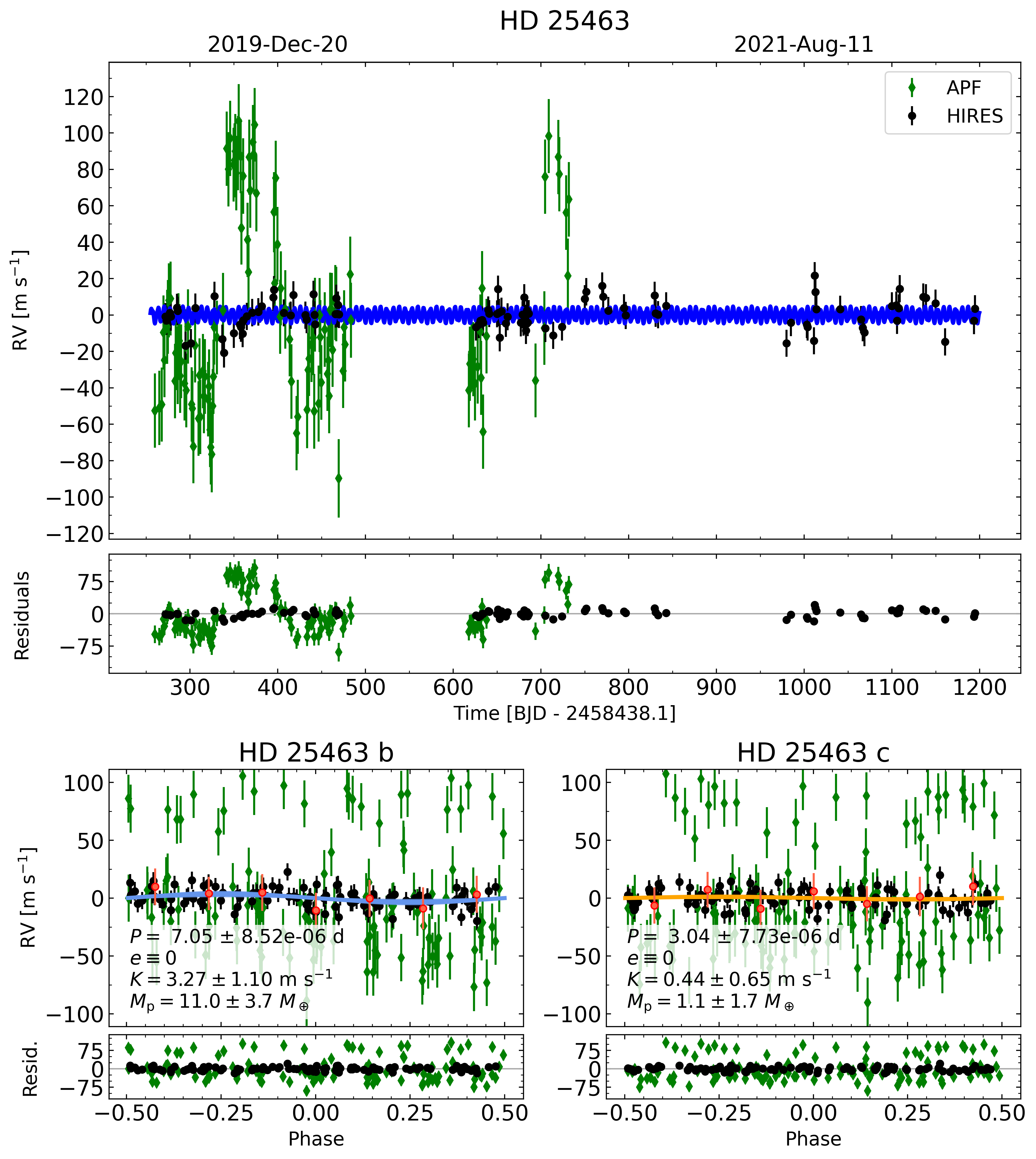}
    \caption{The RV portion of a joint model of the \sysIII \tess photometry and RVs where the \apflevy RVs are measured using the default Doppler pipeline \citep{howard10} as opposed to the \igrand reduction pipeline. This figure is only included for comparison with Figure \ref{fig:hd25463_phot_and_rvs} and the planet properties listed in the phase-folded panels should not be used. The figure description is the same as for Figure \ref{fig:hip8152_phot_and_rvs}. The default \apflevy RVs show unreasonably large scatter, which is a possible symptom of the star's rapid rotation. Mass estimates for planets b and c are consistent with the results of our adopted model, which uses the \igrand \apflevy velocities (Figure \ref{fig:hd25463_phot_and_rvs}). The results of the photometry portion of this joint model (not shown here for brevity) are entirely consistent with those of the model shown on the left in Figure \ref{fig:hd25463_phot_and_rvs}—because both the model here and in Figure \ref{fig:hd25463_phot_and_rvs} assume circular orbits for the planets, there is minimal information sharing between the photometry and RV components (for circular orbits, only the planet's orbital period and time of inferior conjunction, both of which are largely determined by the transits, are shared between the photometry and RV models). As such, we would expect the photometric model results to be essentially the same regardless of which \apflevy velocities are used. \revision{Since they are not listed in an associated table of properties, for use in conjunction with the ``\sysIII Default'' RVs in Table \ref{tab:all_rvs}, the instrumental offsets and jitter terms for this model are: $\gamma_\mathrm{APF} = -7.56 \pm 1.95$ \mps, $\gamma_\mathrm{HIRES} = -1.99 \pm 0.81$ \mps, $\sigma_\mathrm{APF} = 19.97 \pm 0.03$ \mps (which would be larger if not for the joint model's prior on instrumental RV jitter that enforces a maximum of 20 \mps), and $\sigma_\mathrm{HIRES} = 7.31 \pm 0.61$ \mps.}}
    \label{fig:hd25463_default_apf_rvs}
\end{figure*}

As described in \S\ref{sec:rvs_apf_igrand}, the default Doppler reduction pipeline \citep{howard10} fails when computing \apflevy RVs for \sysIII, in part due to the star's rapid rotation (\vsini $=$ \vsiniIII km/s). To circumvent this failure, we slightly alter the default reduction method to fit entire echelle orders of the \apflevy spectra simultaneously, rather than in series using small chunks. In Figure \ref{fig:hd25463_rv_timeseries} we compare the default reduction method's RVs with those from the alternative pipeline (\igrand) and the system's \keckhires RVs. \revision{In addition to containing the \igrand-derived \apflevy velocities for \sysIII, Table \ref{tab:all_rvs} also holds RVs for the system that were derived following the default reduction method (in that table the RVs are found under the label ``\sysIII Default''). We do not suggest using these velocities in future analyses for the reasons described in \S\ref{sec:rvs_apf} and they are only included for completeness.} Figure \ref{fig:hd25463_default_apf_rvs} shows the RV portion of a joint model that is entirely similar to our adopted model for \sysIII (Figure \ref{fig:hd25463_phot_and_rvs}) except the default \apflevy RVs are used.

\end{document}